\documentclass[final,1p,times]{elsarticle}
\usepackage{amssymb}
\usepackage{amsfonts}
\usepackage{amsmath}
\usepackage{geometry}
\usepackage{graphicx}
\usepackage{epsfig}
\usepackage{epstopdf}
\usepackage{bm}% bold math
\numberwithin{equation}{section}
\newtheorem{thm}{Theorem}

\newdefinition{rmk}{Remark}

\newproof{pf}{Proof}
\newproof{pot}{Proof of Theorem  %\ref{thm2}
}
\begin{document}

\begin{frontmatter}
\title{Asymptotic behavior of Heun function and its integral formalism}% Force line breaks with \\
%\thanks{Footnote to title of article.}

\author{Yoon Seok Choun\corref{cor1}}
\ead{Yoon.Choun@baruh.cuny.edu; ychoun@gradcenter.cuny.edu; ychoun@gmail.com}
\cortext[cor1]{Correspondence to: Baruch College, The City University of New York, Natural Science Department, A506, 17 Lexington Avenue, New York, NY 10010} 
\address{Baruch College, The City University of New York, Natural Science Department, A506, 17 Lexington Avenue, New York, NY 10010}
\begin{abstract}

The Heun function generalizes all well-known special functions such as Spheroidal Wave, Lame, Mathieu, and hypergeometric $_2F_1$, $_1F_1$ and $_0F_1$ functions. Heun functions are applicable to diverse areas such as theory of black holes, lattice systems in statistical mechanics, solution of the Schr$\ddot{\mbox{o}}$dinger equation of quantum mechanics, and addition of three quantum spins.

In this paper, applying three term recurrence formula \cite{Chou2012}, we consider asymptotic behaviors of Heun function and its integral formalism including all higher terms of $A_n$'s.\footnote{`` higher terms of $A_n$'s'' means at least two terms of $A_n$'s.} We show how the power series expansion of Heun functions can be converted to closed-form integrals for all cases of infinite series and polynomial.
One interesting observation resulting from the calculations is the fact that a $_2F_1$ function recurs in each of sub-integral forms: the first sub-integral form contains zero term of $A_n's$, the second one contains one term of $A_n$'s, the third one contains two terms of $A_n$'s, etc.   

Applying three term recurrence formula, we consider asymptotic behaviors of Heun functions and their radius of convergences. And we show why Poincar\'{e}-Perron theorem is not always applicable to the Heun equation.

In the appendix, I apply the power series expansion and my integral formalism of Heun function to ``The 192 solutions of the Heun equation'' \cite{Maie2007}. Due to space restriction final equations for all 192 Heun functions is not included in the paper, but feel free to contact me for the final solutions. Section 6 contains two additional examples using integral forms of Huen function. 

This paper is 4th out of 10 in series ``Special functions and three term recurrence formula (3TRF)''. See section 6 for all the papers in the series. The previous paper in series deals with the power series expansion in closed forms of Heun function. The next paper in the series describes the power series expansion of Mathieu function and its integral formalism analytically.
\end{abstract}

\end{frontmatter}  
\section{Introduction}

The Heun function, having three term recurrence relations, are the most outstanding special functions in among every analytic functions.
Due to its complexity Heun function was neglected for almost 100 years\cite{Heun1889}. According to Whittaker's hypothesis, `The Heun function can not be described in form of contour integrals of elementary functions even if it is the simplest class of special functions.'  

Recently Heun function started to appear in theoretical modern physics. For example the Heun functions come out in the hydrogen-molecule ion\cite{Wils1928}, in the Schr$\ddot{\mbox{o}}$dinger equation with doubly anharmonic potential\cite{Ronv1995} (its solution is the confluent forms of Heun function), in the Stark effect\cite{Epst1926}, in perturbations of the Kerr metric\cite{Teuk1973,Leav1985,Bati2006,Bati2007,Bati2010}, in crystalline materials\cite{Slavy2000}, in Collogero-Moser-Sutherland systems\cite{Take2003}, etc., just to mention a few.\cite{Birk2007,Suzu1999,Suzu1998}  Traditionally, we have constructed all physical phenomenons by only using two term recursion relation in power series expansion until 19th century. However, modern physics (quantum gravity, SUSY, general relativity, etc) seem to require at least three or four recurrence relations in power series expansions. Furthermore these type of problems can not be reduced to two term recurrence relations by changing independent variables and coefficients.\cite{Hortacsu:2011rr}

In previous paper we show the analytic solutions of Heun functions for all higher terms of $A_n$'s by applying three term recurrence formula\cite{Chou2012}; power series expansions for an infinite and polynomial cases\cite{Chou2012c}. 

According to Ronveaux (1995 \cite{Ronv1995}), ``Except in some trivial cases, no example has been given of a solution of Heun's equation expressed in the form of a definite integral or contour integral involving only functions which are, in some sense, simpler. It may be reasonably conjectured that no such expressions exist.''

Instead Heun equation is obtained by Fredholm integral equations; such integral relationships express one analytic solution in terms of another analytic solution. More precisely, in earlier literature the integral representations of Heun's equation were constructed by using two types of relations: (1) Linear relations using Fredholm integral equations. \cite{Lamb1934,Erde1942} (2) Non-linear relation (Malurkar-type integral relations) including Fredholm integral equations using two variables. \cite{Slee1969a,Slee1969b,Arsc1964,Schm1979}

Now we consider direct integral representations of Heun functions and their asymptotic behaviors and boundary conditions for the independent variable $x$ by using 3TRF.
 Expressing Heun functions in integral forms resulting in a precise and simplified transformation of Heun functions to other well-known special functions such as hypergeometric functions, Mathieu functions, Lame functions, confluent forms of Heun functions and etc. Also, the orthogonal relations of Heun functions can be obtained from the integral forms.
\vspace{3mm}

In Ref.\cite{Heun1889}, Heun's equation is a second-order linear ordinary differential equation of the form
\begin{equation}
\frac{d^2{y}}{d{x}^2} + \left(\frac{\gamma }{x} +\frac{\delta }{x-1} + \frac{\epsilon }{x-a}\right) \frac{d{y}}{d{x}} +  \frac{\alpha \beta x-q}{x(x-1)(x-a)} y = 0 \label{eq:1}
\end{equation}
With the condition $\epsilon = \alpha +\beta -\gamma -\delta +1$. The parameters play different roles: $a \ne 0 $ is the singularity parameter, $\alpha $, $\beta $, $\gamma $, $\delta $, $\epsilon $ are exponent parameters, $q$ is the accessory parameter. Also, $\alpha $ and $\beta $ are identical to each other. The total number of free parameters is six. It has four regular singular points which are 0, 1, a and $\infty $ with exponents $\{ 0, 1-\gamma \}$, $\{ 0, 1-\delta \}$, $\{ 0, 1-\epsilon \}$ and $\{ \alpha, \beta \}$.Assume that $y(x)$ has a series expansion of the form
\begin{equation}
y(x)= \sum_{n=0}^{\infty } c_n x^{n+\lambda } \label{eq:2}
\end{equation}
where $\lambda $ is an indicial root. Plug (\ref{eq:2})  into (\ref{eq:1}):
\begin{equation}
c_{n+1}=A_n \;c_n +B_n \;c_{n-1} \hspace{1cm};n\geq 1 \label{eq:3}
\end{equation}
where
\begin{subequations}
\begin{eqnarray}
A_n &=& \frac{(n+\lambda )(n-1+\gamma +\epsilon +\lambda + a(n-1+\gamma +\lambda +\delta ))+q}{a(n+1+\lambda )(n+\gamma +\lambda )}\nonumber\\
&=& \frac{(n+\lambda )(n+\alpha +\beta -\delta +\lambda +a(n+\delta +\gamma -1+\lambda ))+q}{a(n+1+\lambda )(n+\gamma +\lambda )} \label{eq:4a}
\end{eqnarray}
\begin{equation}
B_n = -\frac{(n-1+\lambda )(n+\gamma +\delta +\epsilon -2+\lambda )+\alpha \beta }{a(n+1+\lambda )(n+\gamma +\lambda )}= -\frac{(n-1+\lambda +\alpha )(n-1+\lambda +\beta )}{a(n+1+\lambda )(n+\gamma +\lambda )} \label{eq:4b}
\end{equation}
\begin{equation}
c_1= A_0 \;c_0 \label{eq:4c}
\end{equation}
\end{subequations}
We have two indicial roots which are $\lambda = 0$ and $ 1-\gamma $
\section{Asymptotic behavior of the Heun equation}
\subsection{Poincar\'{e}-Perron theorem and its applications for solutions of power series}\label{sec.2}
   
Let's review certain theorems on the asymptotic behavior of solutions of linear difference equations with constant coefficients.
Consider a linear recurrence relation of length $k+1$ with constant coefficients $\alpha _i$ where $i=0,1,2,\cdots,k$ 
\begin{equation}
u(n+1)+ \alpha _1 u(n)+ \alpha _2 u(n-1)+ \alpha _3 u(n-2)+\cdots + \alpha _k u(n-k+1)=0\label{qq:1}
\end{equation}
with $\alpha _k \ne 0$.
The characteristic polynomial equation of recurrence (\ref{qq:1}) is given by
\begin{equation}
t^k + \alpha _1 t^{k-1} +  \alpha _2 t^{k-2}+ \cdots + \alpha _k =0\label{qq:2}
\end{equation}
Denote the roots of the characteristic equation (\ref{qq:2}) by $\lambda _1, . . . ,\lambda _k $.

H. Poincar\'{e}'s suggested that
\begin{equation}
\lim_{n\rightarrow \infty } \frac{u(n+1)}{u(n)}\nonumber
\end{equation}
is equal to one of the roots of the characteristic equation in 1885 \cite{Poin1885}. And a more general theorem has been extended by O. Perron in 1921 \cite{Perr1921}.
\begin{thm}Poincar\'{e}-Perron theorem \cite{Miln1933}: If the coefficient of $u(n)$ in the difference equation of order $k$ be not zero, for $n=0,1,2,\cdots$, and other hypotheses be fulfilled, then the equation possesses $k$ fundamental solutions $u_1(n),\cdots,u_k(n)$, such that
\begin{equation}
\lim_{n\rightarrow \infty } \frac{u_{i}(n+1)}{u_{i}(n)}= \lambda _i \nonumber
\end{equation}
where $i=1,2,\cdots, k$ and $\lambda _i$ is a root of the characteristic equation, and $n\rightarrow \infty $ by positive integral increments.\label{thm.1}
\end{thm}

The recurrence relation of coefficients starts to appear by substituting a series $y(x)=\sum_{n=0}^{\infty }c_n x^n$ into a linear ordinary differential equation (ODE). In general, the 3-term recurrence relation is given by
\begin{equation}
c_{n+1}= \alpha _{1,n} \;c_n + \alpha _{2,n} \;c_{n-1} \hspace{1cm};n\geq 1
\label{qq:3}
\end{equation}
with seed values $c_1= \alpha _{1,0} c_0$. 
For the asymptotic behavior of (\ref{qq:3}), $ \lim_{n\gg 1}\alpha _{j,n}= \alpha _{j}<\infty $ where $j=1,2$ exists. 
Its asymptotic recurrence relation is given by
\begin{equation}
c_{n+1}= \alpha _{1} \;c_n + \alpha _{2} \;c_{n-1} \hspace{1cm};n\geq 1
\label{qq:4}
\end{equation}
where $c_1= \alpha _{1} c_0$. Due to Poincar\'{e}-Perron theorem, we form the characteristic polynomial such as
\begin{equation}
\rho ^2 -\alpha _{1} \rho -\alpha _{2} =0
\label{qq:5}
\end{equation}
The roots of a polynomial (\ref{qq:5}) have two different moduli
\begin{equation}
\rho_1 = \frac{\alpha _{1} -\sqrt{\alpha _{1}^2 +4\alpha _{2}}}{2}\hspace{1cm} \rho_2 = \frac{\alpha _{1} +\sqrt{\alpha _{1}^2 +4\alpha _{2}}}{2} \nonumber
\end{equation}
In general, if $|\rho_1|< |\rho_2|$, then $\lim_{n\rightarrow \infty } |c_{n+1}/c_n| = |\rho_2|$, so that the radius of convergence for a 3-term recursion relation (\ref{qq:3}) is $|\rho_2|^{-1}$. And as if $|\rho_2|< |\rho_1|$, then $\lim_{n\rightarrow \infty } |c_{n+1}/c_n| = |\rho_1|$, and its radius of convergence is increased to $|\rho_1|^{-1}$. For the special case, $\lim_{n\rightarrow \infty } |c_{n+1}/c_n| $ is divergent when $|\rho_1|=|\rho_2|$ and $\rho_1 \ne \rho_2$, and it is convergent when $\rho_1 = \rho_2$. More details are explained in Appendix B of part A \cite{Ronv1995}, Wimp (1984) \cite{Wimp1984}, Kristensson (2010) \cite{Kris2010} or Erd\'{e}lyi (1955) \cite{Erde1955}.

In chapter 3.3 on part A (pp. 34--36) \cite{Ronv1995}, they obtain three-term recursion system by putting a power series with an unknown coefficient into Heun's equation about $x=0$ corresponding to the exponent zero. By applying Poincar\'{e}-Perron theorem, ``We adopt the restriction $|a|>1$ and the series will generally have radius of convergence 1; it will therefore only represent a local solution.'' And its theorem tells us that a Heun function of class I about $\{ 0,1\}$, converging in the circle $|x|<|a|$ as $|a|$ is less than 1 where $a \ne 0$. Table~\ref{cb.1} tells us all possible boundary conditions using Poincar\'{e}-Perron theorem.
%\vspace{1cm}
\begin{table}[h]
\begin{center}
\tabcolsep 5.8pt
\begin{tabular}{l*{6}{c}|r}
Range of the coefficient $a$ & Range of the independent variable $x$ \\
\hline 
As $a=0, -1$ & no solution \\ 
As $|a|>1$  & $|x|<1$  \\ 
As $-1<a<0, 0<a\leq 1$ & $|x|<|a|$ \\
\end{tabular}
\end{center}
\caption{Boundary condition of $x$ of a Heun function about $x=0$ using Poincar\'{e}-Perron theorem}
\label{cb.1}
\end{table}
\begin{figure}[h]
\centering
\includegraphics[scale=.7]{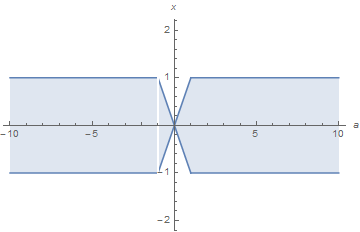}
\caption{Original Poincar\'{e}-Perron theorem}
\label{bc-img1}
\end{figure}

Fig.~\ref{bc-img1} indicates a graph of Table~\ref{cb.1} in the $a$-$x$ plane; the shaded area represents the domain of convergence of the series for a Heun equation around $x=0$ except $a=-1$; it does not include solid lines.

\subsection{Asymptotic behavior for an infinite series of $y(x)$ and the boundary condition for $x$}\label{sec.3}
By rearranging coefficients of $A_n$ and $B_n$ terms in (\ref{eq:3}), let's test for convergence of the Heun function $y(x)$ about $x=0$ for an infinite series. As $n\gg 1$ (for sufficiently large, like an index $n$ is close to infinity, or you can treat as $n\rightarrow \infty $), (\ref{eq:3})--(\ref{eq:4b}) are asymptotically equal to
\begin{subequations} 
\begin{equation}
c_{n+1}= A\;c_n +B\;c_{n-1} \hspace{1cm};n\geq 1 \label{eq:100}
\end{equation}
where
\begin{equation}
\lim_{n\gg 1} A_n = A=  \frac{(1+a)}{a} \hspace{2cm} \lim_{n\gg 1} B_n = B= -\frac{1}{a}\label{eq:5}
\end{equation}
\end{subequations}
Substitute (\ref{eq:5}) into (\ref{eq:100}) by letting $c_1= A c_0$.\footnote{We only have the sense of curiosity about an asymptotic series as  $n\gg 1$ for a given $x$. Actually, $c_1 =  A_0 c_0$. But for a huge value of an index $n$, we treat the coefficient $c_1$ as $ A c_0$ for simple computations.} For $n=0,1,2,\cdots$, it gives
\begin{equation}
\begin{tabular}{ l }
  \vspace{2 mm}
  $c_0$ \\
  \vspace{2 mm}
  $c_1 =  Ac_0 $ \\
  \vspace{2 mm}
  $c_2 = (A^2+ B)c_0 $ \\
  \vspace{2 mm}
  $c_3 = (A^3 + 2AB)c_0 $ \\
  \vspace{2 mm}
  $c_4 = (A^4 + 3 A^2B + B^2)c_0 $ \\
  \vspace{2 mm}
  $c_5 = (A^5 + 4A^3B + 3AB^2)c_0 $ \\
  \vspace{2 mm}
  $c_6 = (A^6+ 5A^4B+ 6A^2B^2+ B^3)c_0 $ \\
  \vspace{2 mm}
  $c_7 = (A^7+ 6A^5B + 10A^3B^2+ 4AB^3)c_0 $ \\
  \vspace{2 mm}
  $c_8 = (A^8+ 7A^6B+ 15A^4B^2+ 10A^2B^3+ B^4)c_0 $ \\
  \vspace{2 mm}                       
 \hspace{2 mm} \vdots \hspace{3cm} \vdots \\
\end{tabular}\label{eq:6}
\end{equation}
If a series solution of a linear differential equation is absolutely convergent, we can rearrange of its terms for the series solution. Indeed, the sum of any arbitrary series is equivalent to the sum of the initial series.

With reminding the above mathematical phenomenon, let assume that a series solution of Heun's equation is absolutely convergent.
The sequence $c_n$ consists of combinations $A$ and $B$ in (\ref{eq:6}). First observe the term inside parentheses of sequence $c_n$ which does not include any $A$'s in (\ref{eq:6}): $c_n$ with even index ($c_0$,$c_2$,$c_4$,$\cdots$). 

\begin{equation}
\begin{tabular}{  l  }
  \vspace{2 mm}
  $c_0$ \\
  \vspace{2 mm}
  $c_2 = B c_0  $ \\
  \vspace{2 mm}
  $c_4 = B^2 c_0  $ \\
  \vspace{2 mm}
  $c_6 = B^3c_0 $ \\
  \vspace{2 mm}
  $c_8 = B^4c_0 $\\
  \vspace{2 mm}
  $c_{10} = B^5c_0 $ \\
  \hspace{2 mm}
   \vdots \hspace{.5cm} \vdots  \\ 
\end{tabular}\label{eq:7}
\end{equation}
When an asymptotic function $y(x)$, analytic at $x=0$, is expanded in a power series, we write
\begin{equation}
y(x)= \sum_{m=0}^{\infty } y_m(x) \label{eq:8}
\end{equation}
where
\begin{equation}
y_m(x)= \sum_{n=0}^{\infty } c_n^m x^{n}\label{eq:9}
\end{equation}
Put(\ref{eq:7}) in (\ref{eq:9}) putting $m=0$. 
\begin{equation}
y_0(x)= c_0 \sum_{n=0}^{\infty } \left( Bx^2\right)^n \label{eq:10}
\end{equation}
Observe the terms inside parentheses of sequence $c_n$ which include one term of $A$'s in (\ref{eq:6}): $c_n$ with odd index ($c_1$, $c_3$, $c_5$,$\cdots$). 

\begin{equation}
\begin{tabular}{  l  }
  \vspace{2 mm}
  $c_1= A c_0$ \\
  \vspace{2 mm}
  $c_3 = 2AB c_0  $ \\
  \vspace{2 mm}
  $c_5 = 3AB^2c_0  $ \\
  \vspace{2 mm}
  $c_7 = 4AB^3c_0  $ \\
  \vspace{2 mm}
  $c_9 = 5AB^4c_0 $\\
  \hspace{2 mm}
   \vdots \hspace{.5cm} \vdots \\ 
\end{tabular}
\end{equation}\label{eq:11}
Put the above sequences $c_n$ in (\ref{eq:9}) putting $m=1$.
\begin{equation}
y_1(x) = c_0 A x\sum_{n=0}^{\infty } \frac{(n+1)}{1!} \left(Bx^2\right)^n\label{eq:12}
\end{equation}
Observe the terms inside parentheses of sequence $c_n$ which include two terms of $A$'s in (\ref{eq:6}): $c_n$ with even index ($c_2$, $c_4$, $c_6$,$\cdots$).  

\begin{equation}
\begin{tabular}{  l  }
  \vspace{2 mm}
  $c_2 = A^2 c_0$ \\
  \vspace{2 mm}
  $c_4 = 3A^2B c_0  $ \\
  \vspace{2 mm}
  $c_6 = 6A^2B^2c_0  $ \\
  \vspace{2 mm}
  $c_8 = 10A^2B^3c_0  $ \\
  \vspace{2 mm}
  $c_{10} = 15A^2B^4c_0 $\\
  \hspace{2 mm}
   \vdots \hspace{1cm} \vdots \\ 
\end{tabular}\label{eq:13}
\end{equation}
Put (\ref{eq:13}) in (\ref{eq:9}) putting $m=2$.
\begin{equation}
y_2(x) = c_0 \left(Ax\right)^2\sum_{n=0}^{\infty } \frac{(n+1)(n+2)}{2!} \left( Bx^2\right)^n \label{eq:14}
\end{equation}
Similarly, the asymptotic function $y_3(x)$ for three terms of $A$'s is given by
\begin{equation}
y_3(x)= c_0 \left( Ax \right)^3 \sum_{n=0}^{\infty } \frac{(n+1)(n+2)(n+3)}{3!}\left( Bx^2 \right)^n
\label{et:14}
\end{equation} 
By repeating this process for all higher terms of $A$'s, we can obtain every $y_m(x)$ terms where $m \geq 4$. Substitute (\ref{eq:10}), (\ref{eq:12}), (\ref{eq:14}), (\ref{et:14}) and including all $y_m(x)$ terms where $m \geq 4$ into (\ref{eq:8}).
\begin{eqnarray}
y(x)&=& \sum_{n=0}^{\infty } c_n x^n = y_0(x)+ y_1(x)+ y_2(x)+y_3(x)+\cdots\nonumber\\
&=& \sum_{n=0}^{\infty } \sum_{m=0}^{\infty } \frac{(n+m)!}{n!\;m!} \tilde{x}^n \tilde{y}^m \hspace{1cm} \mbox{where}\;c_0=1, \tilde{x}= Bx^2 \;\mbox{and} \; \tilde{y}= Ax \label{eq:15}
\end{eqnarray}
By definition, a real or complex series $\sum_{n=0}^{\infty } u_n$ is said to converge absolutely if the series of moduli $\sum_{n=0}^{\infty } |u_n|$  converge. And the series of absolute values (\ref{eq:15}) is
\begin{equation}
\sum_{n=0}^{\infty } \sum_{m=0}^{\infty } \frac{(n+m)!}{n!\;m!} |\tilde{x}|^n |\tilde{y}|^m = \sum_{r=0}^{\infty } (|\tilde{x}| + |\tilde{y}|)^r  \nonumber
\end{equation} 
This double series is absolutely convergent for $|\tilde{x}| + |\tilde{y}| <1$.
(\ref{eq:15}) is simply
\begin{equation}
\lim_{n\gg 1}y(x)= \frac{1}{1-(\tilde{x}+ \tilde{y})}= \frac{1}{1-\left(-\frac{1}{a}x^2 +\frac{1+a}{a}x\right)}  \label{eq:16}
\end{equation}
(\ref{eq:16}) is geometric series. Its condition of an absolute convergence (\ref{eq:16}) is
\begin{equation}
\left|-\frac{1}{a}x^2 \right| + \left| \frac{1+a}{a}x \right|<1 \label{eq:17}
\end{equation} 
The coefficient $a$ decides the range of an independent variable $x$ as we see (\ref{eq:17}). More precisely, 
\begin{table}[h]
\begin{center}
\tabcolsep 5.8pt
\begin{tabular}{l*{6}{c}|r}
Range of the coefficient $a$ & Range of the independent variable $x$ \\
\hline 
As $a=0$ & no solution \\ 
As $a>0$  & $|x|< \frac{1}{2}(-1-a+\sqrt{a^2+6a+1})$  \\ 
As $-1\leq a<0$ & $|x|<|a|$ \\ 
As $a<-1$ & $|x|<1$ \\
\end{tabular}
\end{center}
\caption{Boundary condition of $x$ for the infinite series of a Heun function about $x=0$}
\label{cb.2}
\end{table}
\begin{figure}[h]
\centering
\includegraphics[scale=.7]{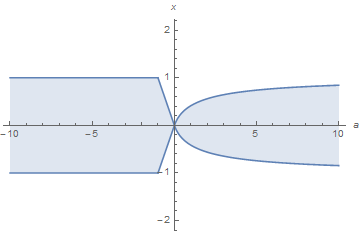} 
\caption{Revised Poincar\'{e}-Perron theorem}
\label{bc-img2}
\end{figure}
 
The corresponding domain of convergence in the real axis, given by (\ref{eq:17}), is shown shaded in Fig.~\ref{bc-img2}; it does not include solid lines, and maximum modulus of $x$ is the unity.

In Table~\ref{cb.2} or the shaded area where $a>0$ in Fig.~\ref{bc-img2},
\begin{equation}
\lim_{a\rightarrow N }\frac{-1-a+\sqrt{a^2+6a+1}}{2} \sim 1 \nonumber
\end{equation}
where $N$ is the sufficiently huge positive real or complex. Then we can argue that $|x|<1$ for $a\rightarrow N$. For examples, if $a=10$, then $|x|< 0.84429$ and as $a=100$, $|x|< 0.98058$. 

In the case of $|a|\gg 1$ assuming $|a|$ is huge numerical values, (\ref{eq:16}) turns to be
\begin{equation}
\lim_{n\gg 1}y(x)\approx  \frac{1}{1- x}  \label{eq:22}
\end{equation}
where $|x|<1$.

\subsection{Original Poincar\'{e}-Perron theorem vs. revised Poincar\'{e}-Perron theorem}\label{sec.4}
 As we compare Table~\ref{cb.2} with Table~\ref{cb.1}, both boundary conditions for radius of convergence are equivalent to each other since $a<0$ except $a=-1$.  Table~\ref{cb.2} allows $a=-1$ for the analytic solution of a Heun function, but there is no solution for a series since $a=-1$ in Table~\ref{cb.1}. As $a\geq 1$, their ranges of $x$ are slightly different: (i) Radius of convergence is the unity in Table~\ref{cb.1} at $a=1$, but we suggest that its radius is approximately 0.414214 in Table~\ref{cb.2}. (ii) If $a$ is quiet huge numerical real or complex values, their radius are almost equal to the unity, i.e., as $a=1000$ in Table~\ref{cb.2}, its range approximates to $|x|<0.998006$, which is really closed to $|x|<1$ in Table~\ref{cb.1}. (iii) In the region at $0<a<1$, maximum absolute value of $x$ in Tables~\ref{cb.1} and ~\ref{cb.2} are quiet different. As we see $x$ where a positive real value in Table~\ref{cb.2}, it is a square root function of $a$ and the range of its slope with respect to $a$ is between 0.207107 and 1. A variable $x$ in Table~\ref{cb.1} is just linearly increasing line with a slope 1 with respect to $a$. Since $x$ is a negative real value in Table~\ref{cb.2}, the slope of a square root function of $a$ is between -1 and -0.207107. And the slope of $x$ in Table~\ref{cb.1} is just -1. (iv) A square root function for a huge value $a$ in Table~\ref{cb.2} is closed to $\pm 1$ which demonstrates strong justification of Poincar\'{e}-Perron theorem, but in the region at $0<a<1$, Poincar\'{e}-Perron theorem is not available to obtain radius of convergence of Heun functions any more. 

Now, let's consider difference between  Tables~\ref{cb.1} and ~\ref{cb.2} with respect to numerical computations. A sequence $c_n$ is derived by putting a power series into a Heun's equation. The boundary condition of $x$ in Table~\ref{cb.1} is obtained by the ratio of sequence $c_{n+1}$ to $c_n$ at the limit $n\rightarrow \infty $. And radius of convergence of $x$ in Table~\ref{cb.2} is constructed by rearranging coefficients $A$ and $B$ in each sequence $c_n$.

For instead, if $a=0.8$ in Table~\ref{cb.2}, its boundary condition is approximately $-0.368858< x< 0.368858$, and the radius convergence in Table~\ref{cb.1} is $-0.8< x< 0.8$. Let allow us that a analytic solution of (\ref{eq:100}) is 
\begin{equation}
y(x) =\sum_{n=0}^{N} c_n x^n \label{qq:6}
\end{equation}
First put (\ref{eq:5}) in (\ref{eq:100}) with $a=0.8$ and substitute the new (\ref{eq:100}) into (\ref{qq:6}) by allowing $x=0.7$ with various positive integer values of $N$ in Mathematica program. Similarly, numerical values of $y(x)$ with $a=0.8$ and $x=0.3$ are given in Tables~\ref{cb.3} and ~\ref{cb.4}. 
\begin{table}[htbp]
\footnotesize
\parbox{.45\linewidth}{
\centering
\begin{center}
\tabcolsep 5.8pt
\begin{tabular}{l*{6}{c}|r}
 $N$ &  $y(x)$ \\
\hline 
$10$ &  $17.722665066666$  \\ 
$50$ &  $26.622563574231$  \\ 
$100$ & $26.666611092450$  \\ 
$200$ & $26.666666666578$ \\
$300$ & $26.666666666667$  \\
$400$ & $26.666666666667$  \\ 
$500$ & $26.666666666667$ \\
$600$ & $26.666666666667$  \\
$700$ & $26.666666666667$ \\
$800$ & $26.666666666667$  \\ 
$900$ & $26.666666666667$ \\
$1000$ & $26.666666666667$  \\ 
\end{tabular}
\end{center}
\caption{ $y(x)$ with $a=0.8$ and $x=0.7$}\label{cb.3}
}
\hfill
\parbox{.45\linewidth}{
\centering
\begin{center}
\tabcolsep 5.8pt
\begin{tabular}{l*{6}{c}|r}
 $N$ &  $y(x)$ \\
\hline      
$10$ &  $2.285559427400$  \\ 
$50$ &  $2.285714285714$  \\ 
$100$ & $2.285714285714$  \\ 
$200$ & $2.285714285714$ \\
$300$ & $2.285714285714$  \\
$400$ & $2.285714285714$  \\ 
$500$ & $2.285714285714$ \\
$600$ & $2.285714285714$  \\
$700$ & $2.285714285714$ \\
$800$ & $2.285714285714$  \\ 
$900$ & $2.285714285714$ \\
$1000$ & $2.285714285714$  \\ 
\end{tabular}
\end{center}
\caption{ $y(x)$ with $a=0.8$ and $x=0.3$}\label{cb.4}
}
\end{table}

%\newpage
Numerical values of $y(x)$ in Tables~\ref{cb.3} and ~\ref{cb.4} are derived by putting a 3-term recursive system into a power series with the specific values of $a$ and $x$.
As we see Table~\ref{cb.3}, $y(x)$ is convergent as $N\rightarrow \infty $, its approximative value is $26.6667$. And Table~\ref{cb.4} tells us that $y(x)\approx 2.2857$ as $N\rightarrow \infty $ is also convergent. It means that the radius of convergence using Poincar\'{e}-Perron theorem and the boundary condition by rearranging of its terms for the series solution are both available for the analytic solutions of Heun functions.

Consider the following summation series such as
\begin{eqnarray}
y(x)= \sum_{n=0}^{N} \sum_{m=0}^{N} \frac{(n+m)!}{n!\;m!} \tilde{x}^n \tilde{y}^m \hspace{1cm} \mbox{where}\; \tilde{x}= -\frac{1}{a}x^2 \;\mbox{and} \; \tilde{y}= \frac{1+a}{a}x \label{qq:7}
\end{eqnarray}
This equation is equivalent to (\ref{eq:15}) as $N\rightarrow \infty $. 
Substitute $a=0.8$ and $x=0.7$ in (\ref{qq:7}) with various positive integer values $N$. And we obtain various numerical values of $y(x)$ where $N= 10,50,100,200,300,\cdots,1000$  by putting $a=0.8$ and $x=0.3$ in (\ref{qq:7}).
\begin{table}[htbp]
\footnotesize
\parbox{.45\linewidth}{
\centering
\begin{center}
\tabcolsep 5.8pt
\begin{tabular}{l*{6}{c}|r}
 $N$ &  $y(x)$ \\
\hline 
$10$ &  $1.00791\times 10^{5}$  \\ 
$50$ &  $1.34009\times 10^{28}$  \\ 
$100$ & $1.99922\times 10^{57}$  \\ 
$200$ & $6.25120\times 10^{115}$ \\
$300$ & $2.25372\times 10^{174}$  \\
$400$ & $8.61497\times 10^{232}$  \\ 
$500$ & $3.40062\times 10^{291}$ \\
$600$ & $1.36992\times 10^{350}$  \\
$700$ & $5.59670\times 10^{408}$ \\
$800$ & $2.31012\times 10^{467}$  \\ 
$900$ & $9.61056\times 10^{525}$ \\
$1000$ &$4.02305\times 10^{584}$  \\ 
\end{tabular}
\end{center}
\caption{ $y(x)$ with $a=0.8$ and $x=0.7$}\label{cb.5}
}
\hfill
\parbox{.45\linewidth}{
\centering
\begin{center}
\tabcolsep 5.8pt
\begin{tabular}{l*{6}{c}|r}
 $N$ &  $y(x)$ \\
\hline      
$10$ &  $2.276337892064$  \\ 
$50$ &  $2.285714285695$  \\ 
$100$ & $2.285714285714$  \\ 
$200$ & $2.285714285714$ \\
$300$ & $2.285714285714$  \\
$400$ & $2.285714285714$  \\ 
$500$ & $2.285714285714$ \\
$600$ & $2.285714285714$  \\
$700$ & $2.285714285714$ \\
$800$ & $2.285714285714$  \\ 
$900$ & $2.285714285714$ \\
$1000$ & $2.285714285714$  \\ 
\end{tabular}
\end{center}
\caption{ $y(x)$ with $a=0.8$ and $x=0.3$}\label{cb.6}
}
\end{table}

A numerical quantities $y(x)$ in Tables~\ref{cb.5} and ~\ref{cb.6} are obtained by rearranging $A_n$ and $B_n$ terms in each sequence $c_n$ with the certain values of $a$ and $x$. Table~\ref{cb.5} tells us that $y(x)$ is divergent as $N\rightarrow \infty $. And Table~\ref{cb.6} informs us that $y(x)\approx 2.2857$ as $N\rightarrow \infty $ is also convergent which is equal to approximative quantities $y(x)$ in Table~\ref{cb.4}.
According to Table~\ref{cb.5}, we notice that the radius of convergence using Poincar\'{e}-Perron theorem is not available in an asymptotic series solution in closed forms which is performed by rearranging coefficients $A$ and $B$ terms in the sequence $c_n$. 

\begin{thm}
We can not use Poincar\'{e}-Perron theorem to obtain the radius of convergence for a power series solution.
And a series solution for an infinite series, obtained by applying Poincar\'{e}-Perron theorem, is not absolute convergent but only conditionally convergent. 
\end{thm}
\begin{pot}
We might have curiosity why we have such errors since we apply one of any values in the interval of convergence of the series, constructed by Poincar\'{e}-Perron theorem, into asymptotic expansion by relating the series to the geometric series. To answer this question, first of all, consider an alternating harmonic series such as
\begin{eqnarray}
\sum_{n=0}^{\infty } \frac{(-1)^n}{n+1}= 1-\frac{1}{2}+\frac{1}{3}-\frac{1}{4}+\frac{1}{5}-\frac{1}{6}+\cdots \nonumber
\end{eqnarray}
This series is well known to have the sum $\ln 2$ since we add terms one by one. However, since we rearrange of its terms for the series solution, its sum can be divergent; if all terms are taken with + signs, it is divergent. Similarly, it is also divergent since we add all terms with -signs. This series is not absolutely convergent but conditionally convergent, based on the Leibniz criterion. 

With reminding this example, let assume that a power series of Heun's equation converges absolutely within its radius of convergence, obtained by applying Poincar\'{e}-Perron theorem. It tells us that even if we rearrange the order of the
terms in series, its solution is also convergent. For instance, consider $a$ and $x$ as real positive numbers. $A$ is real positive number and $B$ is real negative one in (\ref{eq:6}). We observe that all terms in sequence $c_n$ in (\ref{eq:6}) consists of positive and negative real values; any terms having $B^{2m+1}$ where $m=0,1,2,\cdots$ are composed of real negative values, otherwise real positive ones. First, we take all terms with real positive values in each sequence $c_n$ in (\ref{eq:6}) and after that, take every terms having real negative ones. For $n=0,1,2,\cdots$ with $c_0=1$ for simplicity, it gives
%\newpage
\begin{table}[htbp]
\footnotesize
\begin{center}
\tabcolsep 5.8pt
\begin{tabular}{l|l}
 Real positive terms & Real negative terms \\
\hline      
$c_0=1$  &  $c_2 =B$  \\ 
$c_1 =  A  $ &  $c_3 = 2AB$  \\ 
$c_2 = A^2 $ & $c_4 = 3A^2B$  \\ 
$c_3 = A^3 $ & $c_5 = 4A^3B$ \\
$c_4 = A^4 +  B^2 $ & $c_6 = 5A^4B + B^3$  \\
$c_5 = A^5 + 3AB^2 $ & $c_7 = 6A^5B + 4AB^3$  \\ 
$c_6 = A^6 + 6A^2B^2 $  & $c_8 = 7A^6B + 10A^2B^3$ \\
$c_7 = A^7 + 10A^3B^2 $ & $c_9= 8 A^7 B + 20 A^3 B^3$  \\
$c_8 = A^8 + 15A^4B^2 + B^4 $ & $c_{10}= 9 A^8 B + 35 A^4 B^3 + B^5 $ \\
$c_9 = A^9 + 21A^5B^2 + 5AB^4 $ & $c_{11}= 10 A^9 B + 56 A^5 B^3  + 6 A B^5$  \\ 
$c_{10} = A^{10} + 28A^6B^2 + 15A^2B^4 $ & $c_{12}= 11 A^{10} B  + 84 A^6 B^3 + 21 A^2 B^5  $ \\
$c_{11} = A^{11} + 36A^7B^2 + 35A^3B^4 $ & $c_{13}= 12 A^{11} B  + 120 A^7 B^3  + 56 A^3 B^5 $  \\ 
$c_{12} = A^{12} + 45A^8B^2 + 70A^4B^4+ B^6 $ & $c_{14}= 13 A^{12} B + 165 A^8 B^3 + 126 A^4 B^5 + B^7$ \\
$c_{13} = A^{13} + 55A^9B^2 + 126A^5B^4+ 7AB^6 $ & $c_{15}= 14 A^{13} B + 220 A^9 B^3 + 252 A^5 B^5 + 8 A B^7$  \\ 
$\hspace{2 mm} \vdots \hspace{3 cm} \vdots$ & $\hspace{2 mm} \vdots \hspace{3 cm} \vdots$  \\
\end{tabular}
\end{center}
\caption{ all possible terms in sequences $c_n$ from $c_0$ up to $ c_{15}$}
\label{cb.7}
\end{table} 
  
We construct a power series solution of real positive terms, denoted by $y_{+}(x)$ and build a series solution of real negative terms, denominated by $y_{-}(x)$. Since we add $y_{+}(x)$ and $y_{-}(x)$ we get a asymptotic series $y(x)$. 
 
\underline{(A) Series of $y_{+}(x)$} 
%\vspace{.5cm} 

First observe the term of sequence $c_n$ which does not include any $B$'s of real positive terms in Table~\ref{cb.7}: $c_n$ with every index ($c_0$,$c_1$,$c_2$,$\cdots$). 

\begin{equation}
\begin{tabular}{  l  }
  \vspace{2 mm}
  $c_0 =1$ \\
  \vspace{2 mm}
  $c_1 = A  $ \\
  \vspace{2 mm}
  $c_2 = A^2  $ \\
  \vspace{2 mm}
  $c_3 = A^3 $ \\
  \hspace{.5 mm}
   \vdots \hspace{.5cm} \vdots  \\ 
\end{tabular}\label{qq:7a}
\end{equation}
When an asymptotic series $y_{+}(x)$, analytic at $x=0$, is expanded in a power series, we write
\begin{equation}
y_{+}(x)= \sum_{m=0}^{\infty } y_{+}^m(x) \label{qq:8}
\end{equation}
where
\begin{equation}
y_{+}^m(x)= \sum_{n=0}^{\infty } c_n^m x^{n}\label{qq:9}
\end{equation}
Put(\ref{qq:7a}) in (\ref{qq:9}) putting $m=0$. 
\begin{equation}
y_{+}^0(x)= \sum_{n=0}^{\infty } \left( Ax \right)^n \label{qq:10}
\end{equation}
Observe the terms of sequence $c_n$ which include two term of $B$'s of real positive terms in Table~\ref{cb.7}: $c_n$ with every index except $c_0-c_3$ ($c_4$, $c_5$, $c_6$,$\cdots$). 

\begin{equation}
\begin{tabular}{  l  }
  \vspace{2 mm}
  $c_4= B^2$ \\
  \vspace{2 mm}
  $c_5 = 3AB^2  $ \\
  \vspace{2 mm}
  $c_6 = 6A^2B^2   $ \\
  \vspace{2 mm}
  $c_7 = 10A^3B^2  $ \\
  \vspace{2 mm}
  $c_8 = 15A^4B^2 $\\
  \vspace{2 mm}
  $c_9 = 21A^5B^2 $\\
  \hspace{2 mm}
   \vdots \hspace{.5cm} \vdots \\ 
\end{tabular}
\end{equation}\label{qq:11}
Put the above sequences $c_n$ in (\ref{qq:9}) putting $m=1$.
\begin{equation}
y_{+}^1(x) = \left( Bx^2 \right)^2 \sum_{n=0}^{\infty } \frac{ (n+2)!}{2! n!} \left( Ax \right)^n \label{qq:12}
\end{equation}
Observe the terms of sequence $c_n$ which include four terms of $B$'s of real positive terms in Table~\ref{cb.7}: $c_n$ with every index except $c_0-c_7$ ($c_8$, $c_9$, $c_{10}$,$\cdots$). 

\begin{equation}
\begin{tabular}{  l  }
  \vspace{2 mm}
  $c_8 = B^4 $ \\
  \vspace{2 mm}
  $c_9 = 5AB^4 $ \\
  \vspace{2 mm}
  $c_{10} = 15A^2B^4 $ \\
  \vspace{2 mm}
  $c_{11} = 35A^3B^4 $ \\
  \vspace{2 mm}
  $c_{12} = 70A^4B^4 $\\
  \vspace{2 mm}
  $c_{13} = 126A^5B^4 $\\
  \hspace{2 mm}
   \vdots \hspace{1cm} \vdots \\ 
\end{tabular}\label{qq:13}
\end{equation}
Put (\ref{qq:13}) in (\ref{qq:9}) putting $m=2$.
\begin{equation}
y_{+}^2(x) =  \left( Bx^2\right)^4\sum_{n=0}^{\infty } \frac{ (n+4)!}{4! n!} \left( Ax \right)^n \label{qq:14}
\end{equation}
Similarly, the asymptotic series $y_{+}^3(x)$ for six terms of $B$'s is given by
\begin{equation}
y_{+}^3(x)= \left( Bx^2\right)^6\sum_{n=0}^{\infty } \frac{ (n+6)!}{6! n!} \left( Ax \right)^n
\label{qq:15}
\end{equation} 
By mathematical induction, we repeat this process and build series solutions for all higher terms of $B$'s. We construct every $y_{+}^m(x)$ terms where $m \geq 4$. Substitute (\ref{qq:10}), (\ref{qq:12}), (\ref{qq:14}), (\ref{qq:15}) and including all $y_{+}^m(x)$ terms where $m \geq 4$ into (\ref{qq:8}).
\begin{eqnarray}
y_{+}(x)&=& \sum_{n=0}^{\infty } c_n x^n = y_{+}^0(x)+y_{+}^1(x)+ y_{+}^2(x)+y_{+}^3(x)+\cdots\nonumber\\
&=& \sum_{n=0}^{\infty } \sum_{m=0}^{\infty } \frac{(n+2m)!}{n!\;(2m)!} \tilde{x}^{2m} \tilde{y}^n \hspace{1cm} \mbox{where}\;  \tilde{x}= Bx^2 \;\mbox{and} \; \tilde{y}= Ax \label{qq:16}\\
&=& \frac{ 1-\tilde{y} }{1 - \tilde{x}^2 - 2 \tilde{y} + \tilde{y}^2 } \label{qq:17}
\end{eqnarray}
$y_{+}(x)$ gives us a real positive value and the series of absolute values (\ref{qq:16}) is
\begin{equation}
\sum_{n=0}^{\infty } \sum_{m=0}^{\infty } \frac{(n+2m)!}{n!\;(2m)!} |\tilde{x}|^{2m} |\tilde{y}|^n 
= \sum_{n=0}^{\infty } \sum_{l=0}^{\infty } \frac{l!}{n!\;(l-n)!} |\tilde{x}|^{l-n} |\tilde{y}|^n 
= \sum_{r=0}^{\infty } (|\tilde{x}| + |\tilde{y}|)^r  \nonumber
\end{equation} 
This double series is absolutely convergent for $|\tilde{x}| + |\tilde{y}| =\left|-\frac{1}{a}x^2 \right| + \left| \frac{1+a}{a}x \right| <1$ and its boundary condition is same as Table~\ref{cb.2}. It informs that the radius of convergence of $y_{+}(x)$ can not be obtained by applying Poincar\'{e}-Perron theorem
   
\underline{(B) Series of $y_{-}(x)$} 
 
Observe the term of sequence $c_n$ which include one term of $B$'s of real negative terms in Table~\ref{cb.7}: $c_n$ with every index except $c_0$ and $c_1$ ($c_2$, $c_3$, $c_4$,$\cdots$). 

\begin{equation}
\begin{tabular}{  l  }
  \vspace{2 mm}
  $c_2 =B$ \\
  \vspace{2 mm}
  $c_3 = 2AB  $ \\
  \vspace{2 mm}
  $c_4 = 3A^2B  $ \\
  \vspace{2 mm}
  $c_5 = 4A^3B $ \\
  \hspace{.5 mm}
   \vdots \hspace{.5cm} \vdots  \\ 
\end{tabular}\label{qq:18}
\end{equation}
$y_{-}(x)$ will be given by an expression
\begin{equation}
y_{-}(x)= \sum_{m=0}^{\infty } y_{-}^m(x) \label{qq:19}
\end{equation}
where
\begin{equation}
y_{-}^m(x)= \sum_{n=0}^{\infty } c_n^m x^{n}\label{qq:20}
\end{equation}
Put(\ref{qq:18}) in (\ref{qq:20}) putting $m=0$. 
\begin{equation}
y_{-}^0(x)= \left( Bx^2 \right) \sum_{n=0}^{\infty } \frac{(n+1)!}{1! n!}\left( Ax \right)^n \label{qq:21}
\end{equation}
Observe the terms of sequence $c_n$ which include three term of $B$'s of real negative terms in Table~\ref{cb.7}: $c_n$ with every index except $c_0-c_5$ ($c_6$, $c_7$, $c_8$,$\cdots$). 

\begin{equation}
\begin{tabular}{  l  }
  \vspace{2 mm}
  $c_6= B^3$ \\
  \vspace{2 mm}
  $c_7 = 4AB^3  $ \\
  \vspace{2 mm}
  $c_8 = 10A^2B^3   $ \\
  \vspace{2 mm}
  $c_9 = 20A^3B^3  $ \\
  \vspace{2 mm}
  $c_{10} = 35A^4B^3 $\\
  \vspace{2 mm}
  $c_{11} = 56A^5B^3 $\\
  \hspace{2 mm}
   \vdots \hspace{.5cm} \vdots \\ 
\end{tabular}\label{qq:22}
\end{equation}
Put (\ref{qq:22}) in (\ref{qq:20}) putting $m=1$.
\begin{equation}
y_{-}^1(x) = \left( Bx^2 \right)^3 \sum_{n=0}^{\infty } \frac{ (n+3)!}{3! n!} \left( Ax \right)^n \label{qq:23}
\end{equation}

Observe the terms of sequence $c_n$ which include five terms of $B$'s of real negative terms in Table~\ref{cb.7}: $c_n$ with every index except $c_0-c_9$ ($c_{10}$, $c_{11}$, $c_{12}$,$\cdots$). 

\begin{equation}
\begin{tabular}{  l  }
  \vspace{2 mm}
  $c_{10} = B^5 $ \\
  \vspace{2 mm}
  $c_{11} = 6AB^5 $ \\
  \vspace{2 mm}
  $c_{12} = 21A^2B^5 $ \\
  \vspace{2 mm}
  $c_{13} = 56A^3B^5 $ \\
  \vspace{2 mm}
  $c_{14} = 126A^4B^5 $\\
  \vspace{2 mm}
  $c_{15} = 252A^5B^5 $\\
  \hspace{2 mm}
   \vdots \hspace{1cm} \vdots \\ 
\end{tabular}\label{qq:24}
\end{equation}
Put (\ref{qq:24}) in (\ref{qq:20}) putting $m=2$.
\begin{equation}
y_{-}^2(x) =  \left( Bx^2\right)^5\sum_{n=0}^{\infty } \frac{ (n+5)!}{5! n!} \left( Ax \right)^n \label{qq:25}
\end{equation}
And $y_{-}^3(x)$ for seven terms of $B$'s is given by
\begin{equation}
y_{-}^3(x)= \left( Bx^2\right)^7\sum_{n=0}^{\infty } \frac{ (n+7)!}{7! n!} \left( Ax \right)^n
\label{qq:26}
\end{equation} 
In the same way, by repeating this process for all higher terms of $B$'s, we build every $y_{-}^m(x)$ terms where $m \geq 4$. Substitute (\ref{qq:21}), (\ref{qq:23}), (\ref{qq:25}), (\ref{qq:26}) and including all $y_{-}^m(x)$ terms where $m \geq 4$ into (\ref{qq:19}).
\begin{eqnarray}
y_{-}(x)&=& \sum_{n=0}^{\infty } c_n x^n = y_{-}^0(x)+y_{-}^1(x)+ y_{-}^2(x)+y_{-}^3(x)+\cdots\nonumber\\
&=& \sum_{n=0}^{\infty } \sum_{m=0}^{\infty } \frac{(n+2m+1)!}{n!\;(2m+1)!} \tilde{x}^{2m+1} \tilde{y}^n \hspace{1cm} \mbox{where}\;  \tilde{x}= Bx^2 \;\mbox{and} \; \tilde{y}= Ax \label{qq:27}\\
&=& \frac{  \tilde{x} }{1 - \tilde{x}^2 - 2 \tilde{y} + \tilde{y}^2 } \label{qq:28}
\end{eqnarray}
$y_{-}(x)$ provides us a real negative value and the series of absolute values (\ref{qq:27}) is
\begin{eqnarray}
&&\sum_{n=0}^{\infty } \sum_{m=0}^{\infty } \frac{(n+2m+1)!}{n!\;(2m+1)!} |\tilde{x}|^{2m+1} |\tilde{y}|^n \nonumber\\
&=& \sum_{n=0}^{\infty } \sum_{l=1}^{\infty } \frac{l!}{n!\;(l-n)!} |\tilde{x}|^{l-n} |\tilde{y}|^n < \sum_{n=0}^{\infty } \sum_{l=0}^{\infty } \frac{l!}{n!\;(l-n)!} |\tilde{x}|^{l-n} |\tilde{y}|^n
= \sum_{r=0}^{\infty } (|\tilde{x}| + |\tilde{y}|)^r  \nonumber
\end{eqnarray} 
It is also absolutely convergent for $|\tilde{x}| + |\tilde{y}| =\left|-\frac{1}{a}x^2 \right| + \left| \frac{1+a}{a}x \right| <1$ which is equivalent to the boundary condition for $y_{+}(x)$. From this mathematical computations for $y_{-}(x)$, we again notice that Poincar\'{e}-Perron theorem does not give any absolute convergent series solution. Since we add (\ref{qq:17}) and (\ref{qq:28}),
\begin{equation}
y(x)= y_{+}(x)+ y_{-}(x) =\frac{ 1-\tilde{y} }{1 - \tilde{x}^2 - 2 \tilde{y} + \tilde{y}^2 } + \frac{  \tilde{x} }{1 - \tilde{x}^2 - 2 \tilde{y} + \tilde{y}^2 } = \frac{1}{1-(\tilde{x}+\tilde{y})} \label{qq:29}
\end{equation}
(\ref{qq:29}) is same as (\ref{eq:16}). Actually, this is obvious because as long as a solution for a power series is absolutely convergent, we can rearrange any terms for the series solution and its rearranged series is equivalent to the initial series solution. And the radius of convergence for $y(x)$, sum of two partial series $y_{+}(x)$ and $y_{-}(x)$, is also equal to $|\tilde{x}| + |\tilde{y}|<1$. According to the above computations for $y(x)$ which is the rearranged series, we realize that Poincar\'{e}-Perron theorem gives not absolute convergence for a solution of an infinite series but only conditional convergence. 
\qed
\end{pot}

\begin{figure}
\centering
\includegraphics[scale=.7]{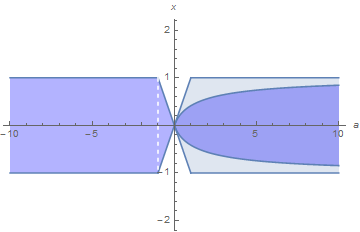} 
\caption{Original and revised Poincar\'{e}-Perron theorems}
\label{bc-img3}
\end{figure}

Fig.\ref{bc-img3} represents two different shaded regions of convergence in Figs.\ref{bc-img1} and \ref{bc-img2}: In the bright shaded area where $a>0$, the domain of absolute convergence of the series for a Heun equation around $x=0$ is not available; it only provides the domain of conditional convergence for it, and the dark shaded region where $a>0$ represents the one of its absolute convergence. 

%\newpage 

Now, you can ask why we can not apply Poincar\'{e}-Perron theorem into a power series in order to obtain its radius of convergence? To answer this question, first remember the following theorem such as
\begin{thm}Suppose we have the series $\sum_{n=0}^{\infty } c_n$ and define
\begin{equation}
L= \lim_{n\rightarrow \infty }\left| \frac{c_{n+1}}{c_n}\right| \label{qq:30}
\end{equation}
Then,

1. if $L<1$, the series is absolutely convergent.

2. if $L>1$, the series is divergent.

3. if $L=1$, the series may converge or diverge.

This test is called the Cauchy ratio test or d'Alembert ratio test.
\end{thm}
Fundamentally, Poincar\'{e}-Perron theorem is obtained by observing a ratio of $c_{n+1}$ to $c_n$ in the difference equation, letting $n\rightarrow \infty $. It demonstrates that its ratio is equivalent to one of roots of the characteristic polynomial of recurrence. And the radius of convergence for a power series $\sum_{n=0}^{\infty } c_n x^n$ is constructed by letting modulus of a root of the characteristic equation times $|x|$ is less than the unity. 

Let us apply (\ref{qq:30}) into a recurrence relation for the radius of convergence of a Heun function about $x=0$, rather than using Thm.\ref{thm.1} directly, in order to review whether Table~\ref{cb.1} is correct or not.
A solution for the recurrence equation for $c_n$ in (\ref{eq:100}) is  
\begin{equation}
c_n =\frac{-\left( A-\sqrt{A^2+4B}\right)^{n+1} + \left( A+ \sqrt{A^2+4B}\right)^{n+1}}{2^{n+1}\sqrt{A^2 + 4B}}
\label{qq:31}
\end{equation}
where $c_1=A$ and $c_0=1$. Substitute (\ref{qq:31}) into (\ref{qq:30}) and multiply the new (\ref{qq:30}) by $|x|$ in order to obtain its radius of convergence.
\begin{equation}
L = \lim_{n\rightarrow \infty } \left| \frac{\frac{A-\sqrt{A^2+4B}}{2}- \frac{A+\sqrt{A^2+4B}}{2}\left(\frac{A+\sqrt{A^2+4B}}{A-\sqrt{A^2+4B}}\right)^{n+1}}{1-\left(\frac{A+\sqrt{A^2+4B}}{A-\sqrt{A^2+4B}}\right)^{n+1}}\right| \left| x\right|<1
\label{qq:32}
\end{equation}
There are two possible solutions of (\ref{qq:32}). 
\begin{equation}
\mbox{If}\; \left| \frac{A+\sqrt{A^2+4B}}{A-\sqrt{A^2+4B}} \right| <1, \;\mbox{then}\;\;L =  \frac{1}{2}\left| A-\sqrt{A^2+4B} \right| \left| x\right| <1
\label{qq:33}
\end{equation} 
Or
\begin{equation}
\mbox{If}\; \left| \frac{A+\sqrt{A^2+4B}}{A-\sqrt{A^2+4B}} \right| >1, \;\mbox{then}\;\;L =  \frac{1}{2}\left| A+\sqrt{A^2+4B} \right| \left| x\right| <1
\label{qq:34}
\end{equation} 
For the special case, if $A=2$ and $B=-1$, (\ref{qq:32}) turns to be
\begin{equation}
L = \left| x\right|<1
\label{qq:35}
\end{equation} 
By putting (\ref{eq:5}) in (\ref{qq:33})--(\ref{qq:35}), we confirm that final solutions of radius of convergence for a Heun function around $x=0$  correspond to Table~\ref{cb.1} accurately.
  I wish that there are some computational errors in Table~\ref{cb.1} in order to agree its solutions to the boundary of the disk of convergence in Table~\ref{cb.2}. But it does not. There are no errors in Table~\ref{cb.1}! What's going on here? Why does boundary conditions for an infinite series of Heun function obtained by applying Poincar\'{e}-Perron theorem disagree with Table~\ref{cb.2}?  
  
We know that the hypergeometric function is defined by the power series 
\begin{equation}
 \sum_{n=0}^{\infty }c_n x^n = 1+ \frac{a b}{c \;1!} x +\frac{a(a+1)b(b+1)}{c(c+1) \;2!} x^2 +\frac{a(a+1)(a+2)b(b+1)(b+2)}{c(c+1)(c+2) \;3!}x^3 +\cdots 
\label{qq:36}
\end{equation} 
And its series consists of 2-term recurrence relation between successive coefficients.
We can decide what condition makes (\ref{qq:36}) as absolute convergent by taking the series of moduli $\sum_{n=0}^{\infty }|c_n| |x|^n$ such as 
\begin{eqnarray}
 \sum_{n=0}^{\infty }|c_n| |x|^n &=& 1+ \left|\frac{a b}{c \;1!}\right| |x| +\left|\frac{a(a+1)b(b+1)}{c(c+1) \;2!}\right| |x|^2 \nonumber\\
&&+\left|\frac{a(a+1)(a+2)b(b+1)(b+2)}{c(c+1)(c+2) \;3!}\right| |x|^3 +\cdots 
\label{qq:37}
\end{eqnarray}
By applying (\ref{qq:30}) into (\ref{qq:37}) to obtain its radius of convergence
\begin{equation}
 \lim_{n\rightarrow \infty }\left|\frac{(n+a)(n+b)}{(n+c)(n+1)}\right| |x| <1 \label{qq:38}
\end{equation}
We know $|x| <1$ for the absolute convergence of its series. 

An asymptotic series expansion of (\ref{eq:100}) is given by
\begin{eqnarray}
 \sum_{n=0}^{\infty }c_n x^n &=& 1+ A x +\left( A^2+ B \right) x^2 +\left( A^3 + 2AB \right) x^3 +\left( A^4 + 3 A^2B + B^2 \right) x^4 \nonumber\\
&&+\left( A^5 + 4A^3B + 3AB^2 \right) x^5 + \cdots 
\label{qq:39}
\end{eqnarray} 
with $c_0=1$ for simplicity. In general, the series of moduli $\sum_{n=0}^{\infty }|c_n| |x|^n$ has been taken to determine that (\ref{qq:39}) is whether absolute convergent or not in the following way. 
\begin{eqnarray}
 \sum_{n=0}^{\infty }|c_n| |x|^n &=& 1+ |A| |x| +\left| A^2+ B \right| |x|^2 +\left| A^3 + 2AB \right| |x|^3 +\left|  A^4 + 3 A^2B + B^2 \right| |x|^4 \nonumber\\
&&+\left| A^5 + 4A^3B + 3AB^2 \right| |x|^5 + \cdots 
\label{qq:40}
\end{eqnarray} 
And Poincar\'{e}-Perron theorem is applied using this basic principle idea into a ratio of $c_{n+1}$ to $c_n$ in the recurrence relation of a Heun function around $x=0$. Actually, this is wrong approach. We should take all absolute values inside parentheses 
of (\ref{qq:39}). Otherwise, we can not obtain any radius of convergence for a Heun function.
\begin{eqnarray}
 \sum_{n=0}^{\infty }|c_n| |x|^n &=& 1+ \big| A\big| |x| +\left( \left| A^2\right| + \big| B \big| \right) |x|^2 +\left( \left| A^3 \right| + \big| 2AB \big| \right) |x|^3 +\left( \left|  A^4\right| + \left| 3 A^2B\right| + \left| B^2 \right| \right) |x|^4 \nonumber\\
&&+\left( \left| A^5 \right| + \left| 4A^3B \right| + \left| 3AB^2 \right| \right) |x|^5 + \cdots 
\label{qq:40}
\end{eqnarray} 
As we compare (\ref{qq:37}) with (\ref{qq:40}), we notice that a 2-term recurrence relation of a hypergeometric function starts to appear and a 3-term recursion relation of a Heun function arises. Each sequence $c_n$ of a hypergeometric function has only one term, but the number of a sum of all coefficients in each sequence $c_n$ in (\ref{eq:6}) for a Heun function follows Fibonacci sequence. We can not take absolute values of whole coefficients in each sequence of a 3-term recursive relation, but we must take absolute values of individual terms, composed of several coefficients, in each sequence. 
Thus we have to take absolute values of $A$ and $B$ in  (\ref{qq:32}) to obtain the radius of convergence.
\begin{equation}
L = \lim_{n\rightarrow \infty } \left| \frac{\frac{|A|-\sqrt{|A|^2+4|B|}}{2}- \frac{|A|+\sqrt{|A|^2+4|B|}}{2}\left(\frac{|A|+\sqrt{|A|^2+4|B|}}{|A|-\sqrt{|A|^2+4|B|}}\right)^{n+1}}{1-\left(\frac{|A|+\sqrt{|A|^2+4|B|}}{|A|-\sqrt{|A|^2+4|B|}}\right)^{n+1}}\right| \left| x\right|<1
\label{qq:41}
\end{equation}
Two possible solutions of (\ref{qq:41}) are given by 
\begin{equation}
\mbox{If}\; \left| \frac{|A|+\sqrt{|A|^2+4|B|}}{|A|-\sqrt{|A|^2+4|B|}} \right| <1, \;\mbox{then}\;\;L =  \frac{1}{2}\left| |A|-\sqrt{|A|^2+4|B|} \right| \left| x\right| <1
\label{qq:42}
\end{equation} 
Or
\begin{equation}
\mbox{If}\; \left| \frac{|A|+\sqrt{|A|^2+4|B|}}{|A|-\sqrt{|A|^2+4|B|}} \right| >1, \;\mbox{then}\;\;L =  \frac{1}{2}\left| |A|+\sqrt{|A|^2+4|B|} \right| \left| x\right| <1
\label{qq:43}
\end{equation}  
Put (\ref{eq:5}) in (\ref{qq:42}) and (\ref{qq:43}), and final solutions of radius of convergence for a Heun function around $x=0$ correspond to Table~\ref{cb.2} except the case of $a=-1$. This is a reason why we obtain errors of the radius of convergence since we apply Poincar\'{e}-Perron theorem directly. Its theorem only verify that a series solution of a Heun function is conditionally convergent. In order to use its theorem, we must take all absolute values of constant coefficients $\alpha _i$ in (\ref{qq:2}) such as 
\begin{equation}
t^k + |\alpha _1| t^{k-1} + |\alpha _2| t^{k-2}+ \cdots + |\alpha _k| =0\label{qq:45}
\end{equation}
The roots of the characteristic equation (\ref{qq:45}) is written by $\lambda _1^{\star }, . . . ,\lambda _k^{\star } $.
And an absolute value of $\frac{u_{i}(n+1)}{u_{i}(n)}$ with limiting $n\rightarrow \infty$ in Thm.\ref{thm.1} is equal to the absolute value of $\lambda _i^{\star }$. With this revision, we also obtain correct radius of convergence for a Heun function which is equivalent to Table~\ref{cb.2}.

However, as we mention the above, we can not obtain the radius of convergence for its series solution in the case of $a=-1$ by applying Poincar\'{e}-Perron theorem. Because as $a=-1$ in $|A|$ and $|B|$, (\ref{qq:41}) turns to be
\begin{equation}
 \lim_{n\rightarrow \infty } \left| \frac{\frac{|A|-\sqrt{|A|^2+4|B|}}{2}- \frac{|A|+\sqrt{|A|^2+4|B|}}{2}\left(\frac{|A|+\sqrt{|A|^2+4|B|}}{|A|-\sqrt{|A|^2+4|B|}}\right)^{n+1}}{1-\left(\frac{|A|+\sqrt{|A|^2+4|B|}}{|A|-\sqrt{|A|^2+4|B|}}\right)^{n+1}}\right| =  \lim_{n\rightarrow \infty } \left| \frac{-1+(-1)^n}{1+(-1)^n}\right| 
\nonumber
\end{equation}
This case is undefined to determine whether the series converge or diverge. Instead, putting $a=-1$ in (\ref{eq:17}) directly, we obtain the interval of convergence for a Heun function around $x=0$.
As we see, even if we use the revised Poincar\'{e}-Perron theorem, we can not decide the series solution for $a=-1$ is the absolute convergent or not accurately. There are no ways to construct asymptotic series solutions in closed forms using its theorem perfectly. Also, it is really hard to obtain the roots of the characteristic polynomial for more than 4 term without using computer simulations. 
Because of these reasons, we develop the new theorem to obtain the radius of convergence and asymptotic series solutions of the multi-term recursive relation in a power series in chapter 3 of Ref.\cite{Choun2014}.
By changing a coefficient $a$ and an variable $x$ in Table~\ref{cb.2}, we can also obtain accurate numerical values of all 192 local solutions of the Heun equation \cite{Maie2007} using machine calculations. 

\section{Integral Formalism}
\subsection{Polynomial which makes $B_n$ term terminated}
There are three types of polynomials in three term recurrence relation of a linear ordinary differential equation: (1) polynomial which makes $B_n$ term terminated: $A_n$ term is not terminated, (2) polynomial which makes $A_n$ term terminated: $B_n$ term is not terminated, (3) polynomial which makes $A_n$ and $B_n$ terms terminated at the same time.\footnote{If $A_n$ and $B_n$ terms are not terminated, it turns to be infinite series.} In general Heun polynomial is defined as type 3 polynomial where $A_n$ and $B_n$ terms terminated. Heun polynomial comes from Heun equation that has a fixed integer value of $\alpha $ or $\beta $, just as it has a fixed value of $q$. In three term recurrence relation, polynomial of type 3 I categorize as complete polynomial. In future papers I will derive type 3 Heun polynomial. 
In Ref.\cite{Choun2013} I construct the power series expansion and an integral form for Heun polynomial of type 2: I treat $\alpha $, $\beta $, $\gamma $ and $\delta $ as free variables and the accessory parameter $q$ as a fixed value.
 In this paper I treat $\alpha $ or/and $\beta $ as a fixed value and $\gamma , \delta $, $q$ as free variables to construct Heun polynomial of type 1 about the singular point at zero. 
\subsubsection{ The case of $\alpha = -2 \alpha _i-i -\lambda $ and $\beta \ne -2 \beta _i -i-\lambda $ where $i, \alpha _i, \beta _i$ = $0,1,2,\cdots$}

Now let's investigate the integral formalism for the polynomial case of $B_n$ term terminated at certain eigenvalue. There is a generalized hypergeometric function which is:  In this paper Pochhammer symbol $(x)_n$ is used to represent the rising factorial: $(x)_n = \frac{\Gamma (x+n)}{\Gamma (x)}$.
\begin{eqnarray}
I_l &=& \sum_{i_l= i_{l-1}}^{\alpha _l} \frac{(-\alpha _l)_{i_l}(\frac{\beta }{2}+\frac{l}{2}+\frac{\lambda }{2})_{i_l}(1+\frac{l}{2}+\frac{\lambda }{2})_{i_{l-1}}(\frac{1}{2}+\frac{\gamma}{2}+\frac{l}{2} +\frac{\lambda }{2})_{i_{l-1}}}{ (-\alpha _l)_{i_{l-1}}(\frac{\beta }{2}+\frac{l}{2}+\frac{\lambda }{2})_{i_{l-1}}(1+\frac{l}{2}+\frac{\lambda }{2})_{i_l}(\frac{1}{2}+\frac{\gamma}{2}+\frac{l}{2} +\frac{\lambda }{2})_{i_l}} z^{i_l}\label{eq:30}\\
&=& z^{i_{l-1}} 
\sum_{j=0}^{\infty } \frac{B(i_{l-1}+\frac{l}{2}+\frac{\lambda }{2},j+1) B(i_{l-1}+\frac{l}{2}-\frac{1}{2}+\frac{\gamma}{2} +\frac{\lambda }{2},j+1)(i_{l-1}-\alpha _l)_j (i_{l-1}+\frac{l}{2}+\frac{\beta }{2}+\frac{\lambda }{2})_j}{(i_{l-1}+\frac{l}{2}+\frac{\lambda }{2})^{-1}(i_{l-1}+\frac{l}{2}-\frac{1}{2}+\frac{\gamma}{2} + \frac{\lambda }{2})^{-1}(1)_j \;j!} z^j\nonumber
\end{eqnarray}
By using integral form of beta function,
\begin{subequations}
\begin{equation}
B\left(i_{l-1}+\frac{l}{2}+\frac{\lambda }{2},j+1\right)= \int_{0}^{1} dt_l\;t_l^{i_{l-1}+\frac{l}{2}-1+\frac{\lambda }{2}} (1-t_l)^j \label{eq:31a}
\end{equation}
\begin{equation}
B\left(i_{l-1}+\frac{l}{2}-\frac{1}{2}+\frac{\gamma}{2} +\frac{\lambda }{2},j+1\right)= \int_{0}^{1} du_l\;u_l^{i_{l-1}+\frac{l}{2}-\frac{3}{2}+\frac{\gamma }{2}+\frac{\lambda }{2}} (1-u_l)^j\label{eq:31b}
\end{equation}
\end{subequations}
Substitute (\ref{eq:31a}) and (\ref{eq:31b}) into (\ref{eq:30}). And divide $(i_{l-1}+\frac{l}{2}+\frac{\lambda }{2})(i_{l-1}+\frac{l}{2}-\frac{1}{2}+\frac{\gamma}{2} + \frac{\lambda }{2})$ into $I_l$.
\begin{eqnarray}
K_l &=& \frac{1}{(i_{l-1}+\frac{l}{2}+\frac{\lambda }{2})(i_{l-1}+\frac{l}{2}-\frac{1}{2}+\frac{\gamma}{2} + \frac{\lambda }{2})}
\sum_{i_l= i_{l-1}}^{\alpha _l} \frac{(-\alpha _l)_{i_l}(\frac{\beta }{2}+\frac{l}{2}+\frac{\lambda }{2})_{i_l}(1+\frac{l}{2}+\frac{\lambda }{2})_{i_{l-1}}(\frac{1}{2}+\frac{\gamma}{2}+\frac{l}{2} +\frac{\lambda }{2})_{i_{l-1}}}{ (-\alpha _l)_{i_{l-1}}(\frac{\beta }{2}+\frac{l}{2}+\frac{\lambda }{2})_{i_{l-1}}(1+\frac{l}{2}+\frac{\lambda }{2})_{i_l}(\frac{1}{2}+\frac{\gamma}{2}+\frac{l}{2} +\frac{\lambda }{2})_{i_l}} z^{i_l}\nonumber\\
&=&  \int_{0}^{1} dt_l\;t_l^{\frac{l}{2}-1+\frac{\lambda }{2}} \int_{0}^{1} du_l\;u_l^{\frac{l}{2}-\frac{3}{2}+\frac{\gamma }{2}+\frac{\lambda }{2}} (z t_l u_l)^{i_{l-1}}\nonumber\\
&&\times \sum_{j=0}^{\infty } \frac{(i_{l-1}-\alpha _l )_j (i_{l-1}+\frac{l}{2}+\frac{\beta }{2}+\frac{\lambda }{2})_j}{(1)_j \;j!} [z(1-t_l)(1-u_l)]^j \label{eq:32}
\end{eqnarray}
The integral form of hypergeometric function is defined by
\begin{eqnarray}
_2F_1 \left( \alpha ,\beta ; \gamma ; z \right) &=& \sum_{n=0}^{\infty } \frac{(\alpha )_n (\beta )_n}{(\gamma )_n (n!)} z^n \nonumber\\
&=& -\frac{1}{2\pi i} \frac{\Gamma(1-\alpha ) \Gamma(\gamma )}{\Gamma (\gamma -\alpha )} \oint dv_l\;(-v_l)^{\alpha -1} (1-v_l)^{\gamma -\alpha -1} (1-zv_l)^{-\beta }\hspace{.2cm}\label{eq:33}\\
&& \mbox{where} \;\mbox{Re}(\gamma -\alpha )>0 \nonumber
\end{eqnarray}
replaced $\alpha $, $\beta $, $\gamma $ and $z$ by $i_{l-1}-\alpha _l$, $i_{l-1}+\frac{l}{2}+\frac{\beta }{2}+\frac{\lambda }{2}$, 1 and $z(1-t_l)(1-u_l)$ in (\ref{eq:33})
\begin{eqnarray}
&& \sum_{j=0}^{\infty } \frac{(i_{l-1}-\alpha _l)_j (i_{l-1}+\frac{l}{2}+\frac{\beta }{2}+\frac{\lambda }{2})_j}{(1)_j \;j!} [z(1-t_l)(1-u_l)]^j \nonumber\\
&=& \frac{1}{2\pi i} \oint dv_l\;\frac{1}{v_l} \left(1- \frac{1}{v_l}\right)^{\alpha_l} (1-z v_l (1-t_l)(1-u_l))^{-\frac{1}{2}(\beta +l+\lambda )}\nonumber\\
&&\times \left(\frac{v_l}{(v_l-1)} \frac{1}{1-z v_l (1-t_l)(1-u_l)}\right)^{i_{l-1}} \label{eq:34}
\end{eqnarray}
Substitute (\ref{eq:34}) into (\ref{eq:32}).
\begin{eqnarray}
K_l &=& \frac{1}{(i_{l-1}+\frac{l}{2}+\frac{\lambda }{2})(i_{l-1}+\frac{l}{2}-\frac{1}{2}+\frac{\gamma}{2} + \frac{\lambda }{2})}
 \sum_{i_l= i_{l-1}}^{\alpha _l} \frac{(-\alpha _l)_{i_l}(\frac{\beta }{2}+\frac{l}{2}+\frac{\lambda }{2})_{i_l}(1+\frac{l}{2}+\frac{\lambda }{2})_{i_{l-1}}(\frac{1}{2}+\frac{\gamma}{2}+\frac{l}{2} +\frac{\lambda }{2})_{i_{l-1}}}{ (-\alpha _l)_{i_{l-1}}(\frac{\beta }{2}+\frac{l}{2}+\frac{\lambda }{2})_{i_{l-1}}(1+\frac{l}{2}+\frac{\lambda }{2})_{i_l}(\frac{1}{2}+\frac{\gamma}{2}+\frac{l}{2} +\frac{\lambda }{2})_{i_l}} z^{i_l}\nonumber\\
&=&  \int_{0}^{1} dt_l\;t_l^{\frac{l}{2}-1+\frac{\lambda }{2}} \int_{0}^{1} du_l\;u_l^{\frac{l}{2}-\frac{3}{2}+\frac{\gamma }{2}+\frac{\lambda }{2}} 
\frac{1}{2\pi i} \oint dv_l\;\frac{1}{v_l} \left(1- \frac{1}{v_l}\right)^{\alpha_l} 
 (1-z v_l (1-t_l)(1-u_l))^{-\frac{1}{2}(\beta +l+\lambda )} \nonumber\\
&&\times \left(\frac{v_l}{(v_l-1)} \frac{z t_l u_l}{1-z v_l (1-t_l)(1-u_l)}\right)^{i_{l-1}}\label{eq:35} 
\end{eqnarray}
In Ref.\cite{Chou2012c}, the general expression of power series of Heun function for polynomial which makes $B_n$ term terminated about $x=0$ where $\alpha = -2 \alpha _i-i -\lambda $ and $\beta \ne -2 \beta _i -i-\lambda $ is
\begin{eqnarray}
 y(x)&=& \sum_{n=0}^{\infty } y_n(x)= y_0(x)+y_1(x)+y_2(x)+y_3(x)\cdots \nonumber\\
&=& c_0 x^{\lambda } \left\{\sum_{i_0=0}^{\alpha _0} \frac{(-\alpha _0)_{i_0} (\frac{\beta }{2}+\frac{\lambda }{2})_{i_0}}{(1+\frac{\lambda }{2})_{i_0}(\frac{1}{2}+ \frac{\gamma}{2} +\frac{\lambda }{2})_{i_0}} z^{i_0}\right.\nonumber\\
&&+ \left\{\sum_{i_0=0}^{\alpha _0}\frac{(i_0+ \frac{\lambda }{2}) \left( i_0+ \Gamma_0^{(S)} \right)+ Q}{(i_0+ \frac{1}{2}+ \frac{\lambda }{2})(i_0 + \frac{\gamma }{2}+ \frac{\lambda }{2})} \frac{(-\alpha _0)_{i_0} (\frac{\beta }{2}+\frac{\lambda }{2})_{i_0}}{(1+\frac{\lambda }{2})_{i_0}(\frac{1}{2}+ \frac{\gamma}{2} +\frac{\lambda }{2})_{i_0}} \right.\nonumber\\
&&\times \left.\sum_{i_1=i_0}^{\alpha _1} \frac{(-\alpha _1)_{i_1}(\frac{1}{2}+\frac{\beta }{2}+ \frac{\lambda }{2})_{i_1}(\frac{3}{2}+\frac{\lambda }{2})_{i_0}(1+\frac{\gamma }{2}+ \frac{\lambda }{2})_{i_0}}{(-\alpha _1)_{i_0}(\frac{1}{2}+\frac{\beta }{2}+ \frac{\lambda }{2})_{i_0}(\frac{3}{2}+\frac{\lambda }{2})_{i_1}(1+ \frac{\gamma}{2} +\frac{\lambda }{2})_{i_1}} z^{i_1} \right\} \eta \nonumber\\
&&+ \sum_{n=2}^{\infty } \left\{ \sum_{i_0=0}^{\alpha _0} \frac{(i_0+ \frac{\lambda }{2}) \left( i_0+ \Gamma_0^{(S)} \right)+ Q}{(i_0+ \frac{1}{2}+ \frac{\lambda }{2})(i_0 + \frac{\gamma }{2}+ \frac{\lambda }{2})}  \frac{(-\alpha _0)_{i_0} (\frac{\beta }{2}+\frac{\lambda }{2})_{i_0}}{(1+\frac{\lambda }{2})_{i_0}(\frac{1}{2}+ \frac{\gamma}{2} +\frac{\lambda }{2})_{i_0}}\right.\nonumber\\
&&\times \prod _{k=1}^{n-1} \left\{ \sum_{i_k=i_{k-1}}^{\alpha _k} \frac{(i_k+\frac{k}{2}+ \frac{\lambda }{2}) \left( i_k+\Gamma_k^{(S)} \right)+ Q}{(i_k+ \frac{k}{2}+\frac{1}{2}+\frac{\lambda }{2})(i_k +\frac{k}{2}+\frac{\gamma }{2}+\frac{\lambda }{2})}  \frac{(-\alpha _k)_{i_k}(\frac{k}{2}+\frac{\beta }{2}+ \frac{\lambda }{2})_{i_k}(1+ \frac{k}{2}+\frac{\lambda }{2})_{i_{k-1}}(\frac{1}{2}+\frac{k}{2}+\frac{\gamma }{2}+ \frac{\lambda }{2})_{i_{k-1}}}{(-\alpha _k)_{i_{k-1}}(\frac{k}{2}+\frac{\beta }{2}+ \frac{\lambda }{2})_{i_{k-1}}(1+\frac{k}{2}+\frac{\lambda }{2})_{i_k}(\frac{1}{2}+ \frac{k}{2}+ \frac{\gamma}{2} +\frac{\lambda }{2})_{i_k}}\right\} \nonumber\\
&&\times \left.\left. \sum_{i_n= i_{n-1}}^{\alpha _n} \frac{(-\alpha _n)_{i_n}(\frac{n}{2}+\frac{\beta }{2}+ \frac{\lambda }{2})_{i_n}(1+ \frac{n}{2}+\frac{\lambda }{2})_{i_{n-1}}(\frac{1}{2}+\frac{n}{2}+\frac{\gamma }{2}+ \frac{\lambda }{2})_{i_{n-1}}}{(-\alpha _n)_{i_{n-1}}(\frac{n}{2}+\frac{\beta }{2}+ \frac{\lambda }{2})_{i_{n-1}}(1+\frac{n}{2}+\frac{\lambda }{2})_{i_n}(\frac{1}{2}+ \frac{n}{2}+ \frac{\gamma}{2} +\frac{\lambda }{2})_{i_n}} z^{i_n} \right\} \eta ^n \right\}\label{eq:36}
\end{eqnarray}
where
\begin{equation}
\begin{cases} z = -\frac{1}{a}x^2 \cr
\eta = \frac{(1+a)}{a} x \cr
\alpha _i\leq \alpha _j \;\;\mbox{only}\;\mbox{if}\;i\leq j\;\;\mbox{where}\;i,j= 0,1,2,\cdots
\end{cases}\nonumber %\label{eq:37}
\end{equation}
and
\begin{equation}
\begin{cases} 
\Gamma_0^{(S)} = \frac{1}{2(1+a)}(-2\alpha _0+ \beta -\delta +a(\delta +\gamma -1+\lambda )) \cr
\Gamma_k^{(S)} = \frac{1}{2(1+a)}(-2\alpha _k+ \beta -\delta +a(\delta +\gamma +\lambda +k-1)) \cr
Q= \frac{q}{4(1+a)}
\end{cases}\nonumber %\label{eq:8}
\end{equation}
Substitute (\ref{eq:35}) into (\ref{eq:36}) where $l=1,2,3,\cdots$; apply $K_1$ into the second summation of sub-power series $y_1(x)$, apply $K_2$ into the third summation and $K_1$ into the second summation of sub-power series $y_2(x)$, apply $K_3$ into the forth summation, $K_2$ into the third summation and $K_1$ into the second summation of sub-power series $y_3(x)$, etc.\footnote{$y_1(x)$ means the sub-power series in (\ref{eq:36}) contains one term of $A_n's$, $y_2(x)$ means the sub-power series in (\ref{eq:36}) contains two terms of $A_n's$, $y_3(x)$ means the sub-power series in (\ref{eq:36}) contains three terms of $A_n's$, etc.}
\begin{thm}
The general representation in the form of integral of Heun polynomial which makes $B_n$ term terminated about $x=0$ as $\alpha = -2 \alpha _i-i -\lambda $ and $\beta \ne -2 \beta _i -i-\lambda $ where $i,\alpha _i,\beta _i \in \mathbb{N}_{0}$ is given by
\begin{eqnarray}
 y(x)&=& \sum_{n=0}^{\infty } y_{n}(x)= y_0(x)+y_1(x)+y_2(x)+y_3(x)+\cdots\nonumber\\
&=& c_0 x^{\lambda } \left\{ \sum_{i_0=0}^{\alpha _0}\frac{(-\alpha _0)_{i_0}(\frac{\beta }{2}+\frac{\lambda }{2})_{i_0}}{(1+\frac{\lambda }{2})_{i_0}(\frac{1}{2}+ \frac{\gamma }{2} +\frac{\lambda }{2})_{i_0}}  z^{i_0}\right.
+ \sum_{n=1}^{\infty } \left\{\prod _{k=0}^{n-1} \Bigg\{ \int_{0}^{1} dt_{n-k}\;t_{n-k}^{\frac{1}{2}(n-k-2+\lambda )} \int_{0}^{1} du_{n-k}\;u_{n-k}^{\frac{1}{2}(n-k-3+\gamma +\lambda )}\right. \nonumber\\
&&\times  \frac{1}{2\pi i}  \oint dv_{n-k} \frac{1}{v_{n-k}} \left( 1-\frac{1}{v_{n-k}}\right)^{\alpha _{n-k}} \left( 1- \overleftrightarrow {w}_{n-k+1,n}v_{n-k}(1-t_{n-k})(1-u_{n-k})\right)^{-\frac{1}{2}(n-k+\beta +\lambda )}\nonumber\\
&&\times  \left( \overleftrightarrow {w}_{n-k,n}^{-\frac{1}{2}(n-k-1+\lambda )}\left(  \overleftrightarrow {w}_{n-k,n} \partial _{ \overleftrightarrow {w}_{n-k,n}}\right) \overleftrightarrow {w}_{n-k,n}^{\frac{1}{2}(n-k-1+\lambda )} \left( \overleftrightarrow {w}_{n-k,n} \partial _{ \overleftrightarrow {w}_{n-k,n}} + \Omega _{n-k-1}^{(S)}\right)+ Q \right) \Bigg\}\nonumber\\
&&\times\left.\left.\sum_{i_0=0}^{\alpha _0}\frac{(-\alpha _0)_{i_0}(\frac{\beta }{2}+\frac{\lambda }{2})_{i_0}}{(1+\frac{\lambda }{2})_{i_0}(\frac{1}{2}+ \frac{\gamma }{2} +\frac{\lambda }{2})_{i_0}}  \overleftrightarrow {w}_{1,n}^{i_0}\right\} \eta ^n \right\} \label{eq:39}
\end{eqnarray}
where
\begin{equation}\overleftrightarrow {w}_{i,j}=
\begin{cases} \displaystyle {\frac{v_i}{(v_i-1)}\; \frac{\overleftrightarrow w_{i+1,j} t_i u_i}{1- \overleftrightarrow w_{i+1,j} v_i (1-t_i)(1-u_i)}}\;\;\mbox{where}\; i\leq j\cr
z \;\;\mbox{only}\;\mbox{if}\; i>j
\end{cases}\nonumber\\ %\label{eq:40}
\end{equation}
and
\begin{equation}
\begin{cases} 
\Omega _{n-k-1}^{(S)} = \frac{1}{2(1+a)}(-2\alpha _{n-k-1}+\beta -\delta +a(\delta +\gamma +n-k-2+\lambda )) \cr
Q= \frac{q}{4(1+a)}
\end{cases}\nonumber %\label{eq:8}
\end{equation}
On the above, the first sub-integral form contains one term of $A_n's$, the second one contains two terms of $A_n$'s, the third one contains three terms of $A_n$'s, etc.
\end{thm}
\begin{pot}
In (\ref{eq:36}) sub-power series $y_0(x) $, $y_1(x)$, $y_2(x)$ and $y_3(x)$ of Heun polynomial which makes $B_n$ term terminated about $x=0$ as $\alpha = -2 \alpha _i-i -\lambda $ and $\beta \ne -2 \beta _i -i-\lambda $ where $i,\alpha _i,\beta _i \in \mathbb{N}_{0}$ are given by
\begin{subequations}
\begin{equation}
 y_0(x)=  c_0 x^{\lambda } \sum_{i_0=0}^{\alpha _0} \frac{(-\alpha _0)_{i_0} (\frac{\beta }{2}+\frac{\lambda }{2})_{i_0}}{(1+\frac{\lambda }{2})_{i_0}(\frac{1}{2}+ \frac{\gamma}{2} +\frac{\lambda }{2})_{i_0}} z^{i_0}\label{eq:300a}
\end{equation}
\begin{eqnarray}
 y_1(x)&=& c_0 x^{\lambda } \left\{\sum_{i_0=0}^{\alpha _0}\frac{(i_0+ \frac{\lambda }{2}) \left( i_0+ \frac{1}{2(1+a)}(-2\alpha _0+ \beta -\delta +a(\delta +\gamma -1+\lambda )) \right)+ \frac{q}{4(1+a)}}{(i_0+ \frac{1}{2}+ \frac{\lambda }{2})(i_0 + \frac{\gamma }{2}+ \frac{\lambda }{2})} \frac{(-\alpha _0)_{i_0} (\frac{\beta }{2}+\frac{\lambda }{2})_{i_0}}{(1+\frac{\lambda }{2})_{i_0}(\frac{1}{2}+ \frac{\gamma}{2} +\frac{\lambda }{2})_{i_0}} \right.\nonumber\\
&\times& \left.\sum_{i_1=i_0}^{\alpha _1} \frac{(-\alpha _1)_{i_1}(\frac{1}{2}+\frac{\beta }{2}+ \frac{\lambda }{2})_{i_1}(\frac{3}{2}+\frac{\lambda }{2})_{i_0}(1+\frac{\gamma }{2}+ \frac{\lambda }{2})_{i_0}}{(-\alpha _1)_{i_0}(\frac{1}{2}+\frac{\beta }{2}+ \frac{\lambda }{2})_{i_0}(\frac{3}{2}+\frac{\lambda }{2})_{i_1}(1+ \frac{\gamma}{2} +\frac{\lambda }{2})_{i_1}} z^{i_1} \right\} \eta \label{eq:300b}
\end{eqnarray}
\begin{eqnarray}
 y_2(x) &=& c_0 x^{\lambda } \left\{\sum_{i_0=0}^{\alpha _0}\frac{(i_0+ \frac{\lambda }{2}) \left( i_0+ \frac{1}{2(1+a)}(-2\alpha _0+ \beta -\delta +a(\delta +\gamma -1+\lambda )) \right)+ \frac{q}{4(1+a)}}{(i_0+ \frac{1}{2}+ \frac{\lambda }{2})(i_0 + \frac{\gamma }{2}+ \frac{\lambda }{2})} \frac{(-\alpha _0)_{i_0} (\frac{\beta }{2}+\frac{\lambda }{2})_{i_0}}{(1+\frac{\lambda }{2})_{i_0}(\frac{1}{2}+ \frac{\gamma}{2} +\frac{\lambda }{2})_{i_0}} \right.\nonumber\\
&\times& \sum_{i_1=i_0}^{\alpha _1}  \frac{(i_1+ \frac{1}{2}+ \frac{\lambda }{2}) \left( i_1+ \frac{1}{2(1+a)}(-2\alpha _1+ \beta -\delta +a(\delta +\gamma +\lambda )) \right)+ \frac{q}{4(1+a)}}{(i_1+ 1+ \frac{\lambda }{2})(i_1 +\frac{1}{2}+ \frac{\gamma }{2}+ \frac{\lambda }{2})}  \nonumber\\
&\times& \left. \frac{(-\alpha _1)_{i_1}(\frac{1}{2}+\frac{\beta }{2}+ \frac{\lambda }{2})_{i_1}(\frac{3}{2}+\frac{\lambda }{2})_{i_0}(1+\frac{\gamma }{2}+ \frac{\lambda }{2})_{i_0}}{(-\alpha _1)_{i_0}(\frac{1}{2}+\frac{\beta }{2}+ \frac{\lambda }{2})_{i_0}(\frac{3}{2}+\frac{\lambda }{2})_{i_1}(1+ \frac{\gamma}{2} +\frac{\lambda }{2})_{i_1}} \sum_{i_2= i_1}^{\alpha _2} \frac{(-\alpha _2)_{i_2}(1+\frac{\beta }{2}+ \frac{\lambda }{2})_{i_2}(2+\frac{\lambda }{2})_{i_1}(\frac{3}{2} +\frac{\gamma }{2}+ \frac{\lambda }{2})_{i_1}}{(-\alpha _2)_{i_1}(1+\frac{\beta }{2}+ \frac{\lambda }{2})_{i_1}(2+\frac{\lambda }{2})_{i_2}(\frac{3}{2} + \frac{\gamma}{2} +\frac{\lambda }{2})_{i_2}} z^{i_2} \right\} \eta ^2 \hspace{1cm}\label{eq:300c}
\end{eqnarray}
\begin{eqnarray}
 y_3(x)&=& c_0 x^{\lambda } \left\{\sum_{i_0=0}^{\alpha _0}\frac{(i_0+ \frac{\lambda }{2}) \left( i_0+ \frac{1}{2(1+a)}(-2\alpha _0+ \beta -\delta +a(\delta +\gamma -1+\lambda )) \right)+ \frac{q}{4(1+a)}}{(i_0+ \frac{1}{2}+ \frac{\lambda }{2})(i_0 + \frac{\gamma }{2}+ \frac{\lambda }{2})} \frac{(-\alpha _0)_{i_0} (\frac{\beta }{2}+\frac{\lambda }{2})_{i_0}}{(1+\frac{\lambda }{2})_{i_0}(\frac{1}{2}+ \frac{\gamma}{2} +\frac{\lambda }{2})_{i_0}} \right.\nonumber\\
&\times& \sum_{i_1=i_0}^{\alpha _1}  \frac{(i_1+ \frac{1}{2}+ \frac{\lambda }{2}) \left( i_1+ \frac{1}{2(1+a)}(-2\alpha _1+ \beta -\delta +a(\delta +\gamma +\lambda )) \right)+ \frac{q}{4(1+a)}}{(i_1+ 1+ \frac{\lambda }{2})(i_1 +\frac{1}{2}+ \frac{\gamma }{2}+ \frac{\lambda }{2})}\nonumber\\
&\times& \frac{(-\alpha _1)_{i_1}(\frac{1}{2}+\frac{\beta }{2}+ \frac{\lambda }{2})_{i_1}(\frac{3}{2}+\frac{\lambda }{2})_{i_0}(1+\frac{\gamma }{2}+ \frac{\lambda }{2})_{i_0}}{(-\alpha _1)_{i_0}(\frac{1}{2}+\frac{\beta }{2}+ \frac{\lambda }{2})_{i_0}(\frac{3}{2}+\frac{\lambda }{2})_{i_1}(1+ \frac{\gamma}{2} +\frac{\lambda }{2})_{i_1}} \nonumber\\
&\times& \sum_{i_2=i_1}^{\alpha _2}  \frac{(i_2+ 1+ \frac{\lambda }{2}) \left( i_2+ \frac{1}{2(1+a)}(-2\alpha _2+ \beta -\delta +a(\delta +\gamma +1+\lambda )) \right)+ \frac{q}{4(1+a)}}{(i_2+ \frac{3}{2}+ \frac{\lambda }{2})(i_2 +1+ \frac{\gamma }{2}+ \frac{\lambda }{2})} 
 \nonumber\\
&\times& \left. \frac{(-\alpha _2)_{i_2}(1+\frac{\beta }{2}+ \frac{\lambda }{2})_{i_2}(2+\frac{\lambda }{2})_{i_1}(\frac{3}{2} +\frac{\gamma }{2}+ \frac{\lambda }{2})_{i_1}}{(-\alpha _2)_{i_1}(1+\frac{\beta }{2}+ \frac{\lambda }{2})_{i_1}(2+\frac{\lambda }{2})_{i_2}(\frac{3}{2} + \frac{\gamma}{2} +\frac{\lambda }{2})_{i_2}} \sum_{i_3= i_2}^{\alpha _3} \frac{(-\alpha _3)_{i_3}(\frac{3}{2}+\frac{\beta }{2}+ \frac{\lambda }{2})_{i_3}(\frac{5}{2}+\frac{\lambda }{2})_{i_2}(2 +\frac{\gamma }{2}+ \frac{\lambda }{2})_{i_2}}{(-\alpha _3)_{i_2}(\frac{3}{2}+\frac{\beta }{2}+ \frac{\lambda }{2})_{i_2}(\frac{5}{2}+\frac{\lambda }{2})_{i_3}(2 + \frac{\gamma}{2} +\frac{\lambda }{2})_{i_3}} z^{i_3} \right\} \eta ^3 \hspace{1cm} \label{eq:300d}
\end{eqnarray}
\end{subequations}
Put $l=1$ in (\ref{eq:35}). Take the new (\ref{eq:35}) into (\ref{eq:300b}).
\begin{eqnarray}
 y_1(x)&=& c_0 x^{\lambda }\int_{0}^{1} dt_1\;t_1^{ \frac{1}{2}(-1+\lambda )} \int_{0}^{1} du_1\;u_1^{ \frac{1}{2}(-2+\gamma +\lambda )}   \frac{1}{2\pi i} \oint dv_1\;\frac{1}{v_1} \left( 1-\frac{1}{v_1} \right)^{\alpha _1} \left( 1- z v_1 (1-t_1)(1-u_1)\right)^{-\frac{1}{2}(\beta +1+\lambda  )}  \nonumber\\
&&\times  \left\{ \sum_{i_0=0}^{\alpha _0}\left( \left( i_0+\frac{\lambda }{2} \right)\left( i_0 +\frac{1}{2(1+a)}\left( -2\alpha _0+\beta -\delta +a(\delta +\gamma -1+\lambda )\right) \right) +\frac{q}{4(1+a)}  \right) \right.\nonumber\\
&&\times \left. \frac{(-\alpha _0)_{i_0} (\frac{\beta }{2}+\frac{\lambda }{2} )_{i_0}}{(1+\frac{\lambda }{2})_{i_0}(\frac{1}{2}+ \frac{\gamma }{2}+\frac{\lambda }{2})_{i_0}} \left( \frac{t_1 u_1 v_1}{(v_1-1)} \frac{z }{1-z v_1 (1-t_1)(1-u_1)}\right)^{i_0} \right\} \eta  \nonumber\\
&=&  c_0 x^{\lambda }\int_{0}^{1} dt_1\;t_1^{ \frac{1}{2}(-1+\lambda )} \int_{0}^{1} du_1\;u_1^{ \frac{1}{2}(-2+\gamma +\lambda )}   \frac{1}{2\pi i} \oint dv_1\;\frac{1}{v_1} \left( 1-\frac{1}{v_1} \right)^{\alpha _1} \left( 1- z v_1 (1-t_1)(1-u_1)\right)^{-\frac{1}{2}(\beta +1+\lambda  )}  \nonumber\\
&&\times  \left(    \overleftrightarrow {w}_{1,1}^{-\frac{\lambda }{2}} \left(  \overleftrightarrow {w}_{1,1} \partial _{ \overleftrightarrow {w}_{1,1}}\right) \overleftrightarrow {w}_{1,1}^{\frac{\lambda }{2}} \left(\overleftrightarrow {w}_{1,1} \partial _{ \overleftrightarrow {w}_{1,1}}+\frac{1}{2(1+a)}\left(  -2\alpha _0+\beta -\delta +a(\delta +\gamma -1+\lambda )\right) \right)+\frac{q}{4(1+a)} \right)\nonumber\\
&&\times \left\{ \sum_{i_0=0}^{\alpha _0} \frac{(-\alpha _0)_{i_0} (\frac{\beta }{2}+\frac{\lambda }{2} )_{i_0}}{(1+\frac{\lambda }{2})_{i_0}(\frac{1}{2}+ \frac{\gamma }{2}+\frac{\lambda }{2})_{i_0}} \overleftrightarrow {w}_{1,1} ^{i_0} \right\} \eta \label{eq:301}
\end{eqnarray}
where 
\begin{equation}
 \overleftrightarrow {w}_{1,1} = \frac{t_1 u_1 v_1}{(v_1-1)}\; \frac{z}{1-z v_1 (1-t_1)(1-u_1)} \nonumber
\end{equation}
Put $l=2$ in (\ref{eq:35}). Take the new (\ref{eq:35}) into (\ref{eq:300c}). 
\begin{eqnarray}
y_2(x) &=& c_0 x^{\lambda } \int_{0}^{1} dt_2\;t_2^{ \frac{\lambda }{2}} \int_{0}^{1} du_2\;u_2^{ \frac{1}{2}(-1+\gamma +\lambda )}   \frac{1}{2\pi i} \oint dv_2\;\frac{1}{v_2} \left( 1-\frac{1}{v_2} \right)^{\alpha _2} \left( 1- z v_2 (1-t_2)(1-u_2)\right)^{-\frac{1}{2}(\beta +2+\lambda  )}  \nonumber\\
&&\times  \left(    \overleftrightarrow {w}_{2,2}^{-\frac{1}{2}(1+\lambda )} \left(  \overleftrightarrow {w}_{2,2} \partial _{ \overleftrightarrow {w}_{2,2}}\right) \overleftrightarrow {w}_{2,2}^{\frac{1}{2}(1+\lambda )} \left(\overleftrightarrow {w}_{2,2} \partial _{ \overleftrightarrow {w}_{2,2}}+\frac{1}{2(1+a)}\left(  -2\alpha _1+\beta -\delta +a(\delta +\gamma +\lambda )\right) \right)+\frac{q}{4(1+a)} \right)\nonumber\\
&&\times \left\{ \sum_{i_0=0}^{\alpha _0}\frac{ (i_0+\frac{\lambda }{2})\left( i_0+\frac{1}{2(1+a)}\left( -2\alpha _0+\beta -\delta +a(\delta +\gamma -1+\lambda )\right) \right) +\frac{q}{4(1+a)} }{(i_0+ \frac{1}{2}+ \frac{\lambda }{2})(i_0 + \frac{\gamma }{2}+ \frac{\lambda }{2})} \frac{(-\alpha _0)_{i_0} ( \frac{\beta }{2}+\frac{\lambda }{2} )_{i_0}}{(1+\frac{\lambda }{2})_{i_0}(\frac{1}{2}+\frac{\gamma }{2}+\frac{\lambda }{2})_{i_0}}\right.\nonumber\\
&&\times \left. \sum_{i_1=i_0}^{\alpha _1} \frac{(-\alpha _1)_{i_1}(\frac{1}{2}+\frac{\beta }{2}+ \frac{\lambda }{2})_{i_1}(\frac{3}{2}+\frac{\lambda }{2})_{i_0}(1+\frac{\gamma }{2}+ \frac{\lambda }{2})_{i_0}}{(-\alpha _1)_{i_0}(\frac{1}{2}+\frac{\beta }{2}+ \frac{\lambda }{2})_{i_0}(\frac{3}{2}+\frac{\lambda }{2})_{i_1}(1+ \frac{\gamma}{2} +\frac{\lambda }{2})_{i_1}} \overleftrightarrow {w}_{2,2}^{i_1} \right\} \eta^2 \label{eq:302}
\end{eqnarray}
where
\begin{equation}
 \overleftrightarrow {w}_{2,2} = \frac{t_2 u_2 v_2}{(v_2-1)} \;\frac{z}{1-z v_2 (1-t_2)(1-u_2)}\nonumber
\end{equation} 
Put $l=1$ and $z =\overleftrightarrow {w}_{2,2}$ in (\ref{eq:35}). Take the new (\ref{eq:35}) into (\ref{eq:302}).
\begin{eqnarray}
y_2(x) &=& c_0 x^{\lambda } \int_{0}^{1} dt_2\;t_2^{ \frac{\lambda }{2}} \int_{0}^{1} du_2\;u_2^{ \frac{1}{2}(-1+\gamma +\lambda )}   \frac{1}{2\pi i} \oint dv_2\;\frac{1}{v_2} \left( 1-\frac{1}{v_2} \right)^{\alpha _2} \left( 1- z v_2 (1-t_2)(1-u_2)\right)^{-\frac{1}{2}(\beta +2+\lambda  )}  \nonumber\\
&&\times  \left(    \overleftrightarrow {w}_{2,2}^{-\frac{1}{2}(1+\lambda )} \left(  \overleftrightarrow {w}_{2,2} \partial _{ \overleftrightarrow {w}_{2,2}}\right) \overleftrightarrow {w}_{2,2}^{\frac{1}{2}(1+\lambda )} \left(\overleftrightarrow {w}_{2,2} \partial _{ \overleftrightarrow {w}_{2,2}}+\frac{1}{2(1+a)}\left(  -2\alpha _1+\beta -\delta +a(\delta +\gamma +\lambda )\right) \right)+\frac{q}{4(1+a)} \right)\nonumber\\
&&\times \int_{0}^{1} dt_1\;t_1^{ \frac{1}{2}(-1+\lambda )} \int_{0}^{1} du_1\;u_1^{ \frac{1}{2}(-2+\gamma +\lambda )}   \frac{1}{2\pi i} \oint dv_1\;\frac{1}{v_1} \left( 1-\frac{1}{v_1} \right)^{\alpha _1} \left( 1- \overleftrightarrow {w}_{2,2} v_1 (1-t_1)(1-u_1)\right)^{-\frac{1}{2}(\beta +1+\lambda  )}  \nonumber\\
&&\times  \left(    \overleftrightarrow {w}_{1,2}^{-\frac{\lambda }{2}} \left(  \overleftrightarrow {w}_{1,2} \partial _{ \overleftrightarrow {w}_{1,2}}\right) \overleftrightarrow {w}_{1,2}^{\frac{\lambda }{2}} \left(\overleftrightarrow {w}_{1,2} \partial _{ \overleftrightarrow {w}_{1,2}}+\frac{1}{2(1+a)}\left(  -2\alpha _0+\beta -\delta +a(\delta +\gamma -1+\lambda )\right) \right)+\frac{q}{4(1+a)} \right)\nonumber\\
&&\times \left\{ \sum_{i_0=0}^{\alpha _0} \frac{(-\alpha _0)_{i_0} (\frac{\beta }{2}+\frac{\lambda }{2})_{i_0}}{(1+\frac{\lambda }{2})_{i_0}(\frac{1}{2}+ \frac{\gamma}{2} +\frac{\lambda }{2})_{i_0}} \overleftrightarrow {w}_{1,2} ^{i_0} \right\} \eta^2 \label{eq:303}
\end{eqnarray}
where
\begin{equation}
 \overleftrightarrow {w}_{1,2}=\frac{t_1 u_1 v_1}{(v_1-1)}\; \frac{\overleftrightarrow {w}_{2,2}}{1-\overleftrightarrow {w}_{2,2} v_1 (1-t_1)(1-u_1)}\nonumber
\end{equation} 
By using similar process for the previous cases of integral forms of $y_1(x)$ and $y_2(x)$, the integral form of sub-power series expansion of $y_3(x)$ is
\begin{eqnarray}
y_3(x)&=& c_0 x^{\lambda } \int_{0}^{1} dt_3\;t_3^{ \frac{1}{2}(1+\lambda )} \int_{0}^{1} du_3\;u_3^{ \frac{1}{2}( \gamma +\lambda )}   \frac{1}{2\pi i} \oint dv_3\;\frac{1}{v_3} \left( 1-\frac{1}{v_3} \right)^{\alpha _3} \left( 1- z v_3 (1-t_3)(1-u_3)\right)^{-\frac{1}{2}(\beta +3+\lambda  )}  \nonumber\\
&&\times  \left(    \overleftrightarrow {w}_{3,3}^{-\frac{1}{2}(2+\lambda )} \left(  \overleftrightarrow {w}_{3,3} \partial _{ \overleftrightarrow {w}_{3,3}}\right) \overleftrightarrow {w}_{3,3}^{\frac{1}{2}(2+\lambda )} \left(\overleftrightarrow {w}_{3,3} \partial _{ \overleftrightarrow {w}_{3,3}}+\frac{1}{2(1+a)}\left(  -2\alpha _2+\beta -\delta +a(\delta +\gamma +1+\lambda )\right) \right)+\frac{q}{4(1+a)} \right)\nonumber\\
&&\times  \int_{0}^{1} dt_2\;t_2^{ \frac{\lambda }{2}} \int_{0}^{1} du_2\;u_2^{ \frac{1}{2}(-1+\gamma +\lambda )}   \frac{1}{2\pi i} \oint dv_2\;\frac{1}{v_2} \left( 1-\frac{1}{v_2} \right)^{\alpha _2} \left( 1- \overleftrightarrow {w}_{3,3} v_2 (1-t_2)(1-u_2)\right)^{-\frac{1}{2}(\beta +2+\lambda  )}  \nonumber\\
&&\times  \left(    \overleftrightarrow {w}_{2,3}^{-\frac{1}{2}(1+\lambda )} \left(  \overleftrightarrow {w}_{2,3} \partial _{ \overleftrightarrow {w}_{2,3}}\right) \overleftrightarrow {w}_{2,3}^{\frac{1}{2}(1+\lambda )} \left(\overleftrightarrow {w}_{2,3} \partial _{ \overleftrightarrow {w}_{2,3}}+\frac{1}{2(1+a)}\left(  -2\alpha _1+\beta -\delta +a(\delta +\gamma +\lambda )\right) \right)+\frac{q}{4(1+a)} \right)\nonumber\\
&&\times \int_{0}^{1} dt_1\;t_1^{ \frac{1}{2}(-1+\lambda )} \int_{0}^{1} du_1\;u_1^{ \frac{1}{2}(-2+\gamma +\lambda )}   \frac{1}{2\pi i} \oint dv_1\;\frac{1}{v_1} \left( 1-\frac{1}{v_1} \right)^{\alpha _1} \left( 1- \overleftrightarrow {w}_{2,3} v_1 (1-t_1)(1-u_1)\right)^{-\frac{1}{2}(\beta +1+\lambda  )}  \nonumber\\
&&\times  \left(    \overleftrightarrow {w}_{1,3}^{-\frac{\lambda }{2}} \left(  \overleftrightarrow {w}_{1,3} \partial _{ \overleftrightarrow {w}_{1,3}}\right) \overleftrightarrow {w}_{1,3}^{\frac{\lambda }{2}} \left(\overleftrightarrow {w}_{1,3} \partial _{ \overleftrightarrow {w}_{1,3}}+\frac{1}{2(1+a)}\left(  -2\alpha _0+\beta -\delta +a(\delta +\gamma -1+\lambda )\right) \right)+\frac{q}{4(1+a)} \right)\nonumber\\
&&\times \left\{ \sum_{i_0=0}^{\alpha _0} \frac{(-\alpha _0)_{i_0} (\frac{\beta }{2}+\frac{\lambda }{2})_{i_0}}{(1+\frac{\lambda }{2})_{i_0}(\frac{1}{2}+ \frac{\gamma}{2} +\frac{\lambda }{2})_{i_0}} \overleftrightarrow{w}_{1,3}^{i_0} \right\} \eta ^3 \label{eq:304}
\end{eqnarray}
where
%\Large
\begin{equation}
\begin{cases} \overleftrightarrow {w}_{3,3} = \frac{t_3 u_3 v_3}{(v_3-1)}\; \frac{z}{1- z v_3 (1-t_3)(1-u_3)}  \cr
\overleftrightarrow {w}_{2,3} = \frac{t_2 u_2 v_2}{(v_2-1)}\; \frac{\overleftrightarrow{w}_{3,3} }{1- \overleftrightarrow{w}_{3,3} v_2 (1-t_2)(1-u_2)} \cr
\overleftrightarrow w_{1,3}= \frac{t_1 u_1 v_1}{(v_1-1)}\frac{\overleftrightarrow w_{2,3} }{1-\overleftrightarrow w_{2,3} v_1 (1-t_1)(1-u_1)}
\end{cases}
\nonumber
\end{equation}
By repeating this process for all higher terms of integral forms of sub-summation $y_m(x)$ terms where $m \geq 4$, we obtain every integral forms of $y_m(x)$ terms. 
Since we substitute (\ref{eq:300a}), (\ref{eq:301}), (\ref{eq:303}), (\ref{eq:304}) and including all integral forms of $y_m(x)$ terms where $m \geq 4$ into (\ref{eq:36}), we obtain (\ref{eq:39}).
\qed
\end{pot}
Put $c_0$= 1 as $\lambda $=0 for the first kind of independent solutions of Heun equation and $\displaystyle{ c_0= \left( a^{-1}(1+a)\right)^{1-\gamma }}$ as $\lambda = 1-\gamma $ for the second one in (\ref{eq:39}). 
\begin{rmk}
The integral representation of Heun equation of the first kind for polynomial which makes $B_n$ term terminated about $x=0$ as $\alpha = -2\alpha_j -j $ where $j, \alpha _j =0,1,2,\cdots$ is 
\begin{eqnarray}
 y(x)&=& HF_{\alpha _j, \beta }\left( \alpha _j =-\frac{1}{2}(\alpha +j)\big|_{j\in \mathbb{N}_{0}}; \eta = \frac{(1+a)}{a} x ; z= -\frac{1}{a} x^2 \right) \nonumber\\
&=& _2F_1 \left(-\alpha _0, \frac{\beta }{2};\frac{1}{2}+\frac{\gamma }{2}; z \right) + \sum_{n=1}^{\infty } \Bigg\{\prod _{k=0}^{n-1} \Bigg\{ \int_{0}^{1} dt_{n-k}\;t_{n-k}^{\frac{1}{2}(n-k-2)} \int_{0}^{1} du_{n-k}\;u_{n-k}^{\frac{1}{2}(n-k-3+\gamma )} \nonumber\\
&&\times \frac{1}{2\pi i}  \oint dv_{n-k} \frac{1}{v_{n-k}} \left( 1-\frac{1}{v_{n-k}}\right)^{\alpha _{n-k}} \left( 1- \overleftrightarrow {w}_{n-k+1,n}v_{n-k}(1-t_{n-k})(1-u_{n-k})\right)^{-\frac{1}{2}(n-k+\beta )}\nonumber\\
&&\times  \left( \overleftrightarrow {w}_{n-k,n}^{-\frac{1}{2}(n-k-1)}\left(  \overleftrightarrow {w}_{n-k,n} \partial _{ \overleftrightarrow {w}_{n-k,n}}\right) \overleftrightarrow {w}_{n-k,n}^{\frac{1}{2}(n-k-1)}\left( \overleftrightarrow {w}_{n-k,n} \partial _{ \overleftrightarrow {w}_{n-k,n}} + \Omega _{n-k-1}^{(S)}\right) +Q\right) \Bigg\}\nonumber\\
&&\times  _2F_1 \left(-\alpha _0, \frac{\beta }{2};\frac{1}{2}+\frac{\gamma }{2}; \overleftrightarrow {w}_{1,n} \right) \Bigg\} \eta ^n \label{eq:41}
\end{eqnarray}
where
\begin{equation}
\begin{cases} 
\Omega _{n-k-1}^{(S)} = \frac{1}{2(1+a)}(-2\alpha _{n-k-1}+\beta -\delta +a(\delta +\gamma +n-k-2)) \cr
Q= \frac{q}{4(1+a)}
\end{cases}\nonumber %\label{eq:8}
\end{equation}
\end{rmk}
\begin{rmk}
The integral representation of Heun equation of the second kind for polynomial which makes $B_n$ term terminated  about $x=0$ as $\alpha = -2\alpha_j -j -1+\gamma $ where $j=0,1,2,\cdots$ is
\begin{eqnarray}
y(x)&=& HS_{\alpha _j, \beta }\left( \alpha _j =-\frac{1}{2}(\alpha +1-\gamma +j)\big|_{j\in \mathbb{N}_{0}}; \eta = \frac{(1+a)}{a} x ; z= -\frac{1}{a} x^2 \right) \nonumber\\
&=& z^{\frac{1}{2}(1-\gamma )} \Bigg\{\; _2F_1 \left(-\alpha _0, \frac{\beta }{2}+\frac{1}{2}-\frac{\gamma }{2};\frac{3}{2}-\frac{\gamma }{2}; z \right)\nonumber\\
&&+ \sum_{n=1}^{\infty } \Bigg\{\prod _{k=0}^{n-1} \Bigg\{ \int_{0}^{1} dt_{n-k}\;t_{n-k}^{\frac{1}{2}(n-k-1-\gamma )} \int_{0}^{1} du_{n-k}\;u_{n-k}^{\frac{1}{2}(n-k-2)}\frac{1}{2\pi i}  \oint dv_{n-k} \frac{1}{v_{n-k}} \left( 1-\frac{1}{v_{n-k}}\right)^{\alpha _{n-k}} \nonumber\\
&&\times \left( 1- \overleftrightarrow {w}_{n-k+1,n}v_{n-k}(1-t_{n-k})(1-u_{n-k})\right)^{-\frac{1}{2}(n-k+1+\beta-\gamma )}\nonumber\\
&&\times  \left( \overleftrightarrow {w}_{n-k,n}^{-\frac{1}{2}(n-k-\gamma )}\left(  \overleftrightarrow {w}_{n-k,n} \partial _{ \overleftrightarrow {w}_{n-k,n}}\right) \overleftrightarrow {w}_{n-k,n}^{\frac{1}{2}(n-k-\gamma )}\left( \overleftrightarrow {w}_{n-k,n} \partial _{ \overleftrightarrow {w}_{n-k,n}}  + \Omega _{n-k-1}^{(S)} \right) + Q \right) \Bigg\}\nonumber\\
&&\times _2F_1 \left(-\alpha _0, \frac{\beta }{2}+\frac{1}{2}-\frac{\gamma }{2};\frac{3}{2}-\frac{\gamma }{2}; \overleftrightarrow {w}_{1,n}\right)  \Bigg\} \eta ^n \Bigg\}\label{eq:42}
\end{eqnarray}
where
\begin{equation}
\begin{cases} 
\Omega _{n-k-1}^{(S)} = \frac{1}{2(1+a)}(-2\alpha _{n-k-1}+\beta -\delta +a(\delta +n-k-1 )) \cr
Q= \frac{q}{4(1+a)}
\end{cases}\nonumber %\label{eq:8}
\end{equation}
\end{rmk}
\subsubsection{The case of $\alpha = -2 \alpha _i-i -\lambda $ and $\beta = -2 \beta _i -i-\lambda $ only if $\alpha _i \leq \beta _i$ where $i, \alpha _i, \beta _i$ = $0,1,2,\cdots$}
%Put $\beta = -2\beta _i -i- \lambda $ and replace the math symbol  $\Omega _{n-k-1}^{(S)}$  by $\Omega _{n-k-1}^{(B)}$ into (\ref{eq:39}): more precisely ${\displaystyle \left(  \beta /2 + \lambda /2 \right)_{i_0} \rightarrow (-\beta _0)_{i_0}}$, ${\displaystyle \left( 1- \overleftrightarrow {w}_{n-k+1,n}v_{n-k}(1-t_{n-k})(1-u_{n-k})\right)^{-\frac{1}{2}(n-k+\beta +\lambda )} \rightarrow \left( 1- \overleftrightarrow {w}_{n-k+1,n}v_{n-k}(1-t_{n-k})(1-u_{n-k})\right)^{\beta _{n-k}}}$ and ${\displaystyle \Omega _{n-k-1}^{(S)} = (2(1+a))^{-1}(-2\alpha _{n-k-1}+\beta -\delta +a(\delta +\gamma +n-k-2+\lambda ))}$ \\ ${\displaystyle \rightarrow  \Omega _{n-k-1}^{(B)}= (2(1+a))^{-1} (-2\alpha _{n-k-1}-2\beta _{n-k-1} -\delta -n+1+k-\lambda +a(\delta +\gamma +n-k-2+\lambda ))}$ in (\ref{eq:39}).
Replace $\beta $ by $ -2\beta _l-l-\lambda $ where $l,\beta _l\in \mathbb{N}_{0}$ into (\ref{eq:35}) .
\begin{eqnarray}
G_l &=& \frac{1}{(i_{l-1}+\frac{l}{2}+\frac{\lambda }{2})(i_{l-1}+\frac{l}{2}-\frac{1}{2}+\frac{\gamma}{2} + \frac{\lambda }{2})}
 \sum_{i_l= i_{l-1}}^{\alpha _l} \frac{(-\alpha _l)_{i_l}(-\beta _l)_{i_l}(1+\frac{l}{2}+\frac{\lambda }{2})_{i_{l-1}}(\frac{1}{2}+\frac{\gamma}{2}+\frac{l}{2} +\frac{\lambda }{2})_{i_{l-1}}}{ (-\alpha _l)_{i_{l-1}}(-\beta _l)_{i_{l-1}}(1+\frac{l}{2}+\frac{\lambda }{2})_{i_l}(\frac{1}{2}+\frac{\gamma}{2}+\frac{l}{2} +\frac{\lambda }{2})_{i_l}} z^{i_l}\nonumber\\
&=&  \int_{0}^{1} dt_l\;t_l^{\frac{l}{2}-1+\frac{\lambda }{2}} \int_{0}^{1} du_l\;u_l^{\frac{l}{2}-\frac{3}{2}+\frac{\gamma }{2}+\frac{\lambda }{2}} 
\frac{1}{2\pi i} \oint dv_l\;\frac{1}{v_l} \left(1- \frac{1}{v_l}\right)^{\alpha_l} 
 (1-z v_l (1-t_l)(1-u_l))^{ \beta _l} \nonumber\\
&&\times \left(\frac{v_l}{(v_l-1)} \frac{z t_l u_l}{1-z v_l (1-t_l)(1-u_l)}\right)^{i_{l-1}}\label{eq:307} 
\end{eqnarray}
In Ref.\cite{Chou2012c}, the general expression of power series of Heun function for polynomial which makes $B_n$ term terminated about $x=0$ where $\alpha = -2 \alpha _i-i -\lambda $ and $\beta = -2 \beta _i -i-\lambda $ only if $\alpha _i \leq \beta _i$ is given by
\begin{eqnarray}
 y(x)&=& \sum_{n=0}^{\infty } y_{n}(x)= y_0(x)+y_1(x)+y_2(x)+y_3(x)+\cdots \nonumber\\
&=& c_0 x^{\lambda } \left\{\sum_{i_0=0}^{\alpha _0} \frac{(-\alpha _0)_{i_0} (-\beta _0)_{i_0}}{(1+\frac{\lambda }{2})_{i_0}(\frac{1}{2}+ \frac{\gamma}{2} +\frac{\lambda }{2})_{i_0}} z^{i_0} \right. \nonumber\\
&&+ \left\{ \sum_{i_0=0}^{\alpha _0}\frac{(i_0+ \frac{\lambda }{2}) \left( i_0+\Gamma_0^{(B)}\right) + Q}{(i_0+ \frac{1}{2}+ \frac{\lambda }{2})(i_0 + \frac{\gamma }{2}+ \frac{\lambda }{2})}   \frac{(-\alpha _0)_{i_0} (-\beta _0)_{i_0}}{(1+\frac{\lambda }{2})_{i_0}(\frac{1}{2}+ \frac{\gamma}{2} +\frac{\lambda }{2})_{i_0}} \sum_{i_1=i_0}^{\alpha _1} \frac{(-\alpha _1)_{i_1}(-\beta _1)_{i_1}(\frac{3}{2}+\frac{\lambda }{2})_{i_0}(1+\frac{\gamma }{2}+ \frac{\lambda }{2})_{i_0}}{(-\alpha _1)_{i_0}(-\beta _1)_{i_0}(\frac{3}{2}+\frac{\lambda }{2})_{i_1}(1+ \frac{\gamma}{2} +\frac{\lambda }{2})_{i_1}} z^{i_1} \right\} \eta \nonumber\\
&&+ \sum_{n=2}^{\infty } \left\{ \sum_{i_0=0}^{\alpha _0} \frac{(i_0+ \frac{\lambda }{2}) \left( i_0+ \Gamma_0^{(B)} \right) + Q}{(i_0+ \frac{1}{2}+ \frac{\lambda }{2})(i_0 + \frac{\gamma }{2}+ \frac{\lambda }{2})}
 \frac{(-\alpha _0)_{i_0} (-\beta _0)_{i_0}}{(1+\frac{\lambda }{2})_{i_0}(\frac{1}{2}+ \frac{\gamma}{2} +\frac{\lambda }{2})_{i_0}}\right. \nonumber\\
&&\times \prod _{k=1}^{n-1} \left\{ \sum_{i_k=i_{k-1}}^{\alpha _k} \frac{(i_k+\frac{k}{2}+ \frac{\lambda }{2}) \left( i_k+\Gamma_k^{(B)}\right) + Q}{(i_k+ \frac{k}{2}+\frac{1}{2}+\frac{\lambda }{2})(i_k +\frac{k}{2}+\frac{\gamma }{2}+\frac{\lambda }{2})}   \frac{(-\alpha _k)_{i_k}(-\beta _k)_{i_k}(1+ \frac{k}{2}+\frac{\lambda }{2})_{i_{k-1}}(\frac{1}{2}+\frac{k}{2}+\frac{\gamma }{2}+ \frac{\lambda }{2})_{i_{k-1}}}{(-\alpha _k)_{i_{k-1}}(-\beta _k)_{i_{k-1}}(1+\frac{k}{2}+\frac{\lambda }{2})_{i_k}(\frac{1}{2}+ \frac{k}{2}+ \frac{\gamma}{2} +\frac{\lambda }{2})_{i_k}}\right\} \nonumber\\
&&\times \left. \left. \sum_{i_n= i_{n-1}}^{\alpha _n} \frac{(-\alpha _n)_{i_n}(-\beta _n)_{i_n}(1+ \frac{n}{2}+\frac{\lambda }{2})_{i_{n-1}}(\frac{1}{2}+\frac{n}{2}+\frac{\gamma }{2}+ \frac{\lambda }{2})_{i_{n-1}}}{(-\alpha _n)_{i_{n-1}}(-\beta _n)_{i_{n-1}}(1+\frac{n}{2}+\frac{\lambda }{2})_{i_n}(\frac{1}{2}+ \frac{n}{2}+ \frac{\gamma}{2} +\frac{\lambda }{2})_{i_n}} z^{i_n} \right\} \eta ^n \right\}\label{eq:305}
\end{eqnarray}
where
\begin{equation}
\begin{cases} z = -\frac{1}{a}x^2 \cr
\eta = \frac{(1+a)}{a} x \cr
\alpha _i\leq \alpha _j \;\;\mbox{only}\;\mbox{if}\;i\leq j\;\;\mbox{where}\;i,j= 0,1,2,\cdots
\end{cases}\nonumber %\label{eq:37}
\end{equation}
and
\begin{equation}
\begin{cases} 
\Gamma_0^{(B)} = \frac{1}{2(1+a)}(-2\alpha _0-2\beta _0-\delta -\lambda +a(\delta +\gamma -1+\lambda )) \cr
\Gamma_k^{(B)} =  \frac{1}{2(1+a)}(-2\alpha _k-2\beta _k-k-\delta -\lambda  +a(\delta +\gamma +k-1+\lambda )) \cr
Q= \frac{q}{4(1+a)}
\end{cases}\nonumber %\label{eq:8}
\end{equation}
Substitute (\ref{eq:307}) into (\ref{eq:305}) where $l=1,2,3,\cdots$; apply $G_1$ into the second summation of sub-power series $y_1(x)$, apply $G_2$ into the third summation and $G_1$ into the second summation of sub-power series $y_2(x)$, apply $G_3$ into the forth summation, $G_2$ into the third summation and $G_1$ into the second summation of sub-power series $y_3(x)$, etc.\footnote{$y_1(x)$ means the sub-power series in (\ref{eq:305}) contains one term of $A_n's$, $y_2(x)$ means the sub-power series in (\ref{eq:305}) contains two terms of $A_n's$, $y_3(x)$ means the sub-power series in (\ref{eq:305}) contains three terms of $A_n's$, etc.}
\begin{thm}
The general representation in the form of integral of Heun polynomial which makes $B_n$ term terminated about $x=0$ as $\alpha = -2 \alpha _i-i -\lambda $ and $\beta = -2 \beta _i -i-\lambda $ only if $\alpha _i \leq \beta _i$ where $i,\alpha _i,\beta _i \in \mathbb{N}_{0}$ is given by 
\begin{eqnarray}
 y(x)&=& \sum_{n=0}^{\infty } y_{n}(x)= y_0(x)+y_1(x)+y_2(x)+y_3(x)+\cdots \nonumber\\
&=& c_0 x^{\lambda } \Bigg\{ \sum_{i_0=0}^{\alpha _0}\frac{(-\alpha _0)_{i_0}(-\beta _0)_{i_0}}{(1+\frac{\lambda }{2})_{i_0}(\frac{1}{2}+ \frac{\gamma }{2} +\frac{\lambda }{2})_{i_0}}  z^{i_0} 
+ \sum_{n=1}^{\infty } \Bigg\{\prod _{k=0}^{n-1} \Bigg\{ \int_{0}^{1} dt_{n-k}\;t_{n-k}^{\frac{1}{2}(n-k-2+\lambda )} \int_{0}^{1} du_{n-k}\;u_{n-k}^{\frac{1}{2}(n-k-3+\gamma +\lambda )} \nonumber\\
&&\times \frac{1}{2\pi i} \oint dv_{n-k} \frac{1}{v_{n-k}} \left( 1-\frac{1}{v_{n-k}}\right)^{\alpha _{n-k}} \left( 1- \overleftrightarrow {w}_{n-k+1,n}v_{n-k}(1-t_{n-k})(1-u_{n-k})\right)^{\beta _{n-k}} \nonumber\\
&&\times \left( \overleftrightarrow {w}_{n-k,n}^{-\frac{1}{2}(n-k-1+\lambda )}\left(  \overleftrightarrow {w}_{n-k,n} \partial _{ \overleftrightarrow {w}_{n-k,n}}\right) \overleftrightarrow {w}_{n-k,n}^{\frac{1}{2}(n-k-1+\lambda )} \left( \overleftrightarrow {w}_{n-k,n} \partial _{ \overleftrightarrow {w}_{n-k,n}}  + \Omega _{n-k-1}^{(B)} \right) +Q\right) \Bigg\} \nonumber\\
&&\times \sum_{i_0=0}^{\alpha _0}\frac{(-\alpha _0)_{i_0}(-\beta _0)_{i_0}}{(1+\frac{\lambda }{2})_{i_0}(\frac{1}{2}+ \frac{\gamma }{2} +\frac{\lambda }{2})_{i_0}}  \overleftrightarrow {w}_{1,n}^{i_0}\Bigg\} \eta ^n \Bigg\} \label{eq:43}
\end{eqnarray}
where
\begin{equation}
\begin{cases} 
\Omega _{n-k-1}^{(B)}= \frac{1}{2(1+a)}(-2\alpha _{n-k-1}-2\beta _{n-k-1} -\delta -n+1+k-\lambda +a(\delta +\gamma +n-k-2+\lambda )) \cr
Q= \frac{q}{4(1+a)}
\end{cases}\nonumber %\label{eq:8}
\end{equation}
\end{thm}
\begin{pot}
In (\ref{eq:305}) sub-power series $y_0(x) $, $y_1(x)$, $y_2(x)$ and $y_3(x)$ of Heun polynomial which makes $B_n$ term terminated about $x=0$ as $\alpha = -2 \alpha _i-i -\lambda $ and $\beta = -2 \beta _i -i-\lambda $ where $i,\alpha _i,\beta _i \in \mathbb{N}_{0}$ are given by
\begin{subequations}
\begin{equation}
 y_0(x)=  c_0 x^{\lambda } \sum_{i_0=0}^{\alpha _0} \frac{(-\alpha _0)_{i_0} (-\beta _0)_{i_0}}{(1+\frac{\lambda }{2})_{i_0}(\frac{1}{2}+ \frac{\gamma}{2} +\frac{\lambda }{2})_{i_0}} z^{i_0}\label{eq:306a}
\end{equation}
\begin{eqnarray}
 y_1(x)&=& c_0 x^{\lambda } \left\{\sum_{i_0=0}^{\alpha _0}\frac{(i_0+ \frac{\lambda }{2}) \left( i_0+ \frac{1}{2(1+a)}(-2\alpha _0-2\beta _0 -\delta -\lambda +a(\delta +\gamma -1+\lambda )) \right)+ \frac{q}{4(1+a)}}{(i_0+ \frac{1}{2}+ \frac{\lambda }{2})(i_0 + \frac{\gamma }{2}+ \frac{\lambda }{2})} \frac{(-\alpha _0)_{i_0} (-\beta _0)_{i_0}}{(1+\frac{\lambda }{2})_{i_0}(\frac{1}{2}+ \frac{\gamma}{2} +\frac{\lambda }{2})_{i_0}} \right.\nonumber\\
&\times& \left.\sum_{i_1=i_0}^{\alpha _1} \frac{(-\alpha _1)_{i_1}(-\beta _1)_{i_1}(\frac{3}{2}+\frac{\lambda }{2})_{i_0}(1+\frac{\gamma }{2}+ \frac{\lambda }{2})_{i_0}}{(-\alpha _1)_{i_0}(-\beta _1)_{i_0}(\frac{3}{2}+\frac{\lambda }{2})_{i_1}(1+ \frac{\gamma}{2} +\frac{\lambda }{2})_{i_1}} z^{i_1} \right\} \eta \label{eq:306b}
\end{eqnarray}
\begin{eqnarray}
 y_2(x) &=& c_0 x^{\lambda } \left\{\sum_{i_0=0}^{\alpha _0}\frac{(i_0+ \frac{\lambda }{2}) \left( i_0+ \frac{1}{2(1+a)}(-2\alpha _0-2\beta _0 -\delta -\lambda +a(\delta +\gamma -1+\lambda )) \right)+ \frac{q}{4(1+a)}}{(i_0+ \frac{1}{2}+ \frac{\lambda }{2})(i_0 + \frac{\gamma }{2}+ \frac{\lambda }{2})} \frac{(-\alpha _0)_{i_0} (-\beta _0)_{i_0}}{(1+\frac{\lambda }{2})_{i_0}(\frac{1}{2}+ \frac{\gamma}{2} +\frac{\lambda }{2})_{i_0}} \right.\nonumber\\
&\times& \sum_{i_1=i_0}^{\alpha _1}  \frac{(i_1+ \frac{1}{2}+ \frac{\lambda }{2}) \left( i_1+ \frac{1}{2(1+a)}(-2\alpha _1-2\beta _1-1-\delta -\lambda +a(\delta +\gamma +\lambda )) \right)+ \frac{q}{4(1+a)}}{(i_1+ 1+ \frac{\lambda }{2})(i_1 +\frac{1}{2}+ \frac{\gamma }{2}+ \frac{\lambda }{2})}  \nonumber\\
&\times& \left. \frac{(-\alpha _1)_{i_1}(-\beta _1)_{i_1}(\frac{3}{2}+\frac{\lambda }{2})_{i_0}(1+\frac{\gamma }{2}+ \frac{\lambda }{2})_{i_0}}{(-\alpha _1)_{i_0}(-\beta _1)_{i_0}(\frac{3}{2}+\frac{\lambda }{2})_{i_1}(1+ \frac{\gamma}{2} +\frac{\lambda }{2})_{i_1}}
 \sum_{i_2= i_1}^{\alpha _2} \frac{(-\alpha _2)_{i_2}(-\beta _2)_{i_2}(2+\frac{\lambda }{2})_{i_1}(\frac{3}{2} +\frac{\gamma }{2}+ \frac{\lambda }{2})_{i_1}}{(-\alpha _2)_{i_1}(-\beta _2)_{i_1}(2+\frac{\lambda }{2})_{i_2}(\frac{3}{2} + \frac{\gamma}{2} +\frac{\lambda }{2})_{i_2}} z^{i_2} \right\} \eta ^2 \hspace{1cm}\label{eq:306c}
\end{eqnarray}
\begin{eqnarray}
 y_3(x)&=& c_0 x^{\lambda } \left\{\sum_{i_0=0}^{\alpha _0}\frac{(i_0+ \frac{\lambda }{2}) \left( i_0+ \frac{1}{2(1+a)}(-2\alpha _0- 2\beta_0 -\delta -\lambda +a(\delta +\gamma -1+\lambda )) \right)+ \frac{q}{4(1+a)}}{(i_0+ \frac{1}{2}+ \frac{\lambda }{2})(i_0 + \frac{\gamma }{2}+ \frac{\lambda }{2})} \frac{(-\alpha _0)_{i_0} (-\beta _0)_{i_0}}{(1+\frac{\lambda }{2})_{i_0}(\frac{1}{2}+ \frac{\gamma}{2} +\frac{\lambda }{2})_{i_0}} \right.\nonumber\\
&\times& \sum_{i_1=i_0}^{\alpha _1}  \frac{(i_1+ \frac{1}{2}+ \frac{\lambda }{2}) \left( i_1+ \frac{1}{2(1+a)}(-2\alpha _1-2\beta _1-1 -\delta -\lambda +a(\delta +\gamma +\lambda )) \right)+ \frac{q}{4(1+a)}}{(i_1+ 1+ \frac{\lambda }{2})(i_1 +\frac{1}{2}+ \frac{\gamma }{2}+ \frac{\lambda }{2})}\nonumber\\
&\times& \frac{(-\alpha _1)_{i_1}(-\beta _1)_{i_1}(\frac{3}{2}+\frac{\lambda }{2})_{i_0}(1+\frac{\gamma }{2}+ \frac{\lambda }{2})_{i_0}}{(-\alpha _1)_{i_0}(-\beta _1)_{i_0}(\frac{3}{2}+\frac{\lambda }{2})_{i_1}(1+ \frac{\gamma}{2} +\frac{\lambda }{2})_{i_1}} \nonumber\\
&\times& \sum_{i_2=i_1}^{\alpha _2}  \frac{(i_2+ 1+ \frac{\lambda }{2}) \left( i_2+ \frac{1}{2(1+a)}(-2\alpha _2-2\beta _2-2-\delta -\lambda +a(\delta +\gamma +1+\lambda )) \right)+ \frac{q}{4(1+a)}}{(i_2+ \frac{3}{2}+ \frac{\lambda }{2})(i_2 +1+ \frac{\gamma }{2}+ \frac{\lambda }{2})}  \nonumber\\
&\times& \left. \frac{(-\alpha _2)_{i_2}(-\beta _2)_{i_2}(2+\frac{\lambda }{2})_{i_1}(\frac{3}{2} +\frac{\gamma }{2}+ \frac{\lambda }{2})_{i_1}}{(-\alpha _2)_{i_1}(-\beta _2)_{i_1}(2+\frac{\lambda }{2})_{i_2}(\frac{3}{2} + \frac{\gamma}{2} +\frac{\lambda }{2})_{i_2}} \sum_{i_3= i_2}^{\alpha _3} \frac{(-\alpha _3)_{i_3}(-\beta _3)_{i_3}(\frac{5}{2}+\frac{\lambda }{2})_{i_2}(2 +\frac{\gamma }{2}+ \frac{\lambda }{2})_{i_2}}{(-\alpha _3)_{i_2}(-\beta _3)_{i_2}(\frac{5}{2}+\frac{\lambda }{2})_{i_3}(2 + \frac{\gamma}{2} +\frac{\lambda }{2})_{i_3}} z^{i_3} \right\} \eta ^3 \hspace{1cm} \label{eq:306d}
\end{eqnarray}
\end{subequations}

Put $l=1$ in (\ref{eq:307}). Take the new (\ref{eq:307}) into (\ref{eq:306b}).
\begin{eqnarray}
 y_1(x)&=& c_0 x^{\lambda }\int_{0}^{1} dt_1\;t_1^{ \frac{1}{2}(-1+\lambda )} \int_{0}^{1} du_1\;u_1^{ \frac{1}{2}(-2+\gamma +\lambda )}   \frac{1}{2\pi i} \oint dv_1\;\frac{1}{v_1} \left( 1-\frac{1}{v_1} \right)^{\alpha _1} \left( 1- z v_1 (1-t_1)(1-u_1)\right)^{\beta _1}  \nonumber\\
&&\times  \left\{ \sum_{i_0=0}^{\alpha _0}\left( \left( i_0+\frac{\lambda }{2} \right)\left( i_0 +\frac{1}{2(1+a)}\left( -2\alpha _0-2\beta _0 -\delta -\lambda +a(\delta +\gamma -1+\lambda )\right) \right) +\frac{q}{4(1+a)}  \right) \right.\nonumber\\
&&\times \left. \frac{(-\alpha _0)_{i_0} (-\beta _0)_{i_0}}{(1+\frac{\lambda }{2})_{i_0}(\frac{1}{2}+ \frac{\gamma }{2}+\frac{\lambda }{2})_{i_0}} \left( \frac{t_1 u_1 v_1}{(v_1-1)} \frac{z }{1-z v_1 (1-t_1)(1-u_1)}\right)^{i_0} \right\} \eta  \nonumber\\
&=&  c_0 x^{\lambda }\int_{0}^{1} dt_1\;t_1^{ \frac{1}{2}(-1+\lambda )} \int_{0}^{1} du_1\;u_1^{ \frac{1}{2}(-2+\gamma +\lambda )}   \frac{1}{2\pi i} \oint dv_1\;\frac{1}{v_1} \left( 1-\frac{1}{v_1} \right)^{\alpha _1} \left( 1- z v_1 (1-t_1)(1-u_1)\right)^{\beta _1}  \nonumber\\
&&\times  \left(    \overleftrightarrow {w}_{1,1}^{-\frac{\lambda }{2}} \left(  \overleftrightarrow {w}_{1,1} \partial _{ \overleftrightarrow {w}_{1,1}}\right) \overleftrightarrow {w}_{1,1}^{\frac{\lambda }{2}} \left(\overleftrightarrow {w}_{1,1} \partial _{ \overleftrightarrow {w}_{1,1}}+\frac{1}{2(1+a)}\left(  -2\alpha _0-2\beta _0 -\delta -\lambda  +a(\delta +\gamma -1+\lambda )\right) \right)+\frac{q}{4(1+a)} \right)\nonumber\\
&&\times \left\{ \sum_{i_0=0}^{\alpha _0} \frac{(-\alpha _0)_{i_0} (-\beta _0)_{i_0}}{(1+\frac{\lambda }{2})_{i_0}(\frac{1}{2}+ \frac{\gamma }{2}+\frac{\lambda }{2})_{i_0}} \overleftrightarrow {w}_{1,1} ^{i_0} \right\} \eta \label{eq:308}
\end{eqnarray}
where 
\begin{equation}
 \overleftrightarrow {w}_{1,1} = \frac{t_1 u_1 v_1}{(v_1-1)}\; \frac{z}{1-z v_1 (1-t_1)(1-u_1)} \nonumber
\end{equation}
Put $l=2$ in (\ref{eq:307}). Take the new (\ref{eq:307}) into (\ref{eq:306c}). 
\begin{eqnarray}
y_2(x) &=& c_0 x^{\lambda } \int_{0}^{1} dt_2\;t_2^{ \frac{\lambda }{2}} \int_{0}^{1} du_2\;u_2^{ \frac{1}{2}(-1+\gamma +\lambda )}   \frac{1}{2\pi i} \oint dv_2\;\frac{1}{v_2} \left( 1-\frac{1}{v_2} \right)^{\alpha _2} \left( 1- z v_2 (1-t_2)(1-u_2)\right)^{\beta _2}  \nonumber\\
&&\times  \left(    \overleftrightarrow {w}_{2,2}^{-\frac{1}{2}(1+\lambda )} \left(  \overleftrightarrow {w}_{2,2} \partial _{ \overleftrightarrow {w}_{2,2}}\right) \overleftrightarrow {w}_{2,2}^{\frac{1}{2}(1+\lambda )} \left(\overleftrightarrow {w}_{2,2} \partial _{ \overleftrightarrow {w}_{2,2}}\right.\right.\nonumber\\
&&+\left.\left. \frac{1}{2(1+a)}\left(  -2\alpha _1-2\beta _1 -1-\delta -\lambda +a(\delta +\gamma +\lambda )\right) \right)+\frac{q}{4(1+a)} \right)\nonumber\\
&&\times \left\{ \sum_{i_0=0}^{\alpha _0}\frac{ (i_0+\frac{\lambda }{2})\left( i_0+\frac{1}{2(1+a)}\left( -2\alpha _0-2\beta _0 -\delta -\lambda  +a(\delta +\gamma -1+\lambda )\right) \right) +\frac{q}{4(1+a)} }{(i_0+ \frac{1}{2}+ \frac{\lambda }{2})(i_0 + \frac{\gamma }{2}+ \frac{\lambda }{2})} \frac{(-\alpha _0)_{i_0} (-\beta _0)_{i_0}}{(1+\frac{\lambda }{2})_{i_0}(\frac{1}{2}+\frac{\gamma }{2}+\frac{\lambda }{2})_{i_0}}\right.\nonumber\\
&&\times \left. \sum_{i_1=i_0}^{\alpha _1} \frac{(-\alpha _1)_{i_1}(-\beta _1)_{i_1}(\frac{3}{2}+\frac{\lambda }{2})_{i_0}(1+\frac{\gamma }{2}+ \frac{\lambda }{2})_{i_0}}{(-\alpha _1)_{i_0}(-\beta _1)_{i_0}(\frac{3}{2}+\frac{\lambda }{2})_{i_1}(1+ \frac{\gamma}{2} +\frac{\lambda }{2})_{i_1}} \overleftrightarrow {w}_{2,2}^{i_1} \right\} \eta^2 \label{eq:309}
\end{eqnarray}
where
\begin{equation}
 \overleftrightarrow {w}_{2,2} = \frac{t_2 u_2 v_2}{(v_2-1)} \;\frac{z}{1-z v_2 (1-t_2)(1-u_2)}\nonumber
\end{equation} 
Put $l=1$ and $z =\overleftrightarrow {w}_{2,2}$ in (\ref{eq:307}). Take the new (\ref{eq:307}) into (\ref{eq:309}).
\begin{eqnarray}
y_2(x) &=& c_0 x^{\lambda } \int_{0}^{1} dt_2\;t_2^{ \frac{\lambda }{2}} \int_{0}^{1} du_2\;u_2^{ \frac{1}{2}(-1+\gamma +\lambda )}   \frac{1}{2\pi i} \oint dv_2\;\frac{1}{v_2} \left( 1-\frac{1}{v_2} \right)^{\alpha _2} \left( 1- z v_2 (1-t_2)(1-u_2)\right)^{\beta _2}  \nonumber\\
&&\times  \left(    \overleftrightarrow {w}_{2,2}^{-\frac{1}{2}(1+\lambda )} \left(  \overleftrightarrow {w}_{2,2} \partial _{ \overleftrightarrow {w}_{2,2}}\right) \overleftrightarrow {w}_{2,2}^{\frac{1}{2}(1+\lambda )} \bigg(\overleftrightarrow {w}_{2,2} \partial _{ \overleftrightarrow {w}_{2,2}} \right. \nonumber\\
&&+\left. \frac{1}{2(1+a)}\left(  -2\alpha _1-2\beta _1-1-\delta -\lambda  +a(\delta +\gamma +\lambda )\right) \bigg)+\frac{q}{4(1+a)} \right)\nonumber\\
&&\times \int_{0}^{1} dt_1\;t_1^{ \frac{1}{2}(-1+\lambda )} \int_{0}^{1} du_1\;u_1^{ \frac{1}{2}(-2+\gamma +\lambda )}   \frac{1}{2\pi i} \oint dv_1\;\frac{1}{v_1} \left( 1-\frac{1}{v_1} \right)^{\alpha _1} \left( 1- \overleftrightarrow {w}_{2,2} v_1 (1-t_1)(1-u_1)\right)^{\beta _1}  \nonumber\\
&&\times  \left(    \overleftrightarrow {w}_{1,2}^{-\frac{\lambda }{2}} \left(  \overleftrightarrow {w}_{1,2} \partial _{ \overleftrightarrow {w}_{1,2}}\right) \overleftrightarrow {w}_{1,2}^{\frac{\lambda }{2}} \left(\overleftrightarrow {w}_{1,2} \partial _{ \overleftrightarrow {w}_{1,2}}+\frac{1}{2(1+a)}\left(  -2\alpha _0-2\beta _0 -\delta -\lambda +a(\delta +\gamma -1+\lambda )\right) \right)+\frac{q}{4(1+a)} \right)\nonumber\\
&&\times \left\{ \sum_{i_0=0}^{\alpha _0} \frac{(-\alpha _0)_{i_0} (-\beta _0)_{i_0}}{(1+\frac{\lambda }{2})_{i_0}(\frac{1}{2}+ \frac{\gamma}{2} +\frac{\lambda }{2})_{i_0}} \overleftrightarrow {w}_{1,2} ^{i_0} \right\} \eta^2 \label{eq:310}
\end{eqnarray}
where
\begin{equation}
 \overleftrightarrow {w}_{1,2}=\frac{t_1 u_1 v_1}{(v_1-1)}\; \frac{\overleftrightarrow {w}_{2,2}}{1-\overleftrightarrow {w}_{2,2} v_1 (1-t_1)(1-u_1)}\nonumber
\end{equation} 
By using similar process for the previous cases of integral forms of $y_1(x)$ and $y_2(x)$, the integral form of sub-power series expansion of $y_3(x)$ is
\begin{eqnarray}
y_3(x)&=& c_0 x^{\lambda } \int_{0}^{1} dt_3\;t_3^{ \frac{1}{2}(1+\lambda )} \int_{0}^{1} du_3\;u_3^{ \frac{1}{2}( \gamma +\lambda )}   \frac{1}{2\pi i} \oint dv_3\;\frac{1}{v_3} \left( 1-\frac{1}{v_3} \right)^{\alpha _3} \left( 1- z v_3 (1-t_3)(1-u_3)\right)^{\beta _3}  \nonumber\\
&&\times  \left(    \overleftrightarrow {w}_{3,3}^{-\frac{1}{2}(2+\lambda )} \left(  \overleftrightarrow {w}_{3,3} \partial _{ \overleftrightarrow {w}_{3,3}}\right) \overleftrightarrow {w}_{3,3}^{\frac{1}{2}(2+\lambda )} \bigg(\overleftrightarrow {w}_{3,3} \partial _{ \overleftrightarrow {w}_{3,3}}\right.\nonumber\\
&&\left.+\frac{1}{2(1+a)}\left(  -2\alpha _2-2\beta _2 -2-\delta -\lambda +a(\delta +\gamma +1+\lambda )\right) \bigg)+\frac{q}{4(1+a)} \right)\nonumber\\
&&\times  \int_{0}^{1} dt_2\;t_2^{ \frac{\lambda }{2}} \int_{0}^{1} du_2\;u_2^{ \frac{1}{2}(-1+\gamma +\lambda )}   \frac{1}{2\pi i} \oint dv_2\;\frac{1}{v_2} \left( 1-\frac{1}{v_2} \right)^{\alpha _2} \left( 1- \overleftrightarrow {w}_{3,3} v_2 (1-t_2)(1-u_2)\right)^{\beta _2}  \nonumber\\
&&\times  \left(    \overleftrightarrow {w}_{2,3}^{-\frac{1}{2}(1+\lambda )} \left(  \overleftrightarrow {w}_{2,3} \partial _{ \overleftrightarrow {w}_{2,3}}\right) \overleftrightarrow {w}_{2,3}^{\frac{1}{2}(1+\lambda )} \bigg(\overleftrightarrow {w}_{2,3} \partial _{ \overleftrightarrow {w}_{2,3}}\right.\nonumber\\
&&\left.+\frac{1}{2(1+a)}\left(  -2\alpha _1 -2\beta _1-1-\delta -\lambda +a(\delta +\gamma +\lambda )\right) \bigg)+\frac{q}{4(1+a)} \right)\nonumber\\
&&\times \int_{0}^{1} dt_1\;t_1^{ \frac{1}{2}(-1+\lambda )} \int_{0}^{1} du_1\;u_1^{ \frac{1}{2}(-2+\gamma +\lambda )}   \frac{1}{2\pi i} \oint dv_1\;\frac{1}{v_1} \left( 1-\frac{1}{v_1} \right)^{\alpha _1} \left( 1- \overleftrightarrow {w}_{2,3} v_1 (1-t_1)(1-u_1)\right)^{\beta _1}  \nonumber\\
&&\times  \left(    \overleftrightarrow {w}_{1,3}^{-\frac{\lambda }{2}} \left(  \overleftrightarrow {w}_{1,3} \partial _{ \overleftrightarrow {w}_{1,3}}\right) \overleftrightarrow {w}_{1,3}^{\frac{\lambda }{2}} \left(\overleftrightarrow {w}_{1,3} \partial _{ \overleftrightarrow {w}_{1,3}}+\frac{1}{2(1+a)}\left(  -2\alpha _0-2\beta _0-\delta -\lambda +a(\delta +\gamma -1+\lambda )\right) \right)+\frac{q}{4(1+a)} \right)\nonumber\\
&&\times \left\{ \sum_{i_0=0}^{\alpha _0} \frac{(-\alpha _0)_{i_0} (-\beta _0)_{i_0}}{(1+\frac{\lambda }{2})_{i_0}(\frac{1}{2}+ \frac{\gamma}{2} +\frac{\lambda }{2})_{i_0}} \overleftrightarrow{w}_{1,3}^{i_0} \right\} \eta ^3 \label{eq:311}
\end{eqnarray}
where
%\Large
\begin{equation}
\begin{cases} \overleftrightarrow {w}_{3,3} = \frac{t_3 u_3 v_3}{(v_3-1)}\; \frac{z}{1- z v_3 (1-t_3)(1-u_3)}  \cr
\overleftrightarrow {w}_{2,3} = \frac{t_2 u_2 v_2}{(v_2-1)}\; \frac{\overleftrightarrow{w}_{3,3} }{1- \overleftrightarrow{w}_{3,3} v_2 (1-t_2)(1-u_2)} \cr
\overleftrightarrow w_{1,3}= \frac{t_1 u_1 v_1}{(v_1-1)}\frac{\overleftrightarrow w_{2,3} }{1-\overleftrightarrow w_{2,3} v_1 (1-t_1)(1-u_1)}
\end{cases}
\nonumber
\end{equation}
By repeating this process for all higher terms of integral forms of sub-summation $y_m(x)$ terms where $m \geq 4$, we obtain every integral forms of $y_m(x)$ terms. 
Since we substitute (\ref{eq:306a}), (\ref{eq:308}), (\ref{eq:310}), (\ref{eq:311}) and including all integral forms of $y_m(x)$ terms where $m \geq 4$ into (\ref{eq:305}), we obtain (\ref{eq:43}).\footnote{Or put $\beta = -2\beta _i -i- \lambda $ into (\ref{eq:39}). Its solution is equivalent to (\ref{eq:43}).}
\qed
\end{pot}
Put $c_0$= 1 as $\lambda $=0 for the first kind of independent solutions of Heun equation and $\displaystyle{ c_0= \left( a^{-1}(1+a)\right)^{1-\gamma }}$ as $\lambda = 1-\gamma $ for the second one in (\ref{eq:43}). 
\begin{rmk}
The integral representation of Heun equation of the first kind for polynomial which makes $B_n$ term terminated about $x=0$ as $\alpha = -2\alpha_j -j $ and $\beta = -2\beta _j -j $ only if $\alpha _j \leq \beta _j$ where $j,\alpha_j, \beta_j =0,1,2,\cdots$ is
\begin{eqnarray}
 y(x)&=& HF_{\alpha _j, \beta_j }\left( \alpha _j =-\frac{1}{2}(\alpha +j), \beta _j =-\frac{1}{2}(\beta +j)\big|_{j\in \mathbb{N}_{0}}; \eta = \frac{(1+a)}{a} x ; z= -\frac{1}{a} x^2 \right) \nonumber\\
&=& _2F_1 \left(-\alpha _0, -\beta _0;\frac{1}{2}+\frac{\gamma }{2}; z \right) + \sum_{n=1}^{\infty } \Bigg\{\prod _{k=0}^{n-1} \Bigg\{ \int_{0}^{1} dt_{n-k}\;t_{n-k}^{\frac{1}{2}(n-k-2)} \int_{0}^{1} du_{n-k}\;u_{n-k}^{\frac{1}{2}(n-k-3+\gamma )} \nonumber\\
&&\times \frac{1}{2\pi i}  \oint dv_{n-k} \frac{1}{v_{n-k}} \left( 1-\frac{1}{v_{n-k}}\right)^{\alpha _{n-k}} \left( 1- \overleftrightarrow {w}_{n-k+1,n}v_{n-k}(1-t_{n-k})(1-u_{n-k})\right)^{\beta _{n-k}}\nonumber\\
&&\times  \left( \overleftrightarrow {w}_{n-k,n}^{-\frac{1}{2}(n-k-1)}\left(  \overleftrightarrow {w}_{n-k,n} \partial _{ \overleftrightarrow {w}_{n-k,n}}\right) \overleftrightarrow {w}_{n-k,n}^{\frac{1}{2}(n-k-1)}\left( \overleftrightarrow {w}_{n-k,n} \partial _{ \overleftrightarrow {w}_{n-k,n}} + \Omega _{n-k-1}^{(B)}\right) +Q\right) \Bigg\}\nonumber\\
&&\times _2F_1 \left(-\alpha _0, -\beta _0 ;\frac{1}{2}+\frac{\gamma }{2}; \overleftrightarrow {w}_{1,n} \right) \Bigg\} \eta ^n \label{eq:44}
\end{eqnarray}
where
\begin{equation}
\begin{cases} 
\Omega _{n-k-1}^{(B)} = \frac{1}{2(1+a)}(-2\alpha _{n-k-1}-2\beta _{n-k-1} -\delta -n+1+k +a(\delta +\gamma +n-k-2)) \cr
Q= \frac{q}{4(1+a)}
\end{cases}\nonumber %\label{eq:8}
\end{equation}
\end{rmk}
\begin{rmk}
The integral representation of Heun equation of the second kind for polynomial which makes $B_n$ term terminated about $x=0$ as $\alpha = -2\alpha_j -j -1+\gamma $ and $\beta = -2\beta _j -j -1+\gamma $ only if $\alpha _j \leq \beta _j$ where $j,\alpha_j, \beta_j =0,1,2,\cdots$ is
\begin{eqnarray}
y(x)&=& HS_{\alpha _j, \beta_j }\Bigg( \alpha _j =-\frac{1}{2}(\alpha +1-\gamma +j), \beta _j =-\frac{1}{2}(\beta +1-\gamma +j)\big|_{j\in \mathbb{N}_{0}}; \eta = \frac{(1+a)}{a} x ; z= -\frac{1}{a} x^2 \Bigg) \nonumber\\
&=& z^{\frac{1}{2}(1-\gamma )} \Bigg\{\; _2F_1 \left(-\alpha _0,-\beta _0;\frac{3}{2}-\frac{\gamma }{2}; z \right)
+ \sum_{n=1}^{\infty } \Bigg\{\prod _{k=0}^{n-1} \Bigg\{ \int_{0}^{1} dt_{n-k}\;t_{n-k}^{\frac{1}{2}(n-k-1-\gamma )} \int_{0}^{1} du_{n-k}\;u_{n-k}^{\frac{1}{2}(n-k-2)} \nonumber\\
&&\times \frac{1}{2\pi i}  \oint dv_{n-k} \frac{1}{v_{n-k}} \left( 1-\frac{1}{v_{n-k}}\right)^{\alpha _{n-k}} \left( 1- \overleftrightarrow {w}_{n-k+1,n}v_{n-k}(1-t_{n-k})(1-u_{n-k})\right)^{\beta _{n-k}}\nonumber\\
&&\times  \left( \overleftrightarrow {w}_{n-k,n}^{-\frac{1}{2}(n-k-\gamma )}\left(  \overleftrightarrow {w}_{n-k,n} \partial _{ \overleftrightarrow {w}_{n-k,n}}\right) \overleftrightarrow {w}_{n-k,n}^{\frac{1}{2}(n-k-\gamma )}\left( \overleftrightarrow {w}_{n-k,n} \partial _{ \overleftrightarrow {w}_{n-k,n}} + \Omega _{n-k-1}^{(B)}\right) +Q \right) \Bigg\}\nonumber\\
&&\times _2F_1 \left(-\alpha _0, -\beta _0;\frac{3}{2}-\frac{\gamma }{2}; \overleftrightarrow {w}_{1,n}\right)  \Bigg\} \eta ^n \Bigg\}\label{eq:45}
\end{eqnarray}
where
\begin{equation}
\begin{cases} 
\Omega _{n-k-1}^{(B)} = \frac{1}{2(1+a)}(-2\alpha _{n-k-1}-2\beta _{n-k-1} -\delta +\gamma -n+k +a(\delta +n-k-1 )) \cr
Q= \frac{q}{4(1+a)}
\end{cases}\nonumber %\label{eq:8}
\end{equation}
\end{rmk}
\subsection{Infinite series}
%For infinite series, replace the finite summation with an interval $[0,\alpha _0]$ by infinite summation with an interval  $[0,\infty ]$ in (\ref{eq:39}). Also, replace $\alpha _0$, $\alpha _{n-k}$, $\alpha _{n-k-1}$ and the math symbol  $\Omega _{n-k-1}^{(S)}$ by $-\frac{1}{2}(\alpha +\lambda )$, $-\frac{1}{2}(\alpha +n-k+\lambda )$, $-\frac{1}{2}(\alpha +n-k-1+\lambda )$ and $\Omega _{n-k-1}^{(I)}$ on it. 
Let's consider the integral representation of Heun equation about $x=0$ for infinite series.
There is a generalized hypergeometric function which is written by
\begin{eqnarray}
M_l &=& \sum_{i_l= i_{l-1}}^{\infty } \frac{(\frac{\alpha }{2}+\frac{l}{2}+\frac{\lambda }{2})_{i_l}(\frac{\beta }{2}+\frac{l}{2}+\frac{\lambda }{2})_{i_l}(1+\frac{l}{2}+\frac{\lambda }{2})_{i_{l-1}}(\frac{1}{2}+\frac{\gamma}{2}+\frac{l}{2} +\frac{\lambda }{2})_{i_{l-1}}}{ (\frac{\alpha }{2}+\frac{l}{2}+\frac{\lambda }{2})_{i_{l-1}}(\frac{\beta }{2}+\frac{l}{2}+\frac{\lambda }{2})_{i_{l-1}}(1+\frac{l}{2}+\frac{\lambda }{2})_{i_l}(\frac{1}{2}+\frac{\gamma}{2}+\frac{l}{2} +\frac{\lambda }{2})_{i_l}} z^{i_l}\label{eq:312}\\
&=& z^{i_{l-1}} 
\sum_{j=0}^{\infty } \frac{B(i_{l-1}+\frac{l}{2}+\frac{\lambda }{2},j+1) B(i_{l-1}+\frac{l}{2}-\frac{1}{2}+\frac{\gamma}{2} +\frac{\lambda }{2},j+1)(i_{l-1}+\frac{l}{2}+\frac{\alpha }{2}+\frac{\lambda }{2})_j (i_{l-1}+\frac{l}{2}+\frac{\beta }{2}+\frac{\lambda }{2})_j}{(i_{l-1}+\frac{l}{2}+\frac{\lambda }{2})^{-1}(i_{l-1}+\frac{l}{2}-\frac{1}{2}+\frac{\gamma}{2} + \frac{\lambda }{2})^{-1}(1)_j \;j!} z^j\nonumber
\end{eqnarray}
Substitute (\ref{eq:31a}) and (\ref{eq:31b}) into (\ref{eq:312}). Divide $(i_{l-1}+\frac{l}{2}+\frac{\lambda }{2})(i_{l-1}+\frac{l}{2}-\frac{1}{2}+\frac{\gamma}{2} + \frac{\lambda }{2})$ into the new (\ref{eq:312}).
\begin{eqnarray}
V_l &=& \frac{1}{(i_{l-1}+\frac{l}{2}+\frac{\lambda }{2})(i_{l-1}+\frac{l}{2}-\frac{1}{2}+\frac{\gamma}{2} + \frac{\lambda }{2})}
\sum_{i_l= i_{l-1}}^{\infty } \frac{(\frac{\alpha }{2}+\frac{l}{2}+\frac{\lambda }{2})_{i_l}(\frac{\beta }{2}+\frac{l}{2}+\frac{\lambda }{2})_{i_l}(1+\frac{l}{2}+\frac{\lambda }{2})_{i_{l-1}}(\frac{1}{2}+\frac{\gamma}{2}+\frac{l}{2} +\frac{\lambda }{2})_{i_{l-1}}}{ (\frac{\alpha }{2}+\frac{l}{2}+\frac{\lambda }{2})_{i_{l-1}}(\frac{\beta }{2}+\frac{l}{2}+\frac{\lambda }{2})_{i_{l-1}}(1+\frac{l}{2}+\frac{\lambda }{2})_{i_l}(\frac{1}{2}+\frac{\gamma}{2}+\frac{l}{2} +\frac{\lambda }{2})_{i_l}} z^{i_l}\nonumber\\
&=&  \int_{0}^{1} dt_l\;t_l^{\frac{l}{2}-1+\frac{\lambda }{2}} \int_{0}^{1} du_l\;u_l^{\frac{l}{2}-\frac{3}{2}+\frac{\gamma }{2}+\frac{\lambda }{2}} (z t_l u_l)^{i_{l-1}}\nonumber\\
&&\times \sum_{j=0}^{\infty } \frac{(i_{l-1}+\frac{l}{2}+\frac{\alpha }{2}+\frac{\lambda }{2} )_j (i_{l-1}+\frac{l}{2}+\frac{\beta }{2}+\frac{\lambda }{2})_j}{(1)_j \;j!} [z(1-t_l)(1-u_l)]^j \label{eq:313}
\end{eqnarray}
The hypergeometric function is defined by
\begin{eqnarray}
_2F_1 \left( \alpha ,\beta ; \gamma ; z \right) &=& \sum_{n=0}^{\infty } \frac{(\alpha )_n (\beta )_n}{(\gamma )_n (n!)} z^n \nonumber\\
&=&  \frac{1}{2\pi i} \frac{\Gamma( 1+\alpha  -\gamma )}{\Gamma (\alpha )} \int_0^{(1+)} dv_l\; (-1)^{\gamma }(-v_l)^{\alpha -1} (1-v_l )^{\gamma -\alpha -1} (1-zv_l)^{-\beta }\hspace{1cm}\label{eq:33.a}\\
&& \mbox{where} \;\gamma -\alpha  \ne 1,2,3,\cdots, \;\mbox{Re}(\alpha )>0 \nonumber
\end{eqnarray}
Replace $\alpha $, $\beta $, $\gamma $ and $z$ by $i_{l-1}+\frac{l}{2}+\frac{\alpha }{2}+\frac{\lambda }{2}$, $i_{l-1}+\frac{l}{2}+\frac{\beta }{2}+\frac{\lambda }{2}$, 1 and $z(1-t_l)(1-u_l)$ in (\ref{eq:33.a}). Take the new (\ref{eq:33.a}) into (\ref{eq:313}).
\begin{eqnarray}
V_l &=& \frac{1}{(i_{l-1}+\frac{l}{2}+\frac{\lambda }{2})(i_{l-1}+\frac{l}{2}-\frac{1}{2}+\frac{\gamma}{2} + \frac{\lambda }{2})}
\sum_{i_l= i_{l-1}}^{\infty } \frac{(\frac{\alpha }{2}+\frac{l}{2}+\frac{\lambda }{2})_{i_l}(\frac{\beta }{2}+\frac{l}{2}+\frac{\lambda }{2})_{i_l}(1+\frac{l}{2}+\frac{\lambda }{2})_{i_{l-1}}(\frac{1}{2}+\frac{\gamma}{2}+\frac{l}{2} +\frac{\lambda }{2})_{i_{l-1}}}{ (\frac{\alpha }{2}+\frac{l}{2}+\frac{\lambda }{2})_{i_{l-1}}(\frac{\beta }{2}+\frac{l}{2}+\frac{\lambda }{2})_{i_{l-1}}(1+\frac{l}{2}+\frac{\lambda }{2})_{i_l}(\frac{1}{2}+\frac{\gamma}{2}+\frac{l}{2} +\frac{\lambda }{2})_{i_l}} z^{i_l}\nonumber\\
&=&  \int_{0}^{1} dt_l\;t_l^{\frac{l}{2}-1+\frac{\lambda }{2}} \int_{0}^{1} du_l\;u_l^{\frac{l}{2}-\frac{3}{2}+\frac{\gamma }{2}+\frac{\lambda }{2}} 
\frac{1}{2\pi i} \oint dv_l\;\frac{1}{v_l} \left(1- \frac{1}{v_l}\right)^{-\frac{1}{2}(\alpha +l+\lambda )} 
 (1-z v_l (1-t_l)(1-u_l))^{-\frac{1}{2}(\beta +l+\lambda )} \nonumber\\
&&\times \left(\frac{v_l}{(v_l-1)} \frac{z t_l u_l}{1-z v_l (1-t_l)(1-u_l)}\right)^{i_{l-1}}\label{eq:314} 
\end{eqnarray}
In Ref.\cite{Chou2012c} the general expression of power series of Heun function for infinite series about $x=0$ is given by
\begin{eqnarray}
 y(x)&=& \sum_{n=0}^{\infty } y_n(x)= y_0(x)+ y_1(x)+ y_2(x)+ y_3(x)+\cdots \nonumber\\
&=& c_0 x^{\lambda } \left\{\sum_{i_0=0}^{\infty } \frac{(\frac{\alpha }{2}+\frac{\lambda }{2})_{i_0} (\frac{\beta }{2}+\frac{\lambda }{2})_{i_0}}{(1+\frac{\lambda }{2})_{i_0}(\frac{1}{2}+ \frac{\gamma}{2} +\frac{\lambda }{2})_{i_0}} z^{i_0}
+ \left\{\sum_{i_0=0}^{\infty }\frac{(i_0+ \frac{\lambda }{2}) \left( i_0+ \Gamma_0^{(I)}\right)+ Q}{(i_0+ \frac{1}{2}+ \frac{\lambda }{2})(i_0 + \frac{\gamma }{2}+ \frac{\lambda }{2})}\right.\right.\nonumber\\
&&\times \left. \frac{(\frac{\alpha }{2}+\frac{\lambda }{2})_{i_0} (\frac{\beta }{2}+\frac{\lambda }{2})_{i_0}}{(1+\frac{\lambda }{2})_{i_0}(\frac{1}{2}+ \frac{\gamma}{2} +\frac{\lambda }{2})_{i_0}} \sum_{i_1=i_0}^{\infty } \frac{(\frac{1}{2}+\frac{\alpha }{2}+ \frac{\lambda }{2})_{i_1}(\frac{1}{2}+\frac{\beta }{2}+ \frac{\lambda }{2})_{i_1}(\frac{3}{2}+\frac{\lambda }{2})_{i_0}(1+\frac{\gamma }{2}+ \frac{\lambda }{2})_{i_0}}{(\frac{1}{2}+\frac{\alpha }{2}+ \frac{\lambda }{2})_{i_0}(\frac{1}{2}+\frac{\beta }{2}+ \frac{\lambda }{2})_{i_0}(\frac{3}{2}+\frac{\lambda }{2})_{i_1}(1+ \frac{\gamma}{2} +\frac{\lambda }{2})_{i_1}} z^{i_1} \right\} \eta \nonumber\\
&&+ \sum_{n=2}^{\infty } \left\{ \sum_{i_0=0}^{\infty } \frac{(i_0+ \frac{\lambda }{2}) \left( i_0+ \Gamma_0^{(I)}\right)+ Q}{(i_0+ \frac{1}{2}+ \frac{\lambda }{2})(i_0 + \frac{\gamma }{2}+ \frac{\lambda }{2})}
 \frac{(\frac{\alpha }{2}+\frac{\lambda }{2})_{i_0} (\frac{\beta }{2}+\frac{\lambda }{2})_{i_0}}{(1+\frac{\lambda }{2})_{i_0}(\frac{1}{2}+ \frac{\gamma}{2} +\frac{\lambda }{2})_{i_0}}\right.\nonumber\\
&&\times \prod _{k=1}^{n-1} \left\{ \sum_{i_k=i_{k-1}}^{\infty } \frac{(i_k+\frac{k}{2}+ \frac{\lambda }{2}) \left( i_k+ \Gamma_k^{(I)}\right)+ Q}{(i_k+ \frac{k}{2}+\frac{1}{2}+\frac{\lambda }{2})(i_k +\frac{k}{2}+\frac{\gamma }{2}+\frac{\lambda }{2})} \right.\nonumber\\
&&\times \left.\frac{(\frac{k}{2}+\frac{\alpha }{2}+ \frac{\lambda }{2})_{i_k}(\frac{k}{2}+\frac{\beta }{2}+ \frac{\lambda }{2})_{i_k}(1+ \frac{k}{2}+\frac{\lambda }{2})_{i_{k-1}}(\frac{1}{2}+\frac{k}{2}+\frac{\gamma }{2}+ \frac{\lambda }{2})_{i_{k-1}}}{(\frac{k}{2}+\frac{\alpha }{2}+ \frac{\lambda }{2})_{i_{k-1}}(\frac{k}{2}+\frac{\beta }{2}+ \frac{\lambda }{2})_{i_{k-1}}(1+\frac{k}{2}+\frac{\lambda }{2})_{i_k}(\frac{1}{2}+ \frac{k}{2}+ \frac{\gamma}{2} +\frac{\lambda }{2})_{i_k}}\right\} \nonumber\\
&&\times \left.\left. \sum_{i_n= i_{n-1}}^{\infty } \frac{(\frac{n}{2}+\frac{\alpha }{2}+ \frac{\lambda }{2})_{i_n}(\frac{n}{2}+\frac{\beta }{2}+ \frac{\lambda }{2})_{i_n}(1+ \frac{n}{2}+\frac{\lambda }{2})_{i_{n-1}}(\frac{1}{2}+\frac{n}{2}+\frac{\gamma }{2}+ \frac{\lambda }{2})_{i_{n-1}}}{(\frac{n}{2}+\frac{\alpha }{2}+ \frac{\lambda }{2})_{i_{n-1}}(\frac{n}{2}+\frac{\beta }{2}+ \frac{\lambda }{2})_{i_{n-1}}(1+\frac{n}{2}+\frac{\lambda }{2})_{i_n}(\frac{1}{2}+ \frac{n}{2}+ \frac{\gamma}{2} +\frac{\lambda }{2})_{i_n}} z^{i_n} \right\} \eta ^n \right\}\label{eq:51}
\end{eqnarray}
where
\begin{equation}
\begin{cases} 
\Gamma_0^{(I)} =  \frac{1}{2(1+a)}(\alpha +\beta -\delta +\lambda +a(\delta +\gamma -1+\lambda ))\cr
\Gamma_k^{(I)} =  \frac{1}{2(1+a)}(\alpha +\beta -\delta +k +\lambda +a(\delta +\gamma -1+k +\lambda )) \cr
Q= \frac{q}{4(1+a)}
\end{cases}\nonumber %\label{eq:8}
\end{equation}
Substitute (\ref{eq:314}) into (\ref{eq:51}) where $l=1,2,3,\cdots$; apply $V_1$ into the second summation of sub-power series $y_1(x)$, apply $V_2$ into the third summation and $V_1$ into the second summation of sub-power series $y_2(x)$, apply $V_3$ into the forth summation, $V_2$ into the third summation and $V_1$ into the second summation of sub-power series $y_3(x)$, etc.\footnote{$y_1(x)$ means the sub-power series in (\ref{eq:51}) contains one term of $A_n's$, $y_2(x)$ means the sub-power series in (\ref{eq:51}) contains two terms of $A_n's$, $y_3(x)$ means the sub-power series in (\ref{eq:51}) contains three terms of $A_n's$, etc.}
\begin{thm}
The general representation in the form of integral of Heun equation for infinite series about $x=0$ is given by 
\begin{eqnarray}
 y(x)&=& \sum_{n=0}^{\infty } y_{n}(x)= y_0(x)+y_1(x)+y_2(x)+y_3(x)+\cdots\nonumber\\
&=& c_0 x^{\lambda } \Bigg\{ \sum_{i_0=0}^{\infty }\frac{(\frac{\alpha }{2}+\frac{\lambda }{2})_{i_0}(\frac{\beta }{2}+\frac{\lambda }{2})_{i_0}}{(1+\frac{\lambda }{2})_{i_0}(\frac{1}{2}+ \frac{\gamma }{2} +\frac{\lambda }{2})_{i_0}}  z^{i_0}+ \sum_{n=1}^{\infty } \Bigg\{\prod _{k=0}^{n-1} \Bigg\{ \int_{0}^{1} dt_{n-k}\;t_{n-k}^{\frac{1}{2}(n-k-2+\lambda )} \int_{0}^{1} du_{n-k}\;u_{n-k}^{\frac{1}{2}(n-k-3+\gamma +\lambda )} \nonumber\\
&&\times \frac{1}{2\pi i}  \oint dv_{n-k} \frac{1}{v_{n-k}} \left( 1-\frac{1}{v_{n-k}}\right)^{-\frac{1}{2}(n-k+\alpha +\lambda )}  \left( 1- \overleftrightarrow {w}_{n-k+1,n}v_{n-k}(1-t_{n-k})(1-u_{n-k})\right)^{-\frac{1}{2}(n-k+\beta +\lambda )}\nonumber\\
&&\times  \left( \overleftrightarrow {w}_{n-k,n}^{-\frac{1}{2}(n-k-1+\lambda )}\left(  \overleftrightarrow {w}_{n-k,n} \partial _{ \overleftrightarrow {w}_{n-k,n}}\right) \overleftrightarrow {w}_{n-k,n}^{\frac{1}{2}(n-k-1+\lambda )}\left( \overleftrightarrow {w}_{n-k,n} \partial _{ \overleftrightarrow {w}_{n-k,n}} + \Omega _{n-k-1}^{(I)} \right) +Q\right) \Bigg\}\nonumber \\
&&\times\sum_{i_0=0}^{\infty }\frac{(\frac{\alpha }{2}+\frac{\lambda }{2})_{i_0}(\frac{\beta }{2}+\frac{\lambda }{2})_{i_0}}{(1+\frac{\lambda }{2})_{i_0}(\frac{1}{2}+ \frac{\gamma }{2} +\frac{\lambda }{2})_{i_0}}  \overleftrightarrow {w}_{1,n}^{i_0}\Bigg\} \eta ^n \Bigg\} \label{eq:46}
\end{eqnarray}
where
\begin{equation}
\begin{cases} 
\Omega _{n-k-1}^{(I)}= \frac{1}{2(1+a)}(\alpha +\beta -\delta +n-k-1+\lambda +a(\delta +\gamma +n-k-2+\lambda )) \cr
Q= \frac{q}{4(1+a)}
\end{cases}\nonumber %\label{eq:8}
\end{equation}
\end{thm}
\begin{pot}
In (\ref{eq:51}) sub-power series $y_0(x) $, $y_1(x)$, $y_2(x)$ and $y_3(x)$ of Heun equation for infinite series about $x=0$ using 3TRF are given by
\begin{subequations}
\begin{equation}
 y_0(x)=  c_0 x^{\lambda } \sum_{i_0=0}^{\infty } \frac{(\frac{\alpha }{2}+\frac{\lambda }{2})_{i_0} (\frac{\beta }{2}+\frac{\lambda }{2})_{i_0}}{(1+\frac{\lambda }{2})_{i_0}(\frac{1}{2}+ \frac{\gamma}{2} +\frac{\lambda }{2})_{i_0}} z^{i_0}\label{eq:315a}
\end{equation}
\begin{eqnarray}
 y_1(x)&=& c_0 x^{\lambda } \left\{\sum_{i_0=0}^{\infty }\frac{(i_0+ \frac{\lambda }{2}) \left( i_0+ \frac{1}{2(1+a)}(\alpha + \beta -\delta +\lambda +a(\delta +\gamma -1+\lambda )) \right)+ \frac{q}{4(1+a)}}{(i_0+ \frac{1}{2}+ \frac{\lambda }{2})(i_0 + \frac{\gamma }{2}+ \frac{\lambda }{2})} \frac{(\frac{\alpha }{2}+\frac{\lambda }{2})_{i_0} (\frac{\beta }{2}+\frac{\lambda }{2})_{i_0}}{(1+\frac{\lambda }{2})_{i_0}(\frac{1}{2}+ \frac{\gamma}{2} +\frac{\lambda }{2})_{i_0}} \right.\nonumber\\
&\times& \left.\sum_{i_1=i_0}^{\infty} \frac{(\frac{1}{2}+\frac{\alpha }{2}+ \frac{\lambda }{2})_{i_1}(\frac{1}{2}+\frac{\beta }{2}+ \frac{\lambda }{2})_{i_1}(\frac{3}{2}+\frac{\lambda }{2})_{i_0}(1+\frac{\gamma }{2}+ \frac{\lambda }{2})_{i_0}}{(\frac{1}{2}+\frac{\alpha }{2}+ \frac{\lambda }{2})_{i_0}(\frac{1}{2}+\frac{\beta }{2}+ \frac{\lambda }{2})_{i_0}(\frac{3}{2}+\frac{\lambda }{2})_{i_1}(1+ \frac{\gamma}{2} +\frac{\lambda }{2})_{i_1}} z^{i_1} \right\} \eta \label{eq:315b}
\end{eqnarray}
\begin{eqnarray}
 y_2(x) &=& c_0 x^{\lambda } \left\{\sum_{i_0=0}^{\infty }\frac{(i_0+ \frac{\lambda }{2}) \left( i_0+ \frac{1}{2(1+a)}(\alpha + \beta -\delta +\lambda +a(\delta +\gamma -1+\lambda )) \right)+ \frac{q}{4(1+a)}}{(i_0+ \frac{1}{2}+ \frac{\lambda }{2})(i_0 + \frac{\gamma }{2}+ \frac{\lambda }{2})} \frac{(\frac{\alpha }{2}+\frac{\lambda }{2})_{i_0} (\frac{\beta }{2}+\frac{\lambda }{2})_{i_0}}{(1+\frac{\lambda }{2})_{i_0}(\frac{1}{2}+ \frac{\gamma}{2} +\frac{\lambda }{2})_{i_0}} \right.\nonumber\\
&\times& \sum_{i_1=i_0}^{\infty }  \frac{(i_1+ \frac{1}{2}+ \frac{\lambda }{2}) \left( i_1+ \frac{1}{2(1+a)}( \alpha + \beta -\delta +1+\lambda  +a(\delta +\gamma +\lambda )) \right)+ \frac{q}{4(1+a)}}{(i_1+ 1+ \frac{\lambda }{2})(i_1 +\frac{1}{2}+ \frac{\gamma }{2}+ \frac{\lambda }{2})}  \nonumber\\
&\times&  \frac{(\frac{1}{2}+\frac{\alpha }{2}+ \frac{\lambda }{2})_{i_1}(\frac{1}{2}+\frac{\beta }{2}+ \frac{\lambda }{2})_{i_1}(\frac{3}{2}+\frac{\lambda }{2})_{i_0}(1+\frac{\gamma }{2}+ \frac{\lambda }{2})_{i_0}}{(\frac{1}{2}+\frac{\alpha }{2}+ \frac{\lambda }{2})_{i_0}(\frac{1}{2}+\frac{\beta }{2}+ \frac{\lambda }{2})_{i_0}(\frac{3}{2}+\frac{\lambda }{2})_{i_1}(1+ \frac{\gamma}{2} +\frac{\lambda }{2})_{i_1}} \nonumber\\
&\times& \left.\sum_{i_2= i_1}^{\infty} \frac{(1+\frac{\alpha }{2}+ \frac{\lambda }{2})_{i_2}(1+\frac{\beta }{2}+ \frac{\lambda }{2})_{i_2}(2+\frac{\lambda }{2})_{i_1}(\frac{3}{2} +\frac{\gamma }{2}+ \frac{\lambda }{2})_{i_1}}{(1+\frac{\alpha }{2}+ \frac{\lambda }{2})_{i_1}(1+\frac{\beta }{2}+ \frac{\lambda }{2})_{i_1}(2+\frac{\lambda }{2})_{i_2}(\frac{3}{2} + \frac{\gamma}{2} +\frac{\lambda }{2})_{i_2}} z^{i_2} \right\} \eta ^2 \hspace{1cm}\label{eq:315c}
\end{eqnarray}
\begin{eqnarray}
 y_3(x)&=& c_0 x^{\lambda } \left\{\sum_{i_0=0}^{\infty }\frac{(i_0+ \frac{\lambda }{2}) \left( i_0+ \frac{1}{2(1+a)}( \alpha + \beta -\delta +\lambda +a(\delta +\gamma -1+\lambda )) \right)+ \frac{q}{4(1+a)}}{(i_0+ \frac{1}{2}+ \frac{\lambda }{2})(i_0 + \frac{\gamma }{2}+ \frac{\lambda }{2})} \frac{(\frac{\alpha }{2}+\frac{\lambda }{2})_{i_0} (\frac{\beta }{2}+\frac{\lambda }{2})_{i_0}}{(1+\frac{\lambda }{2})_{i_0}(\frac{1}{2}+ \frac{\gamma}{2} +\frac{\lambda }{2})_{i_0}} \right.\nonumber\\
&\times& \sum_{i_1=i_0}^{\infty}  \frac{(i_1+ \frac{1}{2}+ \frac{\lambda }{2}) \left( i_1+ \frac{1}{2(1+a)}( \alpha + \beta -\delta +1+\lambda +a(\delta +\gamma +\lambda )) \right)+ \frac{q}{4(1+a)}}{(i_1+ 1+ \frac{\lambda }{2})(i_1 +\frac{1}{2}+ \frac{\gamma }{2}+ \frac{\lambda }{2})}\nonumber\\
&\times& \frac{(\frac{1}{2}+\frac{\alpha }{2}+ \frac{\lambda }{2})_{i_1}(\frac{1}{2}+\frac{\beta }{2}+ \frac{\lambda }{2})_{i_1}(\frac{3}{2}+\frac{\lambda }{2})_{i_0}(1+\frac{\gamma }{2}+ \frac{\lambda }{2})_{i_0}}{(\frac{1}{2}+\frac{\alpha }{2}+ \frac{\lambda }{2})_{i_0}(\frac{1}{2}+\frac{\beta }{2}+ \frac{\lambda }{2})_{i_0}(\frac{3}{2}+\frac{\lambda }{2})_{i_1}(1+ \frac{\gamma}{2} +\frac{\lambda }{2})_{i_1}} \nonumber\\
&\times& \sum_{i_2=i_1}^{\infty}  \frac{(i_2+ 1+ \frac{\lambda }{2}) \left( i_2+ \frac{1}{2(1+a)}( \alpha + \beta -\delta +2+\lambda +a(\delta +\gamma +1+\lambda )) \right)+ \frac{q}{4(1+a)}}{(i_2+ \frac{3}{2}+ \frac{\lambda }{2})(i_2 +1+ \frac{\gamma }{2}+ \frac{\lambda }{2})} 
 \nonumber\\
&\times& \frac{(1+\frac{\alpha }{2}+ \frac{\lambda }{2})_{i_2}(1+\frac{\beta }{2}+ \frac{\lambda }{2})_{i_2}(2+\frac{\lambda }{2})_{i_1}(\frac{3}{2} +\frac{\gamma }{2}+ \frac{\lambda }{2})_{i_1}}{(1+\frac{\alpha }{2}+ \frac{\lambda }{2})_{i_1}(1+\frac{\beta }{2}+ \frac{\lambda }{2})_{i_1}(2+\frac{\lambda }{2})_{i_2}(\frac{3}{2} + \frac{\gamma}{2} +\frac{\lambda }{2})_{i_2}} \nonumber\\
&\times& \left. \sum_{i_3= i_2}^{\infty} \frac{(\frac{3}{2}+\frac{\alpha }{2}+ \frac{\lambda }{2})_{i_3}(\frac{3}{2}+\frac{\beta }{2}+ \frac{\lambda }{2})_{i_3}(\frac{5}{2}+\frac{\lambda }{2})_{i_2}(2 +\frac{\gamma }{2}+ \frac{\lambda }{2})_{i_2}}{(\frac{3}{2}+\frac{\alpha }{2}+ \frac{\lambda }{2})_{i_2}(\frac{3}{2}+\frac{\beta }{2}+ \frac{\lambda }{2})_{i_2}(\frac{5}{2}+\frac{\lambda }{2})_{i_3}(2 + \frac{\gamma}{2} +\frac{\lambda }{2})_{i_3}} z^{i_3} \right\} \eta ^3 \hspace{1cm} \label{eq:315d}
\end{eqnarray}
\end{subequations}  
Put $l=1$ in (\ref{eq:314}). Take the new (\ref{eq:314}) into (\ref{eq:315b}).
\begin{eqnarray}
 y_1(x)&=& c_0 x^{\lambda }\int_{0}^{1} dt_1\;t_1^{ \frac{1}{2}(-1+\lambda )} \int_{0}^{1} du_1\;u_1^{ \frac{1}{2}(-2+\gamma +\lambda )}   \frac{1}{2\pi i} \oint dv_1\;\frac{1}{v_1} \left( 1-\frac{1}{v_1} \right)^{-\frac{1}{2}(\alpha +1+\lambda  )} \left( 1- z v_1 (1-t_1)(1-u_1)\right)^{-\frac{1}{2}(\beta +1+\lambda  )}  \nonumber\\
&&\times  \left\{ \sum_{i_0=0}^{\infty }\left( \left( i_0+\frac{\lambda }{2} \right)\left( i_0 +\frac{1}{2(1+a)}\left( \alpha +\beta -\delta +\lambda +a(\delta +\gamma -1+\lambda )\right) \right) +\frac{q}{4(1+a)}  \right) \right.\nonumber\\
&&\times \left. \frac{(\frac{\alpha }{2}+\frac{\lambda }{2} )_{i_0} (\frac{\beta }{2}+\frac{\lambda }{2} )_{i_0}}{(1+\frac{\lambda }{2})_{i_0}(\frac{1}{2}+ \frac{\gamma }{2}+\frac{\lambda }{2})_{i_0}} \left( \frac{t_1 u_1 v_1}{(v_1-1)} \frac{z }{1-z v_1 (1-t_1)(1-u_1)}\right)^{i_0} \right\} \eta  \nonumber\\
&=&  c_0 x^{\lambda }\int_{0}^{1} dt_1\;t_1^{ \frac{1}{2}(-1+\lambda )} \int_{0}^{1} du_1\;u_1^{ \frac{1}{2}(-2+\gamma +\lambda )}   \frac{1}{2\pi i} \oint dv_1\;\frac{1}{v_1} \left( 1-\frac{1}{v_1} \right)^{-\frac{1}{2}(\alpha +1+\lambda  )} \left( 1- z v_1 (1-t_1)(1-u_1)\right)^{-\frac{1}{2}(\beta +1+\lambda  )}  \nonumber\\
&&\times  \left(    \overleftrightarrow {w}_{1,1}^{-\frac{\lambda }{2}} \left(  \overleftrightarrow {w}_{1,1} \partial _{ \overleftrightarrow {w}_{1,1}}\right) \overleftrightarrow {w}_{1,1}^{\frac{\lambda }{2}} \left(\overleftrightarrow {w}_{1,1} \partial _{ \overleftrightarrow {w}_{1,1}}+\frac{1}{2(1+a)}\left( \alpha  +\beta -\delta +\lambda +a(\delta +\gamma -1+\lambda )\right) \right)+\frac{q}{4(1+a)} \right)\nonumber\\
&&\times \left\{ \sum_{i_0=0}^{\infty } \frac{(\frac{\alpha }{2}+\frac{\lambda }{2})_{i_0} (\frac{\beta }{2}+\frac{\lambda }{2} )_{i_0}}{(1+\frac{\lambda }{2})_{i_0}(\frac{1}{2}+ \frac{\gamma }{2}+\frac{\lambda }{2})_{i_0}} \overleftrightarrow {w}_{1,1} ^{i_0} \right\} \eta \label{eq:316}
\end{eqnarray}
where 
\begin{equation}
 \overleftrightarrow {w}_{1,1} = \frac{t_1 u_1 v_1}{(v_1-1)}\; \frac{z}{1-z v_1 (1-t_1)(1-u_1)} \nonumber
\end{equation}
Put $l=2$ in (\ref{eq:314}). Take the new (\ref{eq:314}) into (\ref{eq:315c}). 
\begin{eqnarray}
y_2(x) &=& c_0 x^{\lambda } \int_{0}^{1} dt_2\;t_2^{ \frac{\lambda }{2}} \int_{0}^{1} du_2\;u_2^{ \frac{1}{2}(-1+\gamma +\lambda )}   \frac{1}{2\pi i} \oint dv_2\;\frac{1}{v_2} \left( 1-\frac{1}{v_2} \right)^{-\frac{1}{2}(\alpha +2+\lambda  )} \left( 1- z v_2 (1-t_2)(1-u_2)\right)^{-\frac{1}{2}(\beta +2+\lambda  )}  \nonumber\\
&&\times  \left(    \overleftrightarrow {w}_{2,2}^{-\frac{1}{2}(1+\lambda )} \left(  \overleftrightarrow {w}_{2,2} \partial _{ \overleftrightarrow {w}_{2,2}}\right) \overleftrightarrow {w}_{2,2}^{\frac{1}{2}(1+\lambda )} \left(\overleftrightarrow {w}_{2,2} \partial _{ \overleftrightarrow {w}_{2,2}}+\frac{1}{2(1+a)}\left( \alpha +\beta -\delta +1+\lambda +a(\delta +\gamma +\lambda )\right) \right)+\frac{q}{4(1+a)} \right)\nonumber\\
&&\times \left\{ \sum_{i_0=0}^{\infty }\frac{ (i_0+\frac{\lambda }{2})\left( i_0+\frac{1}{2(1+a)}\left( \alpha +\beta -\delta +\lambda +a(\delta +\gamma -1+\lambda )\right) \right) +\frac{q}{4(1+a)} }{(i_0+ \frac{1}{2}+ \frac{\lambda }{2})(i_0 + \frac{\gamma }{2}+ \frac{\lambda }{2})} \frac{( \frac{\alpha }{2}+\frac{\lambda }{2} )_{i_0} ( \frac{\beta }{2}+\frac{\lambda }{2} )_{i_0}}{(1+\frac{\lambda }{2})_{i_0}(\frac{1}{2}+\frac{\gamma }{2}+\frac{\lambda }{2})_{i_0}}\right.\nonumber\\
&&\times \left. \sum_{i_1=i_0}^{\infty } \frac{(\frac{1}{2}+\frac{\alpha }{2}+ \frac{\lambda }{2})_{i_1}(\frac{1}{2}+\frac{\beta }{2}+ \frac{\lambda }{2})_{i_1}(\frac{3}{2}+\frac{\lambda }{2})_{i_0}(1+\frac{\gamma }{2}+ \frac{\lambda }{2})_{i_0}}{(\frac{1}{2}+\frac{\alpha }{2}+ \frac{\lambda }{2})_{i_0}(\frac{1}{2}+\frac{\beta }{2}+ \frac{\lambda }{2})_{i_0}(\frac{3}{2}+\frac{\lambda }{2})_{i_1}(1+ \frac{\gamma}{2} +\frac{\lambda }{2})_{i_1}} \overleftrightarrow {w}_{2,2}^{i_1} \right\} \eta^2 \label{eq:317}
\end{eqnarray}
where
\begin{equation}
 \overleftrightarrow {w}_{2,2} = \frac{t_2 u_2 v_2}{(v_2-1)} \;\frac{z}{1-z v_2 (1-t_2)(1-u_2)}\nonumber
\end{equation} 
Put $l=1$ and $z =\overleftrightarrow {w}_{2,2}$ in (\ref{eq:314}). Take the new (\ref{eq:314}) into (\ref{eq:317}).
\begin{eqnarray}
y_2(x) &=& c_0 x^{\lambda } \int_{0}^{1} dt_2\;t_2^{ \frac{\lambda }{2}} \int_{0}^{1} du_2\;u_2^{ \frac{1}{2}(-1+\gamma +\lambda )}   \frac{1}{2\pi i} \oint dv_2\;\frac{1}{v_2} \left( 1-\frac{1}{v_2} \right)^{-\frac{1}{2}(\alpha +2+\lambda  )} \left( 1- z v_2 (1-t_2)(1-u_2)\right)^{-\frac{1}{2}(\beta +2+\lambda  )}  \nonumber\\
&&\times  \left(    \overleftrightarrow {w}_{2,2}^{-\frac{1}{2}(1+\lambda )} \left(  \overleftrightarrow {w}_{2,2} \partial _{ \overleftrightarrow {w}_{2,2}}\right) \overleftrightarrow {w}_{2,2}^{\frac{1}{2}(1+\lambda )} \left(\overleftrightarrow {w}_{2,2} \partial _{ \overleftrightarrow {w}_{2,2}}+\frac{1}{2(1+a)}\left( \alpha +\beta -\delta +1+\lambda +a(\delta +\gamma +\lambda )\right) \right)+\frac{q}{4(1+a)} \right)\nonumber\\
&&\times \int_{0}^{1} dt_1\;t_1^{ \frac{1}{2}(-1+\lambda )} \int_{0}^{1} du_1\;u_1^{ \frac{1}{2}(-2+\gamma +\lambda )}   \frac{1}{2\pi i} \oint dv_1\;\frac{1}{v_1} \left( 1-\frac{1}{v_1} \right)^{-\frac{1}{2}(\alpha +1+\lambda  )} \left( 1- \overleftrightarrow {w}_{2,2} v_1 (1-t_1)(1-u_1)\right)^{-\frac{1}{2}(\beta +1+\lambda  )}  \nonumber\\
&&\times  \left(    \overleftrightarrow {w}_{1,2}^{-\frac{\lambda }{2}} \left(  \overleftrightarrow {w}_{1,2} \partial _{ \overleftrightarrow {w}_{1,2}}\right) \overleftrightarrow {w}_{1,2}^{\frac{\lambda }{2}} \left(\overleftrightarrow {w}_{1,2} \partial _{ \overleftrightarrow {w}_{1,2}}+\frac{1}{2(1+a)}\left( \alpha +\beta -\delta +\lambda  +a(\delta +\gamma -1+\lambda )\right) \right)+\frac{q}{4(1+a)} \right)\nonumber\\
&&\times \left\{ \sum_{i_0=0}^{\infty } \frac{(\frac{\alpha }{2}+\frac{\lambda }{2})_{i_0} (\frac{\beta }{2}+\frac{\lambda }{2})_{i_0}}{(1+\frac{\lambda }{2})_{i_0}(\frac{1}{2}+ \frac{\gamma}{2} +\frac{\lambda }{2})_{i_0}} \overleftrightarrow {w}_{1,2} ^{i_0} \right\} \eta^2 \label{eq:318}
\end{eqnarray}
where
\begin{equation}
 \overleftrightarrow {w}_{1,2}=\frac{t_1 u_1 v_1}{(v_1-1)}\; \frac{\overleftrightarrow {w}_{2,2}}{1-\overleftrightarrow {w}_{2,2} v_1 (1-t_1)(1-u_1)}\nonumber
\end{equation} 
By using similar process for the previous cases of integral forms of $y_1(x)$ and $y_2(x)$, the integral form of sub-power series expansion of $y_3(x)$ is
\begin{eqnarray}
y_3(x)&=& c_0 x^{\lambda } \int_{0}^{1} dt_3\;t_3^{ \frac{1}{2}(1+\lambda )} \int_{0}^{1} du_3\;u_3^{ \frac{1}{2}( \gamma +\lambda )}   \frac{1}{2\pi i} \oint dv_3\;\frac{1}{v_3} \left( 1-\frac{1}{v_3} \right)^{-\frac{1}{2}(\alpha +3+\lambda  )} \left( 1- z v_3 (1-t_3)(1-u_3)\right)^{-\frac{1}{2}(\beta +3+\lambda  )}  \nonumber\\
&&\times  \left(    \overleftrightarrow {w}_{3,3}^{-\frac{1}{2}(2+\lambda )} \left(  \overleftrightarrow {w}_{3,3} \partial _{ \overleftrightarrow {w}_{3,3}}\right) \overleftrightarrow {w}_{3,3}^{\frac{1}{2}(2+\lambda )} \bigg(\overleftrightarrow {w}_{3,3} \partial _{ \overleftrightarrow {w}_{3,3}}\right.\nonumber\\
&&+\left. \frac{1}{2(1+a)}\left( \alpha +\beta -\delta +2+\lambda +a(\delta +\gamma +1+\lambda )\right) \bigg)+\frac{q}{4(1+a)} \right)\nonumber\\
&&\times  \int_{0}^{1} dt_2\;t_2^{ \frac{\lambda }{2}} \int_{0}^{1} du_2\;u_2^{ \frac{1}{2}(-1+\gamma +\lambda )}   \frac{1}{2\pi i} \oint dv_2\;\frac{1}{v_2} \left( 1-\frac{1}{v_2} \right)^{-\frac{1}{2}(\alpha +2+\lambda  )} \left( 1- \overleftrightarrow {w}_{3,3} v_2 (1-t_2)(1-u_2)\right)^{-\frac{1}{2}(\beta +2+\lambda  )}  \nonumber\\
&&\times  \left(    \overleftrightarrow {w}_{2,3}^{-\frac{1}{2}(1+\lambda )} \left(  \overleftrightarrow {w}_{2,3} \partial _{ \overleftrightarrow {w}_{2,3}}\right) \overleftrightarrow {w}_{2,3}^{\frac{1}{2}(1+\lambda )} \bigg(\overleftrightarrow {w}_{2,3} \partial _{ \overleftrightarrow {w}_{2,3}}\right. \nonumber\\
&&+ \left. \frac{1}{2(1+a)}\left(  \alpha +\beta -\delta +1+\lambda +a(\delta +\gamma +\lambda )\right) \bigg)+\frac{q}{4(1+a)} \right)\nonumber\\
&&\times \int_{0}^{1} dt_1\;t_1^{ \frac{1}{2}(-1+\lambda )} \int_{0}^{1} du_1\;u_1^{ \frac{1}{2}(-2+\gamma +\lambda )}   \frac{1}{2\pi i} \oint dv_1\;\frac{1}{v_1} \left( 1-\frac{1}{v_1} \right)^{-\frac{1}{2}(\alpha +1+\lambda  )} \left( 1- \overleftrightarrow {w}_{2,3} v_1 (1-t_1)(1-u_1)\right)^{-\frac{1}{2}(\beta +1+\lambda  )}  \nonumber\\
&&\times  \left(    \overleftrightarrow {w}_{1,3}^{-\frac{\lambda }{2}} \left(  \overleftrightarrow {w}_{1,3} \partial _{ \overleftrightarrow {w}_{1,3}}\right) \overleftrightarrow {w}_{1,3}^{\frac{\lambda }{2}} \left(\overleftrightarrow {w}_{1,3} \partial _{ \overleftrightarrow {w}_{1,3}}+\frac{1}{2(1+a)}\left( \alpha +\beta -\delta +\lambda +a(\delta +\gamma -1+\lambda )\right) \right)+\frac{q}{4(1+a)} \right)\nonumber\\
&&\times \left\{ \sum_{i_0=0}^{\infty } \frac{(\frac{\alpha }{2}+\frac{\lambda }{2})_{i_0} (\frac{\beta }{2}+\frac{\lambda }{2})_{i_0}}{(1+\frac{\lambda }{2})_{i_0}(\frac{1}{2}+ \frac{\gamma}{2} +\frac{\lambda }{2})_{i_0}} \overleftrightarrow{w}_{1,3}^{i_0} \right\} \eta ^3 \label{eq:319}
\end{eqnarray}
where
%\Large
\begin{equation}
\begin{cases} \overleftrightarrow {w}_{3,3} = \frac{t_3 u_3 v_3}{(v_3-1)}\; \frac{z}{1- z v_3 (1-t_3)(1-u_3)}  \cr
\overleftrightarrow {w}_{2,3} = \frac{t_2 u_2 v_2}{(v_2-1)}\; \frac{\overleftrightarrow{w}_{3,3} }{1- \overleftrightarrow{w}_{3,3} v_2 (1-t_2)(1-u_2)} \cr
\overleftrightarrow w_{1,3}= \frac{t_1 u_1 v_1}{(v_1-1)}\frac{\overleftrightarrow w_{2,3} }{1-\overleftrightarrow w_{2,3} v_1 (1-t_1)(1-u_1)}
\end{cases}
\nonumber
\end{equation}
By repeating this process for all higher terms of integral forms of sub-summation $y_m(x)$ terms where $m \geq 4$, we obtain every integral forms of $y_m(x)$ terms. 
Since we substitute (\ref{eq:315a}), (\ref{eq:316}), (\ref{eq:318}), (\ref{eq:319}) and including all integral forms of $y_m(x)$ terms where $m \geq 4$ into (\ref{eq:51}), we obtain (\ref{eq:46}).\footnote{Or replace the finite summation with an interval $[0,\alpha _0]$ by infinite summation with an interval  $[0,\infty ]$ in (\ref{eq:39}). Replace $\alpha _0$, $\alpha _{n-k}$ and $\alpha _{n-k-1}$ by $-\frac{1}{2}(\alpha +\lambda )$, $-\frac{1}{2}(\alpha +n-k+\lambda )$ and $-\frac{1}{2}(\alpha +n-k-1+\lambda )$ into the new (\ref{eq:39}). Its solution is also equivalent to (\ref{eq:46}).} 
\qed
\end{pot}
Put $c_0$= 1 as $\lambda $=0 for the first kind of independent solutions of Heun equation and $\displaystyle{ c_0= \left(a^{-1}(1+a)\right)^{1-\gamma }}$ as $\lambda = 1-\gamma $ for the second one in (\ref{eq:46}). 
\begin{rmk}
The integral representation of Heun equation of the first kind for infinite series about $x=0$ using 3TRF is
\begin{eqnarray}
 y(x)&=& HF_{\alpha , \beta }\left( \eta = \frac{(1+a)}{a} x ; z= -\frac{1}{a} x^2 \right) \nonumber\\
&=& _2F_1 \left( \frac{\alpha }{2}, \frac{\beta }{2};\frac{1}{2}+\frac{\gamma }{2}; z \right) + \sum_{n=1}^{\infty } \Bigg\{\prod _{k=0}^{n-1} \Bigg\{ \int_{0}^{1} dt_{n-k}\;t_{n-k}^{\frac{1}{2}(n-k-2)} \int_{0}^{1} du_{n-k}\;u_{n-k}^{\frac{1}{2}(n-k-3+\gamma )} \nonumber\\
&&\times\frac{1}{2\pi i}  \oint dv_{n-k} \frac{1}{v_{n-k}} \left( 1-\frac{1}{v_{n-k}}\right)^{-\frac{1}{2}(n-k+\alpha )}  \left( 1- \overleftrightarrow {w}_{n-k+1,n}v_{n-k}(1-t_{n-k})(1-u_{n-k})\right)^{-\frac{1}{2}(n-k+\beta )}\nonumber\\
&&\times  \left( \overleftrightarrow {w}_{n-k,n}^{-\frac{1}{2}(n-k-1)}\left(  \overleftrightarrow {w}_{n-k,n} \partial _{ \overleftrightarrow {w}_{n-k,n}}\right) \overleftrightarrow {w}_{n-k,n}^{\frac{1}{2}(n-k-1)}\left( \overleftrightarrow {w}_{n-k,n} \partial _{ \overleftrightarrow {w}_{n-k,n}}  + \Omega _{n-k-1}^{(I)} \right) +Q \right) \Bigg\}\nonumber\\
&&\times _2F_1 \left( \frac{\alpha }{2}, \frac{\beta }{2};\frac{1}{2}+\frac{\gamma }{2}; \overleftrightarrow {w}_{1,n} \right) \Bigg\} \eta ^n \label{eq:47}
\end{eqnarray}
where
\begin{equation}
\begin{cases} 
\Omega _{n-k-1}^{(I)} = \frac{1}{2(1+a)}(\alpha +\beta -\delta +n-k-1 +a(\delta +\gamma +n-k-2)) \cr
Q= \frac{q}{4(1+a)}
\end{cases}\nonumber %\label{eq:8}
\end{equation}
\end{rmk}
\begin{rmk}
The integral representation of Heun equation of the second kind for infinite series about $x=0$ using 3TRF is
\begin{eqnarray}
y(x)&=& HS_{\alpha , \beta }\left( \eta = \frac{(1+a)}{a} x ; z= -\frac{1}{a} x^2 \right) \nonumber\\
&=& z^{\frac{1}{2}(1-\gamma )} \Bigg\{\; _2F_1 \left(\frac{\alpha }{2}+\frac{1}{2}-\frac{\gamma }{2}, \frac{\beta }{2}+\frac{1}{2}-\frac{\gamma }{2};\frac{3}{2}-\frac{\gamma }{2}; z \right) \nonumber\\
&&+ \sum_{n=1}^{\infty } \Bigg\{\prod _{k=0}^{n-1} \Bigg\{ \int_{0}^{1} dt_{n-k}\;t_{n-k}^{\frac{1}{2}(n-k-1-\gamma )} \int_{0}^{1} du_{n-k}\;u_{n-k}^{\frac{1}{2}(n-k-2)}  \nonumber\\
&&\times \frac{1}{2\pi i}  \oint dv_{n-k} \frac{1}{v_{n-k}} \left( 1-\frac{1}{v_{n-k}}\right)^{-\frac{1}{2}(n-k+1+\alpha -\gamma )} \left( 1- \overleftrightarrow {w}_{n-k+1,n}v_{n-k}(1-t_{n-k})(1-u_{n-k})\right)^{-\frac{1}{2}(n-k+1+\beta-\gamma )}\nonumber\\
&&\times  \left( \overleftrightarrow {w}_{n-k,n}^{-\frac{1}{2}(n-k-\gamma )}\left(  \overleftrightarrow {w}_{n-k,n} \partial _{ \overleftrightarrow {w}_{n-k,n}}\right) \overleftrightarrow {w}_{n-k,n}^{\frac{1}{2}(n-k-\gamma )}\left( \overleftrightarrow {w}_{n-k,n} \partial _{ \overleftrightarrow {w}_{n-k,n}} + \Omega _{n-k-1}^{(I)} \right) +Q\right) \Bigg\}\nonumber\\
&&\times _2F_1 \left(\frac{\alpha }{2}+\frac{1}{2}-\frac{\gamma }{2}, \frac{\beta }{2}+\frac{1}{2}-\frac{\gamma }{2};\frac{3}{2}-\frac{\gamma }{2}; \overleftrightarrow {w}_{1,n}\right)  \Bigg\} \eta ^n \Bigg\}\label{eq:48}
\end{eqnarray}
where
\begin{equation}
\begin{cases} 
\Omega _{n-k-1}^{(I)} = \frac{1}{2(1+a)}(\alpha +\beta -\delta -\gamma +n -k +a(\delta +n-k-1 )) \cr
Q= \frac{q}{4(1+a)}
\end{cases}\nonumber %\label{eq:8}
\end{equation}
\end{rmk} 
\section{Summary} 
In my previous paper I show the power series expansion in closed forms of Heun function (infinite series and polynomial) including all higher terms of $A_n$'s. In this paper I derived the integral representation of Heun function and its asymptotic behaviors including all higher terms of $A_n$'s by applying three term recurrence formula.\cite{Chou2012}

 As we see the power series expansions of Heun function for all cases of infinite series and polynomial, denominators and numerators in all $B_n$ terms arise with Pochhammer symbol: the meaning of this is that the analytic solutions of Heun function can be described as hypergoemetric functions in a strict mathematical way. We can express representations in closed form integrals in an easy way since we have power series expansions with Pochhammer symbols in numerators and denominators. We can transform Heun function into all other well-known special functions with two recursive coefficients because a $_2F_1$ function recurs in each of sub-integral forms of Heun function.

Since we get the integral forms of power series expansions in Heun function, we are able to obtain generating functions of it. The generating functions are really helpful in order to derive orthogonal relations, recursion relations and expectation values of physical quantities.
\section{Conclusion}
There are four kinds of confluent forms of Heun equation \cite{Slavy2000,Ronv1995,Slav1998,Wolf1998,Buhr1994} such as the Confluent Heun \cite{Hort2007,Kokk1998,Meix1980}, Doubly-Confluent Heun \cite{Dola2007}, Biconfluent Heun \cite{ChCa2012} and Triconfluent Heun equations \cite{Giac2008}.  We can derive these four confluent forms from Heun equation by combining two or more regular singularities to each other to take form an irregular singularity. Its process, converting Heun equation to other confluent forms, is similar to deriving of confluent hypergeometric equation from the hypergeometric equation. 

We can obtain the analytic solutions of these four confluent forms of Heun function by replacing independent variable $x$ and changing coefficients. Or we are able to have power series expansion, integral forms and generation functions of these four second ordinary differential equations by using three term recurrence formula directly.\cite{Chou2012}: in my future paper I will construct the power series expansion, its integral forms and generating functions of these four confluent forms of Heun equations.

We can apply an integral formalism and power series expansion of Heun functions in many modern physical areas. For example, the Heun functions appear in the solution of Schr$\ddot{\mbox{o}}$dinger equation to the quadratic potentials with inverse even powers of two, four and six.\cite{Figu2005} The solution of the Schr$\ddot{\mbox{o}}$dinger equation to symmetric double Morse potential also need these function.\cite{Figu2007} Also, in ``The stark effect from the point of view of Schr$\ddot{\mbox{o}}$dinger quantum theory"\cite{Epst1926}, the author considers the Schr$\ddot{\mbox{o}}$dinger equation for the hydrogen atom in a constant electric field of magnitude $E$ in the $z$ direction. The Schr$\ddot{\mbox{o}}$dinger equation results into two separated equations by using parabolic coordinates (see (7), (10) in Ref.\cite{Epst1926}). These two equations are of the Biconfluent Heun form.
Biconfluent Heun equation can be obtained from Heun equation by replacing independent variables and changing coefficients. And as we put the new variables and coefficients into integral forms of Heun function on the above for the case of polynomials and infinite series, we might be possible to construct power series expansions and integral forms in closed forms of Biconfluent Heun function. After then, it might be possible to obtain specific eigenvalues for the entire region of $r$ by using the power series expansion of Biconfluent Heun equation. Using the integral forms of Biconfluent Heun equation, it might be possible to construct the normalized wave functions and expectation values of any physical quantity as we want.

In ``The ionized hydrogen molecule"\cite{Wils1928}, the author consider the hydrogen-molecule ion or dihydrogen cation $H_2^+$ in the Born-Oppenheimer approximation. He obtains two individually Confluent Heun equations using the prolate spheroidal coordinates (see (1), (2) in Ref.\cite{Wils1928}). By replacing independent variables and coefficients in Heun equation, we can construct Confluent Heun equation. We might be possible to build power series expansions and integral forms in closed forms of Confluent Heun function putting the new variables and coefficients into integral forms of Heun function on the above for the case of polynomials and infinite series.  
In general, most of wave-functions in physics are quantized with specific eigenvalues. So all solutions on the above examples might be quantized with certain eigenvalues. It means that its analytic wave-functions have polynomial expansions. And there are infinite numbers of eigenvalues surprisingly because of its three term recurrence form\cite{Chou2012}. Also, we can transform representations in the form of integrals in Heun function to other well-known special functions analytically. Because as we see integral forms of Heun function, these functions include $_2F_1$ Hypergeometric function in itself on (\ref{eq:41}), (\ref{eq:42}), (\ref{eq:44}), (\ref{eq:45}), (\ref{eq:47}), (\ref{eq:48}).
\subsection*{Series ``Special functions and three term recurrence formula (3TRF)''} 

This paper is 4th out of 10.
\vspace{3mm}

1. ``Approximative solution of the spin free Hamiltonian involving only scalar potential for the $q-\bar{q}$ system'' \cite{ChCa2012}--in order to solve the spin-free Hamiltonian with light quark masses we are led to develop a totally new kind of special function theory in mathematics that generalize all existing theories of confluent hypergeometric types. We call it the Grand Confluent Hypergeometric Function. Our new solution produces previously unknown extra hidden quantum numbers relevant for the description of supersymmetry and for generating new mass formulas.
\vspace{3mm}

2. ``Generalization of the three-term recurrence formula and its applications'' \cite{Chou2012}--generalize the three term recurrence formula in the linear differential equation.  Obtain the exact solution of the three term recurrence for polynomials and infinite series.
\vspace{3mm}

3. ``The analytic solution for the power series expansion of Heun function'' \cite{Chou2012c}--apply the three term recurrence formula to the power series expansion in closed forms of Heun function (infinite series and polynomials) including all higher terms of $A_n$'s.
\vspace{3mm}

4. ``Asymptotic behavior of Heun function and its integral formalism'' \cite{Chou2012d}--apply the three term recurrence formula, derive the integral formalism, and analyze the asymptotic behavior of Heun function (including all higher terms of $A_n$'s). 
\vspace{3mm}

5. ``The power series expansion of Mathieu function and its integral formalism'' \cite{Chou2012e}--apply the three term recurrence formula, and analyze the power series expansion of Mathieu function and its integral forms.  
\vspace{3mm}

6. ``Lame equation in the algebraic form'' \cite{Chou2012f}--apply the three term recurrence formula, and analyze the power series expansion of Lame function in the algebraic form and its integral forms.
\vspace{3mm}

7. ``Power series and integral forms of Lame equation in Weierstrass's form'' \cite{Chou2012g}--apply the three term recurrence formula, and derive the power series expansion of Lame function in Weierstrass's form and its integral forms. 
\vspace{3mm}

8. ``The generating functions of Lame equation in Weierstrass's form'' \cite{Chou2012h}--derive the generating functions of Lame function in Weierstrass's form (including all higher terms of $A_n$'s).  Apply integral forms of Lame functions in Weierstrass's form.
\vspace{3mm}

9. ``Analytic solution for grand confluent hypergeometric function'' \cite{Chou2012i}--apply the three term recurrence formula, and formulate the exact analytic solution of grand confluent hypergeometric function (including all higher terms of $A_n$'s). Replacing $\mu $ and $\varepsilon \omega $ by 1 and $-q$ transforms the grand confluent hypergeometric function into the Biconfluent Heun function.
\vspace{3mm}

10. ``The integral formalism and the generating function of grand confluent hypergeometric function'' \cite{Chou2012j}--apply the three term recurrence formula, and construct an integral formalism and a generating function of grand confluent hypergeometric function (including all higher terms of $A_n$'s). 

\appendix
\section{Power series expansion of 192 Heun functions}
In this paper the fundamental power series expansion and its integral forms of Heun function about $x=0$ is constructed analytically. The singularity parameter $a \ne 0 $ decides various ranges of an independent variable $x$ according to asymptotic behaviors of a Heun function.

A machine-generated list of 192 (isomorphic to the Coxeter group of the Coxeter diagram $D_4$) local solutions of the Heun equation was obtained by Robert S. Maier(2007) \cite{Maie2007}. We can obtain power series expansion in closed form, asymptotic behaviors and its integral forms of all 192 local solutions of the Heun equation analytically by using three term recurrence formula \cite{Chou2012}.
We derive the analytic solutions of nine out of the 192 local solution of Heun function in Table 2 \cite{Maie2007}.
\subsection{ ${\displaystyle (1-x)^{1-\delta } Hl(a, q - (\delta  - 1)\gamma a; \alpha - \delta  + 1, \beta - \delta + 1, \gamma ,2 - \delta ; x)}$ }
\subsubsection{Polynomial which makes $B_n$ term terminated}
\underline {(1) The case of $\alpha = -2 \alpha _i-i +\delta -1$ and $\beta \ne -2 \beta _i -i+\delta -1$ where $i, \alpha _i, \beta _i$ = $0,1,2,\cdots$.}
\vspace{1mm}

Replace coefficients $q$, $\alpha$, $\beta$, $\delta$, $c_0$ and $\lambda $ by $q - (\delta - 1)\gamma a $, $\alpha - \delta  + 1 $, $\beta - \delta + 1$, $2 - \delta$, 1 and zero into (\ref{eq:36}). Multiply $(1-x)^{1-\delta }$ and the new (\ref{eq:36}) together.
\begin{eqnarray}
& &(1-x)^{1-\delta } y(x)\nonumber\\
&=& (1-x)^{1-\delta } Hl\left(a, q - (\delta  - 1)\gamma a; \alpha - \delta  + 1, \beta - \delta + 1, \gamma ,2 - \delta ; x\right)\nonumber\\
&=& (1-x)^{1-\delta } \left\{\sum_{i_0=0}^{\alpha _0} \frac{(-\alpha _0)_{i_0} (\frac{\beta +1-\delta }{2})_{i_0}}{(1)_{i_0}(\frac{1}{2}+ \frac{\gamma}{2})_{i_0}} z^{i_0} \right. \nonumber\\
&&+ \left\{ \sum_{i_0=0}^{\alpha _0}\frac{i_0 \left( i_0+ \Gamma _0^{(S)} \right)+ Q}{(i_0+ \frac{1}{2})(i_0 + \frac{\gamma }{2})}\frac{(-\alpha _0)_{i_0} (\frac{\beta +1-\delta}{2})_{i_0}}{(1)_{i_0}(\frac{1}{2}+ \frac{\gamma}{2})_{i_0}} \sum_{i_1=i_0}^{\alpha _1} \frac{(-\alpha _1)_{i_1}(1+\frac{\beta -\delta}{2})_{i_1}(\frac{3}{2})_{i_0}(1+\frac{\gamma }{2})_{i_0}}{(-\alpha _1)_{i_0}(1+\frac{\beta -\delta}{2})_{i_0}(\frac{3}{2})_{i_1}(1+ \frac{\gamma}{2})_{i_1}} z^{i_1} \right\} \eta \nonumber\\
&&+ \sum_{n=2}^{\infty } \left\{ \sum_{i_0=0}^{\alpha _0}\frac{i_0\left( i_0+ \Gamma _0^{(S)} \right)+ Q}{(i_0+ \frac{1}{2})(i_0 + \frac{\gamma }{2})} \frac{(-\alpha _0)_{i_0} (\frac{\beta +1-\delta}{2})_{i_0}}{(1)_{i_0}(\frac{1}{2}+ \frac{\gamma}{2})_{i_0}}\right.\nonumber\\
&&\times \prod _{k=1}^{n-1} \left\{ \sum_{i_k=i_{k-1}}^{\alpha _k} \frac{(i_k+\frac{k}{2}) \left( i_k+ \Gamma _k^{(S)}\right)+ Q}{(i_k+ \frac{k}{2}+\frac{1}{2})(i_k +\frac{k}{2}+\frac{\gamma }{2})}  \frac{(-\alpha _k)_{i_k}(\frac{k}{2}+\frac{\beta+1-\delta }{2})_{i_k}(1+ \frac{k}{2})_{i_{k-1}}(\frac{1}{2}+\frac{k}{2}+\frac{\gamma }{2})_{i_{k-1}}}{(-\alpha _k)_{i_{k-1}}(\frac{k}{2}+\frac{\beta+1-\delta }{2})_{i_{k-1}}(1+\frac{k}{2})_{i_k}(\frac{1}{2}+ \frac{k}{2}+ \frac{\gamma}{2})_{i_k}}\right\} \nonumber\\
&&\times \left.\left.\sum_{i_n= i_{n-1}}^{\alpha _n} \frac{(-\alpha _n)_{i_n}(\frac{n}{2}+\frac{\beta+1-\delta }{2})_{i_n}(1+ \frac{n}{2})_{i_{n-1}}(\frac{1}{2}+\frac{n}{2}+\frac{\gamma }{2})_{i_{n-1}}}{(-\alpha _n)_{i_{n-1}}(\frac{n}{2}+\frac{\beta+1-\delta }{2})_{i_{n-1}}(1+\frac{n}{2})_{i_n}(\frac{1}{2}+ \frac{n}{2}+ \frac{\gamma}{2})_{i_n}} z^{i_n} \right\} \eta ^n \right\} \label{eq:49}
\end{eqnarray}
where
\begin{equation}
\begin{cases} z = -\frac{1}{a}x^2 \cr
\eta = \frac{(1+a)}{a} x \cr
\alpha _i\leq \alpha _j \;\;\mbox{only}\;\mbox{if}\;i\leq j\;\;\mbox{where}\;i,j= 0,1,2,\cdots
\end{cases}\nonumber %\label{eq:37}
\end{equation}
and
\begin{equation}
\begin{cases} 
\Gamma _0^{(S)}=  \frac{1}{2(1+a)}(-2\alpha _0+ \beta -1 +a(-\delta +\gamma +1))\cr
\Gamma _k^{(S)}=  \frac{1}{2(1+a)}(-2\alpha _k+ \beta -1 +a(-\delta +\gamma +k+1)) \cr
Q= \frac{q-(\delta -1)\gamma a}{4(1+a)}
\end{cases}\nonumber %\label{eq:8}
\end{equation}
For the minimum value of Heun equation for a polynomial which makes $B_n$ term terminated about $x=0 $, put $\alpha _0=\alpha _1=\alpha _2=\cdots=0$ in (\ref{eq:49}).
\begin{eqnarray}
& &(1-x)^{1-\delta } y(x)\nonumber\\
&=& (1-x)^{1-\delta } Hl\left( a, q - (\delta  - 1)\gamma a; \alpha - \delta  + 1, \beta - \delta + 1, \gamma ,2 - \delta ; x\right)\nonumber\\
&=& (1-x)^{1-\delta } \; _2F_1\left(  \frac{\Lambda_1-\sqrt{ \Lambda _1^2-4a\Omega_1}}{2a}, \frac{\Lambda _1+\sqrt{\Lambda _1^2- 4a\Omega _1}}{2a}; \gamma ; x \right) \nonumber
\end{eqnarray}
where $\Lambda_1= \beta -1+a(\gamma -\delta +1)$ and $\Omega_1=q+a\gamma (1-\delta )$.
It tells us that Heun polynomials in which makes $B_n$ term terminated, for fixed values of $\alpha $, require $|x| < 1$ for the convergence of the radius.

\underline {(2) The case of $\alpha = -2 \alpha _i-i +\delta -1 $ and $\beta = -2 \beta _i -i+\delta -1 $ only if $\alpha _i \leq \beta _i$.}

Put $\beta = -2 \beta _i -i+\delta -1 $ in (\ref{eq:49}).  
\begin{eqnarray}
& &(1-x)^{1-\delta } y(x)\nonumber\\
&=& (1-x)^{1-\delta } Hl\left(a, q - (\delta  - 1)\gamma a; \alpha - \delta  + 1, \beta - \delta + 1, \gamma ,2 - \delta ; x\right)\nonumber\\
&=& (1-x)^{1-\delta } \left\{ \sum_{i_0=0}^{\alpha _0} \frac{(-\alpha _0)_{i_0} (-\beta _0)_{i_0}}{(1)_{i_0}(\frac{1}{2}+ \frac{\gamma}{2})_{i_0}} z^{i_0} \right.\nonumber\\
&&+ \left\{ \sum_{i_0=0}^{\alpha _0}\frac{i_0 \left( i_0+ \Gamma _0^{(B)}\right)+ Q}{(i_0+ \frac{1}{2})(i_0 + \frac{\gamma }{2})}  \frac{(-\alpha _0)_{i_0} (-\beta _0)_{i_0}}{(1)_{i_0}(\frac{1}{2}+ \frac{\gamma}{2})_{i_0}} \sum_{i_1=i_0}^{\alpha _1} \frac{(-\alpha _1)_{i_1}(-\beta _1)_{i_1}(\frac{3}{2})_{i_0}(1+\frac{\gamma }{2})_{i_0}}{(-\alpha _1)_{i_0}(-\beta _1)_{i_0}(\frac{3}{2})_{i_1}(1+ \frac{\gamma}{2})_{i_1}} z^{i_1} \right\} \eta \nonumber\\
&&+ \sum_{n=2}^{\infty } \left\{ \sum_{i_0=0}^{\alpha _0}\frac{i_0\left( i_0+ \Gamma _0^{(B)}\right)+ Q}{(i_0+ \frac{1}{2})(i_0 + \frac{\gamma }{2})} \frac{(-\alpha _0)_{i_0} (-\beta _0)_{i_0}}{(1)_{i_0}(\frac{1}{2}+ \frac{\gamma}{2})_{i_0}}\right.\nonumber\\
&&\times \prod _{k=1}^{n-1} \left\{ \sum_{i_k=i_{k-1}}^{\alpha _k} \frac{(i_k+\frac{k}{2}) \left( i_k+ \Gamma _k^{(B)}\right)+ Q}{(i_k+ \frac{k}{2}+\frac{1}{2})(i_k +\frac{k}{2}+\frac{\gamma }{2})} \frac{(-\alpha _k)_{i_k}(-\beta _k)_{i_k}(1+ \frac{k}{2})_{i_{k-1}}(\frac{1}{2}+\frac{k}{2}+\frac{\gamma }{2})_{i_{k-1}}}{(-\alpha _k)_{i_{k-1}}(-\beta _k)_{i_{k-1}}(1+\frac{k}{2})_{i_k}(\frac{1}{2}+ \frac{k}{2}+ \frac{\gamma}{2} )_{i_k}}\right\}\nonumber\\
&&\times \left.\left.\sum_{i_n= i_{n-1}}^{\alpha _n} \frac{(-\alpha _n)_{i_n}(-\beta _n)_{i_n}(1+ \frac{n}{2})_{i_{n-1}}(\frac{1}{2}+\frac{n}{2}+\frac{\gamma }{2})_{i_{n-1}}}{(-\alpha _n)_{i_{n-1}}(-\beta _n)_{i_{n-1}}(1+\frac{n}{2})_{i_n}(\frac{1}{2}+ \frac{n}{2}+ \frac{\gamma}{2})_{i_n}} z^{i_n} \right\} \eta ^n \right\}  \label{eq:50}
\end{eqnarray}
where
\begin{equation}
\begin{cases} 
\Gamma _0^{(B)}=  \frac{1}{2(1+a)}(-2\alpha _0+ -2\beta _0 -2  +\delta +a(-\delta +\gamma +1))\cr
\Gamma _k^{(B)}=  \frac{1}{2(1+a)}(-2\alpha _k+ -2\beta _k -k -2 +\delta +a(-\delta +\gamma +k+1)) \cr
Q= \frac{q-(\delta -1)\gamma a}{4(1+a)}
\end{cases}\nonumber %\label{eq:8}
\end{equation}
For the minimum value of Heun equation for a polynomial which makes $B_n$ term terminated about $x=0 $, put $\alpha _0=\alpha _1=\alpha _2=\cdots=0$ and $\beta _0=\beta _1=\beta _2=\cdots=0$ in (\ref{eq:50}).
\begin{eqnarray}
& &(1-x)^{1-\delta } y(x)\nonumber\\
&=& (1-x)^{1-\delta } Hl\left( a, q - (\delta  - 1)\gamma a; \alpha - \delta  + 1, \beta - \delta + 1, \gamma ,2 - \delta ; x\right)\nonumber\\
&=& (1-x)^{1-\delta } \; _2F_1\left( \frac{ \Lambda _2-\sqrt{ \Lambda _2^2-4(a-1)\Omega _2}}{2(a-1)}, \frac{ \Lambda _2+\sqrt{ \Lambda _2^2-4(a-1)\Omega _2}}{2(a-1)}; \gamma ; \frac{a-1}{a}x \right) \label{aa:1}
\end{eqnarray}
where $\Lambda_2= \delta -2+a(\gamma -\delta +1)$ and $\Omega_2=q+a\gamma (1-\delta )$.
It tells us that Heun polynomials in which makes $B_n$ term terminated, for fixed values of $\alpha $ and $\beta $, require $\left|\frac{a-1}{a}x\right| < 1$ for the convergence of the radius.

For the special case, if $x=\frac{a}{a-1}$ and $Re\left( \frac{\gamma +\delta -2+a(1-\delta )}{1-a}\right)>0 $ in (\ref{aa:1}),
\begin{eqnarray}
& &(1-a)^{ \delta -1} y\left( \frac{a}{a-1}\right)\nonumber\\
&=& (1-a)^{ \delta -1} Hl\left( a, q - (\delta  - 1)\gamma a; \alpha - \delta  + 1, \beta - \delta + 1, \gamma ,2 - \delta ; x=\frac{a}{a-1}\right)\nonumber\\
&=& (1-a)^{ \delta -1} \frac{ \Gamma \left( \gamma \right) \Gamma \left( \gamma -\frac{\Lambda _2}{a-1} \right)}{\Gamma \left( \gamma - \frac{ \Lambda _2-\sqrt{ \Lambda _2^2-4(a-1)\Omega _2}}{2(a-1)}\right) \Gamma \left( \gamma -\frac{ \Lambda _2+\sqrt{ \Lambda _2^2-4(a-1)\Omega _2}}{2(a-1)} \right)}   \nonumber
\end{eqnarray}
\subsubsection{Infinite series}
Replace coefficients $q$, $\alpha$, $\beta$, $\delta$, $c_0$ and $\lambda $ by $q - (\delta - 1)\gamma a $, $\alpha - \delta  + 1 $, $\beta - \delta + 1$, $2-\delta$, 1 and zero into (\ref{eq:51}). Multiply $(1-x)^{1-\delta }$ and the new (\ref{eq:51}) together.
\begin{eqnarray}
 & &(1-x)^{1-\delta } y(x)\nonumber\\
&=& (1-x)^{1-\delta } Hl\left(a, q - (\delta  - 1)\gamma a; \alpha - \delta + 1, \beta - \delta + 1, \gamma ,2 - \delta ; x\right)\nonumber\\
&=& (1-x)^{1-\delta } \left\{\sum_{i_0=0}^{\infty } \frac{(\frac{\alpha+1-\delta }{2})_{i_0} (\frac{\beta+1-\delta }{2})_{i_0}}{(1)_{i_0}(\frac{1}{2}+ \frac{\gamma}{2})_{i_0}} z^{i_0} \right.\nonumber\\
&&+ \left\{ \sum_{i_0=0}^{\infty }\frac{i_0\left( i_0+ \Gamma_0^{(I)} \right)+ Q}{(i_0+ \frac{1}{2})(i_0 + \frac{\gamma }{2})}\frac{(\frac{\alpha+1-\delta }{2})_{i_0} (\frac{\beta+1-\delta }{2})_{i_0}}{(1)_{i_0}(\frac{1}{2}+ \frac{\gamma}{2})_{i_0}} \sum_{i_1=i_0}^{\infty } \frac{(1+\frac{\alpha-\delta }{2})_{i_1}(1+\frac{\beta -\delta }{2})_{i_1}(\frac{3}{2})_{i_0}(1+\frac{\gamma }{2})_{i_0}}{(1+\frac{\alpha-\delta }{2})_{i_0}(1+\frac{\beta -\delta }{2})_{i_0}(\frac{3}{2})_{i_1}(1+ \frac{\gamma}{2})_{i_1}} z^{i_1} \right\} \eta \nonumber\\
&&+ \sum_{n=2}^{\infty } \left\{ \sum_{i_0=0}^{\infty } \frac{i_0\left( i_0+ \Gamma_0^{(I)} \right)+ Q}{(i_0+ \frac{1}{2})(i_0 + \frac{\gamma }{2})} \frac{(\frac{\alpha+1-\delta }{2})_{i_0} (\frac{\beta +1-\delta }{2})_{i_0}}{(1)_{i_0}(\frac{1}{2}+ \frac{\gamma}{2})_{i_0}}\right.\nonumber\\
&&\times \prod _{k=1}^{n-1} \left\{ \sum_{i_k=i_{k-1}}^{\infty } \frac{(i_k+\frac{k}{2}) \left( i_k+ \Gamma_k^{(I)} \right)+ Q}{(i_k+ \frac{k}{2}+\frac{1}{2})(i_k +\frac{k}{2}+\frac{\gamma }{2})} \frac{(\frac{k}{2}+\frac{\alpha+1-\delta }{2})_{i_k}(\frac{k}{2}+\frac{\beta +1-\delta }{2})_{i_k}(1+ \frac{k}{2})_{i_{k-1}}(\frac{1}{2}+\frac{k}{2}+\frac{\gamma }{2})_{i_{k-1}}}{(\frac{k}{2}+\frac{\alpha +1-\delta }{2})_{i_{k-1}}(\frac{k}{2}+\frac{\beta +1-\delta }{2})_{i_{k-1}}(1+\frac{k}{2})_{i_k}(\frac{1}{2}+ \frac{k}{2}+ \frac{\gamma}{2})_{i_k}}\right\} \nonumber\\
&&\times \left.\left. \sum_{i_n= i_{n-1}}^{\infty } \frac{(\frac{n}{2}+\frac{\alpha +1-\delta }{2})_{i_n}(\frac{n}{2}+\frac{\beta+1-\delta  }{2})_{i_n}(1+ \frac{n}{2})_{i_{n-1}}(\frac{1}{2}+\frac{n}{2}+\frac{\gamma }{2})_{i_{n-1}}}{(\frac{n}{2}+\frac{\alpha +1-\delta }{2})_{i_{n-1}}(\frac{n}{2}+\frac{\beta +1-\delta }{2})_{i_{n-1}}(1+\frac{n}{2})_{i_n}(\frac{1}{2}+ \frac{n}{2}+ \frac{\gamma}{2})_{i_n}} z^{i_n} \right\} \eta ^n \right\} \label{eq:52}
\end{eqnarray}
where
\begin{equation}
\begin{cases} 
\Gamma_0^{(I)}=  \frac{1}{2(1+a)}(\alpha +\beta -\delta +a(-\delta +\gamma +1))\cr
\Gamma_k^{(I)}= \frac{1}{2(1+a)}(\alpha +\beta -\delta +k+a(-\delta +\gamma +1+k )) \cr
Q= \frac{q-(\delta -1)\gamma a}{4(1+a)}
\end{cases}\nonumber %\label{eq:8}
\end{equation}
\subsection{   ${\displaystyle x^{1-\gamma } (1-x)^{1-\delta } Hl(a, q-(\gamma +\delta -2)a-(\gamma -1)(\alpha +\beta -\gamma -\delta +1); \alpha - \gamma -\delta +2}$ \\${\displaystyle, \beta - \gamma -\delta +2, 2-\gamma, 2 - \delta ; x)}$}
\subsubsection{Polynomial which makes $B_n$ term terminated}
\underline {(1) The case of $\alpha = -2 \alpha _i-i -2+\gamma +\delta $ and $\beta \ne -2 \beta _i -i-2+\gamma +\delta$ where $i, \alpha _i, \beta _i$ = $0,1,2,\cdots$.}\vspace{1mm}

Replace coefficients $q$, $\alpha$, $\beta$, $\gamma $, $\delta$, $c_0$ and $\lambda $ by $q-(\gamma +\delta -2)a-(\gamma -1)(\alpha +\beta -\gamma -\delta +1)$, $\alpha - \gamma -\delta +2$, $\beta - \gamma -\delta +2, 2-\gamma$, $2 - \delta$,1 and zero into (\ref{eq:36}). Multiply $x^{1-\gamma } (1-x)^{1-\delta }$ and the new (\ref{eq:36}) together.  
\begin{eqnarray}
& &x^{1-\gamma } (1-x)^{1-\delta } y(x)\nonumber\\
&=& x^{1-\gamma } (1-x)^{1-\delta } Hl(a, q-(\gamma +\delta -2)a-(\gamma -1)(\alpha +\beta -\gamma -\delta +1); \alpha - \gamma -\delta +2 \nonumber\\
&&, \beta - \gamma -\delta +2, 2-\gamma, 2 - \delta ; x)\nonumber\\
&=& x^{1-\gamma } (1-x)^{1-\delta } \left\{\sum_{i_0=0}^{\alpha _0} \frac{(-\alpha _0)_{i_0} (\frac{\beta-\gamma -\delta +2}{2})_{i_0}}{(1)_{i_0}(\frac{3-\gamma }{2})_{i_0}} z^{i_0}\right.\nonumber\\
&&+ \left\{\sum_{i_0=0}^{\alpha _0}\frac{i_0 \left( i_0+ \Gamma_0^{(S)}\right)+ Q_0^{(S)}}{(i_0+ \frac{1}{2})(i_0 + \frac{2-\gamma }{2})} \frac{(-\alpha _0)_{i_0} (\frac{\beta-\gamma -\delta +2 }{2})_{i_0}}{(1)_{i_0}(\frac{3-\gamma}{2})_{i_0}} \sum_{i_1=i_0}^{\alpha _1} \frac{(-\alpha _1)_{i_1}(\frac{\beta-\gamma -\delta +3}{2})_{i_1}(\frac{3}{2})_{i_0}(\frac{4-\gamma }{2})_{i_0}}{(-\alpha _1)_{i_0}(\frac{\beta-\gamma -\delta +3}{2})_{i_0}(\frac{3}{2})_{i_1}(\frac{4-\gamma}{2})_{i_1}} z^{i_1} \right\} \eta \nonumber\\
&&+ \sum_{n=2}^{\infty } \left\{ \sum_{i_0=0}^{\alpha _0} \frac{i_0 \left( i_0+ \Gamma_0^{(S)}\right)+ Q_0^{(S)}}{(i_0+ \frac{1}{2})(i_0 + \frac{2-\gamma }{2})}\frac{(-\alpha _0)_{i_0} (\frac{\beta-\gamma -\delta +2 }{2})_{i_0}}{(1)_{i_0}(\frac{3-\gamma}{2})_{i_0}}\right.\nonumber\\
&&\times \prod _{k=1}^{n-1} \left\{ \sum_{i_k=i_{k-1}}^{\alpha _k} \frac{(i_k+\frac{k}{2}) \left( i_k+ \Gamma_k^{(S)}\right)+ Q_k^{(S)}}{(i_k+ \frac{k}{2}+\frac{1}{2})(i_k +\frac{k+2-\gamma }{2})}\frac{(-\alpha _k)_{i_k}(\frac{k+2+\beta-\gamma -\delta}{2})_{i_k}(1+ \frac{k}{2})_{i_{k-1}}(\frac{k+3-\gamma }{2})_{i_{k-1}}}{(-\alpha _k)_{i_{k-1}}(\frac{k+2+\beta-\gamma -\delta}{2})_{i_{k-1}}(1+\frac{k}{2})_{i_k}(\frac{k+3-\gamma }{2})_{i_k}}\right\} \nonumber\\
&&\times  \left.\left.\sum_{i_n= i_{n-1}}^{\alpha _n} \frac{(-\alpha _n)_{i_n}(\frac{n+2+\beta-\gamma -\delta}{2})_{i_n}(1+ \frac{n}{2})_{i_{n-1}}(\frac{n+3-\gamma }{2})_{i_{n-1}}}{(-\alpha _n)_{i_{n-1}}(\frac{n+2+\beta-\gamma -\delta}{2})_{i_{n-1}}(1+\frac{n}{2})_{i_n}(\frac{n+3-\gamma }{2})_{i_n}} z^{i_n} \right\} \eta ^n \right\}\label{eq:53}
\end{eqnarray}
where
\begin{equation}
\begin{cases} z = -\frac{1}{a}x^2 \cr
\eta = \frac{(1+a)}{a} x \cr
\alpha _i\leq \alpha _j \;\;\mbox{only}\;\mbox{if}\;i\leq j\;\;\mbox{where}\;i,j= 0,1,2,\cdots
\end{cases}\nonumber %\label{eq:37}
\end{equation}
and
\begin{equation}
\begin{cases} 
\Gamma_0^{(S)} = \frac{1}{2(1+a)}(-2\alpha _0+ \beta -\gamma +a(3-\gamma -\delta )) \cr
\Gamma_k^{(S)} = \frac{1}{2(1+a)}(-2\alpha _k+ \beta -\gamma +a(k+3-\gamma -\delta )) \cr
Q_0^{(S)}= \frac{q-(\gamma +\delta -2)a-(\gamma -1)(-2\alpha _0 -1+\beta )}{4(1+a)} \cr
Q_k^{(S)}= \frac{q-(\gamma +\delta -2)a-(\gamma -1)(-2\alpha _k-k-1+\beta )}{4(1+a)}
\end{cases}\nonumber %\label{eq:8}
\end{equation}
For the minimum value of Heun equation for a polynomial which makes $B_n$ term terminated about $x=0 $, put $\alpha _0=\alpha _1=\alpha _2=\cdots=0$ in (\ref{eq:53}).
\begin{eqnarray}
& &x^{1-\gamma } (1-x)^{1-\delta } y(x)\nonumber\\
&=& x^{1-\gamma } (1-x)^{1-\delta } Hl(a, q-(\gamma +\delta -2)a-(\gamma -1)(\alpha +\beta -\gamma -\delta +1); \alpha - \gamma -\delta +2 \nonumber\\
&&, \beta - \gamma -\delta +2, 2-\gamma, 2 - \delta ; x)\nonumber\\
&=& x^{1-\gamma } (1-x)^{1-\delta } \; _2F_1\left(  \frac{\Lambda_3-\sqrt{ \Lambda _3^2-4a\Omega_3}}{2a}, \frac{\Lambda _3+\sqrt{\Lambda _3^2- 4a\Omega _3}}{2a}; 2-\gamma ; x \right) \nonumber
\end{eqnarray}
where $\Lambda_3= \beta -1-a(\gamma +\delta -3)$ and $\Omega_3= q-(\beta -1)(\gamma -1)-a(\gamma +\delta -2)$.
It tells us that Heun polynomials in which makes $B_n$ term terminated, for fixed values of $\alpha $, require $|x| < 1$ for the convergence of the radius.

\underline {(2) The case of $\alpha = -2 \alpha _i-i -2+\gamma +\delta $ and $\beta = -2\beta _i-i -2+\gamma +\delta $ only if $\alpha _i \leq \beta _i$.}

Put $\beta = -2\beta _i-i -2+\gamma +\delta $  where $i=0,1,2,\cdots$ in (\ref{eq:53}).  
\begin{eqnarray}
& &x^{1-\gamma } (1-x)^{1-\delta } y(x)\nonumber\\
&=& x^{1-\gamma } (1-x)^{1-\delta } Hl(a, q-(\gamma +\delta -2)a-(\gamma -1)(\alpha +\beta -\gamma -\delta +1); \alpha - \gamma -\delta +2 \nonumber\\
&&, \beta - \gamma -\delta +2, 2-\gamma, 2 - \delta ; x)\nonumber\\
&=& x^{1-\gamma } (1-x)^{1-\delta } \left\{\sum_{i_0=0}^{\alpha _0} \frac{(-\alpha _0)_{i_0} (-\beta _0)_{i_0}}{(1)_{i_0}(\frac{3-\gamma }{2})_{i_0}} z^{i_0}\right.\nonumber\\
&&+ \left\{\sum_{i_0=0}^{\alpha _0}\frac{i_0 \left( i_0+ \Gamma_0^{(B)}\right)+ Q_0^{(B)}}{(i_0+ \frac{1}{2})(i_0 + \frac{2-\gamma }{2})} \frac{(-\alpha _0)_{i_0} (-\beta _0)_{i_0}}{(1)_{i_0}(\frac{3-\gamma}{2})_{i_0}} \sum_{i_1=i_0}^{\alpha _1} \frac{(-\alpha _1)_{i_1}(-\beta _1)_{i_1}(\frac{3}{2})_{i_0}(\frac{4-\gamma }{2})_{i_0}}{(-\alpha _1)_{i_0}(-\beta _1)_{i_0}(\frac{3}{2})_{i_1}(\frac{4-\gamma}{2})_{i_1}} z^{i_1} \right.\} \eta \nonumber\\
&&+ \sum_{n=2}^{\infty } \left\{ \sum_{i_0=0}^{\alpha _0} \frac{i_0 \left( i_0+ \Gamma_0^{(B)}\right)+ Q_0^{(B)}}{(i_0+ \frac{1}{2})(i_0 + \frac{2-\gamma }{2})} \frac{(-\alpha _0)_{i_0} (-\beta _0)_{i_0}}{(1)_{i_0}(\frac{3-\gamma}{2})_{i_0}}\right.\nonumber\\
&&\times \prod _{k=1}^{n-1} \left\{ \sum_{i_k=i_{k-1}}^{\alpha _k} \frac{(i_k+\frac{k}{2}) \left( i_k+ \Gamma_k^{(B)}\right)+ Q_k^{(B)}}{(i_k+ \frac{k}{2}+\frac{1}{2})(i_k +\frac{k+2-\gamma }{2})} \frac{(-\alpha _k)_{i_k}(-\beta _k)_{i_k}(1+ \frac{k}{2})_{i_{k-1}}(\frac{k+3-\gamma }{2})_{i_{k-1}}}{(-\alpha _k)_{i_{k-1}}(-\beta _k)_{i_{k-1}}(1+\frac{k}{2})_{i_k}(\frac{k+3-\gamma }{2})_{i_k}}\right\}\nonumber\\
&&\times  \left.\left.\sum_{i_n= i_{n-1}}^{\alpha _n} \frac{(-\alpha _n)_{i_n}(-\beta _n)_{i_n}(1+ \frac{n}{2})_{i_{n-1}}(\frac{n+3-\gamma }{2})_{i_{n-1}}}{(-\alpha _n)_{i_{n-1}}(-\beta _n)_{i_{n-1}}(1+\frac{n}{2})_{i_n}(\frac{n+3-\gamma }{2})_{i_n}} z^{i_n} \right\} \eta ^n \right\}\label{eq:54}
\end{eqnarray}
where
\begin{equation}
\begin{cases} 
\Gamma_0^{(B)} = \frac{1}{2(1+a)}(-2\alpha _0-2\beta _0-2+\delta +a(3-\gamma -\delta )) \cr
\Gamma_k^{(B)} = \frac{1}{2(1+a)}(-2\alpha _k-2\beta _k-k-2+\delta +a(k+3-\gamma -\delta )) \cr
Q_0^{(B)}= \frac{q-(\gamma +\delta -2)a-(\gamma -1)(-2\alpha _0-2\beta _0-3+\gamma +\delta)}{4(1+a)} \cr
Q_k^{(B)}= \frac{q-(\gamma +\delta -2)a-(\gamma -1)(-2\alpha _k-2\beta _k-2k-3+\gamma +\delta)}{4(1+a)}
\end{cases}\nonumber %\label{eq:8}
\end{equation}
For the minimum value of Heun equation for a polynomial which makes $B_n$ term terminated about $x=0 $, put $\alpha _0=\alpha _1=\alpha _2=\cdots=0$ and $\beta _0=\beta _1=\beta _2=\cdots=0$ in (\ref{eq:54}).
\begin{eqnarray}
& &x^{1-\gamma } (1-x)^{1-\delta } y(x)\nonumber\\
&=& x^{1-\gamma } (1-x)^{1-\delta } Hl(a, q-(\gamma +\delta -2)a-(\gamma -1)(\alpha +\beta -\gamma -\delta +1); \alpha - \gamma -\delta +2 \nonumber\\
&&, \beta - \gamma -\delta +2, 2-\gamma, 2 - \delta ; x)\nonumber\\
&=& x^{1-\gamma } (1-x)^{1-\delta } \; _2F_1\left( \frac{ \Lambda _4-\sqrt{ \Lambda _4^2-4(a-1)\Omega _4}}{2(a-1)}, \frac{ \Lambda _4+\sqrt{ \Lambda _4^2-4(a-1)\Omega _4}}{2(a-1)}; 2-\gamma ; \frac{a-1}{a}x \right) \hspace{1cm}\label{aa:2}
\end{eqnarray}
where $\Lambda_4= 2(\gamma -2)+\delta -a(\gamma +\delta -3)$ and $\Omega_4=q-(\gamma -1)(\gamma +\delta -3) -a(\gamma +\delta -2)$.
It tells us that Heun polynomials in which makes $B_n$ term terminated, for fixed values of $\alpha $ and $\beta $, require $\left|\frac{a-1}{a}x\right| < 1$ for the convergence of the radius.

For the special case, if $x=\frac{a}{a-1}$ and $Re\left( \frac{\gamma +\delta -2+a(1-\delta )}{1-a}\right)>0 $ in (\ref{aa:2}),
\begin{eqnarray}
& &\left( \frac{a}{a-1}\right)^{1-\gamma } \left( \frac{1}{1-a}\right)^{1-\delta } y\left( \frac{a}{a-1}\right)\nonumber\\
&=& \left( \frac{a}{a-1}\right)^{1-\gamma } \left( \frac{1}{1-a}\right)^{1-\delta } Hl(a, q-(\gamma +\delta -2)a-(\gamma -1)(\alpha +\beta -\gamma -\delta +1); \alpha - \gamma -\delta +2 \nonumber\\
&&, \beta - \gamma -\delta +2, 2-\gamma, 2 - \delta ; a/(a-1))\nonumber\\
&=& \left( \frac{a}{a-1}\right)^{1-\gamma } \left( \frac{1}{1-a}\right)^{1-\delta } \frac{ \Gamma \left( 2-\gamma \right) \Gamma \left( 2-\gamma -\frac{\Lambda _4}{a-1} \right)}{\Gamma \left( 2-\gamma - \frac{ \Lambda _4-\sqrt{ \Lambda _4^2-4(a-1)\Omega _4}}{2(a-1)}\right) \Gamma \left( 2-\gamma - \frac{ \Lambda _4+\sqrt{ \Lambda _4^2-4(a-1)\Omega _4}}{2(a-1)} \right)}   \nonumber
\end{eqnarray}
\subsubsection{Infinite series}
Replace coefficients $q$, $\alpha$, $\beta$, $\gamma $, $\delta$, $c_0$ and $\lambda $ by $q-(\gamma +\delta -2)a-(\gamma -1)(\alpha +\beta -\gamma -\delta +1)$, $\alpha - \gamma -\delta +2$, $\beta - \gamma -\delta +2, 2-\gamma$, $2 - \delta$,1 and zero into (\ref{eq:51}). Multiply $x^{1-\gamma } (1-x)^{1-\delta }$ and the new (\ref{eq:51}) together.
\begin{eqnarray}
 & &x^{1-\gamma } (1-x)^{1-\delta } y(x)\nonumber\\
&=& x^{1-\gamma } (1-x)^{1-\delta } Hl(a, q-(\gamma +\delta -2)a-(\gamma -1)(\alpha +\beta -\gamma -\delta +1); \alpha - \gamma -\delta +2 \nonumber\\
&&, \beta - \gamma -\delta +2, 2-\gamma, 2 - \delta ; x)\nonumber\\
&=& x^{1-\gamma } (1-x)^{1-\delta } \left\{\sum_{i_0=0}^{\infty } \frac{(\frac{\alpha -\gamma -\delta +2}{2})_{i_0} (\frac{\beta-\gamma -\delta +2}{2})_{i_0}}{(1)_{i_0}(\frac{3-\gamma }{2})_{i_0}} z^{i_0}\right.\nonumber\\
&&+ \left\{\sum_{i_0=0}^{\infty }\frac{i_0\left( i_0+ \Gamma_0^{(I)}\right)+ Q}{(i_0+ \frac{1}{2})(i_0 + \frac{2-\gamma }{2})}  \frac{(\frac{\alpha -\gamma -\delta +2 }{2})_{i_0} (\frac{\beta-\gamma -\delta +2 }{2})_{i_0}}{(1)_{i_0}(\frac{3-\gamma}{2})_{i_0}} \sum_{i_1=i_0}^{\infty } \frac{(\frac{\alpha -\gamma -\delta +3}{2})_{i_1}(\frac{\beta-\gamma -\delta +3}{2})_{i_1}(\frac{3}{2})_{i_0}(\frac{4-\gamma }{2})_{i_0}}{(\frac{\alpha -\gamma -\delta +3}{2})_{i_0}(\frac{\beta-\gamma -\delta +3}{2})_{i_0}(\frac{3}{2})_{i_1}(\frac{4-\gamma}{2})_{i_1}} z^{i_1} \right\} \eta \nonumber\\
&&+ \sum_{n=2}^{\infty } \left\{ \sum_{i_0=0}^{\infty } \frac{i_0\left( i_0+ \Gamma_0^{(I)}\right)+ Q}{(i_0+ \frac{1}{2})(i_0 + \frac{2-\gamma }{2})}\frac{(\frac{\alpha -\gamma -\delta +2 }{2})_{i_0} (\frac{\beta-\gamma -\delta +2 }{2})_{i_0}}{(1)_{i_0}(\frac{3-\gamma}{2})_{i_0}}\right.\nonumber\\
&&\times \prod _{k=1}^{n-1} \left\{ \sum_{i_k=i_{k-1}}^{\infty } \frac{(i_k+\frac{k}{2}) \left( i_k+ \Gamma_k^{(I)}\right)+ Q}{(i_k+ \frac{k}{2}+\frac{1}{2})(i_k +\frac{k+2-\gamma }{2})} \frac{(\frac{k+2+\alpha -\gamma -\delta}{2})_{i_k}(\frac{k+2+\beta-\gamma -\delta}{2})_{i_k}(1+ \frac{k}{2})_{i_{k-1}}(\frac{k+3-\gamma }{2})_{i_{k-1}}}{(\frac{k+2+\alpha -\gamma -\delta}{2})_{i_{k-1}}(\frac{k+2+\beta-\gamma -\delta}{2})_{i_{k-1}}(1+\frac{k}{2})_{i_k}(\frac{k+3-\gamma }{2})_{i_k}}\right\}\nonumber\\
&&\times  \left.\left.\sum_{i_n= i_{n-1}}^{\infty } \frac{(\frac{n+2+\alpha -\gamma -\delta}{2})_{i_n}(\frac{n+2+\beta-\gamma -\delta}{2})_{i_n}(1+ \frac{n}{2})_{i_{n-1}}(\frac{n+3-\gamma }{2})_{i_{n-1}}}{(\frac{n+2+\alpha -\gamma -\delta}{2})_{i_{n-1}}(\frac{n+2+\beta-\gamma -\delta}{2})_{i_{n-1}}(1+\frac{n}{2})_{i_n}(\frac{n+3-\gamma }{2})_{i_n}} z^{i_n} \right\} \eta ^n \right\} \label{eq:55}
\end{eqnarray}
where
\begin{equation}
\begin{cases} 
\Gamma_0^{(I)} =  \frac{1}{2(1+a)}(\alpha +\beta -2\gamma -\delta +2 +a(3-\gamma -\delta ))\cr
\Gamma_k^{(I)} =  \frac{1}{2(1+a)}(\alpha +\beta -2\gamma -\delta +2+k +a(k+3-\gamma -\delta )) \cr
Q= \frac{q-(\gamma +\delta -2)a-(\gamma -1)(\alpha +\beta -\gamma -\delta +1)}{4(1+a)}
\end{cases}\nonumber %\label{eq:8}
\end{equation}
\subsection{ ${\displaystyle  Hl(1-a,-q+\alpha \beta; \alpha,\beta, \delta, \gamma; 1-x)}$} 
\subsubsection{Polynomial which makes $B_n$ term terminated}

\underline {(1) The case of $\alpha = -2 \alpha _i-i$ and $\beta \ne -2 \beta _i -i$ where $i, \alpha _i, \beta _i$ = $0,1,2,\cdots$.}\vspace{1mm}

Replace coefficients $a$, $q$, $\gamma $, $\delta$, $x$, $c_0$ and $\lambda $ by $1-a$, $-q +\alpha \beta $, $\delta $, $\gamma $, $1-x$, 1 and zero into (\ref{eq:36}).  
\begin{eqnarray}
y(\xi ) &=& Hl(1-a,-q+\alpha \beta; \alpha,\beta, \delta, \gamma; 1-x)\nonumber\\
&=& \sum_{i_0=0}^{\alpha _0} \frac{(-\alpha _0)_{i_0} (\frac{\beta }{2})_{i_0}}{(1)_{i_0}(\frac{1}{2}+ \frac{\delta }{2})_{i_0}} z^{i_0}\nonumber\\
&&+ \left\{\sum_{i_0=0}^{\alpha _0}\frac{ i_0 \left( i_0+ \Gamma_0^{(S)}\right)+ Q_0^{(S)}}{(i_0+ \frac{1}{2})(i_0 + \frac{\delta }{2})}  \frac{(-\alpha _0)_{i_0} (\frac{\beta }{2})_{i_0}}{(1)_{i_0}(\frac{1}{2}+ \frac{\delta }{2})_{i_0}} \sum_{i_1=i_0}^{\alpha _1} \frac{(-\alpha _1)_{i_1}(\frac{1}{2}+\frac{\beta }{2})_{i_1}(\frac{3}{2})_{i_0}(1+\frac{\delta }{2})_{i_0}}{(-\alpha _1)_{i_0}(\frac{1}{2}+\frac{\beta }{2})_{i_0}(\frac{3}{2})_{i_1}(1+ \frac{\delta }{2})_{i_1}} z^{i_1} \right\} \eta \nonumber\\
&&+ \sum_{n=2}^{\infty } \left\{ \sum_{i_0=0}^{\alpha _0} \frac{i_0 \left( i_0+ \Gamma_0^{(S)}\right)+ Q_0^{(S)}}{(i_0+ \frac{1}{2})(i_0 + \frac{\delta }{2})} \frac{(-\alpha _0)_{i_0} (\frac{\beta }{2})_{i_0}}{(1)_{i_0}(\frac{1}{2}+ \frac{\delta }{2})_{i_0}}\right.\nonumber\\
&&\times \prod _{k=1}^{n-1} \left\{ \sum_{i_k=i_{k-1}}^{\alpha _k} \frac{(i_k+\frac{k}{2}) \left( i_k+ \Gamma_k^{(S)}\right)+ Q_k^{(S)}}{(i_k+ \frac{k}{2}+\frac{1}{2})(i_k +\frac{k}{2}+\frac{\delta }{2})} \frac{(-\alpha _k)_{i_k}(\frac{k}{2}+\frac{\beta }{2})_{i_k}(1+ \frac{k}{2})_{i_{k-1}}(\frac{1}{2}+\frac{k}{2}+\frac{\delta }{2})_{i_{k-1}}}{(-\alpha _k)_{i_{k-1}}(\frac{k}{2}+\frac{\beta }{2})_{i_{k-1}}(1+\frac{k}{2})_{i_k}(\frac{1}{2}+ \frac{k}{2}+ \frac{\delta }{2} )_{i_k}}\right\}\nonumber\\
&&\times  \left. \sum_{i_n= i_{n-1}}^{\alpha _n} \frac{(-\alpha _n)_{i_n}(\frac{n}{2}+\frac{\beta }{2})_{i_n}(1+ \frac{n}{2})_{i_{n-1}}(\frac{1}{2}+\frac{n}{2}+\frac{\delta }{2})_{i_{n-1}}}{(-\alpha _n)_{i_{n-1}}(\frac{n}{2}+\frac{\beta }{2})_{i_{n-1}}(1+\frac{n}{2})_{i_n}(\frac{1}{2}+ \frac{n}{2}+ \frac{\delta }{2})_{i_n}} z^{i_n} \right\} \eta ^n   \label{eq:56}
\end{eqnarray}
where
\begin{equation}
\begin{cases} \xi =1-x \cr
z = \frac{-1}{1-a}\xi^2 \cr
\eta = \frac{2-a}{1-a}\xi \cr
\alpha _i\leq \alpha _j \;\;\mbox{only}\;\mbox{if}\;i\leq j\;\;\mbox{where}\;i,j= 0,1,2,\cdots
\end{cases}\nonumber %\label{eq:37}
\end{equation}
and
\begin{equation}
\begin{cases} 
\Gamma_0^{(S)} =  \frac{1}{2(2-a)}(-2\alpha _0+ \beta -\gamma  +(1-a)(\delta +\gamma -1)) \cr
\Gamma_k^{(S)} =  \frac{1}{2(2-a)}(-2\alpha _k+ \beta -\gamma +(1-a)(\delta +\gamma +k-1))  \cr
Q_0^{(S)}=  \frac{-q -2 \alpha _0 \beta}{4(2-a)} \cr
Q_k^{(S)}=  \frac{-q -(2\alpha _k+k)\beta}{4(2-a)}
\end{cases}\nonumber %\label{eq:8}
\end{equation}
For the minimum value of Heun equation for a polynomial which makes $B_n$ term terminated about $\xi=0 $, put $\alpha _0=\alpha _1=\alpha _2=\cdots=0$ in (\ref{eq:56}).
\begin{eqnarray}
y(\xi ) &=& Hl(1-a,-q+\alpha \beta; \alpha,\beta, \delta, \gamma; 1-x)\nonumber\\
&=&  \; _2F_1\left( \frac{ -\Lambda _5-\sqrt{ \Lambda _5^2-4(a-1)q}}{2(a-1)}, \frac{ -\Lambda _5+\sqrt{ \Lambda _5^2-4(a-1)q}}{2(a-1)}; \delta ; \xi \right) \hspace{1cm}\label{aa:3}
\end{eqnarray}
where $\Lambda_5= \delta -1-a(\gamma +\delta -1)$.
It tells us that Heun polynomials in which makes $B_n$ term terminated, for fixed values of $\alpha $, require $\left|\xi\right| < 1$ for the convergence of the radius.

For the special case, if $\xi=1$ and $Re\left( \frac{ 1+a(\gamma  -1)}{ 1-a }\right)>0 $ in (\ref{aa:3}),
\begin{eqnarray}
y(1) &=& Hl(1-a,-q+\alpha \beta; \alpha,\beta, \delta, \gamma; 1)\nonumber\\
&=& \frac{ \Gamma \left( \delta \right) \Gamma \left( \frac{ 1+a(\gamma  -1)}{ 1-a } \right)}{\Gamma \left( \delta +\frac{  \Lambda _5-\sqrt{ \Lambda _5^2-4(a-1)q}}{2(a-1)}\right) \Gamma \left(\delta + \frac{  \Lambda _5+\sqrt{ \Lambda _5^2-4(a-1)q}}{2(a-1)}\right)}   \nonumber
\end{eqnarray}
\underline {(2) The case of $\alpha = -2 \alpha _i-i$ and $\beta = -2 \beta _i -i$ only if $\alpha _i \leq \beta _i$.}

Put $\beta = -2\beta _i-i $  where $i=0,1,2,\cdots$ in (\ref{eq:56}).  
\begin{eqnarray}
y(\xi ) &=& Hl(1-a,-q+\alpha \beta; \alpha,\beta, \delta, \gamma; 1-x)\nonumber\\
&=& \sum_{i_0=0}^{\alpha _0} \frac{(-\alpha _0)_{i_0} (-\beta _0)_{i_0}}{(1)_{i_0}(\frac{1}{2}+ \frac{\delta }{2})_{i_0}} z^{i_0}\nonumber\\
&&+ \left\{\sum_{i_0=0}^{\alpha _0}\frac{ i_0 \left( i_0+ \Gamma_0^{(B)}\right)+ Q_0^{(B)}}{(i_0+ \frac{1}{2})(i_0 + \frac{\delta }{2})}  \frac{(-\alpha _0)_{i_0} (-\beta _0)_{i_0}}{(1)_{i_0}(\frac{1}{2}+ \frac{\delta }{2})_{i_0}} \sum_{i_1=i_0}^{\alpha _1} \frac{(-\alpha _1)_{i_1}(-\beta _1)_{i_1}(\frac{3}{2})_{i_0}(1+\frac{\delta }{2})_{i_0}}{(-\alpha _1)_{i_0}(-\beta _1)_{i_0}(\frac{3}{2})_{i_1}(1+ \frac{\delta }{2})_{i_1}} z^{i_1} \right\} \eta \nonumber\\
&&+ \sum_{n=2}^{\infty } \left\{ \sum_{i_0=0}^{\alpha _0} \frac{i_0 \left( i_0+ \Gamma_0^{(B)}\right)+ Q_0^{(B)}}{(i_0+ \frac{1}{2})(i_0 + \frac{\delta }{2})} \frac{(-\alpha _0)_{i_0} (-\beta _0)_{i_0}}{(1)_{i_0}(\frac{1}{2}+ \frac{\delta }{2})_{i_0}} \right.\nonumber\\
&&\times \prod _{k=1}^{n-1} \left\{ \sum_{i_k=i_{k-1}}^{\alpha _k} \frac{(i_k+\frac{k}{2}) \left( i_k+ \Gamma_k^{(B)}\right)+ Q_k^{(B)}}{(i_k+ \frac{k}{2}+\frac{1}{2})(i_k +\frac{k}{2}+\frac{\delta }{2})} \frac{(-\alpha _k)_{i_k}(-\beta _k)_{i_k}(1+ \frac{k}{2})_{i_{k-1}}(\frac{1}{2}+\frac{k}{2}+\frac{\delta }{2})_{i_{k-1}}}{(-\alpha _k)_{i_{k-1}}(-\beta _k)_{i_{k-1}}(1+\frac{k}{2})_{i_k}(\frac{1}{2}+ \frac{k}{2}+ \frac{\delta }{2} )_{i_k}}\right\}\nonumber\\
&&\times \left. \sum_{i_n= i_{n-1}}^{\alpha _n} \frac{(-\alpha _n)_{i_n}(-\beta _n)_{i_n}(1+ \frac{n}{2})_{i_{n-1}}(\frac{1}{2}+\frac{n}{2}+\frac{\delta }{2})_{i_{n-1}}}{(-\alpha _n)_{i_{n-1}}(-\beta _n)_{i_{n-1}}(1+\frac{n}{2})_{i_n}(\frac{1}{2}+ \frac{n}{2}+ \frac{\delta }{2})_{i_n}} z^{i_n} \right\} \eta ^n   \label{eq:57}
\end{eqnarray}
where
\begin{equation}
\begin{cases} 
\Gamma_0^{(B)} = \frac{1}{2(2-a)}(-2\alpha _0-2\beta _0 -\gamma  +(1-a)(\delta +\gamma -1))  \cr
\Gamma_k^{(B)} = \frac{1}{2(2-a)}(-2\alpha _k-2\beta _k -k -\gamma +(1-a)(\delta +\gamma +k-1))  \cr
Q_0^{(B)}=  \frac{-q +4 \alpha _0 \beta _0}{4(2-a)} \cr
Q_k^{(B)}=  \frac{-q +(2\alpha _k+k)(2\beta _k+k)}{4(2-a)}
\end{cases}\nonumber %\label{eq:8}
\end{equation}
For the minimum value of Heun equation for a polynomial which makes $B_n$ term terminated about $\xi=0 $, put $\alpha _0=\alpha _1=\alpha _2=\cdots=0$ and $\beta _0=\beta _1=\beta _2=\cdots=0$ in (\ref{eq:57}).
\begin{eqnarray}
y(\xi ) &=& Hl(1-a,-q+\alpha \beta; \alpha,\beta, \delta, \gamma; 1-x)\nonumber\\
&=&  \; _2F_1\left( \frac{ -\Lambda _6-\sqrt{ \Lambda _6^2-4(a-1)q}}{2(a-1)}, \frac{ -\Lambda _6+\sqrt{ \Lambda _6^2-4(a-1)q}}{2(a-1)}; \delta ; \xi \right) \hspace{1cm}\label{aa:4}
\end{eqnarray}
where $\Lambda_6= (1-a)(\gamma +\delta -1)$.
It tells us that Heun polynomials in which makes $B_n$ term terminated, for fixed values of $\alpha $ and $\beta $, require $\left|\xi\right| < 1$ for the convergence of the radius.

For the special case, if $\xi=1$ and $Re\left( \gamma \right)<1 $ in (\ref{aa:4}),
\begin{eqnarray}
y(1) &=& Hl(1-a,-q+\alpha \beta; \alpha,\beta, \delta, \gamma; 1)\nonumber\\
&=& \frac{ \Gamma \left( \delta \right) \Gamma \left( 1-\gamma \right)}{\Gamma \left( \delta +\frac{  \Lambda _6-\sqrt{ \Lambda _6^2-4(a-1)q}}{2(a-1)}\right) \Gamma \left(\delta + \frac{  \Lambda _6+\sqrt{ \Lambda _6^2-4(a-1)q}}{2(a-1)}\right)}   \nonumber
\end{eqnarray}
\subsubsection{Infinite series}
Replace coefficients $a$, $q$, $\gamma $, $\delta$, $x$, $c_0$ and $\lambda $ by $1-a$, $-q +\alpha \beta $, $\delta $, $\gamma $, $1-x$,1 and zero into (\ref{eq:51}).
\begin{eqnarray}
y(\xi ) &=& Hl(1-a,-q+\alpha \beta; \alpha, \beta, \delta, \gamma; 1-x)\nonumber\\
&=& \sum_{i_0=0}^{\infty } \frac{(\frac{\alpha }{2})_{i_0} (\frac{\beta }{2})_{i_0}}{(1)_{i_0}(\frac{1}{2}+ \frac{\delta }{2})_{i_0}} z^{i_0}\nonumber\\
&&+ \left\{\sum_{i_0=0}^{\infty}\frac{ i_0 \left( i_0+ \Gamma_0^{(I)}\right)+ Q}{(i_0+ \frac{1}{2})(i_0 + \frac{\delta }{2})}  \frac{(\frac{\alpha }{2})_{i_0} (\frac{\beta }{2})_{i_0}}{(1)_{i_0}(\frac{1}{2}+ \frac{\delta }{2})_{i_0}} \sum_{i_1=i_0}^{\infty } \frac{(\frac{1}{2}+\frac{\alpha }{2})_{i_1}(\frac{1}{2}+\frac{\beta }{2})_{i_1}(\frac{3}{2})_{i_0}(1+\frac{\delta }{2})_{i_0}}{(\frac{1}{2}+\frac{\alpha }{2})_{i_0}(\frac{1}{2}+\frac{\beta }{2})_{i_0}(\frac{3}{2})_{i_1}(1+ \frac{\delta }{2})_{i_1}} z^{i_1} \right\} \eta \nonumber\\
&&+ \sum_{n=2}^{\infty } \left\{ \sum_{i_0=0}^{\infty } \frac{i_0 \left( i_0+ \Gamma_0^{(I)}\right)+ Q}{(i_0+ \frac{1}{2})(i_0 + \frac{\delta }{2})} \frac{(\frac{\alpha }{2})_{i_0} (\frac{\beta }{2})_{i_0}}{(1)_{i_0}(\frac{1}{2}+ \frac{\delta }{2})_{i_0}}\right.\nonumber\\
&&\times \prod _{k=1}^{n-1} \left\{ \sum_{i_k=i_{k-1}}^{\infty } \frac{(i_k+\frac{k}{2}) \left( i_k+ \Gamma_k^{(I)}\right)+ Q}{(i_k+ \frac{k}{2}+\frac{1}{2})(i_k +\frac{k}{2}+\frac{\delta }{2})} \frac{(\frac{k}{2}+\frac{\alpha }{2})_{i_k}(\frac{k}{2}+\frac{\beta }{2})_{i_k}(1+ \frac{k}{2})_{i_{k-1}}(\frac{1}{2}+\frac{k}{2}+\frac{\delta }{2})_{i_{k-1}}}{(\frac{k}{2}+\frac{\alpha }{2})_{i_{k-1}}(\frac{k}{2}+\frac{\beta }{2})_{i_{k-1}}(1+\frac{k}{2})_{i_k}(\frac{1}{2}+ \frac{k}{2}+ \frac{\delta }{2} )_{i_k}}\right\}\nonumber\\
&&\times \left. \sum_{i_n= i_{n-1}}^{\infty } \frac{(\frac{n}{2}+\frac{\alpha }{2})_{i_n}(\frac{n}{2}+\frac{\beta }{2})_{i_n}(1+ \frac{n}{2})_{i_{n-1}}(\frac{1}{2}+\frac{n}{2}+\frac{\delta }{2})_{i_{n-1}}}{(\frac{n}{2}+\frac{\alpha }{2})_{i_{n-1}}(\frac{n}{2}+\frac{\beta }{2})_{i_{n-1}}(1+\frac{n}{2})_{i_n}(\frac{1}{2}+ \frac{n}{2}+ \frac{\delta }{2})_{i_n}} z^{i_n} \right\} \eta ^n   \label{eq:58}
\end{eqnarray}
where
\begin{equation}
\begin{cases} 
\Gamma_0^{(I)} =   \frac{1}{2(2-a)}(\alpha + \beta -\gamma +(1-a)(\delta +\gamma -1))\cr
\Gamma_k^{(I)} =   \frac{1}{2(2-a)}(\alpha + \beta -\gamma +k+(1-a)(\delta +\gamma +k-1)) \cr
Q=  \frac{-q +\alpha \beta}{4(2-a)}
\end{cases}\nonumber %\label{eq:8}
\end{equation}
\subsection{ ${\displaystyle (1-x)^{1-\delta } Hl(1-a,-q+(\delta -1)\gamma a+(\alpha -\delta +1)(\beta -\delta +1); \alpha-\delta +1,\beta-\delta +1}$\\${\displaystyle, 2-\delta, \gamma; 1-x)}$}
\subsubsection{Polynomial which makes $B_n$ term terminated}

\underline {(1) The case of $\alpha = -2 \alpha _i-i+\delta -1$ and $\beta \ne -2 \beta _i -i+\delta -1$ where $i, \alpha _i, \beta _i$ = $0,1,2,\cdots$.}\vspace{1mm}

Replace coefficients $a$, $q$, $\alpha $, $\beta $, $\gamma $, $\delta$, $x$, $c_0$ and $\lambda $ by $1-a$, $-q+(\delta -1)\gamma a+(\alpha -\delta +1)(\beta -\delta +1)$, $\alpha-\delta +1 $, $\beta-\delta +1 $, $2-\delta$, $\gamma $, $1-x$, 1 and zero into (\ref{eq:36}). Multiply $(1-x)^{1-\delta }$ and the new (\ref{eq:36}) together.  
\begin{eqnarray}
& &(1-x)^{1-\delta } y(\xi)\nonumber\\
&=& (1-x)^{1-\delta } Hl(1-a,-q+(\delta -1)\gamma a+(\alpha -\delta +1)(\beta -\delta +1); \alpha-\delta +1,\beta-\delta +1, 2-\delta, \gamma; 1-x) \nonumber\\
&=& (1-x)^{1-\delta } \left\{\sum_{i_0=0}^{\alpha _0} \frac{(-\alpha _0)_{i_0} (\frac{\beta-\delta +1 }{2})_{i_0}}{(1)_{i_0}(\frac{3-\delta }{2})_{i_0}} z^{i_0}\right.\nonumber\\
&&+ \left\{\sum_{i_0=0}^{\alpha _0}\frac{i_0 \left( i_0+ \Gamma_0^{(S)}\right)+ Q_0^{(S)}}{(i_0+ \frac{1}{2})(i_0 + \frac{2-\delta }{2})} \frac{(-\alpha _0)_{i_0} (\frac{\beta-\delta +1}{2})_{i_0}}{(1)_{i_0}(\frac{3-\delta }{2})_{i_0}} \sum_{i_1=i_0}^{\alpha _1} \frac{(-\alpha _1)_{i_1}(\frac{\beta -\delta +2}{2})_{i_1}(\frac{3}{2})_{i_0}(\frac{4-\delta  }{2})_{i_0}}{(-\alpha _1)_{i_0}(\frac{\beta-\delta +2}{2})_{i_0}(\frac{3}{2})_{i_1}(\frac{4-\delta}{2})_{i_1}} z^{i_1} \right\} \eta \nonumber\\
&&+ \sum_{n=2}^{\infty } \left\{ \sum_{i_0=0}^{\alpha _0} \frac{i_0 \left( i_0+ \Gamma_0^{(S)}\right)+ Q_0^{(S)}}{(i_0+ \frac{1}{2})(i_0 + \frac{2-\delta }{2})}\frac{(-\alpha _0)_{i_0} (\frac{\beta-\delta +1}{2})_{i_0}}{(1)_{i_0}(\frac{3-\delta }{2})_{i_0}}\right.\nonumber\\
&&\times \prod _{k=1}^{n-1} \left\{ \sum_{i_k=i_{k-1}}^{\alpha _k} \frac{(i_k+\frac{k}{2}) \left( i_k+ \Gamma_k^{(S)}\right)+ Q_k^{(S)}}{(i_k+ \frac{k+1}{2})(i_k +\frac{2+k-\delta }{2})} \frac{(-\alpha _k)_{i_k}(\frac{\beta-\delta +k+1 }{2})_{i_k}(\frac{k+2}{2})_{i_{k-1}}(\frac{k+3-\delta }{2})_{i_{k-1}}}{(-\alpha _k)_{i_{k-1}}(\frac{\beta-\delta +k+1}{2})_{i_{k-1}}(\frac{k+2}{2})_{i_k}(\frac{k+3-\delta}{2})_{i_k}}\right\} \nonumber\\
&&\times \left.\left.\sum_{i_n= i_{n-1}}^{\alpha _n} \frac{(-\alpha _n)_{i_n}(\frac{\beta-\delta +n+1}{2})_{i_n}(\frac{n+2}{2})_{i_{n-1}}(\frac{n+3-\delta }{2})_{i_{n-1}}}{(-\alpha _n)_{i_{n-1}}(\frac{\beta-\delta +n+1}{2})_{i_{n-1}}(\frac{n+2}{2})_{i_n}(\frac{n+3-\delta }{2})_{i_n}} z^{i_n} \right\} \eta ^n \right\}\label{eq:59}
\end{eqnarray}
where
\begin{equation}
\begin{cases} \xi =1-x \cr
z = \frac{-1}{1-a}\xi^2 \cr
\eta = \frac{2-a}{1-a}\xi \cr
\alpha _i\leq \alpha _j \;\;\mbox{only}\;\mbox{if}\;i\leq j\;\;\mbox{where}\;i,j= 0,1,2,\cdots
\end{cases}\nonumber %\label{eq:37}
\end{equation}
and
\begin{equation}
\begin{cases} 
\Gamma_0^{(S)} = \frac{1}{2(2-a)}(-2\alpha _0+ \beta-\delta  -\gamma +1 +(1-a)(\gamma -\delta +1)) \cr
\Gamma_k^{(S)} = \frac{1}{2(2-a)}(-2\alpha _k+ \beta-\delta -\gamma +1 +(1-a)(\gamma -\delta  +k+1))  \cr
Q_0^{(S)} = \frac{-q+(\delta -1)\gamma a-2\alpha_0(\beta -\delta +1)}{4(2-a)} \cr
Q_k^{(S)} = \frac{-q+(\delta -1)\gamma a-(2\alpha _k+k)(\beta -\delta +1)}{4(2-a)}
\end{cases}\nonumber %\label{eq:8}
\end{equation}
For the minimum value of Heun equation for a polynomial which makes $B_n$ term terminated about $\xi=0 $, put $\alpha _0=\alpha _1=\alpha _2=\cdots=0$ in (\ref{eq:59}).
\begin{eqnarray}
& &(1-x)^{1-\delta } y(\xi)\nonumber\\
&=& (1-x)^{1-\delta } Hl(1-a,-q+(\delta -1)\gamma a+(\alpha -\delta +1)(\beta -\delta +1); \alpha-\delta +1,\beta-\delta +1, 2-\delta, \gamma; 1-x) \nonumber\\
&=& (1-x)^{1-\delta }  \; _2F_1\left( \frac{ -\Lambda _7-\sqrt{ \Lambda _7^2+4(a-1)\Omega_7}}{2(a-1)}, \frac{ -\Lambda _7+\sqrt{ \Lambda _7^2+4(a-1)\Omega_7}}{2(a-1)}; 2-\delta ; \xi \right) \hspace{1cm}\label{aa:5}
\end{eqnarray}
where $\Lambda_7= -\gamma +(1-a)(\gamma -\delta +1)$ and $\Omega_7=-q+a\gamma ( \delta -1)$.
It tells us that Heun polynomials in which makes $B_n$ term terminated, for fixed values of $\alpha $, require $\left|\xi\right| < 1$ for the convergence of the radius.

For the special case, if $\xi=1$ and $Re\left( \frac{a}{a-1}\gamma \right)<1 $ in (\ref{aa:5}),
\begin{eqnarray}
& & y(1)\nonumber\\
&=& Hl(1-a,-q+(\delta -1)\gamma a+(\alpha -\delta +1)(\beta -\delta +1); \alpha-\delta +1,\beta-\delta +1, 2-\delta, \gamma; 1) \nonumber\\
&=&  \frac{ \Gamma \left( 2-\delta \right) \Gamma \left( 1-\frac{a}{a-1}\gamma  \right)}{\Gamma \left( 2-\delta +\frac{ \Lambda _7-\sqrt{ \Lambda _7^2+4(a-1)\Omega_7}}{2(a-1)} \right) \Gamma \left( 2-\delta +\frac{  \Lambda _7+\sqrt{ \Lambda _7^2+4(a-1)\Omega_7}}{2(a-1)} \right)}   \nonumber
\end{eqnarray}

\underline {(2) The case of $\alpha = -2 \alpha _i-i+\delta -1$ and $\beta = -2 \beta _i -i+\delta -1$ only if $\alpha _i \leq \beta _i$.}

Put $\beta = -2 \beta _i -i+\delta -1$  where $i=0,1,2,\cdots$ in (\ref{eq:59}).  
\begin{eqnarray}
& &(1-x)^{1-\delta } y(\xi)\nonumber\\
&=& (1-x)^{1-\delta } Hl(1-a,-q+(\delta -1)\gamma a+(\alpha -\delta +1)(\beta -\delta +1); \alpha-\delta +1,\beta-\delta +1, 2-\delta, \gamma; 1-x) \nonumber\\
&=& (1-x)^{1-\delta } \left\{\sum_{i_0=0}^{\alpha _0} \frac{(-\alpha _0)_{i_0} (-\beta _0)_{i_0}}{(1)_{i_0}(\frac{3-\delta }{2})_{i_0}} z^{i_0}\right.\nonumber\\
&&+ \left\{\sum_{i_0=0}^{\alpha _0}\frac{i_0 \left( i_0+ \Gamma_0^{(B)} \right)+ Q_0^{(B)}}{(i_0+ \frac{1}{2})(i_0 + \frac{2-\delta }{2})}  \frac{(-\alpha _0)_{i_0} (-\beta _0)_{i_0}}{(1)_{i_0}(\frac{3-\delta }{2})_{i_0}} \sum_{i_1=i_0}^{\alpha _1} \frac{(-\alpha _1)_{i_1}(-\beta _1)_{i_1}(\frac{3}{2})_{i_0}(\frac{4-\delta  }{2})_{i_0}}{(-\alpha _1)_{i_0}(-\beta _1)_{i_0}(\frac{3}{2})_{i_1}(\frac{3-\delta}{2})_{i_1}} z^{i_1} \right\} \eta \nonumber\\
&&+ \sum_{n=2}^{\infty } \left\{ \sum_{i_0=0}^{\alpha _0} \frac{i_0 \left( i_0+ \Gamma_0^{(B)}\right)+ Q_0^{(B)}}{(i_0+ \frac{1}{2})(i_0 + \frac{2-\delta }{2})}\frac{(-\alpha _0)_{i_0} (-\beta _0)_{i_0}}{(1)_{i_0}(\frac{3-\delta }{2})_{i_0}}\right.\nonumber\\
&&\times \prod _{k=1}^{n-1} \left\{ \sum_{i_k=i_{k-1}}^{\alpha _k} \frac{(i_k+\frac{k}{2}) \left( i_k+ \Gamma_k^{(B)}\right)+ Q_k^{(B)}}{(i_k+ \frac{k+1}{2})(i_k +\frac{2+k-\delta }{2})} \frac{(-\alpha _k)_{i_k}(-\beta _k)_{i_k}(\frac{k+2}{2})_{i_{k-1}}(\frac{k+3-\delta }{2})_{i_{k-1}}}{(-\alpha _k)_{i_{k-1}}(-\beta _k)_{i_{k-1}}(\frac{k+2}{2})_{i_k}(\frac{k+3-\delta}{2})_{i_k}}\right\} \nonumber\\
&&\times  \left.\left. \sum_{i_n= i_{n-1}}^{\alpha _n} \frac{(-\alpha _n)_{i_n}(-\beta _n)_{i_n}(\frac{n+2}{2})_{i_{n-1}}(\frac{n+3-\delta }{2})_{i_{n-1}}}{(-\alpha _n)_{i_{n-1}}(-\beta _n)_{i_{n-1}}(\frac{n+2}{2})_{i_n}(\frac{n+3-\delta }{2})_{i_n}} z^{i_n} \right\} \eta ^n \right\}\label{eq:60}
\end{eqnarray}
where
\begin{equation}
\begin{cases} 
\Gamma_0^{(B)} = \frac{1}{2(2-a)}(-2\alpha _0- 2\beta _0- \gamma +(1-a)(\gamma -\delta +1))  \cr
\Gamma_k^{(B)} = \frac{1}{2(2-a)}(-2\alpha _k- 2\beta _k -\gamma -k+(1-a)(\gamma -\delta  +k+1))   \cr
Q_0^{(B)} =  \frac{-q+(\delta -1)\gamma a+4\alpha_0\beta _0}{4(2-a)} \cr
Q_k^{(B)} =  \frac{-q+(\delta -1)\gamma a+(2\alpha _k+k)(2\beta _k+k)}{4(2-a)}
\end{cases}\nonumber %\label{eq:8}
\end{equation}
For the minimum value of Heun equation for a polynomial which makes $B_n$ term terminated about $\xi=0 $, put $\alpha _0=\alpha _1=\alpha _2=\cdots=0$ and $\beta _0=\beta _1=\beta _2=\cdots=0$ in (\ref{eq:60}).
\begin{eqnarray}
& &(1-x)^{1-\delta } y(\xi)\nonumber\\
&=& (1-x)^{1-\delta } Hl(1-a,-q+(\delta -1)\gamma a+(\alpha -\delta +1)(\beta -\delta +1); \alpha-\delta +1,\beta-\delta +1, 2-\delta, \gamma; 1-x) \nonumber\\
&=& (1-x)^{1-\delta }  \; _2F_1\left( \frac{ -\Lambda _8-\sqrt{ \Lambda _8^2+4(a-1)\Omega_8}}{2(a-1)}, \frac{ -\Lambda _8+\sqrt{ \Lambda _8^2+4(a-1)\Omega_8}}{2(a-1)}; 2-\delta ; \xi \right) \hspace{1cm}\label{aa:6}
\end{eqnarray}
where $\Lambda_8= -\delta +1 -a(\gamma -\delta +1)$ and $\Omega_8= qa\gamma ( 1-\delta )$.
It tells us that Heun polynomials in which makes $B_n$ term terminated, for fixed values of $\alpha $ and $\beta $, require $\left|\xi\right| < 1$ for the convergence of the radius.

For the special case, if $\xi=1$ and $Re\left( \frac{a}{a-1}\gamma \right)<1 $ in (\ref{aa:6}),
\begin{eqnarray}
& & y(1)\nonumber\\
&=& Hl(1-a,-q+(\delta -1)\gamma a+(\alpha -\delta +1)(\beta -\delta +1); \alpha-\delta +1,\beta-\delta +1, 2-\delta, \gamma; 1) \nonumber\\
&=&  \frac{ \Gamma \left( 2-\delta \right) \Gamma \left( 1-\frac{a}{a-1}\gamma  \right)}{\Gamma \left( 2-\delta +\frac{ \Lambda _8-\sqrt{ \Lambda _8^2+4(a-1)\Omega_8}}{2(a-1)} \right) \Gamma \left( 2-\delta +\frac{  \Lambda _8+\sqrt{ \Lambda _8^2+4(a-1)\Omega_8}}{2(a-1)} \right)}   \nonumber
\end{eqnarray}
\subsubsection{Infinite series}
Replace coefficients $a$, $q$, $\alpha $, $\beta $, $\gamma $, $\delta$, $x$, $c_0$ and $\lambda $ by $1-a$, $-q+(\delta -1)\gamma a+(\alpha -\delta +1)(\beta -\delta +1)$, $\alpha-\delta +1 $, $\beta-\delta +1 $, $2-\delta$, $\gamma $, $1-x$, 1 and zero into (\ref{eq:51}).  Multiply $(1-x)^{1-\delta }$ and the new (\ref{eq:51}) together.
\begin{eqnarray}
& &(1-x)^{1-\delta } y(\xi)\nonumber\\
&=& (1-x)^{1-\delta } Hl(1-a,-q+(\delta -1)\gamma a+(\alpha -\delta +1)(\beta -\delta +1); \alpha -\delta +1,\beta-\delta +1, 2-\delta, \gamma; 1-x) \nonumber\\
&=& (1-x)^{1-\delta } \Bigg\{\sum_{i_0=0}^{\infty } \frac{(\frac{\alpha -\delta +1 }{2})_{i_0} (\frac{\beta-\delta +1 }{2})_{i_0}}{(1)_{i_0}(\frac{3-\delta }{2})_{i_0}} z^{i_0}\nonumber\\
&&+ \Bigg\{\sum_{i_0=0}^{\infty }\frac{i_0 \left( i_0+ \Gamma_0^{(I)}\right)+ Q}{(i_0+ \frac{1}{2})(i_0 + \frac{2-\delta }{2})}  \frac{(\frac{\alpha -\delta +1}{2})_{i_0} (\frac{\beta-\delta +1}{2})_{i_0}}{(1)_{i_0}(\frac{3-\delta }{2})_{i_0}} \sum_{i_1=i_0}^{\infty } \frac{(\frac{\alpha -\delta +2}{2})_{i_1}(\frac{\beta -\delta +2}{2})_{i_1}(\frac{3}{2})_{i_0}(\frac{4-\delta  }{2})_{i_0}}{(\frac{\alpha -\delta +2}{2})_{i_0}(\frac{\beta-\delta +2}{2})_{i_0}(\frac{3}{2})_{i_1}(\frac{4-\delta}{2})_{i_1}} z^{i_1} \Bigg\} \eta \nonumber\\
&&+ \sum_{n=2}^{\infty } \Bigg\{ \sum_{i_0=0}^{\infty } \frac{i_0 \left( i_0+ \Gamma_0^{(I)}\right)+ Q}{(i_0+ \frac{1}{2})(i_0 + \frac{2-\delta }{2})}\frac{(\frac{\alpha -\delta +1}{2})_{i_0} (\frac{\beta-\delta +1}{2})_{i_0}}{(1)_{i_0}(\frac{3-\delta }{2})_{i_0}} \nonumber\\
&&\times \prod _{k=1}^{n-1} \Bigg\{ \sum_{i_k=i_{k-1}}^{\infty } \frac{(i_k+\frac{k}{2}) \left( i_k+ \Gamma_k^{(I)}\right)+ Q}{(i_k+ \frac{k+1}{2})(i_k +\frac{2+k-\delta }{2})} \frac{(\frac{\alpha -\delta +k+1 }{2})_{i_k}(\frac{\beta-\delta +k+1 }{2})_{i_k}(\frac{k+2}{2})_{i_{k-1}}(\frac{k+3-\delta }{2})_{i_{k-1}}}{(\frac{\alpha -\delta +k+1 }{2})_{i_{k-1}}(\frac{\beta-\delta +k+1}{2})_{i_{k-1}}(\frac{k+2}{2})_{i_k}(\frac{k+3-\delta}{2})_{i_k}}\Bigg\} \nonumber\\
&&\times \sum_{i_n= i_{n-1}}^{\infty } \frac{(\frac{\alpha -\delta +n+1 }{2})_{i_n}(\frac{\beta-\delta +n+1}{2})_{i_n}(\frac{n+2}{2})_{i_{n-1}}(\frac{n+3-\delta }{2})_{i_{n-1}}}{(\frac{\alpha -\delta +n+1 }{2})_{i_{n-1}}(\frac{\beta-\delta +n+1}{2})_{i_{n-1}}(\frac{n+2}{2})_{i_n}(\frac{n+3-\delta }{2})_{i_n}} z^{i_n} \Bigg\} \eta ^n \Bigg\}\label{eq:61}
\end{eqnarray}
where
\begin{equation}
\begin{cases} 
\Gamma_0^{(I)} = \frac{1}{2(2-a)}(\alpha + \beta-2\delta  -\gamma +2 +(1-a)(\gamma -\delta+1))  \cr
\Gamma_k^{(I)} = \frac{1}{2(2-a)}(\alpha + \beta-2\delta -\gamma +k+2+(1-a)(\gamma -\delta  +k+1))   \cr
Q=  \frac{-q+(\delta -1)\gamma a+(\alpha -\delta +1)(\beta -\delta +1)}{4(2-a)} 
\end{cases}\nonumber %\label{eq:8}
\end{equation}
\subsection{ ${\displaystyle x^{-\alpha } Hl\left(\frac{1}{a},\frac{q+\alpha [(\alpha -\gamma -\delta +1)a-\beta +\delta ]}{a}; \alpha , \alpha -\gamma +1, \alpha -\beta +1,\delta ;\frac{1}{x}\right)}$}
\subsubsection{Infinite series}
Replace coefficients $a$, $q$, $\beta $, $\gamma $, $x$, $c_0$ and $\lambda $ by $\frac{1}{a}$, $\frac{q+\alpha [(\alpha -\gamma -\delta +1)a-\beta +\delta ]}{a}$, $\alpha-\gamma +1 $, $\alpha -\beta +1 $, $\frac{1}{x}$, 1 and zero into (\ref{eq:51}). Multiply $x^{-\alpha }$ and the new (\ref{eq:51}) together.
\begin{eqnarray}
& &x^{-\alpha } y(\xi )\nonumber\\
&=& x^{-\alpha }  Hl\left(\frac{1}{a},\frac{q+\alpha [(\alpha -\gamma -\delta +1)a-\beta +\delta ]}{a}; \alpha , \alpha -\gamma +1, \alpha -\beta +1,\delta ;\frac{1}{x}\right) \nonumber\\
&=& x^{-\alpha } \left\{\sum_{i_0=0}^{\infty } \frac{(\frac{\alpha }{2})_{i_0} (\frac{\alpha -\gamma +1}{2})_{i_0}}{(1)_{i_0}(\frac{\alpha -\beta +2}{2})_{i_0}} z^{i_0} \right. \nonumber\\
&&+ \left\{\sum_{i_0=0}^{\infty }\frac{i_0 \left( i_0+ \Gamma_0^{(I)}\right)+ Q}{(i_0+ \frac{1}{2})(i_0 + \frac{\alpha -\beta +1}{2})}  \frac{(\frac{\alpha }{2})_{i_0} (\frac{\alpha -\gamma +1}{2})_{i_0}}{(1)_{i_0}(\frac{\alpha -\beta +2}{2})_{i_0}} \sum_{i_1=i_0}^{\infty } \frac{(\frac{\alpha+1 }{2})_{i_1}(\frac{\alpha -\gamma +2}{2})_{i_1}(\frac{3}{2})_{i_0}(\frac{\alpha -\beta +3}{2})_{i_0}}{(\frac{\alpha +1}{2})_{i_0}(\frac{\alpha -\gamma +2}{2})_{i_0}(\frac{3}{2})_{i_1}(\frac{\alpha -\beta +3}{2} )_{i_1}} z^{i_1} \right\} \eta \nonumber\\
&&+ \sum_{n=2}^{\infty } \left\{ \sum_{i_0=0}^{\infty } \frac{i_0 \left( i_0+ \Gamma_0^{(I)}\right)+ Q}{(i_0+ \frac{1}{2})(i_0 + \frac{\alpha -\beta +1}{2})}  \frac{(\frac{\alpha }{2})_{i_0} (\frac{\alpha -\gamma +1}{2})_{i_0}}{(1)_{i_0}(\frac{\alpha -\beta +2}{2})_{i_0}}\right.\nonumber\\
&&\times \prod _{k=1}^{n-1} \left\{ \sum_{i_k=i_{k-1}}^{\infty } \frac{(i_k+\frac{k}{2}) \left( i_k+ \Gamma_k^{(I)}\right)+ Q}{(i_k+ \frac{k+1}{2})(i_k +\frac{\alpha -\beta +k+1}{2})} \frac{(\frac{\alpha+k }{2})_{i_k}(\frac{\alpha -\gamma +k+1}{2})_{i_k}(\frac{k+2}{2})_{i_{k-1}}(\frac{\alpha -\beta +k+2}{2})_{i_{k-1}}}{(\frac{\alpha +k}{2})_{i_{k-1}}(\frac{\alpha -\gamma +k+1}{2})_{i_{k-1}}(\frac{k+2}{2})_{i_k}(\frac{\alpha -\beta +k+2}{2})_{i_k}}\right\} \nonumber\\
&&\times \left.\left. \sum_{i_n= i_{n-1}}^{\infty } \frac{(\frac{\alpha +n}{2})_{i_n}(\frac{\alpha -\gamma +n+1}{2})_{i_n}(\frac{n+2}{2})_{i_{n-1}}(\frac{\alpha -\beta +n+2}{2})_{i_{n-1}}}{(\frac{\alpha +n}{2})_{i_{n-1}}(\frac{\alpha -\gamma +n+1}{2})_{i_{n-1}}(\frac{n+2}{2})_{i_n}(\frac{\alpha -\beta +n+2}{2})_{i_n}} z^{i_n} \right\} \eta ^n \right\} \label{eq:62}
\end{eqnarray}
where
\begin{equation}
\begin{cases} \xi =\frac{1}{x}\cr
z = -a \xi ^2 \cr
\eta = (1+a)\xi \cr
\end{cases}\nonumber %\label{eq:37}
\end{equation}
and
\begin{equation}
\begin{cases} 
\Gamma_0^{(I)} = \frac{a}{2(1+a)}(2\alpha -\gamma -\delta +1 +\frac{1}{a}(\alpha -\beta +\delta )) \cr
\Gamma_k^{(I)} = \frac{a}{2(1+a)}(2\alpha -\gamma -\delta +k+1 +\frac{1}{a}(\alpha -\beta +\delta +k))  \cr
Q=  \frac{q+\alpha (a(\alpha -\gamma -\delta +1) -\beta +\delta )}{4(1+a)}
\end{cases}\nonumber %\label{eq:8}
\end{equation}
\subsubsection{Polynomial which makes $B_n$ term terminated}
Substitute $\gamma =\alpha +2\gamma _i +i+1$ into (\ref{eq:62}) where $i, \gamma _i= 0,1,2,\cdots$; apply $\gamma =\alpha +2\gamma _0 +1$ into sub-power series $y_0(x)$, apply $\gamma =\alpha +2\gamma _0 +1$ into the first summation and $\gamma =\alpha +2\gamma _1 +2$ into second summation of sub-power series $y_1(x)$, apply $\gamma =\alpha +2\gamma _0 +1$ into the first summation, $\gamma =\alpha +2\gamma _1 +2$ into the second summation and $\gamma =\alpha +2\gamma _2 +3$ into the third summation of sub-power series $y_2(x)$,  etc.\footnote{I treat $\alpha $, $\beta $, $\delta$ and $q$ as free variables and a fixed value of $\gamma $ to construct the polynomial which makes $B_n$ term terminated.} 
\begin{eqnarray}
& &x^{-\alpha } y(\xi )\nonumber\\
&=& x^{-\alpha }  Hl\left(\frac{1}{a},\frac{q+\alpha [(\alpha -\gamma -\delta +1)a-\beta +\delta ]}{a}; \alpha , \alpha -\gamma +1, \alpha -\beta +1,\delta ;\frac{1}{x}\right) \nonumber\\
&=& x^{-\alpha } \left\{\sum_{i_0=0}^{\gamma _0} \frac{(\frac{\alpha }{2})_{i_0} (-\gamma _0)_{i_0}}{(1)_{i_0}(\frac{\alpha -\beta +2}{2})_{i_0}} z^{i_0}
\right.\nonumber\\
&&+ \left\{\sum_{i_0=0}^{\gamma _0}\frac{i_0 \left( i_0+ \Gamma_0^{(S)}\right)+ Q_0^{(S)}}{(i_0+ \frac{1}{2})(i_0 + \frac{\alpha -\beta +1}{2})}  \frac{(\frac{\alpha }{2})_{i_0} (-\gamma _0)_{i_0}}{(1)_{i_0}(\frac{\alpha -\beta +2}{2})_{i_0}} \sum_{i_1=i_0}^{\gamma _1} \frac{(\frac{\alpha+1 }{2})_{i_1}(-\gamma _1)_{i_1}(\frac{3}{2})_{i_0}(\frac{\alpha -\beta +3}{2})_{i_0}}{(\frac{\alpha +1}{2})_{i_0}(-\gamma _1)_{i_0}(\frac{3}{2})_{i_1}(\frac{\alpha -\beta +3}{2} )_{i_1}} z^{i_1} \right\} \eta \nonumber\\
&&+ \sum_{n=2}^{\infty } \left\{ \sum_{i_0=0}^{\gamma _0} \frac{i_0 \left( i_0+ \Gamma_0^{(S)}\right)+ Q_0^{(S)}}{(i_0+ \frac{1}{2})(i_0 + \frac{\alpha -\beta +1}{2})}  \frac{(\frac{\alpha }{2})_{i_0} (-\gamma _0)_{i_0}}{(1)_{i_0}(\frac{\alpha -\beta +2}{2})_{i_0}}\right.\nonumber\\
&&\times \prod _{k=1}^{n-1} \left\{ \sum_{i_k=i_{k-1}}^{\gamma _k} \frac{(i_k+\frac{k}{2}) \left( i_k+ \Gamma_k^{(S)}\right) + Q_k^{(S)}}{(i_k+ \frac{k+1}{2})(i_k +\frac{\alpha -\beta +k+1}{2})} \frac{(\frac{\alpha+k }{2})_{i_k}(-\gamma _k)_{i_k}(\frac{k+2}{2})_{i_{k-1}}(\frac{\alpha -\beta +k+2}{2})_{i_{k-1}}}{(\frac{\alpha +k}{2})_{i_{k-1}}(-\gamma _k)_{i_{k-1}}(\frac{k+2}{2})_{i_k}(\frac{\alpha -\beta +k+2}{2})_{i_k}}\right\}\nonumber\\
&&\times \left.\left. \sum_{i_n= i_{n-1}}^{\gamma _n} \frac{(\frac{\alpha +n}{2})_{i_n}(-\gamma _n)_{i_n}(\frac{n+2}{2})_{i_{n-1}}(\frac{\alpha -\beta +n+2}{2})_{i_{n-1}}}{(\frac{\alpha +n}{2})_{i_{n-1}}(-\gamma _n)_{i_{n-1}}(\frac{n+2}{2})_{i_n}(\frac{\alpha -\beta +n+2}{2})_{i_n}} z^{i_n} \right\} \eta ^n \right\} \label{eq:63}
\end{eqnarray}
where
\begin{equation}
\gamma_i\leq \gamma_j \;\;\mbox{only}\;\mbox{if}\;i\leq j\;\;\mbox{where}\;i,j= 0,1,2,\cdots
\nonumber 
\end{equation}
and
\begin{equation}
\begin{cases} 
\Gamma_0^{(S)} = \frac{a}{2(1+a)}(\alpha -\delta -2\gamma _0 +\frac{1}{a}(\alpha -\beta +\delta ))  \cr
\Gamma_k^{(S)} = \frac{a}{2(1+a)}(\alpha -\delta-2\gamma _k +\frac{1}{a}(\alpha -\beta +\delta +k))   \cr
Q_0^{(S)} =  \frac{q-\alpha (a(\delta +2\gamma _0)+\beta -\delta )}{4(1+a)}  \cr
Q_k^{(S)} =  \frac{q-\alpha (a(\delta +2\gamma _k+k)+\beta -\delta )}{4(1+a)} 
\end{cases}\nonumber %\label{eq:8}
\end{equation}
For the minimum value of Heun equation for a polynomial which makes $B_n$ term terminated about $\xi=0 $, put $\gamma _0=\gamma _1=\gamma _2=\cdots=0$ in (\ref{eq:63}).
\begin{eqnarray}
& &x^{-\alpha } y(\xi )\nonumber\\
&=& x^{-\alpha }  Hl\left(\frac{1}{a},\frac{q+\alpha [(\alpha -\gamma -\delta +1)a-\beta +\delta ]}{a}; \alpha , \alpha -\gamma +1, \alpha -\beta +1,\delta ;\frac{1}{x}\right) \nonumber\\
&=& x^{-\alpha }  \; _2F_1\left( \frac{ \Lambda _9-\sqrt{ \Lambda _9^2-4\Omega_9}}{2}, \frac{ \Lambda _9+\sqrt{ \Lambda _9^2-4\Omega_9}}{2}; \alpha -\beta +1 ; \xi \right) \hspace{1cm}\label{aa:7}
\end{eqnarray}
where $\Lambda_9= \alpha -\beta  -(a-1)\delta $ and $\Omega_9= q-\alpha (\beta +(a-1)\delta )$.
It tells us that Heun polynomials in which makes $B_n$ term terminated, for fixed values of $\gamma $, require $\left| \xi\right| < 1$ for the convergence of the radius.

For the special case, if $\xi=1$ and $Re\left( (a-1)\delta  \right)>-1 $ in (\ref{aa:7}),
\begin{eqnarray}
& & y(1)\nonumber\\
&=&    Hl\left(\frac{1}{a},\frac{q+\alpha [(\alpha -\gamma -\delta +1)a-\beta +\delta ]}{a}; \alpha , \alpha -\gamma +1, \alpha -\beta +1,\delta ;1\right) \nonumber\\
&=&   \frac{ \Gamma \left( \alpha -\beta +1 \right) \Gamma \left( 1+(a-1)\delta \right)}{\Gamma \left( \alpha -\beta +1 - \frac{ \Lambda _9-\sqrt{ \Lambda _9^2-4\Omega_9}}{2}\right) \Gamma \left( \alpha -\beta +1 -\frac{ \Lambda _9+\sqrt{ \Lambda _9^2-4\Omega_9}}{2}\right)}   \nonumber
\end{eqnarray}
\subsection{ ${\displaystyle \left(1-\frac{x}{a} \right)^{-\beta } Hl\left(1-a, -q+\gamma \beta; -\alpha +\gamma +\delta, \beta, \gamma, \delta; \frac{(1-a)x}{x-a} \right)}$}
\subsubsection{Infinite series}
Replace coefficients $a$, $q$, $\alpha $, $x$, $c_0$ and $\lambda $ by $1-a$, $-q+\gamma \beta $, $-\alpha+\gamma +\delta $, $\frac{(1-a)x}{x-a}$, 1 and zero into (\ref{eq:51}). Multiply $\left(1-\frac{x}{a} \right)^{-\beta }$ and the new (\ref{eq:51}) together.
\begin{eqnarray}
&& \left(1-\frac{x}{a} \right)^{-\beta } y(\xi ) \nonumber\\
 &=& \left(1-\frac{x}{a} \right)^{-\beta } Hl\left(1-a, -q+\gamma \beta; -\alpha +\gamma +\delta, \beta, \gamma, \delta; \frac{(1-a)x}{x-a} \right) \nonumber\\
&=& \left(1-\frac{x}{a} \right)^{-\beta } \left\{\sum_{i_0=0}^{\infty } \frac{(\frac{-\alpha +\gamma +\delta }{2})_{i_0} (\frac{\beta }{2})_{i_0}}{(1)_{i_0}(\frac{1}{2}+ \frac{\gamma}{2})_{i_0}} z^{i_0}\right.
\nonumber\\
&&+ \left\{\sum_{i_0=0}^{\infty }\frac{i_0 \left( i_0+ \Gamma_0^{(I)}\right)+ Q}{(i_0+ \frac{1}{2})(i_0 + \frac{\gamma }{2})}  \frac{(\frac{-\alpha +\gamma +\delta }{2})_{i_0} (\frac{\beta }{2})_{i_0}}{(1)_{i_0}(\frac{1}{2}+ \frac{\gamma}{2})_{i_0}} \sum_{i_1=i_0}^{\infty } \frac{(\frac{-\alpha +\gamma +\delta+1}{2})_{i_1}(\frac{1}{2}+\frac{\beta }{2})_{i_1}(\frac{3}{2})_{i_0}(1+\frac{\gamma }{2})_{i_0}}{(\frac{-\alpha +\gamma +\delta+1}{2})_{i_0}(\frac{1}{2}+\frac{\beta }{2})_{i_0}(\frac{3}{2})_{i_1}(1+ \frac{\gamma}{2})_{i_1}} z^{i_1} \right\} \eta \nonumber\\
&&+ \sum_{n=2}^{\infty } \left\{ \sum_{i_0=0}^{\infty } \frac{i_0 \left( i_0+ \Gamma_0^{(I)}\right)+ Q}{(i_0+ \frac{1}{2})(i_0 + \frac{\gamma }{2})}
 \frac{(\frac{-\alpha +\gamma +\delta}{2})_{i_0} (\frac{\beta }{2})_{i_0}}{(1)_{i_0}(\frac{1}{2}+ \frac{\gamma}{2})_{i_0}}\right.\nonumber\\
&&\times \prod _{k=1}^{n-1} \left\{ \sum_{i_k=i_{k-1}}^{\infty } \frac{(i_k+\frac{k}{2}) \left( i_k+ \Gamma_k^{(I)}\right)+ Q}{(i_k+ \frac{k}{2}+\frac{1}{2})(i_k +\frac{k}{2}+\frac{\gamma }{2})}  \frac{(\frac{-\alpha +\gamma +\delta +k}{2})_{i_k}(\frac{k}{2}+\frac{\beta }{2})_{i_k}(1+ \frac{k}{2})_{i_{k-1}}(\frac{1}{2}+\frac{k}{2}+\frac{\gamma }{2})_{i_{k-1}}}{(\frac{-\alpha +\gamma +\delta+k}{2})_{i_{k-1}}(\frac{k}{2}+\frac{\beta }{2})_{i_{k-1}}(1+\frac{k}{2})_{i_k}(\frac{1}{2}+ \frac{k}{2}+ \frac{\gamma}{2})_{i_k}}\right\} \nonumber\\
&&\times \left.\left.\sum_{i_n= i_{n-1}}^{\infty } \frac{(\frac{-\alpha +\gamma +\delta +n}{2})_{i_n}(\frac{n}{2}+\frac{\beta }{2})_{i_n}(1+ \frac{n}{2})_{i_{n-1}}(\frac{1}{2}+\frac{n}{2}+\frac{\gamma }{2})_{i_{n-1}}}{(\frac{-\alpha +\gamma +\delta +n}{2})_{i_{n-1}}(\frac{n}{2}+\frac{\beta }{2})_{i_{n-1}}(1+\frac{n}{2})_{i_n}(\frac{1}{2}+ \frac{n}{2}+ \frac{\gamma}{2})_{i_n}} z^{i_n} \right\} \eta ^n \right\}\label{eq:150}
\end{eqnarray}
where
\begin{equation}
\begin{cases} \xi= \frac{(1-a)x}{x-a} \cr
z = -\frac{1}{1-a} \xi^2 \cr
\eta = \frac{(2-a)}{(1-a)}\xi \cr
\end{cases}\nonumber %\label{eq:37}
\end{equation}
and
\begin{equation}
\begin{cases} 
\Gamma_0^{(I)} = \frac{1}{2(2-a)}(-\alpha +\beta +\gamma +(1-a)(\delta +\gamma -1)) \cr
\Gamma_k^{(I)} = \frac{1}{2(2-a)}(-\alpha +\beta +\gamma +k +(1-a)(\delta +\gamma +k-1)) \cr
Q= \frac{-q+\gamma \beta}{4(2-a)}
\end{cases}\nonumber %\label{eq:8}
\end{equation}
\subsubsection{Polynomial which makes $B_n$ term terminated}
\underline {(1) The case of $\alpha =\gamma +\delta +2\alpha _i +i$ where $i, \alpha _i$ = $0,1,2,\cdots$.}

Substitute $\alpha =\gamma +\delta +2\alpha _i +i$ into (\ref{eq:150}): apply $\alpha =\gamma +\delta +2\alpha _0$ into sub-power series $y_0(x)$, apply $\alpha =\gamma +\delta +2\alpha _0$ into the first summation and $\alpha =\gamma +\delta +2\alpha _1 +1$ into second summation of sub-power series $y_1(x)$, apply $\alpha =\gamma +\delta +2\alpha _0$ into the first summation, $\alpha =\gamma +\delta +2\alpha _1 +1$ into the second summation and $\alpha =\gamma +\delta +2\alpha _2 +2$ into the third summation of sub-power series $y_2(x)$,  etc.\footnote{I treat $\beta $, $\gamma$, $\delta$ and $q$ as free variables and a fixed value of $\alpha $ to construct the polynomial which makes $B_n$ term terminated.} 
\begin{eqnarray}
&& \left(1-\frac{x}{a} \right)^{-\beta } y(\xi ) \nonumber\\
 &=& \left(1-\frac{x}{a} \right)^{-\beta } Hl\left(1-a, -q+\gamma \beta; -\alpha +\gamma +\delta, \beta, \gamma, \delta; \frac{(1-a)x}{x-a} \right) \nonumber\\
&=& \left(1-\frac{x}{a} \right)^{-\beta } \left\{\sum_{i_0=0}^{\alpha _0} \frac{(-\alpha _0)_{i_0} (\frac{\beta }{2})_{i_0}}{(1)_{i_0}(\frac{1}{2}+ \frac{\gamma}{2})_{i_0}} z^{i_0}\right. \nonumber\\
&&+ \left\{\sum_{i_0=0}^{\alpha _0}\frac{i_0 \left( i_0+ \Gamma_0^{(S)}\right)+ Q}{(i_0+ \frac{1}{2})(i_0 + \frac{\gamma }{2})}  \frac{(-\alpha _0)_{i_0} (\frac{\beta }{2})_{i_0}}{(1)_{i_0}(\frac{1}{2}+ \frac{\gamma}{2})_{i_0}} \sum_{i_1=i_0}^{\alpha _1} \frac{(-\alpha _1)_{i_1}(\frac{1}{2}+\frac{\beta }{2})_{i_1}(\frac{3}{2})_{i_0}(1+\frac{\gamma }{2})_{i_0}}{(-\alpha _1)_{i_0}(\frac{1}{2}+\frac{\beta }{2})_{i_0}(\frac{3}{2})_{i_1}(1+ \frac{\gamma}{2})_{i_1}} z^{i_1} \right\} \eta \nonumber\\
&&+ \sum_{n=2}^{\infty } \left\{ \sum_{i_0=0}^{\alpha _0} \frac{i_0 \left( i_0+ \Gamma_0^{(S)}\right)+ Q}{(i_0+ \frac{1}{2})(i_0 + \frac{\gamma }{2})}
 \frac{(-\alpha _0)_{i_0} (\frac{\beta }{2})_{i_0}}{(1)_{i_0}(\frac{1}{2}+ \frac{\gamma}{2})_{i_0}}\right.\nonumber\\
&&\times \prod _{k=1}^{n-1} \left\{ \sum_{i_k=i_{k-1}}^{\alpha _k} \frac{(i_k+\frac{k}{2}) \left( i_k+ \Gamma_k^{(S)}\right)+ Q}{(i_k+ \frac{k}{2}+\frac{1}{2})(i_k +\frac{k}{2}+\frac{\gamma }{2})} \frac{(-\alpha _k)_{i_k}(\frac{k}{2}+\frac{\beta }{2})_{i_k}(1+ \frac{k}{2})_{i_{k-1}}(\frac{1}{2}+\frac{k}{2}+\frac{\gamma }{2})_{i_{k-1}}}{(-\alpha _k)_{i_{k-1}}(\frac{k}{2}+\frac{\beta }{2})_{i_{k-1}}(1+\frac{k}{2})_{i_k}(\frac{1}{2}+ \frac{k}{2}+ \frac{\gamma}{2})_{i_k}}\right\}\nonumber\\
&&\times  \left.\left. \sum_{i_n= i_{n-1}}^{\alpha _n} \frac{(-\alpha _n)_{i_n}(\frac{n}{2}+\frac{\beta }{2})_{i_n}(1+ \frac{n}{2})_{i_{n-1}}(\frac{1}{2}+\frac{n}{2}+\frac{\gamma }{2})_{i_{n-1}}}{(-\alpha _n)_{i_{n-1}}(\frac{n}{2}+\frac{\beta }{2})_{i_{n-1}}(1+\frac{n}{2})_{i_n}(\frac{1}{2}+ \frac{n}{2}+ \frac{\gamma}{2})_{i_n}} z^{i_n} \right\} \eta ^n \right\} \label{eq:151}
\end{eqnarray}
where
\begin{equation}
\alpha _i\leq \alpha _j \;\;\mbox{only}\;\mbox{if}\;i\leq j\;\;\mbox{where}\;i,j= 0,1,2,\cdots
\nonumber 
\end{equation}
and
\begin{equation}
\begin{cases} 
\Gamma_0^{(S)} = \frac{1}{2(2-a)}(\beta -\delta -2\alpha _0 +(1-a)(\delta +\gamma -1)) \cr
\Gamma_k^{(S)} = \frac{1}{2(2-a)}(\beta -\delta -2\alpha _k +(1-a)(\delta +\gamma +k-1)) \cr
Q= \frac{-q+\gamma \beta}{4(2-a)}
\end{cases}\nonumber %\label{eq:8}
\end{equation}
For the minimum value of Heun equation for a polynomial which makes $B_n$ term terminated about $\xi=0 $, put $\alpha _0=\alpha _1=\alpha _2=\cdots=0$ in (\ref{eq:151}).
\begin{eqnarray}
&& \left(1-\frac{x}{a} \right)^{-\beta } y(\xi ) \nonumber\\
 &=& \left(1-\frac{x}{a} \right)^{-\beta } Hl\left(1-a, -q+\gamma \beta; -\alpha +\gamma +\delta, \beta, \gamma, \delta; \frac{(1-a)x}{x-a} \right) \nonumber\\
&=& \left(1-\frac{x}{a} \right)^{-\beta } \; _2F_1\left( \frac{ -\Lambda _{10}-\sqrt{ \Lambda _{10}^2+4(a-1)\Omega_{10}}}{2(a-1)}, \frac{ -\Lambda _{10}+\sqrt{ \Lambda _{10}^2+4(a-1)\Omega_{10}}}{2(a-1)}; \gamma ; \xi \right) \hspace{1cm}\label{aa:8}
\end{eqnarray}
where $\Lambda_{10}= \beta -\delta -(a-1)(\gamma +\delta -1) $ and $\Omega_{10}= -q+\beta \gamma $.
It tells us that Heun polynomials in which makes $B_n$ term terminated, for fixed values of $\alpha $, require $\left| \xi\right| < 1$ for the convergence of the radius.

For the special case, if $\xi=1$ and $Re\left( \frac{\beta -a\delta }{a-1}  \right)>-1 $ in (\ref{aa:8}),
\begin{eqnarray}
&& \left(1-\frac{1}{a} \right)^{-\beta } y(1) \nonumber\\
 &=& \left(1-\frac{1}{a} \right)^{-\beta } Hl\left( 1-a, -q+\gamma \beta; -\alpha +\gamma +\delta, \beta, \gamma, \delta; 1 \right) \nonumber\\
&=& \left(1-\frac{1}{a} \right)^{-\beta }  \frac{ \Gamma \left( \gamma  \right) \Gamma \left( 1+\frac{\beta -a\delta }{a-1} \right)}{\Gamma \left( \gamma +\frac{  \Lambda _{10}-\sqrt{ \Lambda _{10}^2+4(a-1)\Omega_{10}}}{2(a-1)} \right) \Gamma \left( \gamma +\frac{ \Lambda _{10}+\sqrt{ \Lambda _{10}^2+4(a-1)\Omega_{10}}}{2(a-1)} \right)}   \nonumber
\end{eqnarray}

\underline {(2) The case of $\delta =\alpha -\gamma -2\delta _i -i$ where $i, \delta _i$ = $0,1,2,\cdots$.}

Substitute $\delta =\alpha -\gamma -2\delta _i -i$ into (\ref{eq:150}): apply $\delta =\alpha -\gamma -2\delta _0$ into sub-power series $y_0(x)$, apply $\delta =\alpha -\gamma -2\delta _0$ into the first summation and $\delta =\alpha -\gamma -2\delta _1 -1$ into second summation of sub-power series $y_1(x)$, apply $\delta =\alpha -\gamma -2\delta _0$ into the first summation, $\delta =\alpha -\gamma -2\delta _1 -1$ into the second summation and $\delta =\alpha -\gamma -2\delta _2 -2$ into the third summation of sub-power series $y_2(x)$,  etc.\footnote{I treat $\alpha $, $\beta $, $\gamma$ and $q$ as free variables and a fixed value of $\delta $ to construct the polynomial which makes $B_n$ term terminated.}   
\begin{eqnarray}
&& \left(1-\frac{x}{a} \right)^{-\beta } y(\xi ) \nonumber\\
 &=& \left(1-\frac{x}{a} \right)^{-\beta } Hl\left(1-a, -q+\gamma \beta; -\alpha +\gamma +\delta, \beta, \gamma, \delta; \frac{(1-a)x}{x-a} \right) \nonumber\\
&=& \left(1-\frac{x}{a} \right)^{-\beta } \left\{\sum_{i_0=0}^{\delta _0} \frac{(-\delta _0)_{i_0} (\frac{\beta }{2})_{i_0}}{(1)_{i_0}(\frac{1}{2}+ \frac{\gamma}{2})_{i_0}} z^{i_0} \right.\nonumber\\
&&+ \left\{\sum_{i_0=0}^{\delta _0}\frac{i_0 \left( i_0+ \Gamma_0^{(S)}\right)+ Q}{(i_0+ \frac{1}{2})(i_0 + \frac{\gamma }{2})}  \frac{(-\delta _0)_{i_0} (\frac{\beta }{2})_{i_0}}{(1)_{i_0}(\frac{1}{2}+ \frac{\gamma}{2})_{i_0}} \sum_{i_1=i_0}^{\delta _1} \frac{(-\delta _1)_{i_1}(\frac{1}{2}+\frac{\beta }{2})_{i_1}(\frac{3}{2})_{i_0}(1+\frac{\gamma }{2})_{i_0}}{(-\delta _1)_{i_0}(\frac{1}{2}+\frac{\beta }{2})_{i_0}(\frac{3}{2})_{i_1}(1+ \frac{\gamma}{2})_{i_1}} z^{i_1} \right\} \eta \nonumber\\
&&+ \sum_{n=2}^{\infty } \left\{ \sum_{i_0=0}^{\delta _0} \frac{i_0 \left( i_0+ \Gamma_0^{(S)}\right)+ Q}{(i_0+ \frac{1}{2})(i_0 + \frac{\gamma }{2})}
 \frac{(-\delta _0 )_{i_0} (\frac{\beta }{2})_{i_0}}{(1)_{i_0}(\frac{1}{2}+ \frac{\gamma}{2})_{i_0}}\right.\nonumber\\
&&\times \prod _{k=1}^{n-1} \left\{ \sum_{i_k=i_{k-1}}^{\delta _k} \frac{(i_k+\frac{k}{2}) \left( i_k+ \Gamma_k^{(S)}\right)+ Q}{(i_k+ \frac{k}{2}+\frac{1}{2})(i_k +\frac{k}{2}+\frac{\gamma }{2})}  \frac{(-\delta _k)_{i_k}(\frac{k}{2}+\frac{\beta }{2})_{i_k}(1+ \frac{k}{2})_{i_{k-1}}(\frac{1}{2}+\frac{k}{2}+\frac{\gamma }{2})_{i_{k-1}}}{(-\delta _k)_{i_{k-1}}(\frac{k}{2}+\frac{\beta }{2})_{i_{k-1}}(1+\frac{k}{2})_{i_k}(\frac{1}{2}+ \frac{k}{2}+ \frac{\gamma}{2})_{i_k}}\right\} \nonumber\\
&&\times \left.\left.\sum_{i_n= i_{n-1}}^{\delta _n} \frac{(-\delta _n)_{i_n}(\frac{n}{2}+\frac{\beta }{2})_{i_n}(1+ \frac{n}{2})_{i_{n-1}}(\frac{1}{2}+\frac{n}{2}+\frac{\gamma }{2})_{i_{n-1}}}{(-\delta _n)_{i_{n-1}}(\frac{n}{2}+\frac{\beta }{2})_{i_{n-1}}(1+\frac{n}{2})_{i_n}(\frac{1}{2}+ \frac{n}{2}+ \frac{\gamma}{2})_{i_n}} z^{i_n} \right\} \eta ^n \right\}\label{eq:152}
\end{eqnarray}
where
\begin{equation}
\delta _i\leq \delta _j \;\;\mbox{only}\;\mbox{if}\;i\leq j\;\;\mbox{where}\;i,j= 0,1,2,\cdots
\nonumber 
\end{equation}
and
\begin{equation}
\begin{cases} 
\Gamma_0^{(S)} = \frac{1}{2(2-a)}(-\alpha +\beta +\gamma +(1-a)(\alpha -2\delta _0 -1)) \cr
\Gamma_k^{(S)} = \frac{1}{2(2-a)}(-\alpha +\beta +\gamma +k +(1-a)(\alpha -2\delta _k-1))  \cr
Q= \frac{-q+\gamma \beta}{4(2-a)}
\end{cases}\nonumber %\label{eq:8}
\end{equation}
For the minimum value of Heun equation for a polynomial which makes $B_n$ term terminated about $\xi=0 $, put $\delta _0=\delta _1=\delta _2=\cdots=0$ in (\ref{eq:152}).
\begin{eqnarray}
&& \left(1-\frac{x}{a} \right)^{-\beta } y(\xi ) \nonumber\\
 &=& \left(1-\frac{x}{a} \right)^{-\beta } Hl\left( 1-a, -q+\gamma \beta; -\alpha +\gamma +\delta, \beta, \gamma, \delta; \frac{(1-a)x}{x-a} \right) \nonumber\\
&=& \left(1-\frac{x}{a} \right)^{-\beta } \; _2F_1\left( \frac{ \Lambda _{11}-\sqrt{ \Lambda _{11}^2-4 \Omega_{11}}}{2}, \frac{ \Lambda _{11}+\sqrt{ \Lambda _{11}^2-4\Omega_{11}}}{2}; \gamma ; \frac{\xi}{1-a} \right) \hspace{1cm}\label{aa:9}
\end{eqnarray}
where $\Lambda_{11}= \beta +\gamma -1 -a(\alpha -1) $ and $\Omega_{11}= -q+\beta \gamma $.
It tells us that Heun polynomials in which makes $B_n$ term terminated, for fixed values of $\delta $, require $\left| \frac{\xi}{1-a}\right| < 1$ for the convergence of the radius.

For the special case, if $\frac{\xi}{1-a}=1$ and $Re\left( \beta -a(\alpha -1)\right)<1 $ in (\ref{aa:9}),
\begin{eqnarray}
&& \left(1-\frac{1}{a} \right)^{-\beta } y(1-a) \nonumber\\
 &=& \left(1-\frac{1}{a} \right)^{-\beta } Hl\left( 1-a, -q+\gamma \beta; -\alpha +\gamma +\delta, \beta, \gamma, \delta; 1 \right) \nonumber\\
&=& \left(1-\frac{1}{a} \right)^{-\beta } \frac{ \Gamma \left( \gamma  \right) \Gamma \left( -\beta +1+a(\alpha -1) \right)}{\Gamma \left( \gamma - \frac{ \Lambda _{11}-\sqrt{ \Lambda _{11}^2-4 \Omega_{11}}}{2} \right) \Gamma \left( \gamma - \frac{ \Lambda _{11}+\sqrt{ \Lambda _{11}^2-4\Omega_{11}}}{2}\right)}   \nonumber
\end{eqnarray}
\subsection{ ${\displaystyle (1-x)^{1-\delta }\left(1-\frac{x}{a} \right)^{-\beta+\delta -1} Hl\Bigg(1-a, -q+\gamma [(\delta -1)a+\beta -\delta +1]; -\alpha +\gamma +1}$\\ ${\displaystyle, \beta -\delta+1, \gamma, 2-\delta; \frac{(1-a)x}{x-a} \Bigg)}$ }
\subsubsection{Infinite series}
Replace coefficients $a$, $q$, $\alpha $, $\beta $, $\delta $, $x$, $c_0$ and $\lambda $ by $1-a$, $-q+\gamma [(\delta -1)a+\beta -\delta +1]$, $-\alpha +\gamma +1$, $\beta -\delta+1$, $2-\delta $, $\frac{(1-a)x}{x-a}$, 1 and zero into (\ref{eq:51}). Multiply $(1-x)^{1-\delta }\left(1-\frac{x}{a} \right)^{-\beta+\delta -1}$ and the new (\ref{eq:51}) together.
\begin{eqnarray}
&& (1-x)^{1-\delta }\left(1-\frac{x}{a} \right)^{-\beta+\delta -1} y(\xi ) \nonumber\\
&=& (1-x)^{1-\delta }\left(1-\frac{x}{a} \right)^{-\beta+\delta -1} Hl\Bigg(1-a, -q+\gamma [(\delta -1)a+\beta -\delta +1]; -\alpha +\gamma +1, \beta -\delta+1, \gamma \nonumber\\
&&, 2-\delta; \frac{(1-a)x}{x-a} \Bigg) \nonumber\\
&=& (1-x)^{1-\delta }\left(1-\frac{x}{a} \right)^{-\beta+\delta -1} \left\{\sum_{i_0=0}^{\infty } \frac{(\frac{-\alpha+\gamma +1 }{2})_{i_0} (\frac{\beta-\delta +1}{2})_{i_0}}{(1)_{i_0}(\frac{1}{2}+ \frac{\gamma}{2})_{i_0}} z^{i_0}\right.\nonumber\\
&&+ \left\{\sum_{i_0=0}^{\infty }\frac{i_0 \left( i_0+ \Gamma_0^{(I)}\right)+ Q}{(i_0+ \frac{1}{2})(i_0 + \frac{\gamma }{2})}  \frac{(\frac{-\alpha+\gamma +1 }{2})_{i_0} (\frac{\beta-\delta +1 }{2})_{i_0}}{(1)_{i_0}(\frac{1}{2}+ \frac{\gamma}{2})_{i_0}} \sum_{i_1=i_0}^{\infty } \frac{(\frac{-\alpha +\gamma +2}{2})_{i_1}(\frac{\beta-\delta +2 }{2})_{i_1}(\frac{3}{2})_{i_0}(1+\frac{\gamma }{2})_{i_0}}{(\frac{-\alpha +\gamma +2}{2})_{i_0}(\frac{\beta-\delta +2 }{2})_{i_0}(\frac{3}{2})_{i_1}(1+ \frac{\gamma}{2})_{i_1}} z^{i_1} \right\} \eta \nonumber\\
&&+ \sum_{n=2}^{\infty } \left\{ \sum_{i_0=0}^{\infty } \frac{i_0 \left( i_0+ \Gamma_0^{(I)}\right)+ Q}{(i_0+ \frac{1}{2})(i_0 + \frac{\gamma }{2})}
 \frac{(\frac{-\alpha+\gamma +1 }{2})_{i_0} (\frac{\beta -\delta +1}{2})_{i_0}}{(1)_{i_0}(\frac{1}{2}+ \frac{\gamma}{2})_{i_0}}\right. \nonumber\\
&&\times \prod _{k=1}^{n-1} \left\{ \sum_{i_k=i_{k-1}}^{\infty } \frac{(i_k+\frac{k}{2}) \left( i_k+ \Gamma_k^{(I)}\right)+ Q}{(i_k+ \frac{k}{2}+\frac{1}{2})(i_k +\frac{k}{2}+\frac{\gamma }{2})}   \frac{(\frac{-\alpha +\gamma +k+1}{2})_{i_k}(\frac{\beta-\delta +k+1 }{2})_{i_k}(1+ \frac{k}{2})_{i_{k-1}}(\frac{1}{2}+\frac{k}{2}+\frac{\gamma }{2})_{i_{k-1}}}{(\frac{-\alpha +\gamma +k+1}{2})_{i_{k-1}}(\frac{\beta-\delta +k+1 }{2})_{i_{k-1}}(1+\frac{k}{2})_{i_k}(\frac{1}{2}+ \frac{k}{2}+ \frac{\gamma}{2})_{i_k}}\right\} \nonumber\\
&&\times \left.\left.\sum_{i_n= i_{n-1}}^{\infty } \frac{(\frac{-\alpha +\gamma +n+1}{2})_{i_n}(\frac{\beta-\delta +n+1 }{2})_{i_n}(1+ \frac{n}{2})_{i_{n-1}}(\frac{1}{2}+\frac{n}{2}+\frac{\gamma }{2})_{i_{n-1}}}{(\frac{-\alpha +\gamma +n+1}{2})_{i_{n-1}}(\frac{\beta-\delta +n+1 }{2})_{i_{n-1}}(1+\frac{n}{2})_{i_n}(\frac{1}{2}+ \frac{n}{2}+ \frac{\gamma}{2})_{i_n}} z^{i_n} \right\} \eta ^n \right\}\label{eq:153}
\end{eqnarray}
where
\begin{equation}
\begin{cases} \xi= \frac{(1-a)x}{x-a} \cr
z = -\frac{1}{1-a} \xi^2 \cr
\eta = \frac{(2-a)}{(1-a)}\xi \cr
\end{cases}\nonumber %\label{eq:37}
\end{equation}
and
\begin{equation}
\begin{cases} 
\Gamma_0^{(I)} = \frac{1}{2(2-a)}(-\alpha +\beta +\gamma +(1-a)(\gamma -\delta +1)) \cr
\Gamma_k^{(I)} = \frac{1}{2(2-a)}(-\alpha +\beta +\gamma +k +(1-a)(\gamma -\delta +k+1))   \cr
Q=  \frac{-q+\gamma (a(\delta -1) +\beta -\delta +1)}{4(2-a)}
\end{cases}\nonumber %\label{eq:8}
\end{equation}
\subsubsection{Polynomial which makes $B_n$ term terminated}

Substitute $\alpha =\gamma +1 +2\alpha _i +i$ into (\ref{eq:153}): apply $\alpha =\gamma +1 +2\alpha _0$ into sub-power series $y_0(x)$, apply $\alpha =\gamma +1 +2\alpha _0$ into the first summation and $\alpha =\gamma +1 +2\alpha _1 +1$ into second summation of sub-power series $y_1(x)$, apply $\alpha =\gamma +1 +2\alpha _0$ into the first summation, $\alpha =\gamma +1 +2\alpha _1 +1$ into the second summation and $\alpha =\gamma +1 +2\alpha _2 +2$ into the third summation of sub-power series $y_2(x)$,  etc.\footnote{I treat $\beta $, $\gamma$, $\delta$ and $q$ as free variables and a fixed value of $\alpha $ to construct the polynomial which makes $B_n$ term terminated.}  
\begin{eqnarray}
&& (1-x)^{1-\delta }\left(1-\frac{x}{a} \right)^{-\beta+\delta -1} y(\xi ) \nonumber\\
&=& (1-x)^{1-\delta }\left(1-\frac{x}{a} \right)^{-\beta+\delta -1} Hl\Bigg(1-a, -q+\gamma [(\delta -1)a+\beta -\delta +1]; -\alpha +\gamma +1, \beta -\delta+1, \gamma \nonumber\\
&&, 2-\delta; \frac{(1-a)x}{x-a} \Bigg) \nonumber\\
&=& (1-x)^{1-\delta }\left(1-\frac{x}{a} \right)^{-\beta+\delta -1} \left\{\sum_{i_0=0}^{\alpha _0} \frac{(-\alpha _0)_{i_0} (\frac{\beta-\delta +1}{2})_{i_0}}{(1)_{i_0}(\frac{1}{2}+ \frac{\gamma}{2})_{i_0}} z^{i_0}\right. \nonumber\\
&&+ \left\{\sum_{i_0=0}^{\alpha _0}\frac{i_0 \left( i_0+ \Gamma_0^{(S)}\right)+ Q}{(i_0+ \frac{1}{2})(i_0 + \frac{\gamma }{2})} \frac{(-\alpha _0)_{i_0} (\frac{\beta-\delta +1 }{2})_{i_0}}{(1)_{i_0}(\frac{1}{2}+ \frac{\gamma}{2})_{i_0}} \sum_{i_1=i_0}^{\alpha _1} \frac{(-\alpha _1)_{i_1}(\frac{\beta-\delta +2 }{2})_{i_1}(\frac{3}{2})_{i_0}(1+\frac{\gamma }{2})_{i_0}}{(-\alpha _1)_{i_0}(\frac{\beta-\delta +2 }{2})_{i_0}(\frac{3}{2})_{i_1}(1+ \frac{\gamma}{2})_{i_1}} z^{i_1} \right\} \eta \nonumber\\
&&+ \sum_{n=2}^{\infty } \left\{ \sum_{i_0=0}^{\alpha _0} \frac{i_0 \left( i_0+ \Gamma_0^{(S)}\right)+ Q}{(i_0+ \frac{1}{2})(i_0 + \frac{\gamma }{2})}
 \frac{(-\alpha _0)_{i_0} (\frac{\beta -\delta +1}{2})_{i_0}}{(1)_{i_0}(\frac{1}{2}+ \frac{\gamma}{2})_{i_0}}\right.\nonumber\\
&&\times \prod _{k=1}^{n-1} \left\{ \sum_{i_k=i_{k-1}}^{\alpha _k} \frac{(i_k+\frac{k}{2}) \left( i_k+ \Gamma_k^{(S)}\right)+ Q}{(i_k+ \frac{k}{2}+\frac{1}{2})(i_k +\frac{k}{2}+\frac{\gamma }{2})} \frac{(-\alpha _k)_{i_k}(\frac{\beta-\delta +k+1 }{2})_{i_k}(1+ \frac{k}{2})_{i_{k-1}}(\frac{1}{2}+\frac{k}{2}+\frac{\gamma }{2})_{i_{k-1}}}{(-\alpha _k)_{i_{k-1}}(\frac{\beta-\delta +k+1 }{2})_{i_{k-1}}(1+\frac{k}{2})_{i_k}(\frac{1}{2}+ \frac{k}{2}+ \frac{\gamma}{2})_{i_k}}\right\} \nonumber\\
&&\times  \left.\left. \sum_{i_n= i_{n-1}}^{\alpha _n} \frac{(-\alpha _n)_{i_n}(\frac{\beta-\delta +n+1 }{2})_{i_n}(1+ \frac{n}{2})_{i_{n-1}}(\frac{1}{2}+\frac{n}{2}+\frac{\gamma }{2})_{i_{n-1}}}{(-\alpha _n)_{i_{n-1}}(\frac{\beta-\delta +n+1 }{2})_{i_{n-1}}(1+\frac{n}{2})_{i_n}(\frac{1}{2}+ \frac{n}{2}+ \frac{\gamma}{2})_{i_n}} z^{i_n} \right\} \eta ^n \right\} \label{eq:154}
\end{eqnarray}
where
\begin{equation}
\alpha _i\leq \alpha _j \;\;\mbox{only}\;\mbox{if}\;i\leq j\;\;\mbox{where}\;i,j= 0,1,2,\cdots
\nonumber 
\end{equation}
and
\begin{equation}
\begin{cases} 
\Gamma_0^{(S)} = \frac{1}{2(2-a)}(\beta -2\alpha _0-1 +(1-a)(\gamma -\delta +1))  \cr
\Gamma_k^{(S)} = \frac{1}{2(2-a)}(\beta -2\alpha _k-1 +(1-a)(\gamma -\delta +k+1))    \cr
Q=  \frac{-q+\gamma (a(\delta -1) +\beta -\delta +1)}{4(2-a)}
\end{cases}\nonumber %\label{eq:8}
\end{equation}
For the minimum value of Heun equation for a polynomial which makes $B_n$ term terminated about $\xi=0 $, put $\alpha _0=\alpha _1=\alpha _2=\cdots=0$ in (\ref{eq:154}).
\begin{eqnarray}
&& (1-x)^{1-\delta }\left(1-\frac{x}{a} \right)^{-\beta+\delta -1} y(\xi ) \nonumber\\
&=& (1-x)^{1-\delta }\left(1-\frac{x}{a} \right)^{-\beta+\delta -1} Hl\Bigg(1-a, -q+\gamma [(\delta -1)a+\beta -\delta +1]; -\alpha +\gamma +1, \beta -\delta+1, \gamma \nonumber\\
&&, 2-\delta; \frac{(1-a)x}{x-a} \Bigg) \nonumber\\
&=& (1-x)^{1-\delta }\left(1-\frac{x}{a} \right)^{-\beta+\delta -1} \; _2F_1\left( \frac{ -\Lambda _{12}-\sqrt{ \Lambda _{12}^2+4 (a-1)\Omega_{12}}}{2(a-1)}, \frac{ -\Lambda _{12}+\sqrt{ \Lambda _{12}^2+4 (a-1)\Omega_{12}}}{2(a-1)}; \gamma ; \xi \right) \nonumber %\hspace{1cm}\label{aa:10}
\end{eqnarray}
where $\Lambda_{12}= \beta -(a-1)(\gamma -\delta ) $ and $\Omega_{12}= -q+ \gamma (\beta +(a-1)(\delta -1))$.
It tells us that Heun polynomials in which makes $B_n$ term terminated, for fixed values of $\alpha  $, require $\left| \xi \right| < 1$ for the convergence of the radius.
\subsection{ ${\displaystyle x^{-\alpha } Hl\left(\frac{a-1}{a}, \frac{-q+\alpha (\delta a+\beta -\delta )}{a}; \alpha, \alpha -\gamma +1, \delta , \alpha -\beta +1; \frac{x-1}{x} \right)}$}
\subsubsection{Infinite series}
Replace coefficients $a$, $q$, $\beta $, $\gamma $, $\delta $, $x$, $c_0$ and $\lambda $ by $\frac{a-1}{a}$, $\frac{-q+\alpha (\delta a+\beta -\delta )}{a}$, $\alpha -\gamma +1$, $\delta $, $\alpha -\beta +1$, $\frac{x-1}{x}$, 1 and zero into (\ref{eq:51}). Multiply $x^{-\alpha }$ and the new (\ref{eq:51}) together.
\begin{eqnarray}
&& x^{-\alpha } y(\xi ) \nonumber\\
 &=& x^{-\alpha } Hl\left(\frac{a-1}{a}, \frac{-q+\alpha (\delta a+\beta -\delta )}{a}; \alpha, \alpha -\gamma +1, \delta , \alpha -\beta +1; \frac{x-1}{x} \right) \nonumber\\
&=& x^{-\alpha } \left\{\sum_{i_0=0}^{\infty } \frac{(\frac{\alpha }{2})_{i_0} (\frac{\alpha -\gamma +1}{2})_{i_0}}{(1)_{i_0}(\frac{1}{2}+ \frac{\delta }{2})_{i_0}} z^{i_0} \right. \nonumber\\
&&+ \left\{\sum_{i_0=0}^{\infty }\frac{i_0 \left( i_0+ \Gamma_0^{(I)}\right)+ Q}{(i_0+ \frac{1}{2})(i_0 + \frac{\delta }{2})}  \frac{(\frac{\alpha }{2})_{i_0} (\frac{\alpha -\gamma +1}{2})_{i_0}}{(1)_{i_0}(\frac{1}{2}+ \frac{\delta }{2})_{i_0}} \sum_{i_1=i_0}^{\infty } \frac{(\frac{1}{2}+\frac{\alpha }{2})_{i_1}(\frac{\alpha -\gamma +2}{2})_{i_1}(\frac{3}{2})_{i_0}(1+\frac{\delta }{2})_{i_0}}{(\frac{1}{2}+\frac{\alpha }{2})_{i_0}(\frac{\alpha -\gamma +2}{2})_{i_0}(\frac{3}{2})_{i_1}(1+ \frac{\delta }{2})_{i_1}} z^{i_1} \right\} \eta \nonumber\\
&&+ \sum_{n=2}^{\infty } \left\{ \sum_{i_0=0}^{\infty } \frac{i_0 \left( i_0+ \Gamma_0^{(I)}\right)+ Q}{(i_0+ \frac{1}{2})(i_0 + \frac{\delta }{2})}  \frac{(\frac{\alpha }{2})_{i_0} (\frac{\alpha -\gamma +1}{2})_{i_0}}{(1)_{i_0}(\frac{1}{2}+ \frac{\delta }{2})_{i_0}}\right. \nonumber\\
&&\times \prod _{k=1}^{n-1} \left\{ \sum_{i_k=i_{k-1}}^{\infty } \frac{(i_k+\frac{k}{2}) \left( i_k+ \Gamma_k^{(I)}\right)+ Q}{(i_k+ \frac{k}{2}+\frac{1}{2})(i_k +\frac{k}{2}+\frac{\delta }{2})}  \frac{(\frac{k}{2}+\frac{\alpha }{2})_{i_k}(\frac{\alpha -\gamma +k+1}{2})_{i_k}(1+ \frac{k}{2})_{i_{k-1}}(\frac{1}{2}+\frac{k}{2}+\frac{\delta }{2})_{i_{k-1}}}{(\frac{k}{2}+\frac{\alpha }{2})_{i_{k-1}}(\frac{\alpha -\gamma +k+1}{2})_{i_{k-1}}(1+\frac{k}{2})_{i_k}(\frac{1}{2}+ \frac{k}{2}+ \frac{\delta }{2})_{i_k}}\right\} \nonumber\\
&&\times \left.\left. \sum_{i_n= i_{n-1}}^{\infty } \frac{(\frac{n}{2}+\frac{\alpha }{2})_{i_n}(\frac{\alpha -\gamma +n+1}{2})_{i_n}(1+ \frac{n}{2})_{i_{n-1}}(\frac{1}{2}+\frac{n}{2}+\frac{\delta }{2})_{i_{n-1}}}{(\frac{n}{2}+\frac{\alpha }{2})_{i_{n-1}}(\frac{\alpha -\gamma +n+1}{2})_{i_{n-1}}(1+\frac{n}{2})_{i_n}(\frac{1}{2}+ \frac{n}{2}+ \frac{\delta }{2})_{i_n}} z^{i_n} \right\} \eta ^n \right\}\label{eq:155}
\end{eqnarray}
where
\begin{equation}
\begin{cases} \xi= \frac{x-1}{x} \cr
z = \frac{-a}{a-1} \xi^2 \cr
\eta = \frac{2a-1}{a-1}\xi \cr
\end{cases}\nonumber %\label{eq:37}
\end{equation}
and
\begin{equation}
\begin{cases} 
\Gamma_0^{(I)} = \frac{a}{2(2a-1)} (\alpha +\beta -\gamma +\frac{a-1}{a}(\alpha -\beta +\delta ) ) \cr
\Gamma_k^{(I)} = \frac{a}{2(2a-1)} (\alpha +\beta -\gamma +k +\frac{a-1}{a}(\alpha -\beta +\delta +k) )    \cr
Q= \frac{-q+\alpha (\delta a+\beta -\delta )}{4(2a-1)}  
\end{cases}\nonumber %\label{eq:8}
\end{equation}
\subsubsection{Polynomial which makes $B_n$ term terminated}

Substitute $\gamma =\alpha +1+2\gamma _i +i$ into (\ref{eq:155}): apply $\gamma =\alpha +1+2\gamma _0$ into sub-power series $y_0(x)$, apply $\gamma =\alpha +1+2\gamma _0$ into the first summation and $\gamma =\alpha +1+2\gamma _1 +1$ into second summation of sub-power series $y_1(x)$, apply $\gamma =\alpha +1+2\gamma _0$ into the first summation, $\gamma =\alpha +1+2\gamma _1 +1$ into the second summation and $\gamma =\alpha +1+2\gamma _2 +2$ into the third summation of sub-power series $y_2(x)$, etc.\footnote{I treat $\alpha $, $\beta $, $\delta$ and $q$ as free variables and a fixed value of $\gamma $ to construct the polynomial which makes $B_n$ term terminated.}  
\begin{eqnarray}
&& x^{-\alpha } y(\xi ) \nonumber\\
 &=& x^{-\alpha } Hl\left(\frac{a-1}{a}, \frac{-q+\alpha (\delta a+\beta -\delta )}{a}; \alpha, \alpha -\gamma +1, \delta , \alpha -\beta +1; \frac{x-1}{x} \right) \nonumber\\
&=& x^{-\alpha } \left\{\sum_{i_0=0}^{\gamma _0} \frac{(\frac{\alpha }{2})_{i_0} (-\gamma _0)_{i_0}}{(1)_{i_0}(\frac{1}{2}+ \frac{\delta }{2})_{i_0}} z^{i_0} \right. \nonumber\\
&&+ \left\{\sum_{i_0=0}^{\gamma _0}\frac{i_0 \left( i_0+ \Gamma_0^{(S)}\right)+ Q}{(i_0+ \frac{1}{2})(i_0 + \frac{\delta }{2})}  \frac{(\frac{\alpha }{2})_{i_0} (-\gamma _0)_{i_0}}{(1)_{i_0}(\frac{1}{2}+ \frac{\delta }{2})_{i_0}} \sum_{i_1=i_0}^{\gamma _1} \frac{(\frac{1}{2}+\frac{\alpha }{2})_{i_1}(-\gamma _1)_{i_1}(\frac{3}{2})_{i_0}(1+\frac{\delta }{2})_{i_0}}{(\frac{1}{2}+\frac{\alpha }{2})_{i_0}(-\gamma _1)_{i_0}(\frac{3}{2})_{i_1}(1+ \frac{\delta }{2})_{i_1}} z^{i_1} \right\} \eta \nonumber\\
&&+ \sum_{n=2}^{\infty } \left\{\sum_{i_0=0}^{\gamma _0} \frac{i_0 \left( i_0+ \Gamma_0^{(S)}\right)+ Q}{(i_0+ \frac{1}{2})(i_0 + \frac{\delta }{2})}  \frac{(\frac{\alpha }{2})_{i_0} (-\gamma _0)_{i_0}}{(1)_{i_0}(\frac{1}{2}+ \frac{\delta }{2})_{i_0}}\right. \nonumber\\
&&\times \prod _{k=1}^{n-1} \left\{ \sum_{i_k=i_{k-1}}^{\gamma _k} \frac{(i_k+\frac{k}{2}) \left( i_k+ \Gamma_k^{(S)}\right)+ Q}{(i_k+ \frac{k}{2}+\frac{1}{2})(i_k +\frac{k}{2}+\frac{\delta }{2})}   \frac{(\frac{k}{2}+\frac{\alpha }{2})_{i_k}(-\gamma _k)_{i_k}(1+ \frac{k}{2})_{i_{k-1}}(\frac{1}{2}+\frac{k}{2}+\frac{\delta }{2})_{i_{k-1}}}{(\frac{k}{2}+\frac{\alpha }{2})_{i_{k-1}}(-\gamma _k)_{i_{k-1}}(1+\frac{k}{2})_{i_k}(\frac{1}{2}+ \frac{k}{2}+ \frac{\delta }{2})_{i_k}}\right\} \nonumber\\
&&\times \left.\left. \sum_{i_n= i_{n-1}}^{\gamma _n} \frac{(\frac{n}{2}+\frac{\alpha }{2})_{i_n}(-\gamma _n)_{i_n}(1+ \frac{n}{2})_{i_{n-1}}(\frac{1}{2}+\frac{n}{2}+\frac{\delta }{2})_{i_{n-1}}}{(\frac{n}{2}+\frac{\alpha }{2})_{i_{n-1}}(-\gamma _n)_{i_{n-1}}(1+\frac{n}{2})_{i_n}(\frac{1}{2}+ \frac{n}{2}+ \frac{\delta }{2})_{i_n}} z^{i_n} \right\} \eta ^n \right\}\label{eq:156}
\end{eqnarray}
where
\begin{equation}
\gamma_i\leq \gamma_j \;\;\mbox{only}\;\mbox{if}\;i\leq j\;\;\mbox{where}\;i,j= 0,1,2,\cdots
\nonumber 
\end{equation}
and
\begin{equation}
\begin{cases} 
\Gamma_0^{(S)} = \frac{a}{2(2a-1)} (\beta -2\gamma _0-1+\frac{a-1}{a}(\alpha -\beta +\delta ) )  \cr
\Gamma_k^{(S)} =  \frac{a}{2(2a-1)} (\beta -2\gamma _k -1 +\frac{a-1}{a}(\alpha -\beta +\delta +k) ) \cr
Q= \frac{-q+\alpha (\delta a+\beta -\delta )}{4(2a-1)}  
\end{cases}\nonumber %\label{eq:8}
\end{equation}
For the minimum value of Heun equation for a polynomial which makes $B_n$ term terminated about $\xi=0 $, put $\gamma _0=\gamma _1=\gamma _2=\cdots=0$ in (\ref{eq:156}).
\begin{eqnarray}
&& x^{-\alpha } y(\xi ) \nonumber\\
 &=& x^{-\alpha } Hl\left(\frac{a-1}{a}, \frac{-q+\alpha (\delta a+\beta -\delta )}{a}; \alpha, \alpha -\gamma +1, \delta , \alpha -\beta +1; \frac{x-1}{x} \right) \nonumber\\
&=& x^{-\alpha } \; _2F_1\left( \frac{ \Lambda _{13}-\sqrt{ \Lambda _{13}^2-4 (a-1)\Omega_{13}}}{2(a-1)},  \frac{ \Lambda _{13}+\sqrt{ \Lambda _{13}^2-4 (a-1)\Omega_{13}}}{2(a-1)}; \delta  ; \xi \right)  \nonumber %\hspace{1cm}\label{aa:11}
\end{eqnarray}
where $\Lambda_{13}= -\alpha +\beta -\delta +a(\alpha +\delta -1) $ and $\Omega_{13}= -q+ \alpha (\beta +(a-1)\delta )$.
It tells us that Heun polynomials in which makes $B_n$ term terminated, for fixed values of $\gamma  $, require $\left| \xi \right| < 1$ for the convergence of the radius.
\subsection{ ${\displaystyle \left(\frac{x-a}{1-a} \right)^{-\alpha } Hl\left(a, q-(\beta -\delta )\alpha ; \alpha , -\beta+\gamma +\delta , \delta , \gamma; \frac{a(x-1)}{x-a} \right)}$}
\subsubsection{Infinite series}
Replace coefficients $q$, $\beta $, $\gamma $, $\delta $, $x$, $c_0$ and $\lambda $ by $q-(\beta -\delta )\alpha $, $-\beta+\gamma +\delta $, $\delta $,  $\gamma $, $\frac{a(x-1)}{x-a}$, 1 and zero into (\ref{eq:51}). Multiply $\left(\frac{x-a}{1-a} \right)^{-\alpha }$ and the new (\ref{eq:51}) together.
\begin{eqnarray}
&& \left(\frac{x-a}{1-a} \right)^{-\alpha } y(\xi ) \nonumber\\
 &=& \left(\frac{x-a}{1-a} \right)^{-\alpha } Hl\left(a, q-(\beta -\delta )\alpha ; \alpha , -\beta+\gamma +\delta , \delta , \gamma; \frac{a(x-1)}{x-a} \right) \nonumber\\
&=& \left(\frac{x-a}{1-a} \right)^{-\alpha } \left\{\sum_{i_0=0}^{\infty } \frac{(\frac{\alpha }{2})_{i_0} (\frac{-\beta+\gamma +\delta  }{2})_{i_0}}{(1)_{i_0}(\frac{1}{2}+ \frac{\delta }{2})_{i_0}} z^{i_0}\right. \nonumber\\
&&+ \left\{\sum_{i_0=0}^{\infty }\frac{i_0 \left( i_0+ \Gamma_0^{(I)}\right)+ Q}{(i_0+ \frac{1}{2})(i_0 + \frac{\delta }{2})}  \frac{(\frac{\alpha }{2})_{i_0} (\frac{-\beta +\gamma +\delta }{2})_{i_0}}{(1)_{i_0}(\frac{1}{2}+ \frac{\delta }{2})_{i_0}} \sum_{i_1=i_0}^{\infty } \frac{(\frac{1}{2}+\frac{\alpha }{2})_{i_1}(\frac{-\beta +\gamma +\delta+1 }{2})_{i_1}(\frac{3}{2})_{i_0}(1+\frac{\delta }{2})_{i_0}}{(\frac{1}{2}+\frac{\alpha }{2})_{i_0}(\frac{-\beta +\gamma +\delta+1 }{2})_{i_0}(\frac{3}{2})_{i_1}(1+ \frac{\delta }{2})_{i_1}} z^{i_1} \right\} \eta \nonumber\\
&&+ \sum_{n=2}^{\infty } \left\{ \sum_{i_0=0}^{\infty } \frac{i_0 \left( i_0+ \Gamma_0^{(I)}\right)+ Q}{(i_0+ \frac{1}{2})(i_0 + \frac{\delta }{2})}
 \frac{(\frac{\alpha }{2})_{i_0} (\frac{-\beta +\gamma +\delta }{2})_{i_0}}{(1)_{i_0}(\frac{1}{2}+ \frac{\delta }{2})_{i_0}}\right.\nonumber\\
&&\times \prod _{k=1}^{n-1} \left\{ \sum_{i_k=i_{k-1}}^{\infty } \frac{(i_k+\frac{k}{2}) \left( i_k+ \Gamma_k^{(I)}\right)+ Q}{(i_k+ \frac{k}{2}+\frac{1}{2})(i_k +\frac{k}{2}+\frac{\delta }{2})} \frac{(\frac{k}{2}+\frac{\alpha }{2})_{i_k}(\frac{-\beta +\gamma +\delta +k}{2})_{i_k}(1+ \frac{k}{2})_{i_{k-1}}(\frac{1}{2}+\frac{k}{2}+\frac{\delta }{2})_{i_{k-1}}}{(\frac{k}{2}+\frac{\alpha }{2})_{i_{k-1}}(\frac{-\beta +\gamma +\delta +k}{2})_{i_{k-1}}(1+\frac{k}{2})_{i_k}(\frac{1}{2}+ \frac{k}{2}+ \frac{\delta }{2})_{i_k}}\right\} \nonumber\\
&&\times \left.\left. \sum_{i_n= i_{n-1}}^{\infty } \frac{(\frac{n}{2}+\frac{\alpha }{2})_{i_n}(\frac{-\beta +\gamma +\delta +n}{2})_{i_n}(1+ \frac{n}{2})_{i_{n-1}}(\frac{1}{2}+\frac{n}{2}+\frac{\delta }{2})_{i_{n-1}}}{(\frac{n}{2}+\frac{\alpha }{2})_{i_{n-1}}(\frac{-\beta +\gamma +\delta +n}{2})_{i_{n-1}}(1+\frac{n}{2})_{i_n}(\frac{1}{2}+ \frac{n}{2}+ \frac{\delta }{2})_{i_n}} z^{i_n} \right\} \eta ^n \right\}\label{eq:157}
\end{eqnarray}
where
\begin{equation}
\begin{cases} \xi= \frac{a(x-1)}{x-a} \cr
z = -\frac{1}{a} \xi^2 \cr
\eta = \frac{1+a}{a}\xi \cr
\end{cases}\nonumber %\label{eq:37}
\end{equation}
and
\begin{equation}
\begin{cases} 
\Gamma_0^{(I)} = \frac{1}{2(1+a)}(\alpha -\beta +\delta +a(\delta +\gamma -1))  \cr
\Gamma_k^{(I)} = \frac{1}{2(1+a)}(\alpha -\beta +\delta +k+a(\delta +\gamma +k-1))  \cr
Q= \frac{q-(\beta -\delta )\alpha }{4(1+a)} 
\end{cases}\nonumber %\label{eq:8}
\end{equation}
\subsubsection{Polynomial which makes $B_n$ term terminated}
\underline {(1) The case of $\beta =\gamma +\delta +2\beta _i +i$ where $i, \beta _i$ = $0,1,2,\cdots$.}

Substitute $\beta =\gamma +\delta +2\beta _i +i$ into (\ref{eq:157}): apply $\beta =\gamma +\delta +2\beta _0$ into sub-power series $y_0(x)$, apply $\beta =\gamma +\delta +2\beta _0$ into the first summation and $\beta =\gamma +\delta +2\beta _1 +1$ into second summation of sub-power series $y_1(x)$, apply $\beta =\gamma +\delta +2\beta _0$ into the first summation, $\beta =\gamma +\delta +2\beta _1 +1$ into the second summation and $\beta =\gamma +\delta +2\beta _2 +2$ into the third summation of sub-power series $y_2(x)$, etc.\footnote{I treat $\alpha $, $\gamma$, $\delta$ and $q$ as free variables and a fixed value of $\beta $ to construct the polynomial which makes $B_n$ term terminated.}  
\begin{eqnarray}
&& \left(\frac{x-a}{1-a} \right)^{-\alpha } y(\xi ) \nonumber\\
 &=& \left(\frac{x-a}{1-a} \right)^{-\alpha } Hl\left(a, q-(\beta -\delta )\alpha ; \alpha , -\beta+\gamma +\delta , \delta , \gamma; \frac{a(x-1)}{x-a} \right) \nonumber\\
&=& \left(\frac{x-a}{1-a} \right)^{-\alpha } \left\{\sum_{i_0=0}^{\beta _0} \frac{(\frac{\alpha }{2})_{i_0} (-\beta _0)_{i_0}}{(1)_{i_0}(\frac{1}{2}+ \frac{\delta }{2})_{i_0}} z^{i_0} \right. \nonumber\\
&&+ \left\{\sum_{i_0=0}^{\beta _0 }\frac{i_0 \left( i_0+ \Gamma_0^{(S)} \right)+ Q_0^{(S)}}{(i_0+ \frac{1}{2})(i_0 + \frac{\delta }{2})}  \frac{(\frac{\alpha }{2})_{i_0} (-\beta _0)_{i_0}}{(1)_{i_0}(\frac{1}{2}+ \frac{\delta }{2})_{i_0}} \sum_{i_1=i_0}^{\beta _1} \frac{(\frac{1}{2}+\frac{\alpha }{2})_{i_1}(-\beta _1)_{i_1}(\frac{3}{2})_{i_0}(1+\frac{\delta }{2})_{i_0}}{(\frac{1}{2}+\frac{\alpha }{2})_{i_0}(-\beta _1)_{i_0}(\frac{3}{2})_{i_1}(1+ \frac{\delta }{2})_{i_1}} z^{i_1} \right\} \eta \nonumber\\
&&+ \sum_{n=2}^{\infty } \left\{ \sum_{i_0=0}^{\beta _0} \frac{i_0 \left( i_0+ \Gamma_0^{(S)}\right)+ Q_0^{(S)}}{(i_0+ \frac{1}{2})(i_0 + \frac{\delta }{2})}  \frac{(\frac{\alpha }{2})_{i_0} (-\beta _0)_{i_0}}{(1)_{i_0}(\frac{1}{2}+ \frac{\delta }{2})_{i_0}} \right.\nonumber\\
&&\times \prod _{k=1}^{n-1} \left\{ \sum_{i_k=i_{k-1}}^{\beta _k} \frac{(i_k+\frac{k}{2}) \left( i_k+ \Gamma_k^{(S)} \right)+ Q_k^{(S)}}{(i_k+ \frac{k}{2}+\frac{1}{2})(i_k +\frac{k}{2}+\frac{\delta }{2})} \frac{(\frac{k}{2}+\frac{\alpha }{2})_{i_k}(-\beta _k)_{i_k}(1+ \frac{k}{2})_{i_{k-1}}(\frac{1}{2}+\frac{k}{2}+\frac{\delta }{2})_{i_{k-1}}}{(\frac{k}{2}+\frac{\alpha }{2})_{i_{k-1}}(-\beta _k)_{i_{k-1}}(1+\frac{k}{2})_{i_k}(\frac{1}{2}+ \frac{k}{2}+ \frac{\delta }{2})_{i_k}}\right\} \nonumber\\
&&\times \left.\left.  \sum_{i_n= i_{n-1}}^{\beta _n} \frac{(\frac{n}{2}+\frac{\alpha }{2})_{i_n}(-\beta _n)_{i_n}(1+ \frac{n}{2})_{i_{n-1}}(\frac{1}{2}+\frac{n}{2}+\frac{\delta }{2})_{i_{n-1}}}{(\frac{n}{2}+\frac{\alpha }{2})_{i_{n-1}}(-\beta _n)_{i_{n-1}}(1+\frac{n}{2})_{i_n}(\frac{1}{2}+ \frac{n}{2}+ \frac{\delta }{2})_{i_n}} z^{i_n} \right\} \eta ^n \right\} \label{eq:158}
\end{eqnarray}
where
\begin{equation}
\beta _i\leq \beta _j \;\;\mbox{only}\;\mbox{if}\;i\leq j\;\;\mbox{where}\;i,j= 0,1,2,\cdots
\nonumber 
\end{equation}
and
\begin{equation}
\begin{cases} 
\Gamma_0^{(S)} =  \frac{1}{2(1+a)}(\alpha -\gamma -2\beta _0 +a(\delta +\gamma -1))  \cr
\Gamma_k^{(S)} =  \frac{1}{2(1+a)}(\alpha -\gamma -2\beta _k+a(\delta +\gamma +k-1))  \cr
Q_0^{(S)} =  \frac{q-(\gamma +2\beta _0)\alpha }{4(1+a)} \cr
Q_k^{(S)} = \frac{q-(\gamma +2\beta _k+k)\alpha}{4(1+a)}
\end{cases}\nonumber %\label{eq:8}
\end{equation}
For the minimum value of Heun equation for a polynomial which makes $B_n$ term terminated about $\xi=0 $, put $\beta  _0=\beta _1=\beta _2=\cdots=0$ in (\ref{eq:158}).
\begin{eqnarray}
&& \left(\frac{x-a}{1-a} \right)^{-\alpha } y(\xi ) \nonumber\\
 &=& \left(\frac{x-a}{1-a} \right)^{-\alpha } Hl\left(a, q-(\beta -\delta )\alpha ; \alpha , -\beta+\gamma +\delta , \delta , \gamma; \frac{a(x-1)}{x-a} \right) \nonumber\\
&=& \left(\frac{x-a}{1-a} \right)^{-\alpha } \; _2F_1\left( \frac{ \Lambda _{14}-\sqrt{ \Lambda _{14}^2-4a\Omega_{14}}}{2a},  \frac{ \Lambda _{14}+\sqrt{ \Lambda _{14}^2-4a\Omega_{14}}}{2a}; \delta  ; \xi \right)  \hspace{1cm}\label{aa:12}
\end{eqnarray}
where $\Lambda_{14}= -\gamma +a(\gamma +\delta -1) $ and $\Omega_{14}= q-\alpha \gamma  $.
It tells us that Heun polynomials in which makes $B_n$ term terminated, for fixed values of $\beta  $, require $\left| \xi \right| < 1$ for the convergence of the radius.

For the special case, if $\xi=1$ and $Re\left(\frac{1-a}{a}\gamma \right)>-1 $ in (\ref{aa:12}),
\begin{eqnarray}
&& \left(\frac{a}{a-1} \right)^{-\alpha } y(1) \nonumber\\
 &=& \left(\frac{a}{a-1} \right)^{-\alpha } Hl\left( a, q-(\beta -\delta )\alpha ; \alpha , -\beta+\gamma +\delta , \delta , \gamma; 1 \right) \nonumber\\
&=& \left(\frac{a}{a-1} \right)^{-\alpha } \frac{ \Gamma \left( \delta \right) \Gamma \left( 1+\frac{1-a}{a}\gamma  \right)}{\Gamma \left( \delta - \frac{ \Lambda _{14}-\sqrt{ \Lambda _{14}^2-4a\Omega_{14}}}{2a}\right) \Gamma \left( \delta -\frac{ \Lambda _{14}+\sqrt{ \Lambda _{14}^2-4a\Omega_{14}}}{2a}\right)}   \nonumber
\end{eqnarray}

\underline {(2) The case of $\gamma =\beta -\delta -2\gamma _i -i$ where $i, \gamma _i$ = $0,1,2,\cdots$.}

Substitute $\gamma =\beta -\delta -2\gamma _i -i$ into (\ref{eq:157}): apply $\gamma =\beta -\delta -2\gamma _0$ into sub-power series $y_0(x)$, apply $\gamma =\beta -\delta -2\gamma _0$ into the first summation and $\gamma =\beta -\delta -2\gamma _1 -1$ into second summation of sub-power series $y_1(x)$, apply $\gamma =\beta -\delta -2\gamma _0$ into the first summation, $\gamma =\beta -\delta -2\gamma _1 -1$ into the second summation and $\gamma =\beta -\delta -2\gamma _2 -2$ into the third summation of sub-power series $y_2(x)$,  etc. \footnote{I treat $\alpha $, $\beta $, $\delta$ and $q$ as free variables and a fixed value of $\gamma $ to construct the polynomial which makes $B_n$ term terminated.}  
\begin{eqnarray}
&& \left(\frac{x-a}{1-a} \right)^{-\alpha } y(\xi ) \nonumber\\
 &=& \left(\frac{x-a}{1-a} \right)^{-\alpha } Hl\left(a, q-(\beta -\delta )\alpha ; \alpha , -\beta+\gamma +\delta , \delta , \gamma; \frac{a(x-1)}{x-a} \right) \nonumber\\
&=& \left(\frac{x-a}{1-a} \right)^{-\alpha } \left\{\sum_{i_0=0}^{\gamma _0} \frac{(\frac{\alpha }{2})_{i_0} (-\gamma _0)_{i_0}}{(1)_{i_0}(\frac{1}{2}+ \frac{\delta }{2})_{i_0}} z^{i_0} \right.\nonumber\\
&&+ \left\{\sum_{i_0=0}^{\gamma _0}\frac{i_0 \left( i_0+ \Gamma_0^{(S)}\right)+ Q}{(i_0+ \frac{1}{2})(i_0 + \frac{\delta }{2})}  \frac{(\frac{\alpha }{2})_{i_0} (-\gamma _0)_{i_0}}{(1)_{i_0}(\frac{1}{2}+ \frac{\delta }{2})_{i_0}} \sum_{i_1=i_0}^{\gamma _1} \frac{(\frac{1}{2}+\frac{\alpha }{2})_{i_1}(-\gamma _1)_{i_1}(\frac{3}{2})_{i_0}(1+\frac{\delta }{2})_{i_0}}{(\frac{1}{2}+\frac{\alpha }{2})_{i_0}(-\gamma _1)_{i_0}(\frac{3}{2})_{i_1}(1+ \frac{\delta }{2})_{i_1}} z^{i_1} \right\} \eta \nonumber\\
&&+ \sum_{n=2}^{\infty } \left\{ \sum_{i_0=0}^{\gamma _0} \frac{i_0 \left( i_0+ \Gamma_0^{(S)}\right)+ Q}{(i_0+ \frac{1}{2})(i_0 + \frac{\delta }{2})} \frac{(\frac{\alpha }{2})_{i_0} (-\gamma _0)_{i_0}}{(1)_{i_0}(\frac{1}{2}+ \frac{\delta }{2})_{i_0}}
 \right.\nonumber\\
&&\times \prod _{k=1}^{n-1} \left\{ \sum_{i_k=i_{k-1}}^{\gamma _k} \frac{(i_k+\frac{k}{2}) \left( i_k+ \Gamma_k^{(S)}\right)+ Q}{(i_k+ \frac{k}{2}+\frac{1}{2})(i_k +\frac{k}{2}+\frac{\delta }{2})} \frac{(\frac{k}{2}+\frac{\alpha }{2})_{i_k}(-\gamma _k)_{i_k}(1+ \frac{k}{2})_{i_{k-1}}(\frac{1}{2}+\frac{k}{2}+\frac{\delta }{2})_{i_{k-1}}}{(\frac{k}{2}+\frac{\alpha }{2})_{i_{k-1}}(-\gamma _k)_{i_{k-1}}(1+\frac{k}{2})_{i_k}(\frac{1}{2}+ \frac{k}{2}+ \frac{\delta }{2})_{i_k}}\right\} \nonumber\\
&&\times \left.\left. \sum_{i_n= i_{n-1}}^{\gamma _n} \frac{(\frac{n}{2}+\frac{\alpha }{2})_{i_n}(-\gamma _n)_{i_n}(1+ \frac{n}{2})_{i_{n-1}}(\frac{1}{2}+\frac{n}{2}+\frac{\delta }{2})_{i_{n-1}}}{(\frac{n}{2}+\frac{\alpha }{2})_{i_{n-1}}(-\gamma _n)_{i_{n-1}}(1+\frac{n}{2})_{i_n}(\frac{1}{2}+ \frac{n}{2}+ \frac{\delta }{2})_{i_n}} z^{i_n} \right\} \eta ^n \right\} \label{eq:159}
\end{eqnarray}
where
\begin{equation}
\gamma _i\leq \gamma _j \;\;\mbox{only}\;\mbox{if}\;i\leq j\;\;\mbox{where}\;i,j= 0,1,2,\cdots
\nonumber 
\end{equation}
and
\begin{equation}
\begin{cases} 
\Gamma_0^{(S)} =  \frac{1}{2(1+a)}(\alpha -\beta +\delta +a(\beta -2\gamma _0 -1))  \cr
\Gamma_k^{(S)} =  \frac{1}{2(1+a)}(\alpha -\beta +\delta +k+a(\beta -2\gamma _k-1))  \cr
Q  =  \frac{q-(\beta -\delta )\alpha}{4(1+a)} 
\end{cases}\nonumber %\label{eq:8}
\end{equation}
For the minimum value of Heun equation for a polynomial which makes $B_n$ term terminated about $\xi=0 $, put $\gamma _0=\gamma _1=\gamma _2=\cdots=0$ in (\ref{eq:159}).
\begin{eqnarray}
&& \left(\frac{x-a}{1-a} \right)^{-\alpha } y(\xi ) \nonumber\\
 &=& \left(\frac{x-a}{1-a} \right)^{-\alpha } Hl\left(a, q-(\beta -\delta )\alpha ; \alpha , -\beta+\gamma +\delta , \delta , \gamma; \frac{a(x-1)}{x-a} \right) \nonumber\\
&=& \left(\frac{x-a}{1-a} \right)^{-\alpha } \; _2F_1\left( \frac{ \Lambda _{15}-\sqrt{ \Lambda _{15}^2-4\Omega_{15}}}{2},  \frac{ \Lambda _{15}+\sqrt{ \Lambda _{15}^2-4\Omega_{15}}}{2}; \delta  ;\frac{\xi}{a} \right)  \nonumber
\end{eqnarray}
where $\Lambda_{15}= \alpha -\beta +\delta  +a(\beta -1) $ and $\Omega_{15}= q-\alpha (\beta -\delta ) $.
It tells us that Heun polynomials in which makes $B_n$ term terminated, for fixed values of $\gamma  $, require $\left| \frac{\xi}{a} \right| < 1$ for the convergence of the radius.
\section{Asymptotic behaviors of 192 Heun functions} 
\subsection{ ${\displaystyle (1-x)^{1-\delta } Hl(a, q - (\delta  - 1)\gamma a; \alpha - \delta  + 1, \beta - \delta + 1, \gamma ,2 - \delta ; x)}$ \\  \emph{and}\;  ${\displaystyle x^{1-\gamma } (1-x)^{1-\delta } Hl(a, q-(\gamma +\delta -2)a-(\gamma -1)(\alpha +\beta -\gamma -\delta +1); \alpha - \gamma -\delta +2}$ \\ ${\displaystyle, \beta - \gamma -\delta +2, 2-\gamma, 2 - \delta ; x)}$}
The asymptotic behaviors of  $Hl(a, q - (\delta  - 1)\gamma a; \alpha - \delta  + 1, \beta - \delta + 1, \gamma ,2 - \delta ; x)$ and 
$Hl(a, q-(\gamma +\delta -2)a-(\gamma -1)(\alpha +\beta -\gamma -\delta +1); \alpha - \gamma -\delta +2, \beta - \gamma -\delta +2, 2-\gamma, 2 - \delta ; x)$ and those boundary conditions of the independent variable $x$ are same as in Subs.~\ref{sec.3}: see (\ref{eq:16}), (\ref{eq:17}) and (\ref{eq:22}) and Table~\ref{cb.2}.
\subsection{ ${\displaystyle  Hl(1-a,-q+\alpha \beta; \alpha,\beta, \delta, \gamma; 1-x)}$ \\
\emph{and}  ${\displaystyle (1-x)^{1-\delta } Hl(1-a,-q+(\delta -1)\gamma a+(\alpha -\delta +1)(\beta -\delta +1); \alpha-\delta +1,\beta-\delta +1}$\\
${\displaystyle, 2-\delta, \gamma; 1-x)}$} 
Replace a coefficient $a$, $y(x)$ and independent variable $x$ by $1-a$, $Hl(1-a,-q+\alpha \beta; \alpha,\beta, \delta, \gamma; 1-x)$ and $1-x$ in (\ref{eq:16}), (\ref{eq:17}). Repeat same process for the case of $Hl(1-a,-q+(\delta -1)\gamma a+(\alpha -\delta +1)(\beta -\delta +1); \alpha-\delta +1,\beta-\delta +1, 2-\delta, \gamma; 1-x)$.

For an infinite series,
\begin{equation}\footnotesize
\left.
  \begin{array}{l}
    {\displaystyle \lim_{n\gg 1} Hl(1-a,-q+\alpha \beta; \alpha,\beta, \delta, \gamma; 1-x) }\\
   {\displaystyle \lim_{n\gg 1} Hl(1-a,-q+(\delta -1)\gamma a+(\alpha -\delta +1)(\beta -\delta +1); \alpha-\delta +1,\beta-\delta +1, 2-\delta, \gamma; 1-x)}  
  \end{array} \right\}= \frac{1}{1-\left(\frac{-1}{1-a}(1-x)^2 +\frac{2-a}{1-a}(1-x)\right)} \label{xx:1}
\end{equation}  
The condition of convergence of (\ref{xx:1}) is
\begin{equation}
\left|\frac{-1}{1-a}(1-x)^2\right| +\left|\frac{2-a}{1-a}(1-x)\right|<1 \hspace{.5cm}\mbox{where}\;\; a \ne 1 \nonumber %\label{eq:65}
\end{equation}
For $a=2$, (\ref{xx:1}) turns to be   
\begin{equation}\footnotesize
\left.
  \begin{array}{l}
    {\displaystyle \lim_{n\gg 1} Hl(1-a,-q+\alpha \beta; \alpha,\beta, \delta, \gamma; 1-x) }\\
   {\displaystyle \lim_{n\gg 1} Hl(1-a,-q+(\delta -1)\gamma a+(\alpha -\delta +1)(\beta -\delta +1); \alpha-\delta +1,\beta-\delta +1, 2-\delta, \gamma; 1-x)} 
  \end{array} \right\}= \frac{1}{1-(1-x)^2} \label{xx:2}
\end{equation}
The condition of convergence of (\ref{xx:2}) is
\begin{equation}
\left| (1-x)^2 \right|<1 \nonumber %\label{eq:67}
\end{equation}
For $|a|\gg 1$, (\ref{xx:1}) turns to be 
\begin{equation}\footnotesize
\left.
  \begin{array}{l}
    {\displaystyle \lim_{n\gg 1} Hl(1-a,-q+\alpha \beta; \alpha,\beta, \delta, \gamma; 1-x) }\\
   {\displaystyle \lim_{n\gg 1} Hl(1-a,-q+(\delta -1)\gamma a+(\alpha -\delta +1)(\beta -\delta +1); \alpha-\delta +1,\beta-\delta +1, 2-\delta, \gamma; 1-x)}  
  \end{array} \right\}\approx  \frac{1}{1- (1-x)} \label{xx:3}
\end{equation}
The condition of convergence of (\ref{xx:3}) is
\begin{equation}
\left| 1-x \right|<1  \nonumber %\label{eq:69}
\end{equation}
\subsection{ ${\displaystyle x^{-\alpha } Hl\left(\frac{1}{a},\frac{q+\alpha [(\alpha -\gamma -\delta +1)a-\beta +\delta ]}{a}; \alpha , \alpha -\gamma +1, \alpha -\beta +1,\delta ;\frac{1}{x}\right)}$} 
Replace coefficient $a$, $y(x)$ and independent variable x by $\frac{1}{a}$, $ Hl\left(\frac{1}{a},\frac{q+\alpha [(\alpha -\gamma -\delta +1)a-\beta +\delta ]}{a}; \alpha , \alpha -\gamma +1, \alpha -\beta +1,\delta ;\frac{1}{x}\right)$ and $\frac{1}{x}$ in (\ref{eq:16}) and (\ref{eq:17}).

For an infinite series,
\begin{equation}
\lim_{n\gg 1} Hl\left(\frac{1}{a},\frac{q+\alpha [(\alpha -\gamma -\delta +1)a-\beta +\delta ]}{a}; \alpha , \alpha -\gamma +1, \alpha -\beta +1,\delta ;\frac{1}{x}\right) = \frac{1}{1-\left(-a x^{-2} +(1+a)x^{-1}\right)} \label{xx:4}
\end{equation}
 The condition of convergence of (\ref{xx:4}) is
\begin{equation}
\left| a x^{-2}\right| +\left| (1+a)x^{-1} \right|<1 \nonumber
\end{equation}
For $a=-1$, (\ref{xx:4}) turns to be 
\begin{equation}
\lim_{n\gg 1} Hl\left(\frac{1}{a},\frac{q+\alpha [(\alpha -\gamma -\delta +1)a-\beta +\delta ]}{a}; \alpha , \alpha -\gamma +1, \alpha -\beta +1,\delta ;\frac{1}{x}\right)= \frac{1}{1- x^{-2}}  \label{xx:5}
\end{equation}
The condition of convergence of (\ref{xx:5}) is
\begin{equation}
\left| x^{-2}\right|<1 \nonumber
\end{equation}
For $a= 0$, (\ref{xx:4}) turns to be 
\begin{equation}
\lim_{n\gg 1} Hl\left(\frac{1}{a},\frac{q+\alpha [(\alpha -\gamma -\delta +1)a-\beta +\delta ]}{a}; \alpha , \alpha -\gamma +1, \alpha -\beta +1,\delta ;\frac{1}{x}\right)= \frac{1}{1- x^{-1}} \label{xx:6}
\end{equation}
The condition of convergence of (\ref{xx:6}) is
\begin{equation}
\left| x^{-1}\right|<1 \nonumber
\end{equation}
\subsection{ ${\displaystyle \left(1-\frac{x}{a} \right)^{-\beta } Hl\left(1-a, -q+\gamma \beta; -\alpha +\gamma +\delta, \beta, \gamma, \delta; \frac{(1-a)x}{x-a} \right)}$ \\
\emph{and} \small ${\displaystyle (1-x)^{1-\delta }\left(1-\frac{x}{a} \right)^{-\beta+\delta -1} Hl\left(1-a, -q+\gamma [(\delta -1)a+\beta -\delta +1]; -\alpha +\gamma +1, \beta -\delta+1, \gamma, 2-\delta; \frac{(1-a)x}{x-a} \right)}$\normalsize} 
Replace coefficient $a$, $y(x)$ and independent variable $x$ by $1-a$, $Hl\left(1-a, -q+\gamma \beta; -\alpha +\gamma +\delta, \beta, \gamma, \delta; \frac{(1-a)x}{x-a} \right)$ and ${\displaystyle \frac{(1-a)x}{x-a}}$ in (\ref{eq:16}) and (\ref{eq:17}).

Repeat same process for the case of $Hl\left(1-a, -q+\gamma [(\delta -1)a+\beta -\delta +1]; -\alpha +\gamma +1, \beta -\delta+1, \gamma, 2-\delta; \frac{(1-a)x}{x-a} \right)$.

For an infinite series,
\begin{equation}\footnotesize
\left.
  \begin{array}{l}
    {\displaystyle \lim_{n\gg 1} Hl\left(1-a, -q+\gamma \beta; -\alpha +\gamma +\delta, \beta, \gamma, \delta; \frac{(1-a)x}{x-a} \right) }\\
   {\displaystyle \lim_{n\gg 1} Hl\left(1-a, -q+\gamma [(\delta -1)a+\beta -\delta +1]; -\alpha +\gamma +1, \beta -\delta+1, \gamma, 2-\delta; \frac{(1-a)x}{x-a} \right)}  
  \end{array} \right\}= \frac{1}{1-\left( -\frac{(1-a)x^2}{(x-a)^2}+\frac{(2-a)x}{(x-a)}\right)} \label{xx:7}
\end{equation}  
The condition of convergence of (\ref{xx:7}) is
\begin{equation}
\left|-\frac{(1-a)x^2}{(x-a)^2}\right|+\left|\frac{(2-a)x}{(x-a)}\right|<1 \hspace{.5cm}\mbox{where}\;\; x \ne a \nonumber %\label{eq:65}
\end{equation}
For $a=2$, (\ref{xx:7}) turns to be
\begin{equation}\footnotesize
\left.
  \begin{array}{l}
    {\displaystyle \lim_{n\gg 1} Hl\left(1-a, -q+\gamma \beta; -\alpha +\gamma +\delta, \beta, \gamma, \delta; \frac{(1-a)x}{x-a} \right) }\\
   {\displaystyle \lim_{n\gg 1} Hl\left(1-a, -q+\gamma [(\delta -1)a+\beta -\delta +1]; -\alpha +\gamma +1, \beta -\delta+1, \gamma, 2-\delta; \frac{(1-a)x}{x-a} \right)}  
  \end{array} \right\}= \frac{1}{1- \frac{ x^2}{(x-2)^2} } \label{xx:8}
\end{equation}
The condition of convergence of (\ref{xx:8}) is
\begin{equation}
\left|\frac{ x^2}{(x-2)^2}\right|<1 \nonumber %\label{eq:67}
\end{equation}
For $a=1$, (\ref{xx:7}) turns to be
\begin{equation}\footnotesize
\left.
  \begin{array}{l}
    {\displaystyle \lim_{n\gg 1}  Hl\left(1-a, -q+\gamma \beta; -\alpha +\gamma +\delta, \beta, \gamma, \delta; \frac{(1-a)x}{x-a} \right) }\\
   {\displaystyle \lim_{n\gg 1} Hl\left(1-a, -q+\gamma [(\delta -1)a+\beta -\delta +1]; -\alpha +\gamma +1, \beta -\delta+1, \gamma, 2-\delta; \frac{(1-a)x}{x-a} \right)}  
  \end{array} \right\}= \frac{1}{1- \frac{x}{x-1} } \label{xx:9}
\end{equation}
The condition of convergence of (\ref{xx:9}) is
\begin{equation}
\left| \frac{x}{x-1}\right|<1 \nonumber %\label{eq:69}
\end{equation}
For $|a|\gg 1$, (\ref{xx:7}) turns to be
\begin{equation}\footnotesize
\left.
  \begin{array}{l}
    {\displaystyle \lim_{n\gg 1}  Hl\left(1-a, -q+\gamma \beta; -\alpha +\gamma +\delta, \beta, \gamma, \delta; \frac{(1-a)x}{x-a} \right) }\\
   {\displaystyle \lim_{n\gg 1} Hl\left(1-a, -q+\gamma [(\delta -1)a+\beta -\delta +1]; -\alpha +\gamma +1, \beta -\delta+1, \gamma, 2-\delta; \frac{(1-a)x}{x-a} \right)}  
  \end{array} \right\}\approx  \frac{1}{1- x } \label{xx:10}
\end{equation}
The condition of convergence of (\ref{xx:10}) is
\begin{equation}
\left| x\right|<1 \nonumber %\label{eq:69}
\end{equation}
\subsection{ ${\displaystyle x^{-\alpha } Hl\left(\frac{a-1}{a}, \frac{-q+\alpha (\delta a+\beta -\delta )}{a}; \alpha, \alpha -\gamma +1, \delta , \alpha -\beta +1; \frac{x-1}{x} \right)}$} 
Replace coefficient $a$, $y(x)$ and independent variable $x$ by $\frac{a-1}{a}$, $Hl\left(\frac{a-1}{a}, \frac{-q+\alpha (\delta a+\beta -\delta )}{a}; \alpha, \alpha -\gamma +1, \delta , \alpha -\beta +1; \frac{x-1}{x} \right)$ and $\frac{x-1}{x}$ in (\ref{eq:16}) and (\ref{eq:17}).

For an infinite series,
\begin{equation}
\lim_{n\gg 1} Hl\left(\frac{a-1}{a}, \frac{-q+\alpha (\delta a+\beta -\delta )}{a}; \alpha, \alpha -\gamma +1, \delta , \alpha -\beta +1; \frac{x-1}{x} \right) = \frac{1}{1-\left( \frac{a}{(1-a)}\frac{(x-1)^2}{x^2}+\frac{(1-2a)}{(1-a)}\frac{(x-1)}{x}\right)} \label{xx:11}
\end{equation}
 The condition of convergence of (\ref{xx:11}) is
\begin{equation}
\left|\frac{a}{(1-a)}\frac{(x-1)^2}{x^2} \right| +\left|\frac{(1-2a)}{(1-a)}\frac{(x-1)}{x} \right|<1 \hspace{.5cm}\mbox{where}\;\; a \ne 1\nonumber
\end{equation}
For $a= 1/2$, (\ref{xx:11}) turns to be
\begin{equation}
\lim_{n\gg 1} Hl\left(\frac{a-1}{a}, \frac{-q+\alpha (\delta a+\beta -\delta )}{a}; \alpha, \alpha -\gamma +1, \delta , \alpha -\beta +1; \frac{x-1}{x} \right)= \frac{1}{1-\frac{(x-1)^2}{x^2}} \label{xx:12}
\end{equation}
The condition of convergence of (\ref{xx:12}) is
\begin{equation}
\left|\frac{(x-1)^2}{x^2} \right|<1 \nonumber
\end{equation}
For $a=0$, (\ref{xx:11}) turns to be
\begin{equation}
\lim_{n\gg 1} Hl\left(\frac{a-1}{a}, \frac{-q+\alpha (\delta a+\beta -\delta )}{a}; \alpha, \alpha -\gamma +1, \delta , \alpha -\beta +1; \frac{x-1}{x} \right) =  \frac{1}{1- \frac{ x-1 }{x} } \label{xx:13}
\end{equation}
The condition of convergence of (\ref{xx:13}) is
\begin{equation}
\left|\frac{ x-1 }{x}\right|<1 \nonumber
\end{equation}
For $|a|\gg 1$, (\ref{xx:11}) turns to be
\begin{equation}
\lim_{n\gg 1} Hl\left(\frac{a-1}{a}, \frac{-q+\alpha (\delta a+\beta -\delta )}{a}; \alpha, \alpha -\gamma +1, \delta , \alpha -\beta +1; \frac{x-1}{x} \right) \approx  \frac{1}{1-\left( -\frac{(x-1)^2}{x^2}+ \frac{2(x-1)}{x}\right)} \label{xx:14}
\end{equation}
The condition of convergence of (\ref{xx:14}) is
\begin{equation}
\left| \frac{(x-1)^2}{x^2}\right|+ \left|\frac{2(x-1)}{x} \right|<1 \nonumber %\label{eq:69}
\end{equation}
\subsection{ ${\displaystyle \left(\frac{x-a}{1-a} \right)^{-\alpha } Hl\left(a, q-(\beta -\delta )\alpha ; \alpha , -\beta+\gamma +\delta , \delta , \gamma; \frac{a(x-1)}{x-a} \right)}$} 
Replace coefficient $y(x)$ and independent variable $x$ by $Hl\left(a, q-(\beta -\delta )\alpha ; \alpha , -\beta+\gamma +\delta , \delta , \gamma; \frac{a(x-1)}{x-a} \right)$ and $\frac{a(x-1)}{x-a}$ in (\ref{eq:16}) and (\ref{eq:17}).

For an infinite series,
\begin{equation}
\lim_{n\gg 1}  Hl\left(a, q-(\beta -\delta )\alpha ; \alpha , -\beta+\gamma +\delta , \delta , \gamma; \frac{a(x-1)}{x-a} \right) = \frac{1}{1-\left( -\frac{a(x-1)^2}{(x-a)^2}+\frac{(1+a)(x-1)}{(x-a)}\right)}  \label{xx:15}
\end{equation}
 The condition of convergence of (\ref{xx:15}) is
\begin{equation}
\left| \frac{a(x-1)^2}{(x-a)^2}\right| +\left|\frac{(1+a)(x-1)}{(x-a)} \right|<1 \hspace{.5cm}\mbox{where}\;\; x \ne a\nonumber
\end{equation}
For $a= -1$, (\ref{xx:15}) turns to be
\begin{equation}
\lim_{n\gg 1}  Hl\left(a, q-(\beta -\delta )\alpha ; \alpha , -\beta+\gamma +\delta , \delta , \gamma; \frac{a(x-1)}{x-a} \right) = \frac{1}{1- \frac{(x-1)^2}{(x+1)^2} } \label{xx:16}
\end{equation}
The condition of convergence of (\ref{xx:16}) is
\begin{equation}
\left|\frac{(x-1)^2}{(x+1)^2} \right|<1 \nonumber
\end{equation}
For $a= 0$, (\ref{xx:15}) turns to be
\begin{equation}
\lim_{n\gg 1}  Hl\left(a, q-(\beta -\delta )\alpha ; \alpha , -\beta+\gamma +\delta , \delta , \gamma; \frac{a(x-1)}{x-a} \right) = \frac{1}{1- \frac{x-1}{x} } \label{xx:17}
\end{equation}
The condition of convergence of (\ref{xx:17}) is
\begin{equation}
\left|\frac{x-1}{x} \right|<1 \nonumber
\end{equation}
For $|a|\gg 1$, (\ref{xx:15}) turns to be 
\begin{equation}
\lim_{n\gg 1} Hl\left(a, q-(\beta -\delta )\alpha ; \alpha , -\beta+\gamma +\delta , \delta , \gamma; \frac{a(x-1)}{x-a} \right) \approx  \frac{1}{1-(1-x)}  \label{xx:18}
\end{equation}
The condition of convergence of (\ref{xx:18}) is
\begin{equation}
\left| 1-x\right|<1 \nonumber
\end{equation}
\section{Integral formalism of 192 Heun functions} 
\subsection{ ${\displaystyle (1-x)^{1-\delta } Hl(a, q - (\delta  - 1)\gamma a; \alpha - \delta  + 1, \beta - \delta + 1, \gamma ,2 - \delta ; x)}$ }
\subsubsection{Polynomial which makes $B_n$ term terminated}

\underline {(1) The case of $\alpha = -2 \alpha _i-i +\delta -1$ and $\beta \ne -2 \beta _i -i+\delta -1$ where $i, \alpha _i, \beta _i$ = $0,1,2,\cdots$.}\vspace{1mm}

Replace coefficients $q$, $\alpha$, $\beta$, $\delta$, $c_0$ and $\lambda $ by $q - (\delta - 1)\gamma a $, $\alpha - \delta  + 1 $, $\beta - \delta + 1$, $2 - \delta$, 1 and zero into (\ref{eq:39}). Multiply $(1-x)^{1-\delta }$ and the new (\ref{eq:39}) together.
\begin{eqnarray}
& &(1-x)^{1-\delta } y(x)\nonumber\\
&=& (1-x)^{1-\delta } Hl(a, q - (\delta  - 1)\gamma a; \alpha - \delta  + 1, \beta - \delta + 1, \gamma ,2 - \delta ; x)\nonumber\\
&=& (1-x)^{1-\delta } \Bigg\{ \;_2F_1 \left(-\alpha _0, \frac{\beta-\delta +1 }{2};\frac{1}{2}+\frac{\gamma }{2}; z \right) + \sum_{n=1}^{\infty } \Bigg\{\prod _{k=0}^{n-1} \Bigg\{ \int_{0}^{1} dt_{n-k}\;t_{n-k}^{\frac{1}{2}(n-k-2)} \int_{0}^{1} du_{n-k}\;u_{n-k}^{\frac{1}{2}(n-k-3+\gamma )} \nonumber\\
&&\times  \frac{1}{2\pi i}  \oint dv_{n-k} \frac{1}{v_{n-k}} \left( 1-\frac{1}{v_{n-k}}\right)^{\alpha _{n-k}} \left( 1- \overleftrightarrow {w}_{n-k+1,n}v_{n-k}(1-t_{n-k})(1-u_{n-k})\right)^{-\frac{1}{2}(n-k+1+\beta-\delta)}\nonumber\\
&&\times \left( \overleftrightarrow {w}_{n-k,n}^{-\frac{1}{2}(n-k-1)}\left(  \overleftrightarrow {w}_{n-k,n} \partial _{ \overleftrightarrow {w}_{n-k,n}}\right) \overleftrightarrow {w}_{n-k,n}^{\frac{1}{2}(n-k-1)} \left( \overleftrightarrow {w}_{n-k,n} \partial _{ \overleftrightarrow {w}_{n-k,n}} + \Omega _{n-k-1}^{(S)} \right) +Q \right) \Bigg\}\nonumber\\
&&\times \;_2F_1 \left(-\alpha _0, \frac{\beta-\delta +1 }{2};\frac{1}{2}+\frac{\gamma }{2}; \overleftrightarrow {w}_{1,n} \right)\Bigg\} \eta ^n \Bigg\} \label{eq:64}
\end{eqnarray}
where
\begin{equation}
\begin{cases} z = -\frac{1}{a}x^2 \cr
\eta = \frac{(1+a)}{a} x \cr
\alpha _i\leq \alpha _j \;\;\mbox{only}\;\mbox{if}\;i\leq j\;\;\mbox{where}\;i,j= 0,1,2,\cdots
\end{cases}\nonumber %\label{eq:37}
\end{equation}
and
\begin{equation}
\begin{cases} 
\Omega _{n-k-1}^{(S)} = \frac{1}{2(1+a)}(-2\alpha _{n-k-1}+\beta -1 +a(\gamma-\delta+n-k))  \cr
Q=  \frac{q-(\delta-1)\gamma a }{4(1+a)}
\end{cases}\nonumber %\label{eq:8}
\end{equation}
\underline {(2) The case of $\alpha = -2 \alpha _i-i +\delta -1 $ and $\beta = -2 \beta _i -i+\delta -1 $ only if $\alpha _i \leq \beta _i$.}

Put $\beta = -2 \beta _i -i+\delta -1 $ where $i=0,1,2,\cdots$ in (\ref{eq:64}).
\begin{eqnarray}
& &(1-x)^{1-\delta } y(x)\nonumber\\
&=& (1-x)^{1-\delta }Hl(a, q - (\delta  - 1)\gamma a; \alpha - \delta  + 1, \beta - \delta + 1, \gamma ,2 - \delta ; x)\nonumber\\
&=& (1-x)^{1-\delta } \Bigg\{ \;_2F_1 \left(-\alpha _0, -\beta _0;\frac{1}{2}+\frac{\gamma }{2}; z \right) + \sum_{n=1}^{\infty } \Bigg\{\prod _{k=0}^{n-1} \Bigg\{ \int_{0}^{1} dt_{n-k}\;t_{n-k}^{\frac{1}{2}(n-k-2)} \int_{0}^{1} du_{n-k}\;u_{n-k}^{\frac{1}{2}(n-k-3+\gamma )} \nonumber\\
&&\times  \frac{1}{2\pi i}  \oint dv_{n-k} \frac{1}{v_{n-k}} \left( 1-\frac{1}{v_{n-k}}\right)^{\alpha _{n-k}} \left( 1- \overleftrightarrow {w}_{n-k+1,n}v_{n-k}(1-t_{n-k})(1-u_{n-k})\right)^{\beta _{n-k}}\nonumber\\
&&\times \left( \overleftrightarrow {w}_{n-k,n}^{-\frac{1}{2}(n-k-1)}\left(  \overleftrightarrow {w}_{n-k,n} \partial _{ \overleftrightarrow {w}_{n-k,n}}\right) \overleftrightarrow {w}_{n-k,n}^{\frac{1}{2}(n-k-1)} \left( \overleftrightarrow {w}_{n-k,n} \partial _{ \overleftrightarrow {w}_{n-k,n}}  + \Omega _{n-k-1}^{(B)}\right) +Q\right) \Bigg\}\nonumber\\
&&\times \;_2F_1 \left(-\alpha _0, -\beta _0; \frac{1}{2}+\frac{\gamma }{2}; \overleftrightarrow {w}_{1,n} \right)\Bigg\} \eta ^n \Bigg\} \label{eq:65}
\end{eqnarray}
where
\begin{equation}
\begin{cases} 
\Omega _{n-k-1}^{(B)} =  \frac{1}{2(1+a)}(-2\alpha _{n-k-1}-2\beta _{n-k-1}-n+k-1+\delta +a(\gamma-\delta+n-k ) )  \cr
Q=  \frac{q-(\delta-1)\gamma a }{4(1+a)}
\end{cases}\nonumber %\label{eq:8}
\end{equation}
\subsubsection{Infinite series}
Replace coefficients q, $\alpha$, $\beta$, $\delta$, $c_0$ and $\lambda $ by $q - (\delta - 1)\gamma a $, $\alpha - \delta  + 1 $, $\beta - \delta + 1$, $2-\delta$, 1 and zero into (\ref{eq:46}). Multiply $(1-x)^{1-\delta }$ and the new (\ref{eq:46}) together.
\begin{eqnarray}
 & &(1-x)^{1-\delta } y(x)\nonumber\\
&=& (1-x)^{1-\delta } Hl\left(a, q - (\delta  - 1)\gamma a; \alpha - \delta + 1, \beta - \delta + 1, \gamma ,2 - \delta ; x\right)\nonumber\\
&=& (1-x)^{1-\delta } \Bigg\{ \;_2F_1 \left(\frac{\alpha -\delta +1 }{2}, \frac{\beta-\delta +1 }{2};\frac{1}{2}+\frac{\gamma }{2}; z \right) + \sum_{n=1}^{\infty } \Bigg\{\prod _{k=0}^{n-1} \Bigg\{ \int_{0}^{1} dt_{n-k}\;t_{n-k}^{\frac{1}{2}(n-k-2)} \int_{0}^{1} du_{n-k}\;u_{n-k}^{\frac{1}{2}(n-k-3+\gamma )} \nonumber\\
&&\times  \frac{1}{2\pi i}  \oint dv_{n-k} \frac{1}{v_{n-k}} \left( 1-\frac{1}{v_{n-k}}\right)^{-\frac{1}{2}(n-k+1+\alpha -\delta)} \left( 1- \overleftrightarrow {w}_{n-k+1,n}v_{n-k}(1-t_{n-k})(1-u_{n-k})\right)^{-\frac{1}{2}(n-k+1+\beta-\delta)}\nonumber\\
&&\times \left( \overleftrightarrow {w}_{n-k,n}^{-\frac{1}{2}(n-k-1)}\left(  \overleftrightarrow {w}_{n-k,n} \partial _{ \overleftrightarrow {w}_{n-k,n}}\right) \overleftrightarrow {w}_{n-k,n}^{\frac{1}{2}(n-k-1)} \left( \overleftrightarrow {w}_{n-k,n} \partial _{ \overleftrightarrow {w}_{n-k,n}} + \Omega _{n-k-1}^{(I)} \right) + Q\right) \Bigg\} \nonumber\\
&&\times \;_2F_1 \left(\frac{\alpha -\delta +1 }{2}, \frac{\beta-\delta +1 }{2};\frac{1}{2}+\frac{\gamma }{2}; \overleftrightarrow {w}_{1,n} \right)\Bigg\} \eta ^n \Bigg\}  \label{eq:66}
\end{eqnarray}
where
\begin{equation}
\begin{cases} 
\Omega _{n-k-1}^{(I)} =  \frac{1}{2(1+a)}(\alpha +\beta -\delta +n-k-1 +a(\gamma-\delta+n-k ) )   \cr
Q=  \frac{q-(\delta-1)\gamma a }{4(1+a)}
\end{cases}\nonumber %\label{eq:8}
\end{equation}
\subsection{ ${\displaystyle x^{1-\gamma } (1-x)^{1-\delta } Hl(a, q-(\gamma +\delta -2)a-(\gamma -1)(\alpha +\beta -\gamma -\delta +1); \alpha - \gamma -\delta +2}$ \\${\displaystyle, \beta - \gamma -\delta +2, 2-\gamma, 2 - \delta ; x)}$}
\subsubsection{Polynomial which makes $B_n$ term terminated}
Replace coefficients $q$, $\alpha$, $\beta$, $\gamma $, $\delta$, $c_0$ and $\lambda $ by $q-(\gamma +\delta -2)a-(\gamma -1)(\alpha +\beta -\gamma -\delta +1)$, $\alpha - \gamma -\delta +2$, $\beta - \gamma -\delta +2, 2-\gamma$, $2 - \delta$,1 and zero into (\ref{eq:39}). Multiply $x^{1-\gamma } (1-x)^{1-\delta }$ and the new (\ref{eq:39}) together.

\underline {(1) The case of $\alpha = -2 \alpha _i-i -2+\gamma +\delta $ and $\beta \ne -2 \beta _i -i-2+\gamma +\delta$ where $i, \alpha _i, \beta _i$ = $0,1,2,\cdots$.}
\begin{eqnarray}
& &x^{1-\gamma } (1-x)^{1-\delta } y(x)\nonumber\\
&=& x^{1-\gamma } (1-x)^{1-\delta } Hl(a, q-(\gamma +\delta -2)a-(\gamma -1)(\alpha +\beta -\gamma -\delta +1); \alpha - \gamma -\delta +2\nonumber\\
&&, \beta - \gamma -\delta +2, 2-\gamma, 2 - \delta ; x)\nonumber\\
&=& x^{1-\gamma } (1-x)^{1-\delta } \Bigg\{ \;_2F_1 \left(-\alpha _0, \frac{\beta-\gamma -\delta +2}{2};\frac{3-\gamma }{2}; z \right)+ \sum_{n=1}^{\infty } \Bigg\{\prod _{k=0}^{n-1} \Bigg\{ \int_{0}^{1} dt_{n-k}\;t_{n-k}^{\frac{1}{2}(n-k-2)} \int_{0}^{1} du_{n-k}\;u_{n-k}^{\frac{1}{2}(n-k-1-\gamma)} \nonumber\\
&&\times  \frac{1}{2\pi i}  \oint dv_{n-k} \frac{1}{v_{n-k}} \left( 1-\frac{1}{v_{n-k}}\right)^{\alpha _{n-k}} \left( 1- \overleftrightarrow {w}_{n-k+1,n}v_{n-k}(1-t_{n-k})(1-u_{n-k})\right)^{-\frac{1}{2}(n-k+2+\beta-\gamma -\delta)}\nonumber\\
&&\times \left( \overleftrightarrow {w}_{n-k,n}^{-\frac{1}{2}(n-k-1)}\left(  \overleftrightarrow {w}_{n-k,n} \partial _{ \overleftrightarrow {w}_{n-k,n}}\right) \overleftrightarrow {w}_{n-k,n}^{\frac{1}{2}(n-k-1)} \left( \overleftrightarrow {w}_{n-k,n} \partial _{ \overleftrightarrow {w}_{n-k,n}}  + \Omega _{n-k-1}^{(S)}\right) + Q _{n-k-1}^{(S)}\right) \Bigg\}\nonumber\\
&&\times \;_2F_1 \left(-\alpha _0, \frac{\beta-\gamma -\delta +2}{2};\frac{3-\gamma }{2}; \overleftrightarrow {w}_{1,n}\right) \Bigg\} \eta ^n \Bigg\} \label{eq:67}
\end{eqnarray}
where
\begin{equation}
\begin{cases} z = -\frac{1}{a}x^2 \cr
\eta = \frac{(1+a)}{a} x \cr
\alpha _i\leq \alpha _j \;\;\mbox{only}\;\mbox{if}\;i\leq j\;\;\mbox{where}\;i,j= 0,1,2,\cdots
\end{cases}\nonumber %\label{eq:37}
\end{equation}
and
\begin{equation}
\begin{cases} 
\Omega _{n-k-1}^{(S)} =  \frac{1}{2(1+a)}(-2\alpha _{n-k-1}+\beta -\gamma +a(-\delta -\gamma +n-k+2))   \cr
Q _{n-k-1}^{(S)} =  \frac{q-(\gamma +\delta -2)a-(\gamma -1)(-2\alpha _{n-k-1}+\beta -n+k)}{4(1+a)} 
\end{cases}\nonumber %\label{eq:8}
\end{equation}
\underline {(2) The case of $\alpha = -2 \alpha _i-i -2+\gamma +\delta $ and $\beta = -2\beta _i-i -2+\gamma +\delta $ only if $\alpha _i \leq \beta _i$.}

Put $\beta = -2\beta _i-i -2+\gamma +\delta $  where $i=0,1,2,\cdots$ in (\ref{eq:67}).
\begin{eqnarray}
 & &x^{1-\gamma } (1-x)^{1-\delta } y(x)\nonumber\\
&=& x^{1-\gamma } (1-x)^{1-\delta } Hl(a, q-(\gamma +\delta -2)a-(\gamma -1)(\alpha +\beta -\gamma -\delta +1); \alpha - \gamma -\delta +2\nonumber\\
&&, \beta - \gamma -\delta +2, 2-\gamma, 2 - \delta ; x)\nonumber\\
&=& x^{1-\gamma } (1-x)^{1-\delta } \Bigg\{ \;_2F_1 \left(-\alpha _0, -\beta _0;\frac{3-\gamma }{2}; z \right)+ \sum_{n=1}^{\infty } \Bigg\{\prod _{k=0}^{n-1} \Bigg\{ \int_{0}^{1} dt_{n-k}\;t_{n-k}^{\frac{1}{2}(n-k-2)} \int_{0}^{1} du_{n-k}\;u_{n-k}^{\frac{1}{2}(n-k-1-\gamma)} \nonumber\\
&&\times  \frac{1}{2\pi i}  \oint dv_{n-k} \frac{1}{v_{n-k}} \left( 1-\frac{1}{v_{n-k}}\right)^{\alpha _{n-k}} \left( 1- \overleftrightarrow {w}_{n-k+1,n}v_{n-k}(1-t_{n-k})(1-u_{n-k})\right)^{\beta _{n-k}}\nonumber\\
&&\times \left( \overleftrightarrow {w}_{n-k,n}^{-\frac{1}{2}(n-k-1)}\left(  \overleftrightarrow {w}_{n-k,n} \partial _{ \overleftrightarrow {w}_{n-k,n}}\right) \overleftrightarrow {w}_{n-k,n}^{\frac{1}{2}(n-k-1)} \left( \overleftrightarrow {w}_{n-k,n} \partial _{ \overleftrightarrow {w}_{n-k,n}}  + \Omega _{n-k-1}^{(B)}\right) +Q _{n-k-1}^{(B)}\right) \Bigg\} \nonumber\\
&&\times \;_2F_1 \left(-\alpha _0, -\beta _0;\frac{3-\gamma }{2}; \overleftrightarrow {w}_{1,n}\right) \Bigg\} \eta ^n \Bigg\}  \label{eq:68}
\end{eqnarray}
where
\begin{equation}
\begin{cases} 
\Omega _{n-k-1}^{(B)} = \frac{1}{2(1+a)}(-2\alpha _{n-k-1}-2\beta _{n-k-1}-n+k-1+\delta +a(-\delta -\gamma +n-k+2))     \cr
Q _{n-k-1}^{(B)} =  \frac{q-(\gamma +\delta -2)a-(\gamma -1)(-2\alpha _{n-k-1}-2\beta _{n-k-1}-2n+2k-1+\gamma +\delta)}{4(1+a)}  
\end{cases}\nonumber %\label{eq:8}
\end{equation}

\subsubsection{Infinite series}
Replace coefficients q, $\alpha$, $\beta$, $\gamma $, $\delta$, $c_0$ and $\lambda $ by $q-(\gamma +\delta -2)a-(\gamma -1)(\alpha +\beta -\gamma -\delta +1)$, $\alpha - \gamma -\delta +2$, $\beta - \gamma -\delta +2, 2-\gamma$, $2 - \delta$,1 and zero into (\ref{eq:46}). Multiply $x^{1-\gamma } (1-x)^{1-\delta }$ and the new (\ref{eq:46}) together.
\begin{eqnarray}
& &x^{1-\gamma } (1-x)^{1-\delta } y(x)\nonumber\\
&=& x^{1-\gamma } (1-x)^{1-\delta } Hl(a, q-(\gamma +\delta -2)a-(\gamma -1)(\alpha +\beta -\gamma -\delta +1); \alpha - \gamma -\delta +2\nonumber\\
&&, \beta - \gamma -\delta +2, 2-\gamma, 2 - \delta ; x)\nonumber\\
&=& x^{1-\gamma } (1-x)^{1-\delta } \Bigg\{ \;_2F_1 \left(\frac{\alpha -\gamma -\delta +2}{2}, \frac{\beta-\gamma -\delta +2}{2};\frac{3-\gamma }{2}; z \right)\nonumber\\
&&+ \sum_{n=1}^{\infty } \Bigg\{\prod _{k=0}^{n-1} \Bigg\{ \int_{0}^{1} dt_{n-k}\;t_{n-k}^{\frac{1}{2}(n-k-2)} \int_{0}^{1} du_{n-k}\;u_{n-k}^{\frac{1}{2}(n-k-1-\gamma)} \nonumber\\
&&\times  \frac{1}{2\pi i}  \oint dv_{n-k} \frac{1}{v_{n-k}} \left( 1-\frac{1}{v_{n-k}}\right)^{-\frac{1}{2}(n-k+2+\alpha -\gamma -\delta)} \left( 1- \overleftrightarrow {w}_{n-k+1,n}v_{n-k}(1-t_{n-k})(1-u_{n-k})\right)^{-\frac{1}{2}(n-k+2+\beta-\gamma -\delta)}\nonumber\\
&&\times \left( \overleftrightarrow {w}_{n-k,n}^{-\frac{1}{2}(n-k-1)}\left(  \overleftrightarrow {w}_{n-k,n} \partial _{ \overleftrightarrow {w}_{n-k,n}}\right) \overleftrightarrow {w}_{n-k,n}^{\frac{1}{2}(n-k-1)} \left( \overleftrightarrow {w}_{n-k,n} \partial _{ \overleftrightarrow {w}_{n-k,n}}  + \Omega _{n-k-1}^{(I)}\right) + Q\right) \Bigg\} \nonumber\\
&&\times \;_2F_1 \left(\frac{\alpha -\gamma -\delta +2}{2}, \frac{\beta-\gamma -\delta +2}{2};\frac{3-\gamma }{2}; \overleftrightarrow {w}_{1,n}\right) \Bigg\} \eta ^n \Bigg\} \label{eq:69}
\end{eqnarray}
where
\begin{equation}
\begin{cases} 
\Omega _{n-k-1}^{(I)} = \frac{1}{2(1+a)}(\alpha +\beta -2\gamma -\delta +n-k+1 +a(-\delta -\gamma +n-k+2))   \cr
Q  =  \frac{q-(\gamma +\delta -2)a-(\gamma -1)(\alpha +\beta -\gamma -\delta +1)}{4(1+a)}   
\end{cases}\nonumber %\label{eq:8}
\end{equation}
\subsection{ ${\displaystyle  Hl(1-a,-q+\alpha \beta; \alpha,\beta, \delta, \gamma; 1-x)}$} 
\subsubsection{Polynomial which makes $B_n$ term terminated}
Replace coefficients $a$, $q$, $\gamma $, $\delta$, $x$, $c_0$ and $\lambda $ by $1-a$, $-q +\alpha \beta $, $\delta $, $\gamma $, $1-x$, 1 and zero into (\ref{eq:39}).

\underline {(1) The case of $\alpha = -2 \alpha _i-i$ and $\beta \ne -2 \beta _i -i$ where $i, \alpha _i, \beta _i$ = $0,1,2,\cdots$.}
\begin{eqnarray}
y(\xi ) &=& Hl(1-a,-q+\alpha \beta; \alpha,\beta, \delta, \gamma; 1-x)\nonumber\\
&=&  \;_2F_1 \left(-\alpha _0, \frac{\beta }{2}; \frac{1}{2}+ \frac{\delta }{2}; z \right) + \sum_{n=1}^{\infty } \Bigg\{\prod _{k=0}^{n-1} \Bigg\{ \int_{0}^{1} dt_{n-k}\;t_{n-k}^{\frac{1}{2}(n-k-2)} \int_{0}^{1} du_{n-k}\;u_{n-k}^{\frac{1}{2}(n-k-3+\delta )} \nonumber\\
&&\times  \frac{1}{2\pi i}  \oint dv_{n-k} \frac{1}{v_{n-k}} \left( 1-\frac{1}{v_{n-k}}\right)^{\alpha _{n-k}} \left( 1- \overleftrightarrow {w}_{n-k+1,n}v_{n-k}(1-t_{n-k})(1-u_{n-k})\right)^{-\frac{1}{2}(n-k+\beta )}\nonumber\\
&&\times \left( \overleftrightarrow {w}_{n-k,n}^{-\frac{1}{2}(n-k-1)}\left(  \overleftrightarrow {w}_{n-k,n} \partial _{ \overleftrightarrow {w}_{n-k,n}}\right) \overleftrightarrow {w}_{n-k,n}^{\frac{1}{2}(n-k-1)} \left( \overleftrightarrow {w}_{n-k,n} \partial _{ \overleftrightarrow {w}_{n-k,n}}  + \Omega _{n-k-1}^{(S)}\right) + Q_{n-k-1}^{(S)}\right) \Bigg\}\nonumber\\
&&\times\;_2F_1 \left( -\alpha _0, \frac{\beta }{2}; \frac{1}{2}+ \frac{\delta }{2}; \overleftrightarrow {w}_{1,n} \right) \Bigg\} \eta ^n \label{eq:70}
\end{eqnarray}
where
\begin{equation}
\begin{cases} \xi =1-x \cr
z = \frac{-1}{1-a}\xi^2 \cr
\eta = \frac{2-a}{1-a}\xi \cr
\alpha _i\leq \alpha _j \;\;\mbox{only}\;\mbox{if}\;i\leq j\;\;\mbox{where}\;i,j= 0,1,2,\cdots
\end{cases}\nonumber %\label{eq:37}
\end{equation}
and
\begin{equation}
\begin{cases} 
\Omega _{n-k-1}^{(S)} =  \frac{1}{2(2-a)}(-2\alpha _{n-k-1}+\beta -\gamma  +(1-a)(\delta +\gamma +n-k-2))   \cr
Q_{n-k-1}^{(S)}  = \frac{-q-(2\alpha _{n-k-1}+n-k-1)\beta }{4(2-a)}   
\end{cases}\nonumber %\label{eq:8}
\end{equation}
\underline {(2) The case of $\alpha = -2 \alpha _i-i$ and $\beta = -2 \beta _i -i$ only if $\alpha _i \leq \beta _i$.}

Put $\beta = -2\beta _i-i $  where $i=0,1,2,\cdots$ in (\ref{eq:70}).
\begin{eqnarray}
y(\xi ) &=& Hl(1-a,-q+\alpha \beta; \alpha,\beta, \delta, \gamma; 1-x)\nonumber\\
&=&  \;_2F_1 \left(-\alpha _0, -\beta _0; \frac{1}{2}+ \frac{\delta }{2}; z \right) + \sum_{n=1}^{\infty } \Bigg\{\prod _{k=0}^{n-1} \Bigg\{ \int_{0}^{1} dt_{n-k}\;t_{n-k}^{\frac{1}{2}(n-k-2)} \int_{0}^{1} du_{n-k}\;u_{n-k}^{\frac{1}{2}(n-k-3+\delta )} \nonumber\\
&&\times  \frac{1}{2\pi i}  \oint dv_{n-k} \frac{1}{v_{n-k}} \left( 1-\frac{1}{v_{n-k}}\right)^{\alpha _{n-k}} \left( 1- \overleftrightarrow {w}_{n-k+1,n}v_{n-k}(1-t_{n-k})(1-u_{n-k})\right)^{\beta _{n-k}}\nonumber\\
&&\times \left( \overleftrightarrow {w}_{n-k,n}^{-\frac{1}{2}(n-k-1)}\left(  \overleftrightarrow {w}_{n-k,n} \partial _{ \overleftrightarrow {w}_{n-k,n}}\right) \overleftrightarrow {w}_{n-k,n}^{\frac{1}{2}(n-k-1)} \left( \overleftrightarrow {w}_{n-k,n} \partial _{ \overleftrightarrow {w}_{n-k,n}}  + \Omega _{n-k-1}^{(B)}\right) +Q_{n-k-1}^{(B)}\right) \Bigg\} \nonumber\\
&&\times \;_2F_1 \left( -\alpha _0, -\beta _0; \frac{1}{2}+ \frac{\delta }{2}; \overleftrightarrow {w}_{1,n} \right) \Bigg\} \eta ^n  \label{eq:71}
\end{eqnarray}
where
\begin{equation}
\begin{cases} 
\Omega _{n-k-1}^{(B)} = \frac{1}{2(2-a)}(-2\alpha _{n-k-1}-2\beta _{n-k-1}-\gamma -n+k+1 +(1-a)(\delta +\gamma +n-k-2)) \cr
Q_{n-k-1}^{(B)}  =  \frac{-q+(2\alpha _{n-k-1}+n-k-1)(2\beta _{n-k-1}+n-k-1)}{4(2-a)} 
\end{cases}\nonumber %\label{eq:8}
\end{equation}
\subsubsection{Infinite series}
Replace coefficients a, q, $\gamma $, $\delta$, $x$, $c_0$ and $\lambda $ by $1-a$, $-q +\alpha \beta $, $\delta $, $\gamma $, $1-x$,1 and zero into (\ref{eq:46}).
\begin{eqnarray}
y(\xi ) &=& Hl(1-a,-q+\alpha \beta; \alpha, \beta, \delta, \gamma; 1-x)\nonumber\\
&=&  \;_2F_1 \left(\frac{\alpha }{2}, \frac{\beta }{2}; \frac{1}{2}+ \frac{\delta }{2}; z \right) + \sum_{n=1}^{\infty } \Bigg\{\prod _{k=0}^{n-1} \Bigg\{ \int_{0}^{1} dt_{n-k}\;t_{n-k}^{\frac{1}{2}(n-k-2)} \int_{0}^{1} du_{n-k}\;u_{n-k}^{\frac{1}{2}(n-k-3+\delta )} \nonumber\\
&&\times  \frac{1}{2\pi i}  \oint dv_{n-k} \frac{1}{v_{n-k}} \left( 1-\frac{1}{v_{n-k}}\right)^{-\frac{1}{2}(n-k+\alpha )} \left( 1- \overleftrightarrow {w}_{n-k+1,n}v_{n-k}(1-t_{n-k})(1-u_{n-k})\right)^{-\frac{1}{2}(n-k+\beta )}\nonumber\\
&&\times \left( \overleftrightarrow {w}_{n-k,n}^{-\frac{1}{2}(n-k-1)}\left(  \overleftrightarrow {w}_{n-k,n} \partial _{ \overleftrightarrow {w}_{n-k,n}}\right) \overleftrightarrow {w}_{n-k,n}^{\frac{1}{2}(n-k-1)} \left( \overleftrightarrow {w}_{n-k,n} \partial _{ \overleftrightarrow {w}_{n-k,n}} + \Omega _{n-k-1}^{(I)}\right) + Q\right) \Bigg\}\nonumber\\
&&\times \;_2F_1 \left(\frac{\alpha }{2}, \frac{\beta }{2}; \frac{1}{2}+ \frac{\delta }{2}; \overleftrightarrow {w}_{1,n} \right) \Bigg\} \eta ^n \label{eq:72}
\end{eqnarray}
where
\begin{equation}
\begin{cases} 
\Omega _{n-k-1}^{(I)} = \frac{1}{2(2-a)}(\alpha +\beta -\gamma +n-k-1+(1-a)(\delta +\gamma +n-k-2))  \cr
Q  =  \frac{-q+\alpha \beta }{4(2-a)}  
\end{cases}\nonumber %\label{eq:8}
\end{equation}
\subsection{ ${\displaystyle (1-x)^{1-\delta } Hl(1-a,-q+(\delta -1)\gamma a+(\alpha -\delta +1)(\beta -\delta +1); \alpha-\delta +1,\beta-\delta +1, 2-\delta, \gamma; 1-x)}$}
\subsubsection{Polynomial which makes $B_n$ term terminated}

\underline {(1) The case of $\alpha = -2 \alpha _i-i+\delta -1$ and $\beta \ne -2 \beta _i -i+\delta -1$ where $i, \alpha _i, \beta _i$ = $0,1,2,\cdots$.}\vspace{1mm}

Replace coefficients $a$, $q$, $\alpha $, $\beta $, $\gamma $, $\delta$, $x$, $c_0$ and $\lambda $ by $1-a$, $-q+(\delta -1)\gamma a+(\alpha -\delta +1)(\beta -\delta +1)$, $\alpha-\delta +1 $, $\beta-\delta +1 $, $2-\delta$, $\gamma $, $1-x$, 1 and zero into (\ref{eq:39}). Multiply $(1-x)^{1-\delta }$ and the new (\ref{eq:39}) together.
\begin{eqnarray}
& &(1-x)^{1-\delta } y(\xi)\nonumber\\
&=& (1-x)^{1-\delta } Hl(1-a,-q+(\delta -1)\gamma a+(\alpha -\delta +1)(\beta -\delta +1); \alpha-\delta +1,\beta-\delta +1, 2-\delta, \gamma; 1-x) \nonumber\\
&=& (1-x)^{1-\delta } \Bigg\{ \;_2F_1 \left(-\alpha _0, \frac{\beta -\delta +1}{2}; \frac{3-\delta }{2}; z\right)  + \sum_{n=1}^{\infty } \Bigg\{\prod _{k=0}^{n-1} \Bigg\{ \int_{0}^{1} dt_{n-k}\;t_{n-k}^{\frac{1}{2}(n-k-2)} \int_{0}^{1} du_{n-k}\;u_{n-k}^{\frac{1}{2}(n-k-1-\delta)} \nonumber\\
&&\times  \frac{1}{2\pi i}  \oint dv_{n-k} \frac{1}{v_{n-k}} \left( 1-\frac{1}{v_{n-k}}\right)^{\alpha _{n-k}} \left( 1- \overleftrightarrow {w}_{n-k+1,n}v_{n-k}(1-t_{n-k})(1-u_{n-k})\right)^{-\frac{1}{2}(n-k+1+\beta-\delta )}\nonumber\\
&&\times \left( \overleftrightarrow {w}_{n-k,n}^{-\frac{1}{2}(n-k-1)}\left(  \overleftrightarrow {w}_{n-k,n} \partial _{ \overleftrightarrow {w}_{n-k,n}}\right) \overleftrightarrow {w}_{n-k,n}^{\frac{1}{2}(n-k-1)} \left( \overleftrightarrow {w}_{n-k,n} \partial _{ \overleftrightarrow {w}_{n-k,n}} + \Omega _{n-k-1}^{(S)} \right) +Q_{n-k-1}^{(S)}\right) \Bigg\} \nonumber\\
&&\times\;_2F_1 \left(-\alpha _0, \frac{\beta -\delta +1}{2}; \frac{3-\delta }{2}; \overleftrightarrow {w}_{1,n}\right) \Bigg\} \eta ^n \Bigg\} \label{eq:73}
\end{eqnarray}
where
\begin{equation}
\begin{cases} \xi =1-x \cr
z = \frac{-1}{1-a}\xi^2 \cr
\eta = \frac{2-a}{1-a}\xi \cr
\alpha _i\leq \alpha _j \;\;\mbox{only}\;\mbox{if}\;i\leq j\;\;\mbox{where}\;i,j= 0,1,2,\cdots
\end{cases}\nonumber %\label{eq:37}
\end{equation}
and
\begin{equation}
\begin{cases} 
\Omega _{n-k-1}^{(S)} = \frac{1}{2(2-a)}(-2\alpha _{n-k-1}+\beta -\gamma -\delta +1 +(1-a)(\gamma -\delta +n-k))  \cr
Q_{n-k-1}^{(S)}  =   \frac{-q+(\delta -1)\gamma a-(2\alpha_{n-k-1}+n-k-1)(\beta -\delta +1)}{4(2-a)}
\end{cases}\nonumber %\label{eq:8}
\end{equation}
\underline {(2) The case of $\alpha = -2 \alpha _i-i+\delta -1$ and $\beta = -2 \beta _i -i+\delta -1$ only if $\alpha _i \leq \beta _i$.}

Put $\beta = -2 \beta _i -i+\delta -1$  where $i=0,1,2,\cdots$ in (\ref{eq:73}).
\begin{eqnarray}
& &(1-x)^{1-\delta } y(\xi)\nonumber\\
&=& (1-x)^{1-\delta } Hl(1-a,-q+(\delta -1)\gamma a+(\alpha -\delta +1)(\beta -\delta +1); \alpha-\delta +1,\beta-\delta +1, 2-\delta, \gamma; 1-x) \nonumber\\
&=& (1-x)^{1-\delta } \Bigg\{ \;_2F_1 \left(-\alpha _0, -\beta _0; \frac{3-\delta }{2}; z\right)  + \sum_{n=1}^{\infty } \Bigg\{\prod _{k=0}^{n-1} \Bigg\{ \int_{0}^{1} dt_{n-k}\;t_{n-k}^{\frac{1}{2}(n-k-2)} \int_{0}^{1} du_{n-k}\;u_{n-k}^{\frac{1}{2}(n-k-1-\delta)} \nonumber\\
&&\times  \frac{1}{2\pi i}  \oint dv_{n-k} \frac{1}{v_{n-k}} \left( 1-\frac{1}{v_{n-k}}\right)^{\alpha _{n-k}} \left( 1- \overleftrightarrow {w}_{n-k+1,n}v_{n-k}(1-t_{n-k})(1-u_{n-k})\right)^{\beta _{n-k}} \nonumber\\
&&\times \left( \overleftrightarrow {w}_{n-k,n}^{-\frac{1}{2}(n-k-1)}\left(  \overleftrightarrow {w}_{n-k,n} \partial _{ \overleftrightarrow {w}_{n-k,n}}\right) \overleftrightarrow {w}_{n-k,n}^{\frac{1}{2}(n-k-1)} \left( \overleftrightarrow {w}_{n-k,n} \partial _{ \overleftrightarrow {w}_{n-k,n}}  + \Omega _{n-k-1}^{(B)} \right) +Q_{n-k-1}^{(B)}\right) \Bigg\} \nonumber\\
&&\times \;_2F_1 \left(-\alpha _0, -\beta _0; \frac{3-\delta }{2}; \overleftrightarrow {w}_{1,n}\right) \Bigg\} \eta ^n \Bigg\}  \label{eq:74}
\end{eqnarray}
where
\begin{equation}
\begin{cases} 
\Omega _{n-k-1}^{(B)} = \frac{1}{2(2-a)}(-2\alpha _{n-k-1}-2\beta _{n-k-1}-\gamma -n+k+1 +(1-a)(\gamma -\delta +n-k))  \cr
Q_{n-k-1}^{(B)}  = \frac{-q+(\delta -1)\gamma a+(2\alpha_{n-k-1}+n-k-1)(2\beta _{n-k-1}+n-k-1)}{4(2-a)}   
\end{cases}\nonumber %\label{eq:8}
\end{equation}
\subsubsection{Infinite series}
Replace coefficients a, q, $\alpha $, $\beta $, $\gamma $, $\delta$, $x$, $c_0$ and $\lambda $ by $1-a$, $-q+(\delta -1)\gamma a+(\alpha -\delta +1)(\beta -\delta +1)$, $\alpha-\delta +1 $, $\beta-\delta +1 $, $2-\delta$, $\gamma $, $1-x$, 1 and zero into (\ref{eq:46}).  Multiply $(1-x)^{1-\delta }$ and the new (\ref{eq:46}) together.
\begin{eqnarray}
& &(1-x)^{1-\delta } y(\xi)\nonumber\\
&=& (1-x)^{1-\delta } Hl(1-a,-q+(\delta -1)\gamma a+(\alpha -\delta +1)(\beta -\delta +1); \alpha -\delta +1,\beta-\delta +1, 2-\delta, \gamma; 1-x) \nonumber\\
&=& (1-x)^{1-\delta } \Bigg\{ \;_2F_1 \left(\frac{\alpha -\delta +1}{2}, \frac{\beta -\delta +1}{2}; \frac{3-\delta }{2}; z\right)  + \sum_{n=1}^{\infty } \Bigg\{\prod _{k=0}^{n-1} \Bigg\{ \int_{0}^{1} dt_{n-k}\;t_{n-k}^{\frac{1}{2}(n-k-2)} \int_{0}^{1} du_{n-k}\;u_{n-k}^{\frac{1}{2}(n-k-1-\delta)} \nonumber\\
&&\times  \frac{1}{2\pi i}  \oint dv_{n-k} \frac{1}{v_{n-k}} \left( 1-\frac{1}{v_{n-k}}\right)^{-\frac{1}{2}(n-k+1+\alpha -\delta )} \left( 1- \overleftrightarrow {w}_{n-k+1,n}v_{n-k}(1-t_{n-k})(1-u_{n-k})\right)^{-\frac{1}{2}(n-k+1+\beta-\delta )}\nonumber\\
&&\times \left( \overleftrightarrow {w}_{n-k,n}^{-\frac{1}{2}(n-k-1)}\left(  \overleftrightarrow {w}_{n-k,n} \partial _{ \overleftrightarrow {w}_{n-k,n}}\right) \overleftrightarrow {w}_{n-k,n}^{\frac{1}{2}(n-k-1)} \left( \overleftrightarrow {w}_{n-k,n} \partial _{ \overleftrightarrow {w}_{n-k,n}}  + \Omega _{n-k-1}^{(I)}\right) +Q \right) \Bigg\} \nonumber\\
&&\times \;_2F_1 \left(\frac{\alpha -\delta +1}{2}, \frac{\beta -\delta +1}{2}; \frac{3-\delta }{2}; \overleftrightarrow {w}_{1,n}\right) \Bigg\} \eta ^n \Bigg\} \label{eq:75}
\end{eqnarray}
where
\begin{equation}
\begin{cases} 
\Omega _{n-k-1}^{(I)} = \frac{1}{2(2-a)}(\alpha +\beta -\gamma -2\delta +n-k+1 +(1-a)(\gamma -\delta +n-k))   \cr
Q =   \frac{-q+(\delta -1)\gamma a+(\alpha -\delta +1)(\beta -\delta +1)}{4(2-a)}  
\end{cases}\nonumber %\label{eq:8}
\end{equation}
\subsection{ ${\displaystyle x^{-\alpha } Hl\left(\frac{1}{a},\frac{q+\alpha [(\alpha -\gamma -\delta +1)a-\beta +\delta ]}{a}; \alpha , \alpha -\gamma +1, \alpha -\beta +1,\delta ;\frac{1}{x}\right)}$}
\subsubsection{Infinite series}
Replace coefficients $a$, $q$, $\beta $, $\gamma $, $x$, $c_0$ and $\lambda $ by $\frac{1}{a}$, $\frac{q+\alpha [(\alpha -\gamma -\delta +1)a-\beta +\delta ]}{a}$, $\alpha-\gamma +1 $, $\alpha -\beta +1 $, $\frac{1}{x}$, 1 and zero into (\ref{eq:46}). Multiply $x^{-\alpha }$ and the new (\ref{eq:46}) together.
\begin{eqnarray}
& &x^{-\alpha } y(\xi )\nonumber\\
&=& x^{-\alpha } Hl\left(\frac{1}{a},\frac{q+\alpha [(\alpha -\gamma -\delta +1)a-\beta +\delta ]}{a}; \alpha , \alpha -\gamma +1, \alpha -\beta +1,\delta ;\frac{1}{x}\right) \nonumber\\
&=& x^{-\alpha } \Bigg\{ \;_2F_1 \left(\frac{\alpha }{2}, \frac{\alpha -\gamma +1}{2}; \frac{\alpha -\beta +2}{2}; z\right) + \sum_{n=1}^{\infty } \Bigg\{\prod _{k=0}^{n-1} \Bigg\{ \int_{0}^{1} dt_{n-k}\;t_{n-k}^{\frac{1}{2}(n-k-2)} \int_{0}^{1} du_{n-k}\;u_{n-k}^{\frac{1}{2}(n-k-2+\alpha -\beta )} \nonumber\\
&&\times \frac{1}{2\pi i}  \oint dv_{n-k} \frac{1}{v_{n-k}} \left( 1-\frac{1}{v_{n-k}}\right)^{-\frac{1}{2}(n-k+\alpha)}  \left( 1- \overleftrightarrow {w}_{n-k+1,n}v_{n-k}(1-t_{n-k})(1-u_{n-k})\right)^{-\frac{1}{2}(n-k+1+\alpha -\gamma )}\nonumber\\
&&\times  \left( \overleftrightarrow {w}_{n-k,n}^{-\frac{1}{2}(n-k-1)}\left(  \overleftrightarrow {w}_{n-k,n} \partial _{ \overleftrightarrow {w}_{n-k,n}}\right) \overleftrightarrow {w}_{n-k,n}^{\frac{1}{2}(n-k-1)}\left( \overleftrightarrow {w}_{n-k,n} \partial _{ \overleftrightarrow {w}_{n-k,n}} + \Omega _{n-k-1}^{(I)}\right) +Q\right) \Bigg\} \nonumber\\
&&\times \;_2F_1 \left(\frac{\alpha }{2}, \frac{\alpha -\gamma +1}{2}; \frac{\alpha -\beta +2}{2}; \overleftrightarrow {w}_{1,n}\right) \Bigg\} \eta ^n \Bigg\} \label{eq:76}
\end{eqnarray}
where
\begin{equation}
\begin{cases} \xi = \frac{1}{x} \cr
z = -a \xi ^2 \cr
\eta = (1+a)\xi 
\end{cases}\nonumber %\label{eq:37}
\end{equation}
and
\begin{equation}
\begin{cases} 
\Omega _{n-k-1}^{(I)} = \frac{a}{2(1+a)}(2\alpha -\gamma -\delta +n-k +\frac{1}{a}(\alpha -\beta +\delta +n-k-1))  \cr
Q =  \frac{q+\alpha [(\alpha -\gamma -\delta +1)a-\beta +\delta ]}{4(1+a)}
\end{cases}\nonumber %\label{eq:8}
\end{equation}
\subsubsection{Polynomial which makes $B_n$ term terminated}
Substitute $\gamma =\alpha +2\gamma _i +i+1$ into (\ref{eq:76}) where $i, \gamma _i= 0,1,2,\cdots$.
\begin{eqnarray}
& &x^{-\alpha } y(\xi )\nonumber\\
&=& x^{-\alpha } Hl\left(\frac{1}{a},\frac{q+\alpha [(\alpha -\gamma -\delta +1)a-\beta +\delta ]}{a}; \alpha , \alpha -\gamma +1, \alpha -\beta +1,\delta ;\frac{1}{x}\right) \nonumber\\
&=& x^{-\alpha } \Bigg\{ \;_2F_1 \left(\frac{\alpha }{2}, -\gamma _0; \frac{\alpha -\beta +2}{2}; z\right) + \sum_{n=1}^{\infty } \Bigg\{\prod _{k=0}^{n-1} \Bigg\{ \int_{0}^{1} dt_{n-k}\;t_{n-k}^{\frac{1}{2}(n-k-2)} \int_{0}^{1} du_{n-k}\;u_{n-k}^{\frac{1}{2}(n-k-2+\alpha -\beta )} \nonumber\\
&&\times \frac{1}{2\pi i}  \oint dv_{n-k} \frac{1}{v_{n-k}} \left( 1-\frac{1}{v_{n-k}}\right)^{-\frac{1}{2}(n-k+\alpha)}  \left( 1- \overleftrightarrow {w}_{n-k+1,n}v_{n-k}(1-t_{n-k})(1-u_{n-k})\right)^{\gamma _{n-k}}\nonumber\\
&&\times  \left( \overleftrightarrow {w}_{n-k,n}^{-\frac{1}{2}(n-k-1)}\left(  \overleftrightarrow {w}_{n-k,n} \partial _{ \overleftrightarrow {w}_{n-k,n}}\right) \overleftrightarrow {w}_{n-k,n}^{\frac{1}{2}(n-k-1)}\left( \overleftrightarrow {w}_{n-k,n} \partial _{ \overleftrightarrow {w}_{n-k,n}} + \Omega _{n-k-1}^{(S)}\right) + Q _{n-k-1}^{(S)}\right) \Bigg\} \nonumber\\
&&\times \;_2F_1 \left(\frac{\alpha }{2}, -\gamma _0; \frac{\alpha -\beta +2}{2}; \overleftrightarrow {w}_{1,n}\right) \Bigg\} \eta ^n \Bigg\} \label{eq:77}
\end{eqnarray}
where
\begin{equation}
\gamma _i\leq \gamma _j \;\;\mbox{only}\;\mbox{if}\;i\leq j\;\;\mbox{where}\;i,j= 0,1,2,\cdots \nonumber %\label{eq:37}
\end{equation}
and
\begin{equation}
\begin{cases} 
\Omega _{n-k-1}^{(S)} = \frac{a}{2(1+a)}(-2\gamma _{n-k-1}+\alpha -\delta +\frac{1}{a}(\alpha -\beta +\delta +n-k-1)) \cr
Q _{n-k-1}^{(S)} =   \frac{q+\alpha [(-2\gamma _{n-k-1}-\delta -n+k+1)a-\beta +\delta ]}{4(1+a)}
\end{cases}\nonumber %\label{eq:8}
\end{equation}
\subsection{ ${\displaystyle \left(1-\frac{x}{a} \right)^{-\beta } Hl\left(1-a, -q+\gamma \beta; -\alpha +\gamma +\delta, \beta, \gamma, \delta; \frac{(1-a)x}{x-a} \right)}$}
\subsubsection{Infinite series}
Replace coefficients $a$, $q$, $\alpha $, $x$, $c_0$ and $\lambda $ by $1-a$, $-q+\gamma \beta $, $-\alpha+\gamma +\delta $, $\frac{(1-a)x}{x-a}$, 1 and zero into (\ref{eq:46}). Multiply $\left(1-\frac{x}{a} \right)^{-\beta }$ and the new (\ref{eq:46}) together.
\begin{eqnarray}
&& \left(1-\frac{x}{a} \right)^{-\beta } y(\xi ) \nonumber\\
 &=& \left(1-\frac{x}{a} \right)^{-\beta } Hl\left(1-a, -q+\gamma \beta; -\alpha +\gamma +\delta, \beta, \gamma, \delta; \frac{(1-a)x}{x-a} \right) \nonumber\\
&=& \left(1-\frac{x}{a} \right)^{-\beta } \Bigg\{ \;_2F_1 \left(\frac{-\alpha+\gamma +\delta  }{2}, \frac{\beta }{2}; \frac{1}{2}+ \frac{\gamma }{2}; z\right) + \sum_{n=1}^{\infty } \Bigg\{\prod _{k=0}^{n-1} \Bigg\{ \int_{0}^{1} dt_{n-k}\;t_{n-k}^{\frac{1}{2}(n-k-2)} \int_{0}^{1} du_{n-k}\;u_{n-k}^{\frac{1}{2}(n-k-3+\gamma)} \nonumber\\
&&\times \frac{1}{2\pi i}  \oint dv_{n-k} \frac{1}{v_{n-k}} \left( 1-\frac{1}{v_{n-k}}\right)^{-\frac{1}{2}(n-k-\alpha+\gamma +\delta)}  \left( 1- \overleftrightarrow {w}_{n-k+1,n}v_{n-k}(1-t_{n-k})(1-u_{n-k})\right)^{-\frac{1}{2}(n-k+\beta)}\nonumber\\
&&\times  \left( \overleftrightarrow {w}_{n-k,n}^{-\frac{1}{2}(n-k-1)}\left(  \overleftrightarrow {w}_{n-k,n} \partial _{ \overleftrightarrow {w}_{n-k,n}}\right) \overleftrightarrow {w}_{n-k,n}^{\frac{1}{2}(n-k-1)}\left( \overleftrightarrow {w}_{n-k,n} \partial _{ \overleftrightarrow {w}_{n-k,n}} + \Omega _{n-k-1}^{(I)}\right) +Q \right) \Bigg\} \nonumber\\
&&\times \;_2F_1 \left(\frac{-\alpha+\gamma +\delta  }{2}, \frac{\beta }{2}; \frac{1}{2}+ \frac{\gamma }{2}; \overleftrightarrow {w}_{1,n}\right) \Bigg\} \eta ^n \Bigg\} \label{eq:160}
\end{eqnarray}
where
\begin{equation}
\begin{cases} \xi= \frac{(1-a)x}{x-a} \cr
z = -\frac{1}{1-a} \xi^2 \cr
\eta = \frac{(2-a)}{(1-a)}\xi \cr
\end{cases}\nonumber %\label{eq:37}
\end{equation}
and
\begin{equation}
\begin{cases} 
\Omega _{n-k-1}^{(I)} = \frac{1}{2(2-a)}(-\alpha +\beta +\gamma +n-k-1+(1-a)(\delta +\gamma +n-k-2)) \cr
Q =   \frac{-q+\gamma \beta}{4(2-a)}
\end{cases}\nonumber %\label{eq:8}
\end{equation}
\subsubsection{Polynomial which makes $B_n$ term terminated}
\underline {(1) The case of $\alpha =\gamma +\delta +2\alpha _i +i$ where $i, \alpha _i$ = $0,1,2,\cdots$.}

Substitute $\alpha =\gamma +\delta +2\alpha _i +i$ into (\ref{eq:160}).
\begin{eqnarray}
&& \left(1-\frac{x}{a} \right)^{-\beta } y(\xi ) \nonumber\\
 &=& \left(1-\frac{x}{a} \right)^{-\beta } Hl\left(1-a, -q+\gamma \beta; -\alpha +\gamma +\delta, \beta, \gamma, \delta; \frac{(1-a)x}{x-a} \right) \nonumber\\
&=& \left(1-\frac{x}{a} \right)^{-\beta } \Bigg\{ \;_2F_1 \left(-\alpha _0, \frac{\beta }{2}; \frac{1}{2}+ \frac{\gamma }{2}; z\right) + \sum_{n=1}^{\infty } \Bigg\{\prod _{k=0}^{n-1} \Bigg\{ \int_{0}^{1} dt_{n-k}\;t_{n-k}^{\frac{1}{2}(n-k-2)} \int_{0}^{1} du_{n-k}\;u_{n-k}^{\frac{1}{2}(n-k-3+\gamma)} \nonumber\\
&&\times \frac{1}{2\pi i}  \oint dv_{n-k} \frac{1}{v_{n-k}} \left( 1-\frac{1}{v_{n-k}}\right)^{\alpha _{n-k}}  \left( 1- \overleftrightarrow {w}_{n-k+1,n}v_{n-k}(1-t_{n-k})(1-u_{n-k})\right)^{-\frac{1}{2}(n-k+\beta)}\nonumber\\
&&\times  \left( \overleftrightarrow {w}_{n-k,n}^{-\frac{1}{2}(n-k-1)}\left(  \overleftrightarrow {w}_{n-k,n} \partial _{ \overleftrightarrow {w}_{n-k,n}}\right) \overleftrightarrow {w}_{n-k,n}^{\frac{1}{2}(n-k-1)}\left( \overleftrightarrow {w}_{n-k,n} \partial _{ \overleftrightarrow {w}_{n-k,n}} + \Omega _{n-k-1}^{(S)}\right) +Q\right) \Bigg\} \nonumber\\
&&\times  \;_2F_1 \left(-\alpha _0, \frac{\beta }{2}; \frac{1}{2}+ \frac{\gamma }{2}; \overleftrightarrow {w}_{1,n}\right)\Bigg\} \eta ^n \Bigg\} \label{eq:161}
\end{eqnarray}
where
\begin{equation}
\alpha _i\leq \alpha _j \;\;\mbox{only}\;\mbox{if}\;i\leq j\;\;\mbox{where}\;i,j= 0,1,2,\cdots
\nonumber 
\end{equation}
and
\begin{equation}
\begin{cases} 
\Omega _{n-k-1}^{(S)} =  \frac{1}{2(2-a)}(\beta -\delta -2\alpha _{n-k-1} +(1-a)(\delta +\gamma +n-k-2)) \cr
Q =   \frac{-q+\gamma \beta}{4(2-a)}
\end{cases}\nonumber %\label{eq:8}
\end{equation}
\underline {(2) The case of $\delta =\alpha -\gamma -2\delta _i -i$ where $i, \delta _i$ = $0,1,2,\cdots$.}

Substitute $\delta =\alpha -\gamma -2\delta _i -i$ into (\ref{eq:160}).
\begin{eqnarray}
&& \left(1-\frac{x}{a} \right)^{-\beta } y(\xi ) \nonumber\\
 &=& \left(1-\frac{x}{a} \right)^{-\beta } Hl\left(1-a, -q+\gamma \beta; -\alpha +\gamma +\delta, \beta, \gamma, \delta; \frac{(1-a)x}{x-a} \right) \nonumber\\
&=& \left(1-\frac{x}{a} \right)^{-\beta } \Bigg\{ \;_2F_1 \left(-\delta _0, \frac{\beta }{2}; \frac{1}{2}+ \frac{\gamma }{2}; z\right) + \sum_{n=1}^{\infty } \Bigg\{\prod _{k=0}^{n-1} \Bigg\{ \int_{0}^{1} dt_{n-k}\;t_{n-k}^{\frac{1}{2}(n-k-2)} \int_{0}^{1} du_{n-k}\;u_{n-k}^{\frac{1}{2}(n-k-3+\gamma)} \nonumber\\
&&\times \frac{1}{2\pi i}  \oint dv_{n-k} \frac{1}{v_{n-k}} \left( 1-\frac{1}{v_{n-k}}\right)^{\delta _{n-k}}  \left( 1- \overleftrightarrow {w}_{n-k+1,n}v_{n-k}(1-t_{n-k})(1-u_{n-k})\right)^{-\frac{1}{2}(n-k+\beta)}\nonumber\\
&&\times  \left( \overleftrightarrow {w}_{n-k,n}^{-\frac{1}{2}(n-k-1)}\left(  \overleftrightarrow {w}_{n-k,n} \partial _{ \overleftrightarrow {w}_{n-k,n}}\right) \overleftrightarrow {w}_{n-k,n}^{\frac{1}{2}(n-k-1)}\left( \overleftrightarrow {w}_{n-k,n} \partial _{ \overleftrightarrow {w}_{n-k,n}}  + \Omega _{n-k-1}^{(S)}\right) +Q \right) \Bigg\} \nonumber\\
&&\times \;_2F_1 \left(-\delta _0, \frac{\beta }{2}; \frac{1}{2}+ \frac{\gamma }{2}; \overleftrightarrow {w}_{1,n}\right) \Bigg\} \eta ^n \Bigg\} \label{eq:162}
\end{eqnarray}
where
\begin{equation}
\delta _i\leq \delta _j \;\;\mbox{only}\;\mbox{if}\;i\leq j\;\;\mbox{where}\;i,j= 0,1,2,\cdots
\nonumber 
\end{equation}
and
\begin{equation}
\begin{cases} 
\Omega _{n-k-1}^{(S)} =  \frac{1}{2(2-a)}(-\alpha +\beta +\gamma +n-k-1+(1-a)(\alpha -2\delta _{n-k-1}-1))  \cr
Q =   \frac{-q+\gamma \beta}{4(2-a)}
\end{cases}\nonumber %\label{eq:8}
\end{equation}
\subsection{  ${\displaystyle (1-x)^{1-\delta }\left(1-\frac{x}{a} \right)^{-\beta+\delta -1} Hl\Bigg(1-a, -q+\gamma [(\delta -1)a+\beta -\delta +1]; -\alpha +\gamma +1, \beta -\delta+1}$\\ ${\displaystyle, \gamma, 2-\delta; \frac{(1-a)x}{x-a} \Bigg)}$ }
\subsubsection{Infinite series}
Replace coefficients $a$, $q$, $\alpha $, $\beta $, $\delta $, $x$, $c_0$ and $\lambda $ by $1-a$, $-q+\gamma [(\delta -1)a+\beta -\delta +1]$, $-\alpha +\gamma +1$, $\beta -\delta+1$, $2-\delta $, $\frac{(1-a)x}{x-a}$, 1 and zero into (\ref{eq:46}). Multiply $(1-x)^{1-\delta }\left(1-\frac{x}{a} \right)^{-\beta+\delta -1}$ and the new (\ref{eq:46}) together.
\begin{eqnarray}
&& (1-x)^{1-\delta }\left(1-\frac{x}{a} \right)^{-\beta+\delta -1} y(\xi ) \nonumber\\
&=& (1-x)^{1-\delta }\left(1-\frac{x}{a} \right)^{-\beta+\delta -1} Hl\Bigg(1-a, -q+\gamma [(\delta -1)a+\beta -\delta +1]; -\alpha +\gamma +1, \beta -\delta+1\nonumber\\
&&, \gamma, 2-\delta; \frac{(1-a)x}{x-a} \Bigg) \nonumber\\
&=& (1-x)^{1-\delta }\left(1-\frac{x}{a} \right)^{-\beta+\delta -1} \Bigg\{ \;_2F_1 \left(\frac{-\alpha +\gamma +1}{2}, \frac{\beta-\delta +1 }{2}; \frac{1}{2}+ \frac{\gamma }{2}; z\right) \nonumber\\
&&+ \sum_{n=1}^{\infty } \Bigg\{\prod _{k=0}^{n-1} \Bigg\{ \int_{0}^{1} dt_{n-k}\;t_{n-k}^{\frac{1}{2}(n-k-2)} \int_{0}^{1} du_{n-k}\;u_{n-k}^{\frac{1}{2}(n-k-3+\gamma)}  \frac{1}{2\pi i}  \oint dv_{n-k} \frac{1}{v_{n-k}} \left( 1-\frac{1}{v_{n-k}}\right)^{-\frac{1}{2}(n-k+1-\alpha+\gamma )} \nonumber\\
&&\times \left( 1- \overleftrightarrow {w}_{n-k+1,n}v_{n-k}(1-t_{n-k})(1-u_{n-k})\right)^{-\frac{1}{2}(n-k+1+\beta-\delta)}  \nonumber\\
&&\times \left( \overleftrightarrow {w}_{n-k,n}^{-\frac{1}{2}(n-k-1 )}\left(  \overleftrightarrow {w}_{n-k,n} \partial _{ \overleftrightarrow {w}_{n-k,n}}\right) \overleftrightarrow {w}_{n-k,n}^{\frac{1}{2}(n-k-1)}\left( \overleftrightarrow {w}_{n-k,n} \partial _{ \overleftrightarrow {w}_{n-k,n}} + \Omega _{n-k-1}^{(I)}\right) +Q\right) \Bigg\}\nonumber\\
&&\times \;_2F_1 \left(\frac{-\alpha +\gamma +1}{2}, \frac{\beta-\delta +1 }{2}; \frac{1}{2}+ \frac{\gamma }{2}; \overleftrightarrow {w}_{1,n}\right) \Bigg\} \eta ^n \Bigg\} \label{eq:163}
\end{eqnarray}
where
\begin{equation}
\begin{cases} \xi= \frac{(1-a)x}{x-a} \cr
z = -\frac{1}{1-a} \xi^2 \cr
\eta = \frac{(2-a)}{(1-a)}\xi \cr
\end{cases}\nonumber %\label{eq:37}
\end{equation}
and
\begin{equation}
\begin{cases} 
\Omega _{n-k-1}^{(I)} = \frac{1}{2(2-a)}(-\alpha +\beta +\gamma +n-k-1 +(1-a)(\gamma-\delta +n-k ))   \cr
Q =   \frac{-q+\gamma [(\delta -1)a+\beta -\delta +1]}{4(2-a)}
\end{cases}\nonumber %\label{eq:8}
\end{equation}
\subsubsection{Polynomial which makes $B_n$ term terminated}

Substitute $\alpha =\gamma +1 +2\alpha _i +i$ into (\ref{eq:163}) where $i, \alpha _i= 0,1,2,\cdots$.
\begin{eqnarray}
&& (1-x)^{1-\delta }\left(1-\frac{x}{a} \right)^{-\beta+\delta -1} y(\xi ) \nonumber\\
&=& (1-x)^{1-\delta }\left(1-\frac{x}{a} \right)^{-\beta+\delta -1} Hl\Bigg(1-a, -q+\gamma [(\delta -1)a+\beta -\delta +1]; -\alpha +\gamma +1, \beta -\delta+1\nonumber\\
&&, \gamma, 2-\delta; \frac{(1-a)x}{x-a} \Bigg) \nonumber\\
&=& (1-x)^{1-\delta }\left(1-\frac{x}{a} \right)^{-\beta+\delta -1} \Bigg\{ \;_2F_1 \left(-\alpha _0, \frac{\beta-\delta +1 }{2}; \frac{1}{2}+ \frac{\gamma }{2}; z\right) \nonumber\\
&&+ \sum_{n=1}^{\infty } \Bigg\{\prod _{k=0}^{n-1} \Bigg\{ \int_{0}^{1} dt_{n-k}\;t_{n-k}^{\frac{1}{2}(n-k-2)} \int_{0}^{1} du_{n-k}\;u_{n-k}^{\frac{1}{2}(n-k-3+\gamma)}  \frac{1}{2\pi i}  \oint dv_{n-k} \frac{1}{v_{n-k}} \left( 1-\frac{1}{v_{n-k}}\right)^{\alpha _{n-k}} \nonumber\\
&&\times \left( 1- \overleftrightarrow {w}_{n-k+1,n}v_{n-k}(1-t_{n-k})(1-u_{n-k})\right)^{-\frac{1}{2}(n-k+1+\beta-\delta)} \nonumber\\
&&\times \left( \overleftrightarrow {w}_{n-k,n}^{-\frac{1}{2}(n-k-1 )}\left(  \overleftrightarrow {w}_{n-k,n} \partial _{ \overleftrightarrow {w}_{n-k,n}}\right) \overleftrightarrow {w}_{n-k,n}^{\frac{1}{2}(n-k-1)}\left( \overleftrightarrow {w}_{n-k,n} \partial _{ \overleftrightarrow {w}_{n-k,n}}  + \Omega _{n-k-1}^{(S)}\right) +Q\right) \Bigg\}\nonumber\\
&&\times \;_2F_1 \left(-\alpha _0, \frac{\beta-\delta +1 }{2}; \frac{1}{2}+ \frac{\gamma }{2}; \overleftrightarrow {w}_{1,n}\right) \Bigg\} \eta ^n \Bigg\} \label{eq:164}
\end{eqnarray}
where
\begin{equation}
\alpha _i\leq \alpha _j \;\;\mbox{only}\;\mbox{if}\;i\leq j\;\;\mbox{where}\;i,j= 0,1,2,\cdots
\nonumber 
\end{equation}
and
\begin{equation}
\begin{cases} 
\Omega _{n-k-1}^{(S)} = \frac{1}{2(2-a)}(\beta -2\alpha _{n-k-1}-1 +(1-a)(\gamma-\delta +n-k )) \cr
Q =   \frac{-q+\gamma [(\delta -1)a+\beta -\delta +1]}{4(2-a)}
\end{cases}\nonumber %\label{eq:8}
\end{equation}
\subsection{  ${\displaystyle x^{-\alpha } Hl\left(\frac{a-1}{a}, \frac{-q+\alpha (\delta a+\beta -\delta )}{a}; \alpha, \alpha -\gamma +1, \delta , \alpha -\beta +1; \frac{x-1}{x} \right)}$}
\subsubsection{Infinite series}
Replace coefficients $a$, $q$, $\beta $, $\gamma $, $\delta $, $x$, $c_0$ and $\lambda $ by $\frac{a-1}{a}$, $\frac{-q+\alpha (\delta a+\beta -\delta )}{a}$, $\alpha -\gamma +1$, $\delta $, $\alpha -\beta +1$, $\frac{x-1}{x}$, 1 and zero into (\ref{eq:46}). Multiply $x^{-\alpha }$ and the new (\ref{eq:46}) together.
\begin{eqnarray}
&& x^{-\alpha } y(\xi ) \nonumber\\
&=& x^{-\alpha } Hl\left(\frac{a-1}{a}, \frac{-q+\alpha (\delta a+\beta -\delta )}{a}; \alpha, \alpha -\gamma +1, \delta , \alpha -\beta +1; \frac{x-1}{x} \right)\nonumber\\
&=& x^{-\alpha } \Bigg\{ \;_2F_1 \left(\frac{\alpha }{2}, \frac{\alpha -\gamma +1}{2}; \frac{1}{2}+ \frac{\delta }{2}; z\right)\nonumber\\
&&+ \sum_{n=1}^{\infty } \Bigg\{\prod _{k=0}^{n-1} \Bigg\{ \int_{0}^{1} dt_{n-k}\;t_{n-k}^{\frac{1}{2}(n-k-2)} \int_{0}^{1} du_{n-k}\;u_{n-k}^{\frac{1}{2}(n-k-3+\delta )} \frac{1}{2\pi i}  \oint dv_{n-k} \frac{1}{v_{n-k}} \left( 1-\frac{1}{v_{n-k}}\right)^{-\frac{1}{2}(n-k+\alpha )} \nonumber\\
&&\times \left( 1- \overleftrightarrow {w}_{n-k+1,n}v_{n-k}(1-t_{n-k})(1-u_{n-k})\right)^{-\frac{1}{2}(n-k+1+\alpha -\gamma  )} \nonumber\\
&&\times \left( \overleftrightarrow {w}_{n-k,n}^{-\frac{1}{2}(n-k-1)}\left(  \overleftrightarrow {w}_{n-k,n} \partial _{ \overleftrightarrow {w}_{n-k,n}}\right) \overleftrightarrow {w}_{n-k,n}^{\frac{1}{2}(n-k-1)}\left( \overleftrightarrow {w}_{n-k,n} \partial _{ \overleftrightarrow {w}_{n-k,n}}  + \Omega _{n-k-1}^{(I)}\right) +Q\right) \Bigg\} \nonumber\\
&&\times   \;_2F_1 \left(\frac{\alpha }{2}, \frac{\alpha -\gamma +1}{2}; \frac{1}{2}+ \frac{\delta }{2}; \overleftrightarrow {w}_{1,n}\right)\Bigg\} \eta ^n \Bigg\} \label{eq:165}
\end{eqnarray}
where
\begin{equation}
\begin{cases} \xi= \frac{x-1}{x} \cr
z = \frac{-a}{a-1} \xi^2 \cr
\eta = \frac{2a-1}{a-1}\xi \cr
\end{cases}\nonumber %\label{eq:37}
\end{equation}
and
\begin{equation}
\begin{cases} 
\Omega _{n-k-1}^{(I)} = \frac{a}{2(2a-1)}\left(\alpha +\beta -\gamma +n-k-1 +\frac{(a-1)}{a}(\alpha -\beta +\delta +n-k-1)\right)  \cr
Q =  \frac{-q+\alpha (\delta a+\beta -\delta )}{4(2a-1)}  
\end{cases}\nonumber %\label{eq:8}
\end{equation}
\subsubsection{Polynomial which makes $B_n$ term terminated}

Substitute $\gamma =\alpha +1+2\gamma _i +i$ into (\ref{eq:165}) where $i, \gamma _i= 0,1,2,\cdots$.
\begin{eqnarray}
&& x^{-\alpha } y(\xi ) \nonumber\\
 &=& x^{-\alpha } Hl\left(\frac{a-1}{a}, \frac{-q+\alpha (\delta a+\beta -\delta )}{a}; \alpha, \alpha -\gamma +1, \delta , \alpha -\beta +1; \frac{x-1}{x} \right) \nonumber\\
&=& x^{-\alpha } \Bigg\{ \;_2F_1 \left(\frac{\alpha }{2}, -\gamma _0; \frac{1}{2}+ \frac{\delta }{2}; z\right)\nonumber\\
&&+ \sum_{n=1}^{\infty } \Bigg\{\prod _{k=0}^{n-1} \Bigg\{ \int_{0}^{1} dt_{n-k}\;t_{n-k}^{\frac{1}{2}(n-k-2)} \int_{0}^{1} du_{n-k}\;u_{n-k}^{\frac{1}{2}(n-k-3+\delta )} \frac{1}{2\pi i}  \oint dv_{n-k} \frac{1}{v_{n-k}} \left( 1-\frac{1}{v_{n-k}}\right)^{-\frac{1}{2}(n-k+\alpha )} \nonumber\\
&&\times \left( 1- \overleftrightarrow {w}_{n-k+1,n}v_{n-k}(1-t_{n-k})(1-u_{n-k})\right)^{\gamma _{n-k}} \nonumber\\
&&\times \left( \overleftrightarrow {w}_{n-k,n}^{-\frac{1}{2}(n-k-1)} \left(  \overleftrightarrow {w}_{n-k,n} \partial _{ \overleftrightarrow {w}_{n-k,n}}\right) \overleftrightarrow {w}_{n-k,n}^{\frac{1}{2}(n-k-1)}\left( \overleftrightarrow {w}_{n-k,n} \partial _{ \overleftrightarrow {w}_{n-k,n}}  + \Omega _{n-k-1}^{(S)}\right) +Q\right) \Bigg\}\nonumber\\
&&\times   \;_2F_1 \left(\frac{\alpha }{2}, -\gamma _0; \frac{1}{2}+ \frac{\delta }{2}; \overleftrightarrow {w}_{1,n}\right)\Bigg\} \eta ^n \Bigg\} \label{eq:166}
\end{eqnarray}
where
\begin{equation}
\gamma_i\leq \gamma_j \;\;\mbox{only}\;\mbox{if}\;i\leq j\;\;\mbox{where}\;i,j= 0,1,2,\cdots
\nonumber 
\end{equation}
and
\begin{equation}
\begin{cases} 
\Omega _{n-k-1}^{(S)} =  \frac{a}{2(2a-1)}\left(\beta -2\gamma _{n-k-1}-1 +\frac{(a-1)}{a}(\alpha -\beta +\delta +n-k-1)\right)  \cr
Q =  \frac{-q+\alpha (\delta a+\beta -\delta )}{4(2a-1)}  
\end{cases}\nonumber %\label{eq:8}
\end{equation}
\subsection{ ${\displaystyle \left(\frac{x-a}{1-a} \right)^{-\alpha } Hl\left(a, q-(\beta -\delta )\alpha ; \alpha , -\beta+\gamma +\delta , \delta , \gamma; \frac{a(x-1)}{x-a} \right)}$}
\subsubsection{Infinite series}
Replace coefficients $q$, $\beta $, $\gamma $, $\delta $, $x$, $c_0$ and $\lambda $ by $q-(\beta -\delta )\alpha $, $-\beta+\gamma +\delta $, $\delta $,  $\gamma $, $\frac{a(x-1)}{x-a}$, 1 and zero into (\ref{eq:46}). Multiply $\left(\frac{x-a}{1-a} \right)^{-\alpha }$ and the new (\ref{eq:46}) together.
\begin{eqnarray}
&& \left(\frac{x-a}{1-a} \right)^{-\alpha } y(\xi ) \nonumber\\
 &=& \left(\frac{x-a}{1-a} \right)^{-\alpha } Hl\left(a, q-(\beta -\delta )\alpha ; \alpha , -\beta+\gamma +\delta , \delta , \gamma; \frac{a(x-1)}{x-a} \right) \nonumber\\
&=& \left(\frac{x-a}{1-a} \right)^{-\alpha } \Bigg\{ \;_2F_1 \left(\frac{\alpha }{2}, \frac{-\beta +\gamma +\delta }{2}; \frac{1}{2}+ \frac{\delta }{2}; z\right) \nonumber\\
&&+ \sum_{n=1}^{\infty } \Bigg\{\prod _{k=0}^{n-1} \Bigg\{ \int_{0}^{1} dt_{n-k}\;t_{n-k}^{\frac{1}{2}(n-k-2 )} \int_{0}^{1} du_{n-k}\;u_{n-k}^{\frac{1}{2}(n-k-3+\delta )}  \frac{1}{2\pi i}  \oint dv_{n-k} \frac{1}{v_{n-k}} \left( 1-\frac{1}{v_{n-k}}\right)^{-\frac{1}{2}(n-k+\alpha )} \nonumber\\
&&\times \left( 1- \overleftrightarrow {w}_{n-k+1,n}v_{n-k}(1-t_{n-k})(1-u_{n-k})\right)^{-\frac{1}{2}(n-k-\beta+\gamma +\delta )} \nonumber\\
&&\times \left( \overleftrightarrow {w}_{n-k,n}^{-\frac{1}{2}(n-k-1)}\left(  \overleftrightarrow {w}_{n-k,n} \partial _{ \overleftrightarrow {w}_{n-k,n}}\right) \overleftrightarrow {w}_{n-k,n}^{\frac{1}{2}(n-k-1 )}\left( \overleftrightarrow {w}_{n-k,n} \partial _{ \overleftrightarrow {w}_{n-k,n}}  + \Omega _{n-k-1}^{(I)}\right) +Q\right) \Bigg\} \nonumber\\
&&\times \;_2F_1 \left(\frac{\alpha }{2}, \frac{-\beta +\gamma +\delta }{2}; \frac{1}{2}+ \frac{\delta }{2}; \overleftrightarrow {w}_{1,n}\right) \Bigg\} \eta ^n \Bigg\} \label{eq:167}
\end{eqnarray}
where
\begin{equation}
\begin{cases} \xi= \frac{a(x-1)}{x-a} \cr
z = -\frac{1}{a} \xi^2 \cr
\eta = \frac{1+a}{a}\xi \cr
\end{cases}\nonumber %\label{eq:37}
\end{equation}
and
\begin{equation}
\begin{cases} 
\Omega _{n-k-1}^{(I)} = \frac{1}{2(1+a)}(\alpha -\beta +\delta +n-k-1 +a(\delta +\gamma +n-k-2 ))   \cr
Q =   \frac{q-(\beta -\delta )\alpha}{4(1+a)}
\end{cases}\nonumber %\label{eq:8}
\end{equation}
\subsubsection{Polynomial which makes $B_n$ term terminated}
\underline {(1) The case of $\beta =\gamma +\delta +2\beta _i +i$ where $i, \beta _i$ = $0,1,2,\cdots$.}

Substitute $\beta =\gamma +\delta +2\beta _i +i$ into (\ref{eq:167}) where $i, \beta _i= 0,1,2,\cdots$.
\begin{eqnarray}
&& \left(\frac{x-a}{1-a} \right)^{-\alpha } y(\xi ) \nonumber\\
 &=& \left(\frac{x-a}{1-a} \right)^{-\alpha } Hl\left(a, q-(\beta -\delta )\alpha ; \alpha , -\beta+\gamma +\delta , \delta , \gamma; \frac{a(x-1)}{x-a} \right) \nonumber\\
&=& \left(\frac{x-a}{1-a} \right)^{-\alpha } \Bigg\{ \;_2F_1 \left(\frac{\alpha }{2}, -\beta _0; \frac{1}{2}+ \frac{\delta }{2}; z\right) \nonumber\\
&&+ \sum_{n=1}^{\infty } \Bigg\{\prod _{k=0}^{n-1} \Bigg\{ \int_{0}^{1} dt_{n-k}\;t_{n-k}^{\frac{1}{2}(n-k-2 )} \int_{0}^{1} du_{n-k}\;u_{n-k}^{\frac{1}{2}(n-k-3+\delta )}  \frac{1}{2\pi i}  \oint dv_{n-k} \frac{1}{v_{n-k}} \left( 1-\frac{1}{v_{n-k}}\right)^{-\frac{1}{2}(n-k+\alpha )} \nonumber\\
&&\times \left( 1- \overleftrightarrow {w}_{n-k+1,n}v_{n-k}(1-t_{n-k})(1-u_{n-k})\right)^{\beta _{n-k}}  \nonumber\\
&&\times  \left( \overleftrightarrow {w}_{n-k,n}^{-\frac{1}{2}(n-k-1)}\left(  \overleftrightarrow {w}_{n-k,n} \partial _{ \overleftrightarrow {w}_{n-k,n}}\right) \overleftrightarrow {w}_{n-k,n}^{\frac{1}{2}(n-k-1 )}\left( \overleftrightarrow {w}_{n-k,n} \partial _{ \overleftrightarrow {w}_{n-k,n}}  + \Omega _{n-k-1}^{(S)} \right) +Q_{n-k-1}^{(S)}\right) \Bigg\} \nonumber\\
&&\times \;_2F_1 \left(\frac{\alpha }{2}, -\beta _0; \frac{1}{2}+ \frac{\delta }{2}; \overleftrightarrow {w}_{1,n}\right) \Bigg\} \eta ^n \Bigg\} \label{eq:168}
\end{eqnarray}
where
\begin{equation}
\beta _i\leq \beta _j \;\;\mbox{only}\;\mbox{if}\;i\leq j\;\;\mbox{where}\;i,j= 0,1,2,\cdots
\nonumber 
\end{equation}
and
\begin{equation}
\begin{cases} 
\Omega _{n-k-1}^{(S)} =  \frac{1}{2(1+a)}(\alpha -\gamma  -2\beta _{n-k-1} +a(\delta +\gamma +n-k-2 )) \cr
Q_{n-k-1}^{(S)} =  \frac{q-(\gamma +2\beta _{n-k-1}+n-k-1)\alpha}{4(1+a)}
\end{cases}\nonumber %\label{eq:8}
\end{equation}
\underline {(2) The case of $\gamma =\beta -\delta -2\gamma _i -i$ where $i, \gamma _i$ = $0,1,2,\cdots$.}

Substitute $\gamma =\beta -\delta -2\gamma _i -i$ into (\ref{eq:167}) where $i, \gamma _i= 0,1,2,\cdots$.
\begin{eqnarray}
&& \left(\frac{x-a}{1-a} \right)^{-\alpha } y(\xi ) \nonumber\\
 &=& \left(\frac{x-a}{1-a} \right)^{-\alpha } Hl\left(a, q-(\beta -\delta )\alpha ; \alpha , -\beta+\gamma +\delta , \delta , \gamma; \frac{a(x-1)}{x-a} \right) \nonumber\\
&=& \left(\frac{x-a}{1-a} \right)^{-\alpha } \Bigg\{ \;_2F_1 \left(\frac{\alpha }{2}, -\gamma _0; \frac{1}{2}+ \frac{\delta }{2}; z\right) \nonumber\\
&&+ \sum_{n=1}^{\infty } \Bigg\{\prod _{k=0}^{n-1} \Bigg\{ \int_{0}^{1} dt_{n-k}\;t_{n-k}^{\frac{1}{2}(n-k-2 )} \int_{0}^{1} du_{n-k}\;u_{n-k}^{\frac{1}{2}(n-k-3+\delta )}  \frac{1}{2\pi i}  \oint dv_{n-k} \frac{1}{v_{n-k}} \left( 1-\frac{1}{v_{n-k}}\right)^{-\frac{1}{2}(n-k+\alpha )} \nonumber\\
&&\times \left( 1- \overleftrightarrow {w}_{n-k+1,n}v_{n-k}(1-t_{n-k})(1-u_{n-k})\right)^{\gamma _{n-k}} \nonumber\\
&&\times \left( \overleftrightarrow {w}_{n-k,n}^{-\frac{1}{2}(n-k-1)}\left(  \overleftrightarrow {w}_{n-k,n} \partial _{ \overleftrightarrow {w}_{n-k,n}}\right) \overleftrightarrow {w}_{n-k,n}^{\frac{1}{2}(n-k-1 )}\left( \overleftrightarrow {w}_{n-k,n} \partial _{ \overleftrightarrow {w}_{n-k,n}}  + \Omega _{n-k-1}^{(S)}\right) +Q\right) \Bigg\} \nonumber\\
&&\times \;_2F_1 \left(\frac{\alpha }{2}, -\gamma _0; \frac{1}{2}+ \frac{\delta }{2}; \overleftrightarrow {w}_{1,n}\right) \Bigg\} \eta ^n \Bigg\} \label{eq:169}
\end{eqnarray}
where
\begin{equation}
\gamma _i\leq \gamma _j \;\;\mbox{only}\;\mbox{if}\;i\leq j\;\;\mbox{where}\;i,j= 0,1,2,\cdots
\nonumber 
\end{equation}
and
\begin{equation}
\begin{cases} 
\Omega _{n-k-1}^{(S)} =  \frac{1}{2(1+a)}(\alpha -\beta +\delta +n-k-1 +a(\beta -2\gamma _{n-k-1}-1))  \cr
Q  =   \frac{q-(\beta -\delta )\alpha}{4(1+a)}
\end{cases}\nonumber %\label{eq:8}
\end{equation}

\section*{Acknowledgment}
I thank Bogdan Nicolescu. The discussions I had with him on number theory was of great joy. 
\vspace{3mm}

\bibliographystyle{model1a-num-names}
\bibliography{<your-bib-database>}
 
\end{document}